\documentclass[preprint,3p,12pt,sort&compress]{elsarticle}
\usepackage{mathrsfs}
\usepackage{amsmath}
\usepackage{stmaryrd}
\usepackage{bbding}
\usepackage{dcolumn}
\usepackage{graphicx}
\usepackage{amsfonts}
\usepackage{amssymb}
\usepackage{psfrag}
\usepackage{wrapfig}
\usepackage{subfigure}
\usepackage{makeidx}
\usepackage{bm}
\usepackage{epsf}
\usepackage{epsfig}
\usepackage{setspace}
\usepackage{graphicx}
\usepackage{epstopdf}
\usepackage{psfrag}
\usepackage{subfigure}
\usepackage{color}
\usepackage{enumerate}
\usepackage{verbatim} 
\usepackage{booktabs}

\usepackage{tikz}
\usetikzlibrary{positioning,calc}
\usetikzlibrary{arrows,shapes}
\tikzstyle{block} = [rectangle,rounded corners,thin,align=center,fill=green!20,draw=black!20]
\tikzstyle{line} = [-latex]

\begin{document}

\title{A stochastic kinetic scheme for multi-scale plasma transport with uncertainty quantification}

\author[KIT]{Tianbai Xiao\corref{cor}}
\ead{tianbaixiao@gmail.com}

\author[KIT]{Martin Frank}
\ead{martin.frank@kit.edu}

\address[KIT]{Karlsruhe Institute of Technology, Karlsruhe, Germany}

\cortext[cor]{Corresponding author}

\begin{abstract}

Plasmas present a diverse set of behaviors in different regimes. Given the intrinsic multi-scale nature of plasma dynamics, classical theoretical and numerical methods are often employed at separate scales with corresponding assumptions and approximations. Clearly, the coarse-grained modeling may introduce considerable uncertainties between the field solutions of flow and electromagnetic variables, and the real plasma physics. 
To study the emergence, propagation and evolution of randomness from gyrations of charged particles to magnetohydrodynamics poses great opportunities and challenges to develop both sound theories and reliable numerical algorithms. 
In this paper, a physics-oriented stochastic kinetic scheme will be developed that includes random inputs from both flow and electromagnetic fields via a hybridization of stochastic Galerkin and collocation methods. 
Based on the BGK-type relaxation model of the multi-component Boltzmann equation, a scale-dependent kinetic central-upwind flux function is designed in both physical and particle velocity space, and the governing equations in the discrete temporal-spatial-random domain are constructed.
By solving Maxwell's equations with the wave-propagation method, the evolutions of ions, electrons and electromagnetic field are coupled throughout the simulation.
We prove that the scheme is formally asymptotic-preserving in the Vlasov, magnetohydrodynamical, and neutral Euler regimes with the inclusion of random variables.
Therefore, it can be used for the study of multi-scale and multi-physics plasma system under the effects of uncertainties,
and provide scale-adaptive physical solutions under different ratios among numerical cell size, particle mean free path and gyroradius (or time step, local particle collision time and plasma period).
Numerical experiments including one-dimensional Landau Damping, the two-stream instability and the Brio-Wu shock tube problem with one- to three-dimensional velocity settings, and each under stochastic initial conditions with one-dimensional uncertainty, will be presented to validate the scheme. 

\end{abstract}

\begin{keyword}
	Boltzmann equation, plasma physics, kinetic theory, uncertainty quantification, multi-scale methods
\end{keyword}

\maketitle

\section{Introduction}

\begin{figure}[htb!]
	\centering
	\includegraphics[width=0.9\textwidth]{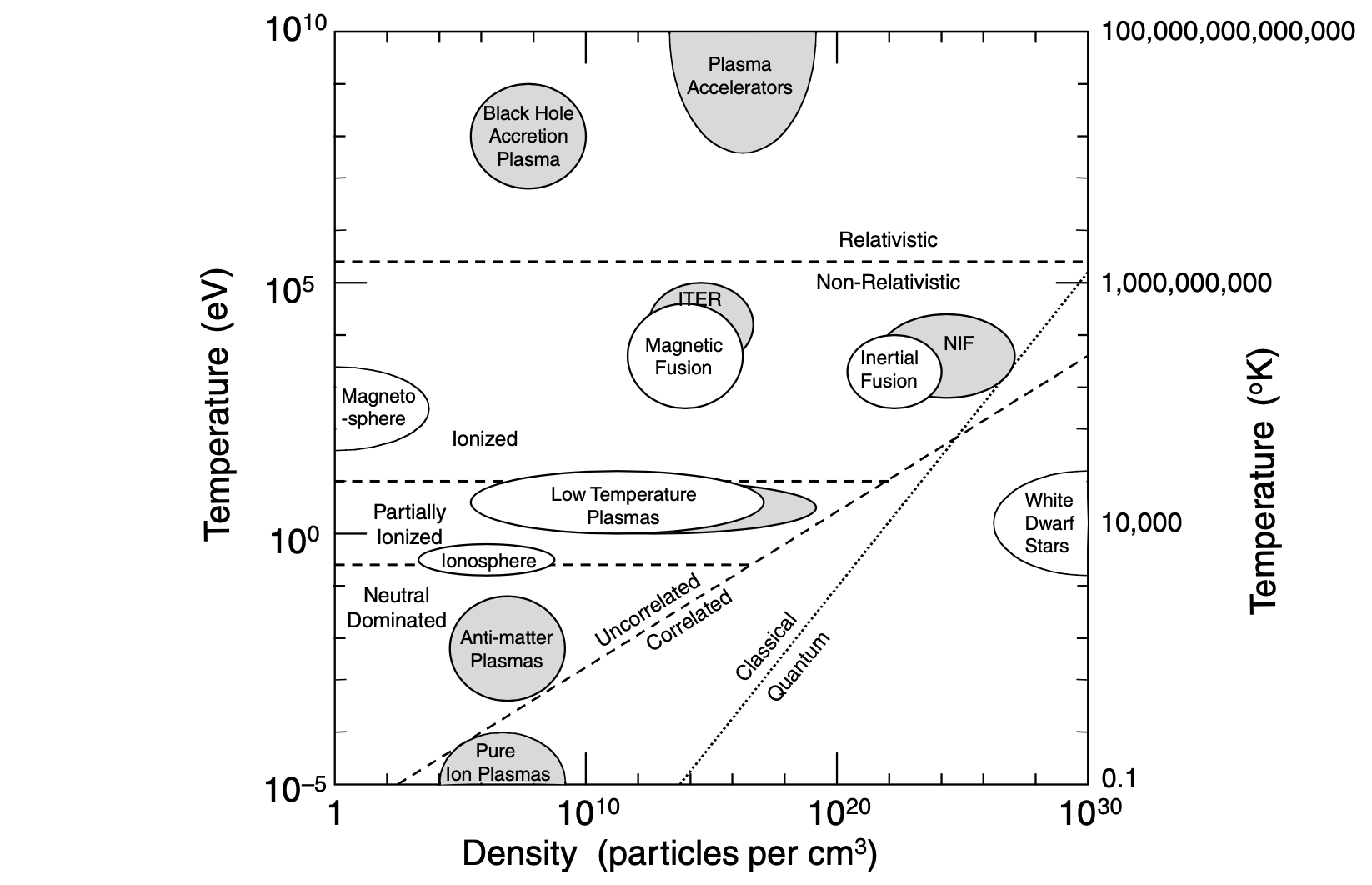}
	\caption{An illustrative demonstration of different plasma regimes and related phenomena \cite{national2008plasma}.}
	\label{pic:plasma regime}
\end{figure}

Plasma applications cover an extremely wide range of density $n$ from $ 10^{6}$ to $10^{34} \ \rm m^{-3}$ and temperature $k_BT$ from $10^{-1}$ to $10^{10} \ \rm eV$ \cite{national2008plasma}.
As F. Chen wrote in his famous monograph \cite{chen2012introduction}, "What makes plasmas particularly difficult to analyze is the fact that the densities fall in an intermediate range. They behave sometimes like fluids, and sometimes like a collection of individual particles."
A qualitative classification of typical plasma regimes with respect to density and temperature is presented in Fig. \ref{pic:plasma regime}.
In this paper, we confine ourselves to the study of classical, non-relativistic and weakly coupled plasmas, e.g.\ the magnetosphere, whose behaviors can be well described by the kinetic theory of gases.

Given the intrinsic multi-scale nature in plasma physics, classical theories are devoted to different governing equations at different hierarchies.
For example, at a particle mean free path which is larger than the Debye length $\lambda_D=(\varepsilon_0 k_B T_{ele}/ne^2)^{1/2}$, and a mean collision time which is larger than the reciprocal of the plasma frequency $\omega_p=(ne^2/\varepsilon_0 m_{ion})^{1/2}$, the motions of ions and electrons can be depicted statistically through kinetic equations, e.g. the Vlasov equation and the Fokker-Planck-Landau equation.
On the other hand, at the macroscopic level with intensive intermolecular collisions, the fluid dynamic equations are routinely used to model the collective behaviors of charged particles, i.e. the magnetohydrodynamics (MHD) equations.

Since the 1950s, rapid development has been made in deterministic numerical methods for plasma simulations \cite{vahedi1995monte,filbet2001conservative,crouseilles2009forward,qiu2010conservative,brio1988upwind,powell1999solution,degond2010asymptotic,degond2017asymptotic}.
However, given the coarse-grained approximation in the field theories of plasmas and errors inherited from numerical simulations, considerable uncertainties may be introduced inevitably. 
One typical example are the uncertain inputs of the initial/boundary value problem.
Furthermore, for the evaluation of collision kernel in the kinetic equations, the phenomenological model parameters often need to be calibrated by experiments to reproduce correct transport coefficients, which introduce errors into the simulations.
Another example goes to the vacuum permeability employed in the Maxwell's equations for electromagnetic fields.
In the SI system which has gone into force in 2019 \cite{nistrefer2019}, this value is measured experimentally as $\mu_0=8.854 187 8128 \times 10^{-12} \rm F\cdot m^{-1} $, with a relative standard deviation being $1.5 \times 10^{-10}$.

Uncertainty quantification (UQ) is a thriving subject that quantifies one's lack of knowledge concerning a physical reality.
It applies itself to answer the challenging questions, e.g.\ how predictive are the simulation results from the idealized models, and how can one explicitly assess the effects of uncertainties on the quality of model predictions.
Depending on the methodology to model the random variables, the methods for UQ study can be classified into intrusive and non-intrusive ones. 
In the former case, a series of realizations of random inputs are generated based on a prescribed probability distribution.
Each realization is solved by a deterministic solver, and then a post-processing is employed to estimated the uncertainties.
In contrast, intrusive methods work in a way such that we reformulate the original deterministic system.

One prevalent intrusive strategy is the stochastic Galerkin (SG) method \cite{xiu2010numerical}, in which the solutions are expressed into orthogonal polynomials of the input random parameters. 
It promises spectral convergence in random space when the solution depends smoothly on the stochastic parameters \cite{canuto1982approximation}. 
In a nonlinear Galerkin system, all the expansion coefficients are essentially coupled, which becomes cumbersome in massive computations.
The stochastic collocation (SC) method \cite{xiu2005high}, although a non-intrusive method, can be seen as a middle way.
It combines the strengths of non-intrusive sampling and SG by evaluating the generalized polynomial chaos (gPC) expansions on quadrature points in random space. 
As a result, a set of decoupled equations can be derived and solved with deterministic solvers on each quadrature point. 
Provided the solutions posses sufficient smoothness over random space, the SC methods maintain similar convergence as SG, but suffers from aliasing errors due to limited number of quadrature points.

Although the UQ field has undergone rapid development over the past few years, its applications on plasma physics mainly focus on the two limits of Vlasov \cite{hu2016stochastic, hu2017uncertainty,jin2018hypocoercivity,ding2019random} and MHD \cite{phillips2015stochastic,yamazaki2017stochastic} with standard stochastic settings.
To the best of the authors' knowledge, limited work has been conducted on the evolutionary process of uncertainty in multi-scale physics. 
Given the nonlinear system including intermolecular collisions, initial
inputs, fluid-surface interactions and geometric complexities, uncertainties may emerge from molecular-level nature, develop upwards, affect macroscopic collective behaviors, and vice versa. 
To study the emergence, propagation and evolution of uncertainty poses great opportunities and challenges to develop both sound theories and reliable multi-scale numerical algorithms.

It is noticeable that tracking the evolution of stochastic variables with either polynomial chaos or quadrature rules is similar in spirit to solving kinetic equations in phase space with moment or discrete velocity methods.
The advantages of SG and SC methods can be combined when the integrals that are necessary for SG inside the algorithm are computed numerically using SC.
In this paper, we follow the strategy proposed in \cite{xiao2020stochastic} and develop a stochastic kinetic scheme for multi-scale plasma transport and we couple it to Maxwell's equations.
Based on the Bhatnagar-Gross-Krook (BGK) type relaxation model for multi-component plasmas, a scale-dependent central-upwind flux function is constructed in both physical and particle velocity space, which considers simultaneously the individual particle transports and their collective behaviors.
The update of source terms of plasma and electromagnetic fields are solved in a coupled way implicitly.
We thus combine the advantages of SG
and SC methods with the construction principle of kinetic schemes, and obtain an efficient and accurate scheme for cross-scale BGK-Maxwell system with uncertainties. The 
randomly initial inputs of both flow and electromagnetic fields are considered.

The rest of this paper is organized as follows. 
Sec.\ 2 is a brief introduction of kinetic theory of plasma and its stochastic formulation. 
Sec.\ 3 presents the numerical implementation of the current scheme and detailed solution algorithm. 
Sec.\ 4 includes numerical experiments to demonstrate the performance of the current scheme and analyze some new physical observations. 
The last section is the conclusion.

\section{Deterministic and stochastic theories}\label{sec:kinetic theory}

\subsection{Kinetic theory of plasmas}

The gas kinetic theory describes the time-space evolution of particle distribution function.
With a separate modeling of particle transport and collision processes, the evolution equation of monatomic plasmas writes as
\begin{equation}
\frac{\partial f_{\alpha}}{\partial t}+\mathbf{u} \cdot \nabla_{\mathbf{x}} f_{\alpha}+\frac{q_{\alpha}}{m_{\alpha}}\left(\mathbf{E}+\mathbf{u} \times \mathbf{B}\right) \cdot \nabla_{\mathbf{u}} f_{\alpha}=Q_{\alpha},
\label{eqn:boltzmann}
\end{equation}
where $\alpha = ion,ele$ denotes a specific ion or electron, $(q_\alpha, m_\alpha)$ are particle charge and mass, and $(\bf E, B)$ are electric and magnetic fields respectively.
For the Coulomb collisions between charged particles, the limiting case of Boltzmann collision integral leads to the Fokker-Planck-Landau operator,
\begin{equation}
Q_\alpha = \sum_\beta^N \left\{ \nabla_\mathbf u \cdot \int \mathbf \Phi(\mathbf u - \mathbf u') \left[ f_\beta(\mathbf u') \nabla_\mathbf u f_\alpha(\mathbf u) - f_\alpha(\mathbf u) \nabla_\mathbf u f_\beta(\mathbf u') \right] d\mathbf u' \right\},
\label{eqn:fokker-planck collision operator}
\end{equation}
where $(\mathbf{u},\ \mathbf{u}')$ are the velocities of two classes of particles and $\mathbf \Phi(\mathbf u) = (|\mathbf u|^2 \mathbf I_3-\mathbf u \otimes \mathbf u)/|\mathbf u|^3$ is a $3\times3$ matrix.

Instead of solving the above collision integral directly, here we employ a multi-component BGK-type relaxation model proposed by Andries, Aoki and Perthame \cite{andries2002consistent} in the current work to mimic the collision process, 
\begin{equation}
Q_\alpha = \nu_\alpha ( \mathcal M_\alpha-f_\alpha ),
\label{eqn:aap}
\end{equation}
where $\nu_{\alpha}$ is the collision frequency.
The equilibrium distribution is defined based on the local modified macroscopic variables, i.e.,
\begin{equation}
\mathcal M_\alpha=n_\alpha\left(\frac{m_\alpha}{2\pi k_B \bar T_\alpha}\right)^\frac{3}{2}\exp \left(-\frac{m_\alpha}{2k_B \bar T_\alpha}(\mathbf u -\mathbf {\bar U}_\alpha)^2\right),
\label{eqn:maxwellian}
\end{equation}
where $n_\alpha$ is number density, $m_\alpha$ is molecular mass and $k_B$ is the Boltzmann constant.
The design of modified temperature $\bar T_\alpha$ and velocity $\mathbf {\bar U}_\alpha$ is based on the idea that the macroscopic transfer rates in the moments equations derived from BGK model should be consistent with that from multi-component Boltzmann equation.
For elastic scattering, the evaluation of modified variables for Maxwell and hard sphere molecules can be written as,
\begin{equation}
\begin{aligned}
&\mathbf {\bar{U}}_\alpha=\mathbf U_\alpha+\tau_\alpha\sum_{r}\frac{2 m_r}{m_\alpha+m_r}\nu_{\alpha r}(\mathbf U_r-\mathbf U_\alpha), \\
&\frac{3}{2}k_B \bar T_\alpha=\frac{3}{2}k_BT_\alpha-\frac{m_\alpha}{2}(\mathbf {\bar U}_\alpha-\mathbf U_\alpha)^2\\
&+\tau_\alpha\sum_{r}\frac{4 m_\alpha m_r}{(m_\alpha+m_r)^2}\nu_{\alpha r}\left[ \frac{3}{2}k_BT_r-\frac{3}{2}k_BT_\alpha+\frac{m_r}{2}(\mathbf U_r-\mathbf U_\alpha)^2 \right],
\label{eqn:modified variable}
\end{aligned}
\end{equation}
where $\nu_{\alpha r}$ is the frequency of intermolecular interactions which can be derived through specific molecule models \cite{morse1963energy}, and it determines the relaxation time by $\tau_\alpha=1/\sum_r^N \nu_{\alpha r}$.
Here we adopt the hard-sphere molecules, i.e.,
\begin{equation}
\nu_{\alpha r}=\frac{4 \sqrt{\pi} n_r}{3}\left(\frac{2 k_{B} T_{\alpha}}{m_{\alpha}}+\frac{2 k_{B} T_{r}}{m_{r}}\right)^{1 / 2}\left(\frac{d_{\alpha}+d_{r}}{2}\right)^{2},
\label{eqn:collision frequency}
\end{equation}
where $d$ is the kinetic molecule diameter.

Macroscopic conservative flow variables are related to the moments of the particle distribution function,
\begin{equation}
\mathbf{W}_\alpha =\left(
\begin{matrix}
\rho_\alpha \\
\rho_\alpha \mathbf U_\alpha \\
\rho_\alpha \mathscr E_\alpha
\end{matrix}
\right)=\int m_\alpha f_\alpha \varpi d\mathbf u,
\label{eqn:moment relation}
\end{equation}
where $\mathscr E_\alpha=(\mathbf U_\alpha)^2/2+3k_BT_\alpha/2m_\alpha$ is total energy density, and $\varpi=\left(1,\mathbf u,\frac{1}{2} \mathbf u^2 \right)^T$ is the vector of collision invariants.
Hence, macroscopic transport equations can be derived by taking moments of the kinetic equation with respect to the collision invariants, i.e.,
\begin{equation}
\begin{aligned}
&\frac{\partial n_\alpha}{\partial t}+\nabla_\mathbf x \cdot (n_\alpha \mathbf U_\alpha)=0, \\
&\frac{\partial (\rho_\alpha \mathbf U_\alpha) }{\partial t}+\nabla_\mathbf x \cdot (\rho_\alpha \mathbf U_\alpha \mathbf U_\alpha)=\nabla_\mathbf x \cdot \mathbf P_\alpha+n_\alpha q_\alpha (\mathbf E+\mathbf U_\alpha \times \mathbf B)+\mathbf R_\alpha, \\
&\frac{\partial (\rho_\alpha \mathscr E_\alpha) }{\partial t}+\nabla_\mathbf x \cdot (\rho_\alpha \mathscr E_\alpha \mathbf U_\alpha) = \nabla_\mathbf x \cdot (\mathbf P_\alpha\mathbf U_\alpha ) - \nabla_\mathbf x \cdot \mathbf q_\alpha + n_\alpha q_\alpha \mathbf U_\alpha \cdot \mathbf E + H_\alpha,
\end{aligned}
\label{eqn:two-fluid}
\end{equation}
where the source terms $\mathbf R_\alpha$ and $H_\alpha$ in the balance laws come from the moments of collision term respectively,
\begin{equation}
\begin{aligned} 
\mathbf R _ { \alpha } & = \int \mathbf { u } m_\alpha { \nu _ { \alpha } } ( \mathcal M _ { \alpha } - f _ { \alpha } ) d \mathbf { u } = \sum _ { r } \frac { 2 m _ { \alpha } m _ { r } } { m _ { \alpha } + m _ { r } } n _ { \alpha } v _ { \alpha r } \left( \mathbf { U } _ { r } - \mathbf { U } _ { \alpha } \right), \\ 
H _ { \alpha } & = \int \frac { 1 } { 2 } ( \mathbf { u } - \mathbf { U } ) ^ { 2 } m_\alpha { \nu _ { \alpha } } ( \mathcal M _ { \alpha } - f _ { \alpha } ) d \mathbf { u } \\ & = \sum _ { r } \frac { 4 m _ { \alpha } m _ { r } } { \left( m _ { \alpha } + m _ { r } \right) ^ { 2 } } n _ { \alpha } v _ { \alpha r } \left[ \frac { 3 } { 2 } k _ { B } T _ { r } - \frac { 3 } { 2 } k _ { B } T _ { \alpha } + \frac { m _ { r } } { 2 } \left( \mathbf { U } _ { r } - \mathbf { U } _ { \alpha } \right) ^ { 2 } \right].
\end{aligned}
\end{equation}
Eq.(\ref{eqn:two-fluid}) is consistent with the well-known Braginskii's two-fluid model \cite{braginskii1965transport}.

\subsection{Generalized polynomial chaos approximation of kinetic equation with uncertainties}
Several sources of uncertainty can be considered in the BGK equation. Here we consider uncertain initial and boundary conditions, which turn the deterministic system into stochastic case.
We employ the generalized polynomial chaos (gPC) expansion of particle distribution with degree $N$, i.e.,
\begin{equation}
f_\alpha(t,\mathbf x,\mathbf u,\mathbf z) \simeq f_{\alpha N} = \sum_{|i|=0}^N \hat f_{\alpha i} (t,\mathbf x,\mathbf u) \Phi_i (\mathbf z) = \hat{\boldsymbol f}_\alpha^T \boldsymbol \Phi,
\label{eqn:polynomial chaos}
\end{equation}
where i could be a scalar or a $K$-dimensional vector $i=(i_1,i_2,\cdots,i_K)$ with $|i|=i_1+i_2+\cdots+i_K$.
The $\hat f_{\alpha \mathbf i}$ is the coefficient of $i$-th polynomial chaos expansion,
and the basis functions used are orthogonal polynomials \{$\Phi_ i(\mathbf z)$\} satisfying the following constraints,
\begin{equation}
\mathbb{E}[\Phi_ j (\mathbf z) \Phi_ k (\mathbf z)] = \gamma_ k \delta_{ j  k}, \quad 0 \leq | j|, | k| \leq N,
\end{equation}
where 
\begin{equation}
\gamma_ k=\mathbb{E}[\Phi_ k^2 (\mathbf z)], \quad 0 \leq |\mathbf k| \leq N,
\end{equation}
are the normalization factors.
The expectation value defines a scalar product,
\begin{equation}
\mathbb{E}[\Phi_ j (\mathbf z) \Phi_ k (\mathbf z)] = \int_{I_{\mathbf z}} \Phi_{ j}(\mathbf z) \Phi_{ k}(\mathbf z) \varrho(\mathbf z) d \mathbf z,
\label{eqn:continuous expectation value}
\end{equation}
where $\varrho(\mathbf z)$ is the probability density function.
In practice, it can be evaluated theoretically or with numerical quadrature rule, i.e.,
\begin{equation}
\mathbb{E}[\Phi_ j (\mathbf z) \Phi_ k (\mathbf z)] = \sum_i \Phi_{ j}(\mathbf z_i) \Phi_{ k}(\mathbf z_i) w(\mathbf z_i) ,
\label{eqn:discrete expectation value}
\end{equation}
where $w(\mathbf z_i)$ is the corresponding quadrature weight function in random space.
In the following we adopt a uniform notation $\langle \Phi_ j \Phi_ k \rangle$ to denote the integrals over random space from Eq.(\ref{eqn:continuous expectation value}) and (\ref{eqn:discrete expectation value}).

Given the correspondence between macroscopic and mesoscopic variables, from Eq.(\ref{eqn:moment relation}) we can derive,
\begin{equation}
\begin{aligned}
\mathbf W_\alpha & \simeq \int f_{\alpha N} \varpi d\mathbf u = \int \sum_{ i}^N \hat f_{\alpha  i} (t,\mathbf x,\mathbf u) \Phi_ i (\mathbf z) \varpi d\mathbf u = \sum_{ i} \left( \int \hat f_{\alpha  i} \varpi d\mathbf u \right) \Phi_ i\\
& \simeq \mathbf W_{\alpha N} = \sum_{ i}^N \hat w_{\alpha  i} \Phi_ i = \hat{\boldsymbol w}_\alpha^T \boldsymbol \Phi.
\end{aligned}
\end{equation}

After substituting the Eq.(\ref{eqn:polynomial chaos}) into the kinetic equation (\ref{eqn:boltzmann}) and (\ref{eqn:aap}), and performing a Galerkin projection, we then obtain
\begin{equation}
\frac{\partial \hat{\boldsymbol f}_{\alpha}}{\partial t}+\mathbf{u} \cdot \nabla_{\mathbf{x}} \hat {\boldsymbol f}_{\alpha} + \hat{\boldsymbol{G}}_\alpha = \hat{\boldsymbol Q}_{\alpha} = \nu_{\alpha} ( \hat {\boldsymbol m}_{\alpha} - \hat {\boldsymbol f}_{\alpha} ),
\label{eqn:stochastic bgk equation}
\end{equation}
where $\hat{\boldsymbol Q}_{\alpha}$ is the gPC coefficient vector of the projection from collision operator to the polynomial basis, 
\begin{equation}
\hat{\boldsymbol Q}_{\alpha} = \nu_{\alpha} ( \hat {\boldsymbol m}_{\alpha} - \hat {\boldsymbol f}_{\alpha} ),
\label{eqn:galerkin collision term}
\end{equation}
with $\hat {\mathbf m}_{\alpha}$ being the vector of gPC coefficients of Maxwellian distribution (which depends implicitly on the $\hat {\boldsymbol f}_{\alpha}$) and $\nu_\alpha$ being a deterministic collision frequency. 
The electromagnetic forcing term $\hat{\boldsymbol{G}}_\alpha$ is
\begin{equation}
\hat{\boldsymbol{G}}_\alpha = \hat{\boldsymbol{G}}_\alpha^T \boldsymbol{\Phi}= \sum_{ i}^{N} \hat G_ i \Phi_ i,\ 
\hat{G}_{{i}} =
\frac{q_\alpha}{m_\alpha}
\frac{\sum_{{j}}^{N} \sum_{{k}}^{N} 
\left(	
\hat{\mathbf E}_ j+\mathbf{u} \times \hat{\mathbf B}_ j
\right)
\nabla_\mathbf u \hat f_{\alpha k}
	\left\langle\Phi_{{j}} \Phi_{{k}} \Phi_{{i}}\right\rangle}{\left\langle\Phi_{{i}}^{2}\right\rangle}.
\label{eqn:galerkin force term}
\end{equation}
Notice that the gPC coefficients of both $\hat{\boldsymbol{G}}_\alpha$ and $\hat{\boldsymbol{Q}}_\alpha$ are nonlinear functions of the state variables
(cf.\ Eq.(\ref{eqn:modified variable})).

\subsection{Maxwellian distribution in generalized polynomial chaos}\label{sec:maxwellian evaluation}

For a deterministic system, the evaluation of the Maxwellian distribution given in Eq.(\ref{eqn:maxwellian}) is straight-forward.
However, given a generalized polynomial chaos (gPC) system, the multiplication and division can't be operated directly on the stochastic moments without modifying the orthogonal basis.
Starting from a known particle distribution function in Eq.(\ref{eqn:polynomial chaos}), here we draw a brief outline to approximately evaluate the Maxwellian distribution function in the gPC expansion.

1. Derive the macroscopic conservative variables from particle distribution function with gPC expansion,
\begin{equation}
\mathbf{W}_{\alpha N}  =\left(
\begin{matrix}
\rho_{\alpha N} \\
(\rho_\alpha \mathbf U_\alpha)_{N} \\
(\rho_\alpha \mathscr E_\alpha)_{N}
\end{matrix}
\right)=\sum_{i}^N \left( \int \hat f_{\alpha i} \psi d\mathbf u \right) \Phi_ i ;
\end{equation}

2. Locate conservative variables on quadrature points $\mathbf z_j$ of random space and calculate primitive variables, e.g. flow velocity
\begin{equation}
\mathbf U_\alpha(\mathbf z_j)
= \frac{(\rho_\alpha \mathbf U_\alpha)_N(\mathbf z_j)}{\rho_{\alpha N} (\mathbf z_j)},
\end{equation}
and
\begin{equation}
T_\alpha(\mathbf z_j) = \frac{ (\rho_\alpha E_\alpha)_N(\mathbf z_j) - (\rho_\alpha U_\alpha)_N^2(\mathbf z_j)/2\rho_{\alpha N}(\mathbf z_j) }
{3 k_B \rho_{\alpha N} / 2m_\alpha},
\end{equation}
and then calculate the modified velocity and temperature via Eq.(\ref{eqn:modified variable}).

3. Calculate Maxwellian distribution on quadrature points
\begin{equation}
\mathcal M_\alpha(\mathbf z_j)=n_{\alpha}(\mathbf z_j) \left( \frac{m_\alpha}{2\pi k_B \bar{T}_\alpha(\mathbf z_j)} \right)^{\frac{3}{2}} e^{-\bar \lambda_\alpha(\mathbf z_j)(\mathbf u-\bar{\mathbf U}_\alpha(\mathbf z_j))^2},
\end{equation}
and decompose it into a gPC expansion
\begin{equation}
\mathcal M_{\alpha N} = \sum_{i}^ N \hat m_{\alpha i} \Phi_ i,
\end{equation}
with each coefficient in the expansion being given by a quadrature rule
\begin{equation}
\hat m_{\alpha i} = \frac{\langle \mathcal M_{\alpha},\Phi_ i \rangle}{\langle \Phi_{ i}^2 \rangle} = \frac{\sum_j \mathcal M_\alpha(\mathbf z_j) \Phi_{\mathbf i}(\mathbf z_j) \varrho(\mathbf z_j)}{\int_{I_{\mathbf z}}  ( \Phi_{ i}(\mathbf z) )^2 \varrho(\mathbf z) d \mathbf z}.
\end{equation}

\subsection{Maxwell's equations}

For the self-consistent problem of plasma dynamics, the evolutions of electric and magnetic fields $(\mathbf E, \mathbf B)$ are coupled with the motions of charged particles, which can be described by the linear Maxwell's equations in vacuum,
\begin{equation}
\begin{aligned}
&\frac{\partial \mathbf E} {\partial t} - c^2 \nabla_\mathbf x \times \mathbf B = - \frac{1}{\varepsilon_0} \mathbf J, \\
&\frac{\partial \mathbf B}{\partial t} + \nabla_\mathbf x \times \mathbf E = 0, \\
&\nabla_\mathbf x \cdot \mathbf E = \frac{\sigma}{\varepsilon_0}, \\
&\nabla_\mathbf x \cdot \mathbf B = 0. \\
\end{aligned}
\end{equation}
Here $\sigma=e(n_i-n_e)$ is the net charge density, $\mathbf J$ is the current, and the speed of light is related to the permeability and permittivity of vacuum with $c=(\mu_0 \varepsilon_0)^{-1/2}$.
To ensure the divergence constraints in numerical simulations, some techniques can be used in solving Maxwell's equations. 
Here we employ the perfectly hyperbolic Maxwell's equations (PHM) \cite{munz2000divergence},
\begin{equation}
\begin{aligned}
&\frac{\partial \mathbf E} {\partial t} - c^2 \nabla_\mathbf x \times \mathbf B + \chi c^2 \nabla_\mathbf x \phi = - \frac{1}{\varepsilon_0} \mathbf J, \\
&\frac{\partial \mathbf B}{\partial t} + \nabla_\mathbf x \times \mathbf E + \gamma \nabla_\mathbf x \psi= 0, \\
&\frac{1}{\chi} \frac{\partial \phi}{\partial t} + \nabla_\mathbf x \cdot \mathbf E = \frac{\sigma}{\varepsilon_0}, \\
&\frac{\varepsilon_0 \mu_0}{\gamma} \frac{\partial \psi}{\partial t} + \nabla_\mathbf x \cdot \mathbf B = 0, \\
\end{aligned}
\label{eqn:phm}
\end{equation}
where $\phi, \psi$ are two additional correction potentials, and the propagation speed of errors for the divergence of magnetic and electric fields are $\gamma c$ and $\chi c$ correspondingly.
With the stochastic Galerkin formulation, the PHM system can be rewritten as
\begin{equation}
\begin{aligned}
&\frac{\partial \hat{\boldsymbol E}} {\partial t} - c^2 \nabla_\mathbf x \times \hat{\boldsymbol B} + \chi c^2 \nabla_\mathbf x \hat{\boldsymbol \phi} = -  (\hat{\boldsymbol J / \varepsilon_0}), \\
&\frac{\partial \hat{\boldsymbol B}}{\partial t} + \nabla_\mathbf x \times \hat{\boldsymbol E} + \gamma \nabla_\mathbf x \hat {\boldsymbol \psi}= 0, \\
&\frac{1}{\chi} \frac{\partial \hat{\boldsymbol \phi}}{\partial t} + \nabla_\mathbf x \cdot \hat{\boldsymbol E} = (\hat{ \sigma/\varepsilon_0}), \\
&\frac{\varepsilon_0 \mu_0}{\gamma} \frac{\partial \hat {\boldsymbol \psi}}{\partial t} + \nabla_\mathbf x \cdot \hat{\boldsymbol B} = 0. \\
\end{aligned}
\label{eqn:sg phm}
\end{equation}
The parameters in Maxwell's equations are assumed to be deterministic.

\section{Solution algorithm} \label{sec:solution algorithm}

\subsection{Update algorithm}

The current numerical algorithm is implemented within the finite volume framework. We adopt the notation of cell averaged macroscopic conservative variables and particle distribution function in a control volume,
\begin{equation*}
\begin{aligned}
&\mathbf W_\alpha({t^n,\mathbf x_i,\mathbf z_k})=(\mathbf W_\alpha)_{i,k}^n=\frac{1}{\Omega_{i}(\mathbf x) \Omega_{k}(\mathbf z)} \int_{\Omega_{i}}\int_{\Omega_{k}} \mathbf W_\alpha(t^n, \mathbf x, \mathbf z)d\mathbf x d\mathbf z,\\
&f_\alpha({t^n,\mathbf x_i,\mathbf u_j,\mathbf z_k})=(f_\alpha)_{i,j,k}^n=\frac{1}{\Omega_{i}(\mathbf x)\Omega_{j}(\mathbf u)\Omega_{k}(\mathbf z)} \int_{\Omega_{i}} \int_{\Omega_{j}} \int_{\Omega_{k}} f_\alpha(t^n,\mathbf x,\mathbf u,\mathbf z) d\mathbf xd\mathbf u d\mathbf z,
\end{aligned}
\end{equation*}
along with the coefficient vector in the gPC expansions,
\begin{equation*}
\begin{aligned}
&\hat {\boldsymbol W}_\alpha (t^n,\mathbf x_i)=(\hat {\boldsymbol W}_\alpha)_{i}^n=\frac{1}{\Omega_{i}(\mathbf x)} \int_{\Omega_{i}} \hat {\boldsymbol W}_\alpha(t^n, \mathbf x)d\mathbf x ,\\
&\hat {\boldsymbol f}_\alpha(t^n,\mathbf x_i,\mathbf u_j)=(\hat {\boldsymbol f}_\alpha)_{i,j}^n=\frac{1}{\Omega_{i}(\mathbf x)\Omega_{j}(\mathbf u)} \int_{\Omega_{i}} \int_{\Omega_{j}}\hat {\boldsymbol f}_\alpha (t^n,\mathbf x,\mathbf u) d\mathbf xd\mathbf u,
\end{aligned}
\end{equation*}
where $\Omega_{i}$, $\Omega_{j}$ and $\Omega_{k}$ are the cell area in the discretized physical, velocity and random space.

The update of the stochastic Galerkin coefficients for particle distribution function can be written as,
\begin{equation}
\begin{aligned}
(\hat {\boldsymbol f}_\alpha)_{i,j}^{n+1}
=&(\hat {\boldsymbol f}_\alpha)_{i,j}^n+\frac{1}{\Omega_{i}}\int_{t^n}^{t^{n+1}} \sum_{S_{r} \in \partial \Omega_{i}} S_r (\hat {\boldsymbol F}_\alpha)_{r,j}^f dt \\
&+\frac{1}{\Omega_{j}}\int_{t^n}^{t^{n+1}} \sum_{S_{r} \in \partial \Omega_{j}} S_r (\hat {\boldsymbol F}_\alpha)_{i,r}^f dt
+ \int_{t^n}^{t^{n+1}} (\hat {\boldsymbol Q}_\alpha)_{i,j}^f dt .
\label{eqn:galerkin micro update}
\end{aligned}
\end{equation}
where $(\hat{\boldsymbol{F}}_\alpha)^f_r$ is the time-dependent fluxes for distribution function at interface $r$ in physical and velocity space, $S_r$ is the interface area, and $\hat{\boldsymbol Q}_\alpha^f$ is the collision term.
Taking velocity moments of Eq.(\ref{eqn:galerkin micro update}), we obtain the corresponding macroscopic system,
\begin{equation}
\begin{aligned}
(\hat {\boldsymbol{W}}_\alpha)_{i}^{n+1}
=&(\hat {\boldsymbol{W}}_\alpha)_{i}^n+\frac{1}{\Omega_{i}}\int_{t^n}^{t^{n+1}}\sum_{\mathbf S_{r} \in \partial \Omega_{i}} \mathbf{S}_{r} \cdot (\hat{\boldsymbol{F}}_{\alpha})_{r}^{W} dt \\
&+\int_{t^{n}}^{t^{n+1}}(\hat{\boldsymbol{G}}_{\alpha})_{i}^{W} d t
+\int_{t^{n}}^{t^{n+1}}(\hat{\boldsymbol{Q}}_{\alpha})_{i}^{W} d t,
\label{eqn:galerkin macro update}
\end{aligned}
\end{equation}
where $(\hat{\boldsymbol{F}}_\alpha)^W_r$ is the flux functions for macroscopic conservative variables, $\mathbf{S}_{r} =\boldsymbol{n} S_r $ is the interface area vector, and $\hat{\boldsymbol{G}}_{\alpha}^W$ is the external force terms related to Eq.(\ref {eqn:two-fluid}).

Notice that the gPC coefficients of different orders are coupled through the external force term $\hat{\boldsymbol G}_\alpha$ (\ref{eqn:galerkin collision term}) and the collision term $\hat{\boldsymbol Q}_\alpha$ (\ref{eqn:galerkin force term}). Instead of solving this large nonlinear system, we can locate the gPC system onto the collocation points in random space.
It results a decoupled system that can be solved efficiently.
Therefore, we combine the advantages of stochastic Galerkin and collcation methods, which is one of the novelties of this paper.
To make use of it, in the solution algorithm, we first update the gPC coefficients to an intermediate step $t^*$,
\begin{equation}
\begin{aligned}
&(\hat {\boldsymbol{W}}_\alpha)_{i}^{*}
=(\hat {\boldsymbol{W}}_\alpha)_{i}^n+\frac{1}{\Omega_{i}}\int_{t^n}^{t^{n+1}}\sum_{\mathbf S_{r} \in \partial \Omega_{i}} \mathbf{S}_{r} \cdot (\hat{\boldsymbol{F}}_{\alpha})_{r}^{W} dt, \\
&(\hat {\boldsymbol f}_\alpha)_{i,j}^{*}
=(\hat {\boldsymbol f}_\alpha)_{i,j}^n+\frac{1}{\Omega_{i}}\int_{t^n}^{t^{n+1}} \sum_{S_{r} \in \partial \Omega_{i}} S_r (\hat {\boldsymbol F}_\alpha)_{r,j}^f dt,
\end{aligned}
\end{equation}
which are then evaluated on random quadrature cell $\Omega_{k}$,
\begin{equation}
\begin{aligned}
\mathbf W_{i, k}^{*}=\mathbf W_{N i}^{*}\left(\mathbf z_{k}\right)=\sum_{m}^{N} \hat{w}_{i, m}^{*}\left(\mathbf z_{k}\right) \Phi_{m}\left(\mathbf z_{k}\right), \\
f_{i, j, k}^{*}=f_{N i, j}^{*}\left(\mathbf z_{k}\right)=\sum_{m}^{N} \hat{f}_{i, j, m}^{*}\left(\mathbf z_{k}\right) \Phi_{m}\left(\mathbf z_{k}\right).
\end{aligned}
\end{equation}
Afterwards, the collision and forcing term are evaluated on collocation points via
\begin{equation}
\left(\mathbf{W}_{\alpha}\right)_{i,k}^{n+1}
=\left(\mathbf{W}_{\alpha}\right)_{i,k}^{*}
+\int_{t^{n}}^{t^{n+1}}\left(\mathbf{G}_{\alpha}\right)_{i,k}^{W} d t
+\int_{t^{n}}^{t^{n+1}}\left(\mathbf{Q}_{\alpha}\right)_{i,k}^{W} d t,
\label{eqn:macro update}
\end{equation}
and
\begin{equation}
\left(f_{\alpha}\right)_{i, j, k}^{n+1}
=\left(f_{\alpha}\right)_{i, j, k}^{*}
+\frac{1}{\Omega_{j}} \int_{t^{n}}^{t^{n+1}} \sum_{S_{r} \in \partial \Omega_{j}} S_{r}\left(F_{\alpha}\right)_{i,r,k}^{f} d t
+\int_{t^{n}}^{t^{n+1}}\left(Q_{\alpha}\right)_{i, j, k}^{f} d t,
\label{eqn:micro update}
\end{equation}
where $(F_\alpha)^f_r$ is the numerical flux at interface $r$ in velocity space.
In the solution algorithm loop, Eq.(\ref{eqn:macro update}) can be solved first, and then the updated variables at $t^{n+1}$ can be employed to evaluate the Maxwellian distribution in Eq.(\ref{eqn:micro update}) implicitly.

For plasma transport, the evolution of electromagnetic field should be solved in a coupled way with flow field.
The hybrid Galerkin-collocation method is employed as well, where the gPC coefficients are stepped to the intermediate state first,
\begin{equation}
\hat{\boldsymbol{M}}_{i}^{*}=\hat{\boldsymbol{M}}_{i}^{n}+\frac{1}{\Omega_{i}} \int_{t^{n}}^{t^{n+1}} \sum_{r} \Delta \mathbf{S}_{r} \cdot \hat{\boldsymbol{F}}_{r}^{M} d t,
\label{eqn:galerkin em update}
\end{equation}
where $\hat{\boldsymbol{F}}^M$ is the flux functions for the electromagnetic fields $(\hat{\boldsymbol E}, \hat{\boldsymbol B}, \hat{\boldsymbol \phi}, \hat{\boldsymbol \psi})$.
Then the source terms are solved via,
\begin{equation}
\begin{aligned}
	&{\mathbf{E}}_{i}^{n+1}={\mathbf{E}}_{i}^{*}
	-\frac{e}{\varepsilon_0} \int_{t^n}^{t^{n+1}} \left( {( n \mathbf U)}_{ion} - {( n \mathbf U)}_{ele} \right) dt, \\
	&{{\phi}}_{i}^{n+1}={{\phi}}_{i}^{*}
	+\frac{e}{\varepsilon_0} \int_{t^n}^{t^{n+1}} \left(n_{ion}-n_{ele} \right) dt. \\
\end{aligned}
\end{equation}

\subsection{Fluxes computed using Stochastic Galerkin}\label{sec:sg treatment}

\subsubsection{Plasma flux}

Based on the finite volume framework, a scale-dependent interface flux function is needed in multi-scale modeling and simulation.
Different from a purely upwind flux which loses efficiency in the collisional limit, we here develop a kinetic central-upwind flux function based on an integral solution of the kinetic model equation.
The integral solution originates from Kogan's monograph on rarefied gas dynamics \cite{koganrarefied} and has been employed by a series of gas-kinetic schemes \cite{xu2010unified,xiao2017well,liu2017unified,xiao2019unified, xiao2020velocity}.
Let us rewrite the stochastic BGK equation (\ref{eqn:stochastic bgk equation}) for the gPC coefficients vector along the characteristics,
\begin{equation}
\frac{D \hat {\boldsymbol f}_\alpha}{D t} + \nu_\alpha \hat {\boldsymbol f}_\alpha = \nu_\alpha \hat {\boldsymbol m}_\alpha.
\end{equation}
We assume that the collision frequency $\nu_\alpha$ is kept fixed at the value that can be computed from the macroscopic variables at the previous timestep.
Then the following integral solution holds along the characteristics,
\begin{equation}
\hat{\boldsymbol f}_\alpha(t,\mathbf x,\mathbf u)=\nu_\alpha \int_{0}^t \hat{ \boldsymbol m}_\alpha(t',\mathbf x',\mathbf u')e^{-\nu_\alpha(t-t')}dt' +e^{-\nu_\alpha t} \hat {\boldsymbol f}_\alpha(0,\mathbf x^0,\mathbf u^0),
\label{eqn:integral solution}
\end{equation}
where $\mathbf x'=\mathbf x-\mathbf u'(t-t')-\frac{1}{2}\mathbf a (t-t')^2$ is the particle trajectory, and $\mathbf x^0=\mathbf x-\mathbf u't-\frac{1}{2} \mathbf a t^2$ is the location at initial time $t=0$. The above solution indicates a self-conditioned mechanism for multi-scale gas dynamics.
For example, when the evolving time $t$ is much less than the mean collision time $\tau=1/\nu_\alpha$, the latter term in Eq.(\ref{eqn:integral solution}) dominates and describes the free transport of particles.
And if $t$ is much larger than $\tau$, the second term approaches to zero, and then the distribution function will be an accumulation of equilibrium state along the characteristic lines, which provides the underlying wave-propagation mechanism for hydrodynamic solutions.
In the following, we present a detailed strategy for the construction of  numerical fluxes.
Since the electromagnetic force term $\mathbf a$ is a stochastic variable, we do an operator splitting to evaluate fluxes in physical and velocity space.

With the simplified notations of physical cell interface $\mathbf x_{i+1/2}=0$ and initial time $t^n=0$, Eq.(\ref{eqn:integral solution}) along physical trajectories of particles can be rewritten into the following form,
\begin{equation}
\hat{\boldsymbol f}_\alpha(t,0,\mathbf u_j)=\nu_\alpha \int_{0}^{t} \hat{\boldsymbol m}_\alpha(t',\mathbf x',\mathbf u_j)e^{-\nu_\alpha (t-t')}dt' +e^{-\nu_\alpha t}\hat{\boldsymbol f}_\alpha(0,-\mathbf u_j t,\mathbf u_j),
\label{eqn:interface integral solution}
\end{equation}
where $\hat{\boldsymbol f}_\alpha(0,-\mathbf u_j t,\mathbf u_j)$ is the initial distribution at each time step.

In the numerical scheme,
the initial distribution function at cell interface can be obtained through reconstruction, i.e.,
\begin{equation}
(\hat{\boldsymbol f}_\alpha)(0,\pm 0,\mathbf u_j)=\left\{
\begin{aligned}
&(\hat{\boldsymbol f}_\alpha)_{i+1/2,j}^L, \quad x= 0^-, \\
&(\hat{\boldsymbol f}_\alpha)_{i+1/2,j}^R, \quad x= 0^+.
\end{aligned}
\right.
\label{eqn:f0 reconstruct 1st}
\end{equation}
The initial distributions $(\hat{\boldsymbol f}_\alpha)_{i+1/2,j}^{L,R}$ at the left and right hand sides of a cell interface are obtained through the van-Leer limiter.

The macroscopic conservative variables in the gPC expansions at the interface can be evaluated by taking moments over velocity space,
\begin{equation*}
\hat{\boldsymbol w}=\sum_{u_j>0} \hat{\boldsymbol f}_{i+1/2,j}^L\psi \Delta u_j +\sum_{u_j<0} \hat{\boldsymbol f}_{i+1/2,j}^R\psi \Delta u_j.
\end{equation*}
The equilibrium distribution at interface can be determined as illustrated in Sec.\ \ref{sec:maxwellian evaluation}.

After all coefficients are obtained, the interface distribution function becomes 
\begin{equation}
\begin{aligned}
\hat{\boldsymbol f}_\alpha(t,0,u_j)=&\left(1-e^{-\nu_\alpha t}\right) (\hat{\boldsymbol m}_{\alpha})_{j} \\
&+e^{-\nu_\alpha t}\left[(\hat{\boldsymbol f}_\alpha)_{i+1/2,j}^L H\left[\mathbf u_j\right] + (\hat{\boldsymbol f}_\alpha)_{i+1/2,j}^R (1-H\left[\mathbf u_j\right])\right],
\end{aligned}
\label{eqn:interface distribution gPC 1st}
\end{equation}
where $H(\mathbf u)$ is the heaviside step function.
The above interface distribution function can be regarded as a combination of central difference and upwind methods.
With the variation of the ratio between evolving time $t$ (i.e., the time step in the computation) and collision time $\tau_\alpha=1/\nu_\alpha$, the factor $e^{-\nu_\alpha t}$ plays as a modulator and provides a self-adaptive solution between equilibrium and non-equilibrium physics.

After the coefficients of distribution function at all orders are determined, the corresponding gPC expansion can be expressed as,
\begin{equation}
f_{\alpha N}(t,0,\mathbf u_j) = \sum_{m=0}^N \hat f_{m}(t,0,\mathbf u_j) \Phi_m(\mathbf z) , 
\end{equation}
and the corresponding fluxes of particle distribution function and conservative flow variables can be evaluated via
\begin{equation}
\begin{aligned}
&{{F}}_{\alpha N}^f(t,0,\mathbf u_j,\mathbf z)= \mathbf u_j f_{\alpha N}(t,0,\mathbf u_j,\mathbf z) , \\
&{\mathbf{F}}_{\alpha N}^W(t,0,\mathbf z)=\int \mathbf u f_{\alpha N}(t,0,\mathbf u,\mathbf z) \varpi d\mathbf u \simeq \sum w_j \mathbf u_j f_{\alpha N}(t,0,\mathbf u_j,\mathbf z) \varpi_j  ,
\end{aligned}
\end{equation}
where $\mathbf u_j$ denotes quadrature points in particle velocity space, and $w_j$ is its integral weight in velocity space,
and the time-integrated fluxes in Eq.(\ref{eqn:galerkin macro update}) and (\ref{eqn:galerkin micro update}) can be evaluated with respect to time in Eq.(\ref{eqn:interface distribution gPC 1st}).

\subsubsection{Electromagnetic flux}

Besides the flow variables, the numerical fluxes of electromagnetic fields in Eq.(\ref{eqn:sg phm}) are calculated by the wave-propagation method developed by Hakim et al. \cite{hakim2006high}.

\subsection{Fluxes and sources computed using stochastic collocation} \label{sec:sc treatment}

Besides the construction of the interface flux, the source terms need to be evaluated inside each control volume within each time step. 
In this part, we show the detailed update algorithm for the collision and external force term with stochastic collocation method.

\subsubsection{Macroscopic source terms}

Given the intermediate macroscopic variables in the discrete cell $(\Omega_{i}, \Omega_{j}, \Omega_{k})$, the macroscopic system writes as follows,
\begin{equation}
\begin{aligned}
& (\rho_\alpha \mathbf U_\alpha)^{n+1}_{i,k} = (\rho_\alpha \mathbf U_\alpha)^{*}_{i,k}
+ \Delta t q_\alpha \left( n_\alpha (\mathbf E+\mathbf U_\alpha \times \mathbf B) \right)_{i,k}^{n+1} 
+\int_{t^n}^{t^{n+1}} \left( \nu_{\alpha}\rho_{\alpha} ( \overline{\mathbf{U}}^*_{\alpha}- \mathbf{U}^*_{\alpha}) \right)_{i,k} dt, \\
& (\rho_\alpha \mathscr E_\alpha)_{i,k}^{n+1}  = (\rho_\alpha \mathscr E_\alpha)_{i,k}^{*}
+ \Delta t q_\alpha (n_\alpha \mathbf E \cdot \mathbf U_\alpha)_{i,k}^{n+1}
+ \int_{t^n}^{t^{n+1}} \left( \nu_{\alpha} (\rho_{\alpha} \overline{\mathscr{E}}^*_{\alpha} -\rho_{\alpha} \mathscr{E}^*_{\alpha} ) \right)_{i,k} dt, \\
&{\mathbf{E}}_{i,k}^{n+1}={\mathbf{E}}_{i,k}^{*}
-\frac{e \Delta t}{\varepsilon_0}\left( {( n \mathbf U)}_{ion} - {( n \mathbf U)}_{ele} \right)^{n+1}_{i,k}, \\
&{{\phi}}_{i,k}^{n+1}={{\phi}}_{i,k}^{*} 
+\frac{e \Delta t}{\varepsilon_0}\left({n}_{ion}-{n}_{ele} \right)^{n+1}_{i,k} .
\end{aligned}
\label{eqn:macro source}
\end{equation}
We split the above system into two parts.
First, the flow variables are evolved with respect to mixture source terms,
\begin{equation}
\begin{aligned}
& (\rho_\alpha \mathbf U_\alpha)^{**}_{i,k} = (\rho_\alpha \mathbf U_\alpha)^{*}_{i,k}
+\int_{t^n}^{t^{n+1}} \left( \nu_{\alpha}\rho^{n+1}_{\alpha} ( \overline{\mathbf{U}}^*_{\alpha}- \mathbf{U}^*_{\alpha}) \right)_{i,k} dt, \\
& (\rho_\alpha \mathscr E_\alpha)_{i,k}^{**}  = (\rho_\alpha \mathscr E_\alpha)_{i,k}^{*}
+ \int_{t^n}^{t^{n+1}} \left( \nu_{\alpha}\rho_{\alpha}^{n+1} \left( \overline{\mathscr{E}}_{\alpha}^{*}- \mathscr{E}_{\alpha}^* \right) \right)_{i,k} dt .
\end{aligned}
\end{equation}
The relaxation integrals are evaluated through the Rosenbrock method \cite{shampine1982implementation} to overcome possible stiffness.

Then we solve the electromagnetic sources implicitly,
in which a linear system can be solved, i.e.,
\begin{equation}
\begin{aligned}
& (\rho_\alpha \mathbf U_\alpha)^{n+1}_{i,k} = (\rho_\alpha \mathbf U_\alpha)^{**}_{i,k}
+ \Delta t q_\alpha \left( n_\alpha (\mathbf E+\mathbf U_\alpha \times \mathbf B) \right)_{i,k}^{n+1}, \\
& (\rho_\alpha \mathscr E_\alpha)_{i,k}^{n+1}  = (\rho_\alpha \mathscr E_\alpha)_{i,k}^{**}
+ \Delta t q_\alpha (n_\alpha \mathbf E \cdot \mathbf U_\alpha)_{i,k}^{n+1}, \\
&{\mathbf{E}}_{i,k}^{n+1}={\mathbf{E}}_{i,k}^{*}
-\frac{e \Delta t}{\varepsilon_0}\left( {( n \mathbf U)}_{ion} - {( n \mathbf U)}_{ele} \right)^{n+1}_{i,k}, \\
&{{\phi}}_{i,k}^{n+1}={{\phi}}_{i,k}^{*} 
+\frac{e \Delta t}{\varepsilon_0}\left({n}_{ion}-{n}_{ele} \right)^{n+1}_{i,k} .
\end{aligned}
\label{eqn:em source}
\end{equation}

\subsubsection{Particle distribution function}

The updated macroscopic variables can be used to step the particle distribution functions of ion and electron implicitly.
Let us consider the kinetic equation,
\begin{equation}
\frac{\partial f_{\alpha}}{\partial t}+ \mathbf a_\alpha \cdot \nabla_{\mathbf{u}} f_{\alpha}=\nu_\alpha (\mathcal M_\alpha - f_\alpha),
\label{eqn:force bgk}
\end{equation}
where the electromagnetic force is 
\begin{equation*}
\mathbf a_\alpha=\frac{q_{\alpha}}{m_{\alpha}}\left(\mathbf{E}^{n+1}+\mathbf{u}_{\alpha} \times \mathbf{B}^{n+1}\right).
\end{equation*}

Making the simplified notations of velocity cell interface $\mathbf u_{j+1/2}=0$ and initial time $t^n=0$ again, we write the integral solution of Eq.(\ref{eqn:force bgk}) as,
\begin{equation}
\begin{aligned}
f_\alpha(t,\mathbf x_i,0,\mathbf z_k)
&=\nu_\alpha \int_{0}^{t} \mathcal M^*_\alpha(t',\mathbf x_i,\mathbf u',\mathbf z_k)e^{-\nu_\alpha (t-t')}dt' +e^{-\nu_\alpha t} f^*_\alpha(0,\mathbf x_i,-\mathbf a_\alpha t,\mathbf z_k) \\
&=\mathcal M^*_\alpha(0,\mathbf{x}, \mathbf{u}-\mathbf a_\alpha t,\mathbf z_k)\left(1-\mathbf{e}^{-\nu_\alpha t}\right)+f^*_{\alpha}(0,\mathbf{x}_i, \mathbf{u}-\mathbf a_\alpha t,\mathbf z_k) \mathbf{e}^{-\nu_\alpha t}.
\end{aligned}
\label{eqn:velocity interface distribution}
\end{equation}
Similar to physical space, the above interface distribution function can also be regarded as a combination of central difference and upwind methods in particle velocity space.
With the variation of the ratio between evolving time $t$ (i.e., the time step in the computation) and collision time $\tau_\alpha=1/\nu_\alpha$, it provides a self-adaptive solution from equilibrium to non-equilibrium. 
Based the above solution, the interface flux in particle velocity space can be constructed as
\begin{equation}
{{F}}_{\alpha}^f(t,\mathbf x_i,0,\mathbf z_k)= \mathbf a_\alpha f_{\alpha}(t,\mathbf x_i,0,\mathbf z_k),
\end{equation}
and the time-integrated flux can be evaluated directly with respect to $t$.

For the collision term, with the updated flow variables at $t^{n+1}$, an implicit update can be arranged. 
Therefore, the update algorithm of particle distribution function can be written as,
\begin{equation}
\left(f_{\alpha}\right)_{i, j, k}^{n+1}
=\left(f_{\alpha}\right)_{i, j, k}^{*}
+\frac{1}{\Omega_{j}} \int_{t^{n}}^{t^{n+1}} \sum_{S_{r} \in \partial \Omega_{j}} S_{r}\left(F_{\alpha}\right)_{i,r,k}^{f} d t
+\Delta t \nu_\alpha^{n+1} \left( (\mathcal M_\alpha)_{i,j,k}^{n+1} - (f_\alpha)_{i,j,k}^{n+1} \right).
\end{equation}

\subsection{Time step}

In the current scheme, the time step is determined by the Courant-Friedrichs-Lewy condition in phase space,
\begin{equation}
\Delta t=\mathcal{C} \min \left( 
\frac{\min (|\Delta \mathbf x|)}{\max (|\mathbf u|)+\max (|\mathbf U|)}, 
\frac{\min (|\Delta \mathbf x|)}{c} ,
\frac{\min (|\Delta \mathbf u|)}{\max (|\mathbf a|)}
\right),
\end{equation}
where $\mathcal{C}$ is the CFL number, $\mathbf u=(\mathbf u_{ion}, \mathbf u_{ele})$ is particle velocity, $\mathbf U=(\mathbf U_{ion}, \mathbf U_{ele})$ is fluid velocity, and $c$ is speed of light.
In the computation, $c$ usually takes a pseudo value which is less than $3.0 \times 10^8 m/s$ but a few orders larger than particle velocity.

\subsection{Asymptotic analysis}

In this part we will present theoretical analysis on the current kinetic model and numerical algorithm, with special focus on their asymptotic limits.

\subsubsection{Asymptotic limits of the BGK-Maxwell system}

Let us return to the BGK equation (\ref{eqn:boltzmann}).
From Eq.(\ref{eqn:collision frequency}), we see that the collision frequency is positively correlated with plasma density and temperature.
When the plasma gets rarefied, the intensity of collision term decreases correspondingly.
As the collision frequency $\nu_\alpha$ goes to zero, the BGK equation automatically reduces to the Vlasov equation, i.e.,
\begin{equation}
\frac{\partial f_{\alpha}}{\partial t}+\mathbf{u} \cdot \nabla_{\mathbf{x}} f_{\alpha}+\frac{q_{\alpha}}{m_{\alpha}}\left(\mathbf{E}+\mathbf{u}_{\alpha} \times \mathbf{B}\right) \cdot \nabla_{\mathbf{u}} f_{\alpha}=0.
\label{eqn:vlasov}
\end{equation}

On the other hand, as the collision frequency increases, then the two-fluid system (\ref{eqn:two-fluid}) is equivalent with the BGK equation, which describes the motions of plasma as fluid.
Under the quasineutral assumption $n_i\simeq n_e$ with $\lambda_D\ll L_0$ ($L_0$ is the characteristic length of system), $m_i \gg m_e$ and $\nu_i,\ \nu_e \gg 0$, it can be further degenerated to the single-fluid Hall-MHD equations,
\begin{equation}
\begin{aligned}
&\frac{\partial \rho}{\partial t} + \nabla_{\mathbf{x}} \cdot(\rho \mathbf{U})=0,\\
&\frac{\partial (\rho \mathbf U)}{\partial t} + \nabla_{\mathbf{x}} \cdot\left(\rho \mathbf{U} \mathbf{U}\right)=-\nabla_\mathbf x p + \mathbf{J} \times \mathbf{B},\\
&\frac{\partial (\rho \mathscr E) }{\partial t}+\nabla_\mathbf x \cdot (\rho \mathscr E \mathbf U) = \nabla_\mathbf x \cdot (p\mathbf U ) + \mathbf J \cdot \mathbf E ,\\
&\mathbf{E}+\mathbf{U} \times \mathbf{B}=\eta \mathbf{J}+\frac{1}{en} (\mathbf{J} \times \mathbf{B}+\nabla_{\mathbf{x}} p),\\
&\frac{\partial \sigma}{\partial t} + \nabla_{\mathbf{x}} \cdot \mathbf{J}=0.
\end{aligned}
\end{equation}
where $\sigma$ is electric charge density and $\mathbf J$ is net current.
Together with Maxwell’s equations, this set is able to describe the equilibrium state of the plasma.

The specific resistivity $\eta$ is proportional to the interaction frequency between ions and electrons, i.e.,
\begin{equation}
\eta=\frac{m_e}{ne^2}\nu_{ie}.
\end{equation}
Therefore, given the fully conductive conditions with $\nu_{ie} \to 0$ and ignore the Hall current term and pressure gradient in the generalized Ohm's law, we get the ideal MHD equations,
\begin{equation}
\begin{aligned}
&\frac{\partial \rho}{\partial t} + \nabla_{\mathbf{x}} \cdot(\rho \mathbf{U})=0,\\
&\frac{\partial (\rho \mathbf U)}{\partial t} + \nabla_{\mathbf{x}} \cdot\left(\rho \mathbf{U} \mathbf{U}\right)=-\nabla_\mathbf x p + \frac{\left(\mathbf{B} \cdot \nabla_{\mathbf{x}}\right) \mathbf{B}}{\mu_{0}}-\nabla_{\mathbf{x}}\left(\frac{\mathbf{B}^{2}}{2 \mu_{0}}\right),\\
&\frac{\partial (\rho \mathscr E) }{\partial t}+\nabla_\mathbf x \cdot (\rho \mathscr E \mathbf U) = \nabla_\mathbf x \cdot (p\mathbf U ) + \frac{1}{\mu_{0}} \rho \mathbf{U} \cdot\left(\nabla_{\mathbf{x}} \times \mathbf{B} \times \mathbf{B}\right) ,\\
&\frac{\partial \mathbf B}{\partial t}+\nabla_\mathbf x \times (\mathbf U \times \mathbf{B})=0.
\end{aligned}
\end{equation}

In contrast to the ideal MHD regime, we can also consider the non-conductive limit where the interspecies molecular interaction is intensive with $\nu_{ie} \to \infty$. 
As a result, the plasma now behaves like dielectric material, and the two-fluid system deduces to the Euler equations,
\begin{equation}
\begin{aligned}
&\frac{\partial \rho}{\partial t} + \nabla_{\mathbf{x}} \cdot(\rho \mathbf{U})=0,\\
&\frac{\partial (\rho \mathbf U)}{\partial t} + \nabla_{\mathbf{x}} \cdot\left(\rho \mathbf{U} \mathbf{U}\right)=-\nabla_\mathbf x p ,\\
&\frac{\partial (\rho \mathscr E) }{\partial t}+\nabla_\mathbf x \cdot (\rho \mathscr E \mathbf U) = \nabla_\mathbf x \cdot (p\mathbf U ) .
\end{aligned}
\end{equation}

\subsubsection{Interface fluxes}

Besides the theoretical modeling, the asymptotic preserving property for the limiting solutions is also a preferred nature for the kinetic scheme. 
Let us consider the interface fluxes constructed in Sec. \ref{sec:sg treatment} and \ref{sec:sc treatment} first.
In the collisionless limit where $\nu_\alpha$ approaches zero, the relation $\nu_\alpha \Delta t \ll 1$ holds naturally, and the collision term in Eq.(\ref{eqn:boltzmann}) disappears. 
In this case, the interface distribution function in both Eq.(\ref{eqn:interface distribution gPC 1st}) and (\ref{eqn:velocity interface distribution}) go to the non-equilibrium distribution parts.
For brevity, we write them down into collocation form, i.e.,
\begin{equation}
\begin{aligned}
&(f_\alpha)_{i+1/2,j,k} = 
(f_\alpha)_{i+1/2,j}^L H(\mathbf u_j) + (f_\alpha)_{i+1/2,j}^R (1-H(\mathbf u_j)), \\
&(f_\alpha)_{i,j+1/2,k}=
f^*_{\alpha}(\mathbf{x}_i, \mathbf{u}-\mathbf a_\alpha t,\mathbf z_k),
\end{aligned}
\end{equation}
which constitute a fully upwind scheme for the Vlasov equation.

On the other hand, in the hydrodynamic regime with intensive collisions, the interface distribution becomes,
\begin{equation}
\begin{aligned}
&(f_\alpha)_{i+1/2,j,k} = 
(\mathcal M_\alpha)_{i+1/2,j} , \\
&(f_\alpha)_{i,j+1/2,k}=
\mathcal M^*_{\alpha}(\mathbf{x}_i, \mathbf{u}-\mathbf a_\alpha t,\mathbf z_k),
\end{aligned}
\end{equation}
leading to near-equilibrium MHD or Euler solutions.

\subsubsection{Source terms}

Besides the flux functions, the treatment of source terms plays an important role for the asymptotic property of scheme.
Let us rewrite the electromagnetic sources in Eq.(\ref{eqn:em source}) as follows,
\begin{equation}
\begin{aligned}
	& (\rho_{ion} \mathbf U_{ion})^{n+1}_{i,k} = (\rho_{ion} \mathbf U_{ion})^{**}_{i,k}
	+ \Delta t e \left( n_{ion} (\mathbf E+\mathbf U_{ion} \times \mathbf B) \right)_{i,k}^{n+1}, \\
	& (\rho_{ele} \mathbf U_{ele})^{n+1}_{i,k} = (\rho_{ele} \mathbf U_{ele})^{**}_{i,k}
	- \Delta t e \left( n_{ele} (\mathbf E+\mathbf U_{ele} \times \mathbf B) \right)_{i,k}^{n+1}, \\
	& (\rho_{ion} \mathscr E_{ion})_{i,k}^{n+1} = (\rho_{ion} \mathscr E_{ion})_{i,k}^{**}
	+ \Delta t e (n_{ion} \mathbf E \cdot \mathbf U_{ion})_{i,k}^{n+1}, \\
	& (\rho_{ele} \mathscr E_{ele})_{i,k}^{n+1} = (\rho_{ele} \mathscr E_{ele})_{i,k}^{**}
	- \Delta t e (n_{ele} \mathbf E \cdot \mathbf U_{ele})_{i,k}^{n+1}, \\
	&{\mathbf{E}}_{i,k}^{n+1}={\mathbf{E}}_{i,k}^{*}
	-\frac{e \Delta t}{\varepsilon_0}\left( { n }_{ion} \mathbf U_{ion} - n_{ele} { \mathbf U}_{ele} \right)^{n+1}_{i,k}, \\
	&{{\phi}}_{i,k}^{n+1}={{\phi}}_{i,k}^{*} 
	+\frac{e \Delta t}{\varepsilon_0}\left({n}_{ion}-{n}_{ele} \right)^{n+1}_{i,k} .
\end{aligned}
\label{eqn:cell em source}
\end{equation}

Adding the momentum and energy equations of ion and electron, we get
\begin{equation}
\begin{aligned}
&(\rho \mathbf U)^{n+1}_{i,k} - (\rho \mathbf U)^{**}_{i,k}
= \Delta t e \left((n_{ion}-n_{ele}) \mathbf{E}\right)^{n+1}_{i,k}
+ \Delta t \left((\mathbf J_{ion} + \mathbf J_{ele}) \times \mathbf{B}\right)^{n+1}_{i,k}, \\
&(\rho \mathscr E)^{n+1}_{i,k} - (\rho \mathscr E)^{**}_{i,k}
=\Delta t \left((\mathbf J_{ion} + \mathbf J_{ele}) \cdot \mathbf{E} \right)^{n+1}_{i,k}. \\
\end{aligned}
\end{equation}
With the quasineutral assumption, it can be simplified as
\begin{equation}
\begin{aligned}
&(\rho \mathbf U)^{n+1}_{i,k} - (\rho \mathbf U)^{**}_{i,k}
= \Delta t \left(\mathbf J \times \mathbf{B}\right)^{n+1}_{i,k}, \\
&(\rho \mathscr E)^{n+1}_{i,k} - (\rho \mathscr E)^{**}_{i,k}
=\Delta t \left(\mathbf J \cdot \mathbf{E} \right)^{n+1}_{i,k}. \\
\end{aligned}
\end{equation}
At the same time, let us multiply the first equation in Eq.(\ref{eqn:cell em source}) by $m_{ele}$ and second by $m_{ion}$ and subtract the latter from the former, which results,
\begin{equation}
\begin{aligned}
m_{ion} m_{ele} \left(n\left(\mathbf U_{ion}-\mathbf U_{ele}\right)\right)^{n+1}_{i,k}
-m_{ion} m_{ele} \left(n\left(\mathbf U_{ion}-\mathbf U_{ele}\right)\right)^{**}_{i,k} \\
=\Delta t e \left(m_{ion}+m_{ele}\right) (n \mathbf E)^{n+1}_{i,k}
+\Delta t e \left(n \left(m_{ele} \mathbf U_{ion}+m_{ion} \mathbf U_{ele}\right) \times \mathbf{B}\right)^{n+1}_{i,k}.
\end{aligned}
\end{equation}
Given $m_{ion} \gg m_{ele}$, the above the equation can be degenerated to
\begin{equation}
\mathbf{E}_{i,k}^{n+1} + \mathbf{U}_{i,k}^{n+1} \times \mathbf{B}_{i,k}^{n+1}
=\frac{1}{e \rho_{i,k}^{n+1}}
\left(
\frac{m_{ion} m_{ele}}{e} (\mathbf J^{n+1}_{i,k} - \mathbf J^{**}_{i,k})
+m_{ion} \mathbf{J}^{n+1}_{i,k} \times \mathbf{B}^{n+1}_{i,k}
\right),
\end{equation}
which is the generalized Ohm's law.
At a time scale where the inertial effects (i.e., cyclotron frequency) are unimportant, the time difference of current is ignorable.
The second term is the Hall current term, which becomes unimportant in the ideal MHD limit.
Let $\nu_{ie} \to 0$ with fully conductive assumption, then we get the ideal Ohm's law,
\begin{equation}
\mathbf{E}_{i,k}^{n+1} + \mathbf{U}_{i,k}^{n+1} \times \mathbf{B}_{i,k}^{n+1}
=0.
\end{equation}

\subsection{Summary}

\begin{figure}[htb!]
	\centering
	\includegraphics[width=0.7\textwidth]{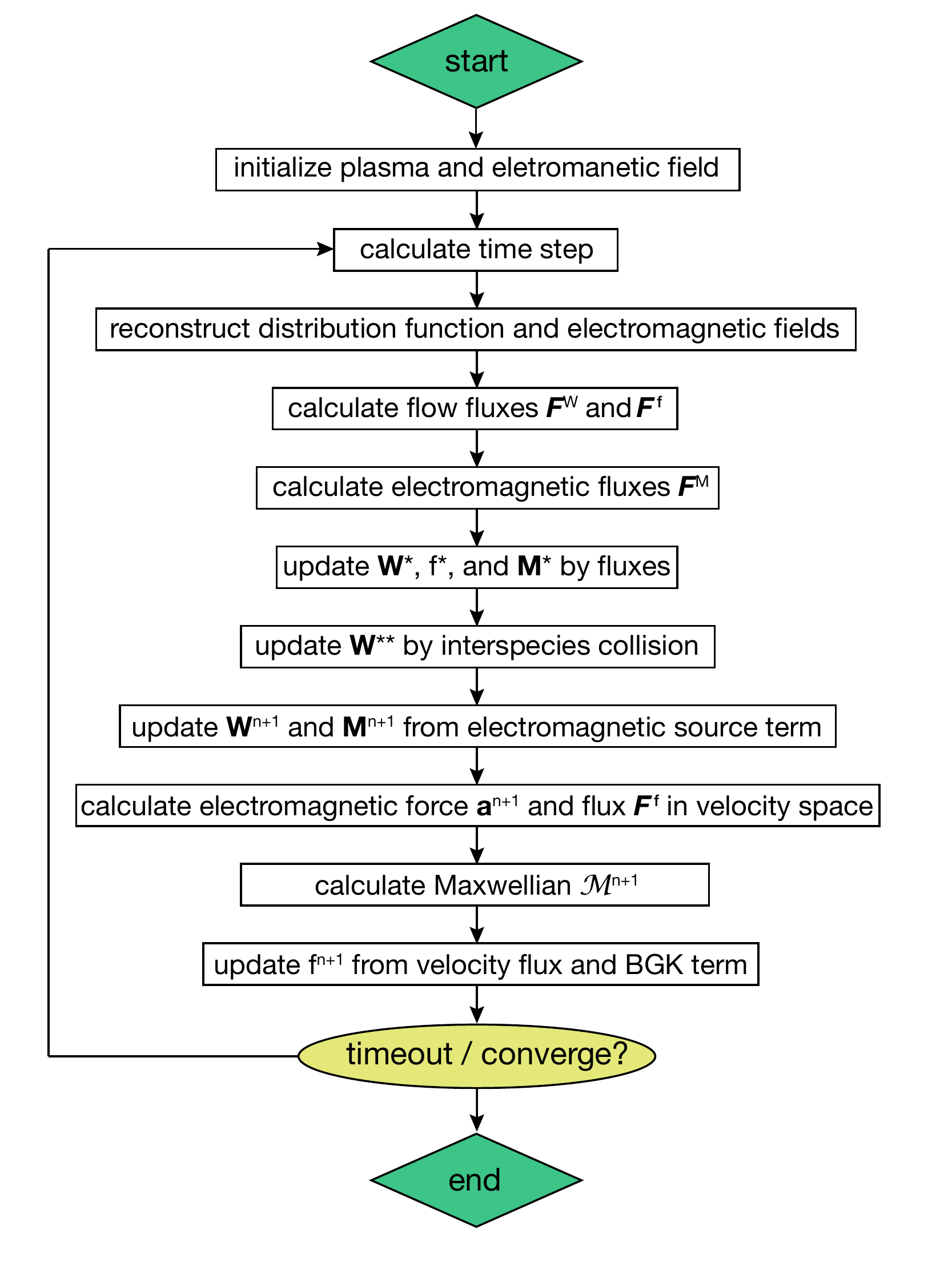}
	\caption{Flowchart of solution algorithm.}
	\label{pic:flowchart}
\end{figure}

The flowchart of the current solution algorithm is summarized in Fig. \ref{pic:flowchart}.

\section{Numerical experiments} \label{sec:numerical experiment}

In this section, we will present some numerical results.
The goal of numerical experiments is not simply to validate the performance of the current scheme, but also to present and analyze new physical observations.
In order to demonstrate the multi-scale nature of the algorithm, simulations from Vlasov to magnetohydrodynamics (MHD) regimes are presented. The following dimensionless flow variables are introduced in the simulations,
\begin{equation*}
\begin{aligned}
& \tilde{\mathbf x}=\frac{\mathbf x}{L_0},
\tilde{t}=\frac{t}{L_0 / U_0},
\tilde{m}=\frac{m}{m_{ion}},
\tilde{n}=\frac{n}{n_0},
\tilde{\mathbf U}=\frac{\mathbf U}{U_0}, 
\tilde{T}=\frac{T}{T_0}, \\ 
& \tilde{\mathbf P}=\frac{\mathbf P}{m_{ion} n_0 U_0^2}, 
\tilde{\mathbf q}=\frac{\mathbf q}{m_{ion} n_0 U_0^{3}}, 
\tilde{f}=\frac{f}{n_0 U_0^{3}}, 
\tilde{\mathbf u}=\frac{\mathbf u}{U_0}, \\
&\tilde{\mathbf B}=\frac{\mathbf B}{B_0}, 
\tilde{\mathbf E}=\frac{\mathbf E}{B_0 U_0}, 
\tilde{\sigma}=\frac{\sigma}{e n_0},
\tilde{\mathbf J}=\frac{\mathbf J}{e n_0 U_0}, \tilde{\lambda_D}=\frac{\lambda_D}{r_g},
\end{aligned}
\end{equation*}
where $U_0=\sqrt{k_B T_0/m_{ion}}$ is the thermal velocity of ions, $\lambda_D=\sqrt{\varepsilon_0 k_B T_0/n_0 e^2}$ is the Debye length, and $r_{g}=m_{ion} U_0/e B_0$ is the gyroradius of ion in the reference state.  
For brevity, the tilde notation for dimensionless variables will be removed henceforth. 
Therefore, the dimensionless BGK-Maxwell system becomes,
\begin{equation}
\begin{aligned}
	&\frac{\partial f_{\alpha}}{\partial t}+\mathbf{u} \cdot \nabla_{\mathbf{x}} f_{\alpha}+\frac{1}{r_{g} m_{\alpha}}(\mathbf{E}+\mathbf{u} \times \mathbf{B}) \cdot \nabla_{\mathbf{u}} f_{\alpha}={\nu_{\alpha}}({\mathcal M_{\alpha}-f_{\alpha}}), \\
	&\frac{\partial \mathbf{E}}{\partial t}-c^{2} \nabla_{\mathbf{x}} \times \mathbf{B}=-\frac{1}{\lambda_{D}^{2} r_{g}} \mathbf{J}, \\
	&\frac{\partial \mathbf{B}}{\partial t}+\nabla_{\mathbf{x}} \times \mathbf{E}=0, \\
	&\frac{1}{\chi} \frac{\partial \phi}{\partial t} + \nabla_\mathbf x \cdot \mathbf E = \frac{\sigma}{\lambda_D^2 r_g}, \\
	&\frac{1}{c^2 \gamma} \frac{\partial \psi}{\partial t} + \nabla_\mathbf x \cdot \mathbf B = 0. \\
\end{aligned}
\end{equation}

\subsection{Landau damping}

The Landau damping is a physical phenomenon first predicted by Landau \cite{landau1946vibration} based on theoretical derivation, namely the exponential decay effect of electromagnetic waves in collisionless plasmas.
Here, we use the example of Landau damping to verify the performance of the current stochastic scheme in the Vlasov limit. 
For brevity, we consider the one-dimensional case first, 
and thus the Maxwell's equations for the electrostatic field degenerates into the Poisson equation for the electric potential.

\subsubsection{Linear case}

Consider the following macroscopic system
\begin{equation*}
\left[\begin{array}{c}
	n_{ion} \\
	n_{ele} \\
	U \\
	T \\
\end{array}\right]_{\text {t=0}}=\left[\begin{array}{c}
	1 \\
	1 + \alpha \xi \cos(kx) \\
	0 \\
	1 \\
\end{array}\right],
\end{equation*}
where $\alpha$ is the amplitude of electromagnetic wave, $k$ is the wave number, and $\xi$ is a random parameter.
The corresponding particle distribution functions at $t=0$ are the Maxwellian, 
\begin{equation*}
f_\alpha=n_\alpha \left(\frac{m_\alpha}{2\pi T}\right)^{1/2} \exp\left(-\frac{m_\alpha}{2 T} (u-U)^2\right).
\end{equation*}

The computational setup is listed in Table \ref{tab:linear damping}.
Given the minor amplitude $\alpha$, the Vlasov-Poisson system can be regarded as the Maxwellian plus lineazised perturbation near equilibrium.
With the large mass ratio, we fix the slow motions of ions as background, and solve the evolution of electrons.
The standard non-intrusive stochastic collocation method is also employed to provide reference solution.
Besides, the damping rate of electric field energy can be derived by linear theory \cite{crouseilles2009forward}.
We hereby combine $\alpha$ and $\xi$ into a new perturbation amplitude $\beta$, and derive the following relation as benchmark,
\begin{equation}
\mathcal E(x, t) \simeq 4 \beta \times 0.3677 e^{-0.1533 t} \sin (k x) \cos (1.4156 t-0.536245).
\label{eqn:damping theory}
\end{equation}
\begin{table}[htbp]
	\caption{Computational setup of linear Landau damping.} 
	\centering
	\begin{tabular}{lllllll} 
		\toprule 
		$t$ & $x$ & $N_x$ & $m_{ion}/m_{ele}$ & $u_{ele}$ & $N_u$ & $\alpha$ \\ 
		\midrule 
		$(0,40]$ & $[-\pi/k,\pi/k]$ & 128 & $1836$ & $[-5,5]$ & 128 & 0.01 \\ 
		\midrule 
		$k$ & $\xi$ & Polynomial & $N_z$(gPC) & $N_z$(quad) & Kn & cfl \\ 
		\midrule 
		0.5 & $\mathcal U(0,1)$ & Legendre & $5$ & $9$ & $100$  & 0.2   \\ 
		\bottomrule 
	\end{tabular} 
	\label{tab:linear damping}
\end{table}

\begin{figure}[htb!]
	\centering
	\subfigure[Expectation value]{
		\includegraphics[width=0.45\textwidth]{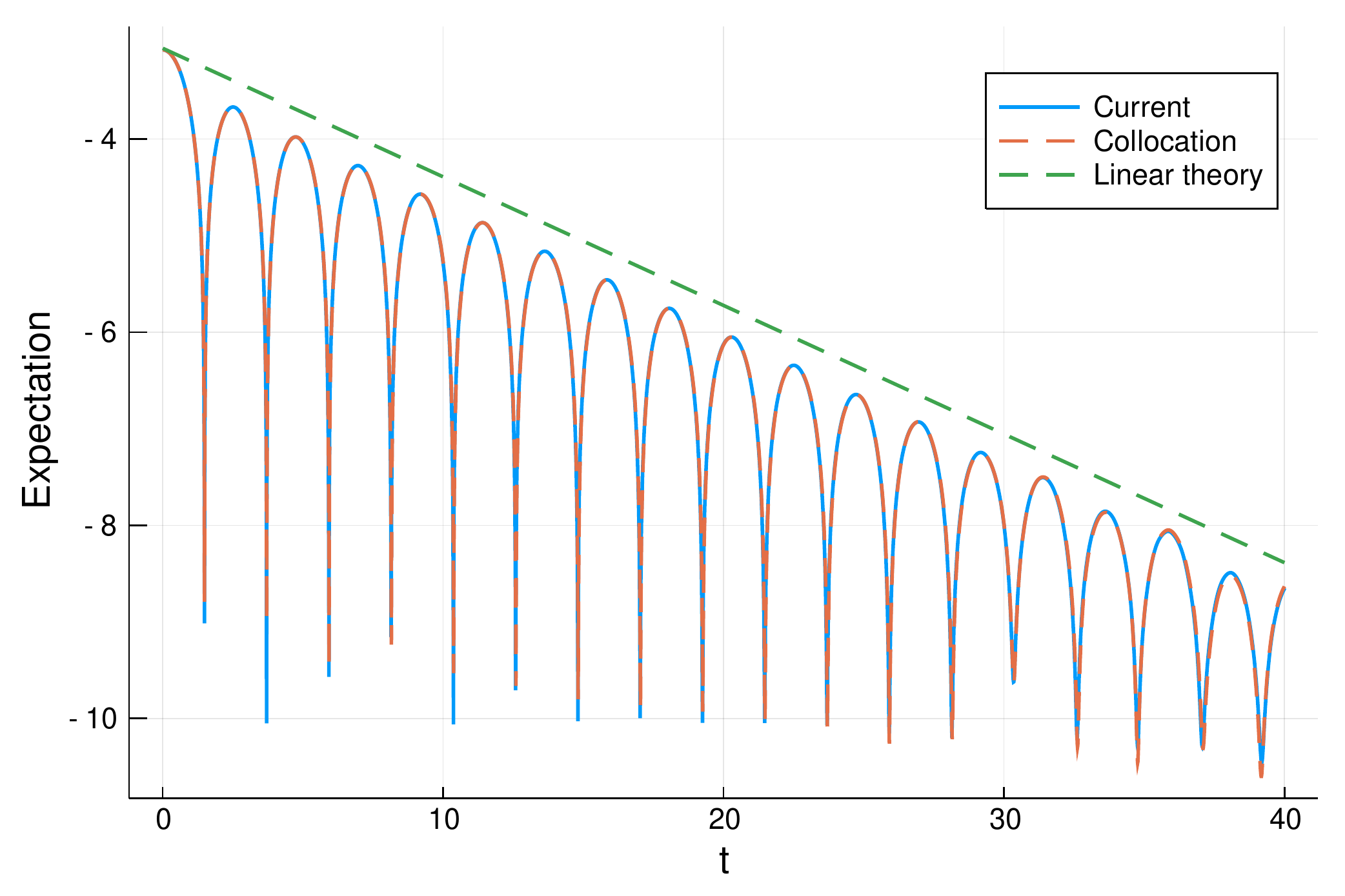}
	}
	\subfigure[Standard deviation]{
		\includegraphics[width=0.45\textwidth]{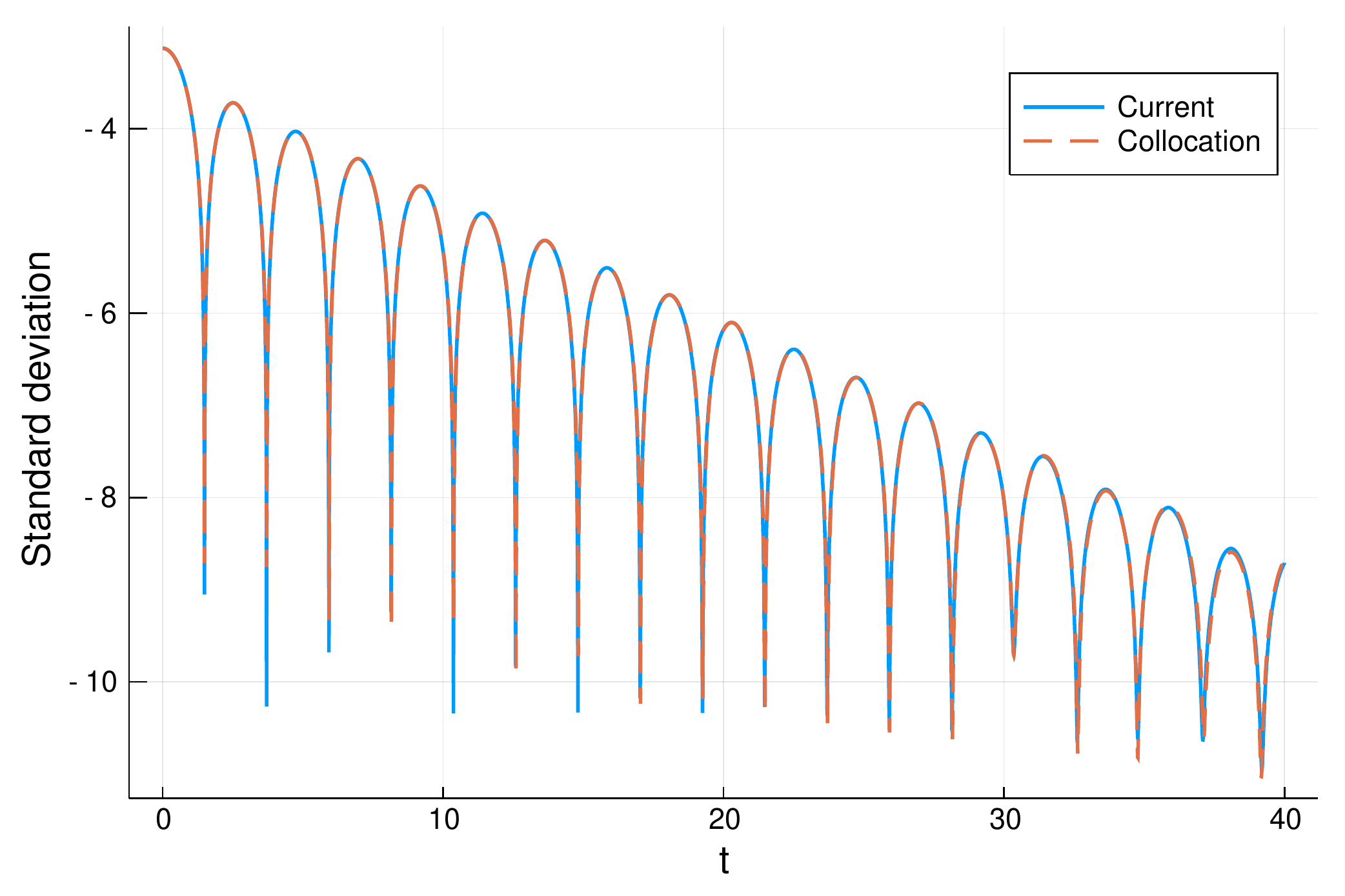}
	}
	\caption{Expectation value and standard deviation of electric field energy (logarithmic) in linear landau damping.}
	\label{pic:damping linear}
\end{figure}

Fig. \ref{pic:damping linear} presents the time evolution of electric energy.
As is shown, the current stochastic kinetic scheme provides equivalent numerical solution as the standard collocation method.
The expected energy damping rate is consistent with the theoretical damping rate $-0.1533$, and the oscillation frequency of the electromagnetic wave corresponds well to the theoretical value $\omega = 1.4156$.

Besides the evolution of expectation value, the stochastic scheme provides the opportunity to study the uncertainty propagation modes simultaneously.
As shown in Fig. \ref{pic:damping linear}(b), the uncertainties travel along with the structure of electromagnetic wave and present similar evolution patterns with mean field. 
The magnitudes of expectation and variance are on the same order due to the small perturbation strength.

\begin{figure}[htb!]
	\centering
	\subfigure[Expectation value]{
		\includegraphics[width=0.45\textwidth]{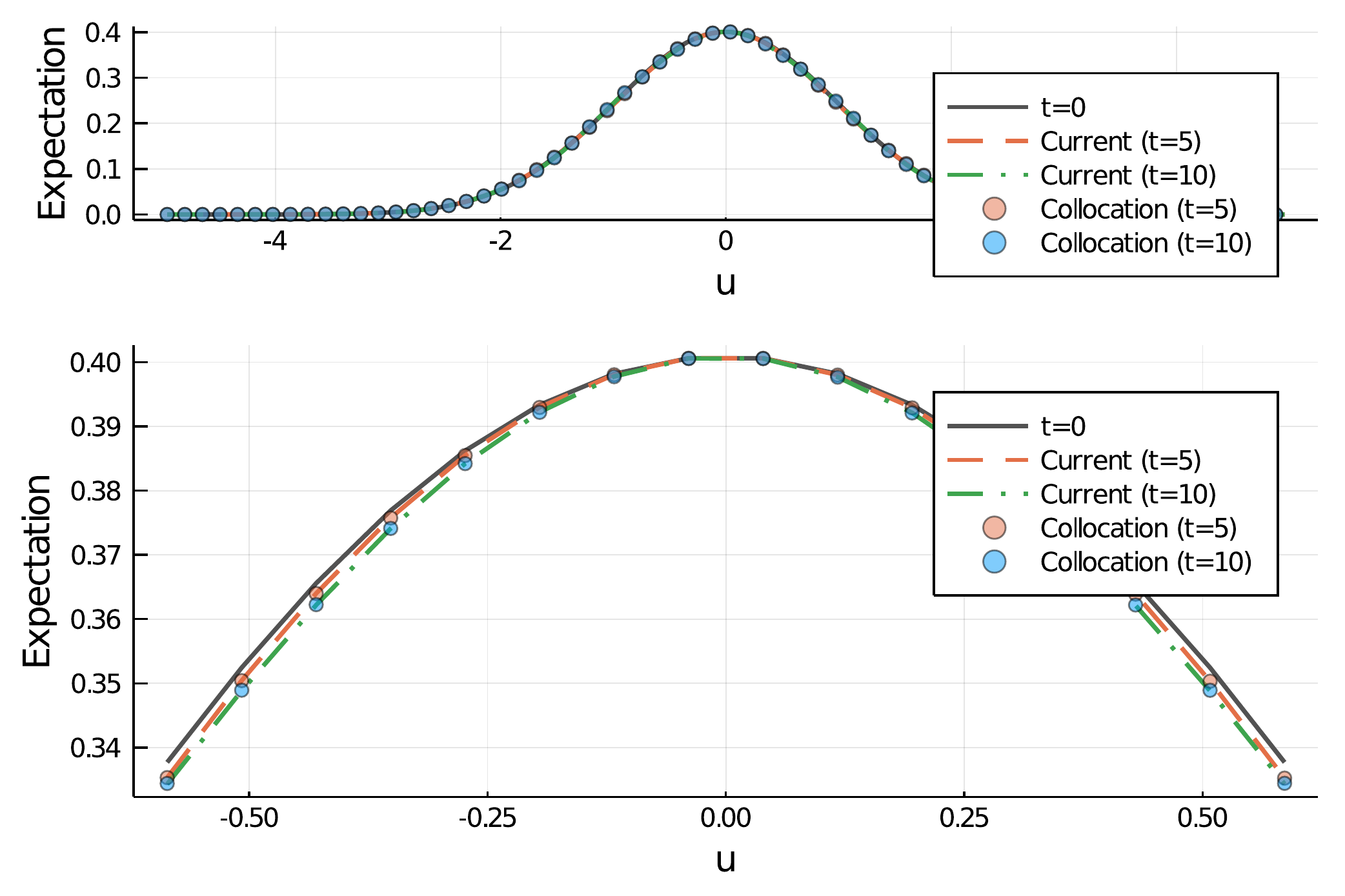}
	}
	\subfigure[Standard deviation]{
		\includegraphics[width=0.45\textwidth]{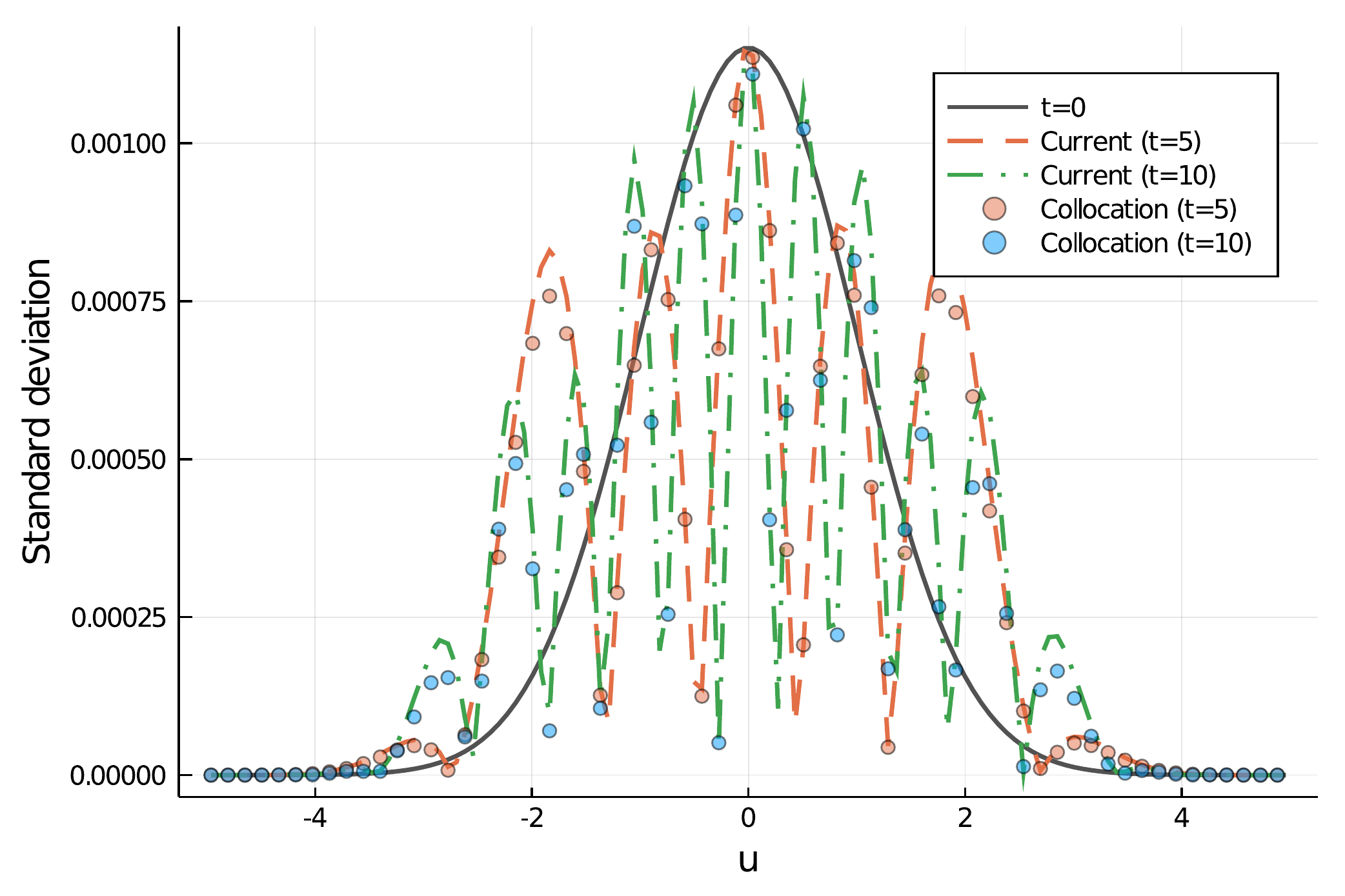}
	}
	\caption{Time evolved expectation value and standard deviation of particle distribution function at $x=0$ in linear landau damping.}
	\label{pic:damping linear pdf}
\end{figure}

Fig. \ref{pic:damping linear pdf} presents the time evolution of particle distribution function at the domain center $x=0$.
From the zoom-in expected distribution function around $u=0$, we see that the low-speed particles resonate with the electromagnetic wave and gradually absorb energy, resulting in decreased value of particle distribution.
In spite of the minor variation of expected value around the Maxwellian for the linear damping case, the standard deviation of solutions shows an increasingly symmetric oscillation in the velocity space during the resonance process.
It indicates a more significant sensitivity compared to the mean field, and
provides a quantitative description of the non-equilibrium effects triggered by the particle-wave resonance.

\subsubsection{Nonlinear case}

\begin{figure}[htb!]
	\centering
	\subfigure[Expectation value]{
		\includegraphics[width=0.45\textwidth]{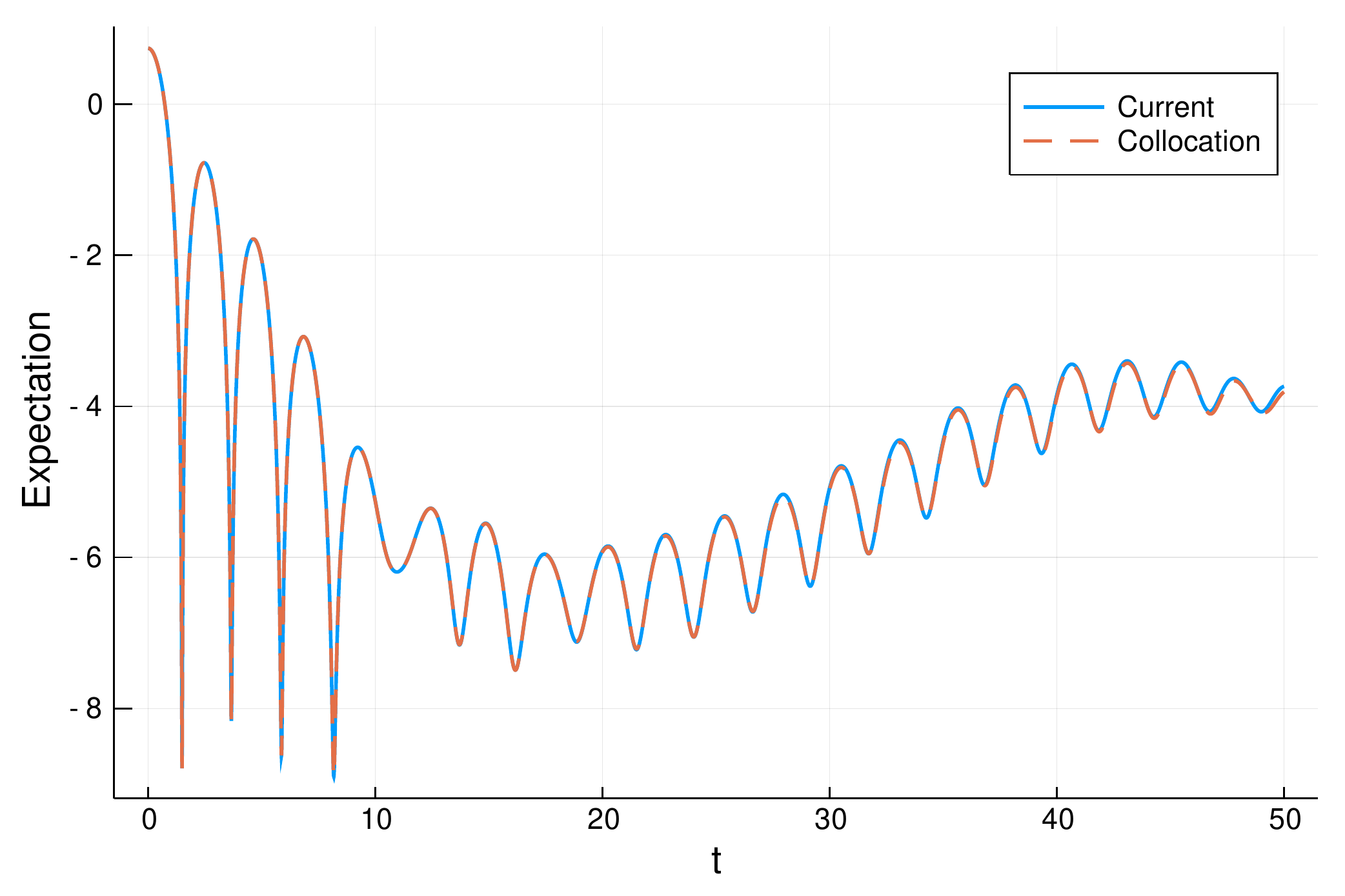}
	}
	\subfigure[Standard deviation]{
		\includegraphics[width=0.45\textwidth]{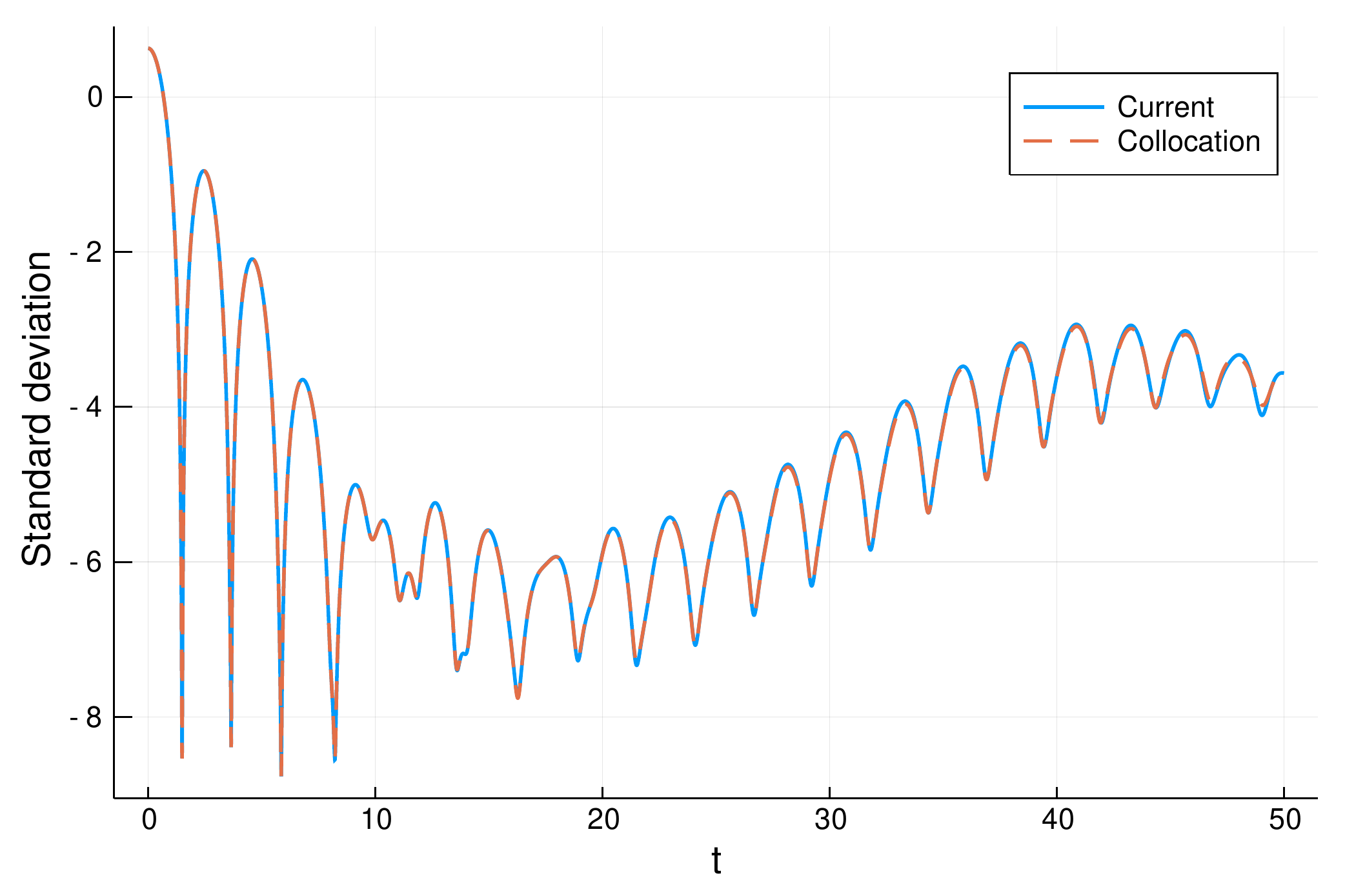}
	}
	\caption{Expectation value and standard deviation of electric field energy (logarithmic) in nonlinear landau damping.}
	\label{pic:damping nonlinear}
\end{figure}

The same initial conditions and computational setups as linear case are followed, except for the enhanced amplitude $\alpha=0.5$. 
The evolution of electric energy is presented in Fig. \ref{pic:damping nonlinear}.
As the amplitude of the electromagnetic wave increases, nonlinear effects would emerge correspondingly, resulting in a rise in energy after the initial damping.

\begin{figure}[htb!]
	\centering
	\subfigure[$t=0$]{
		\includegraphics[width=0.3\textwidth]{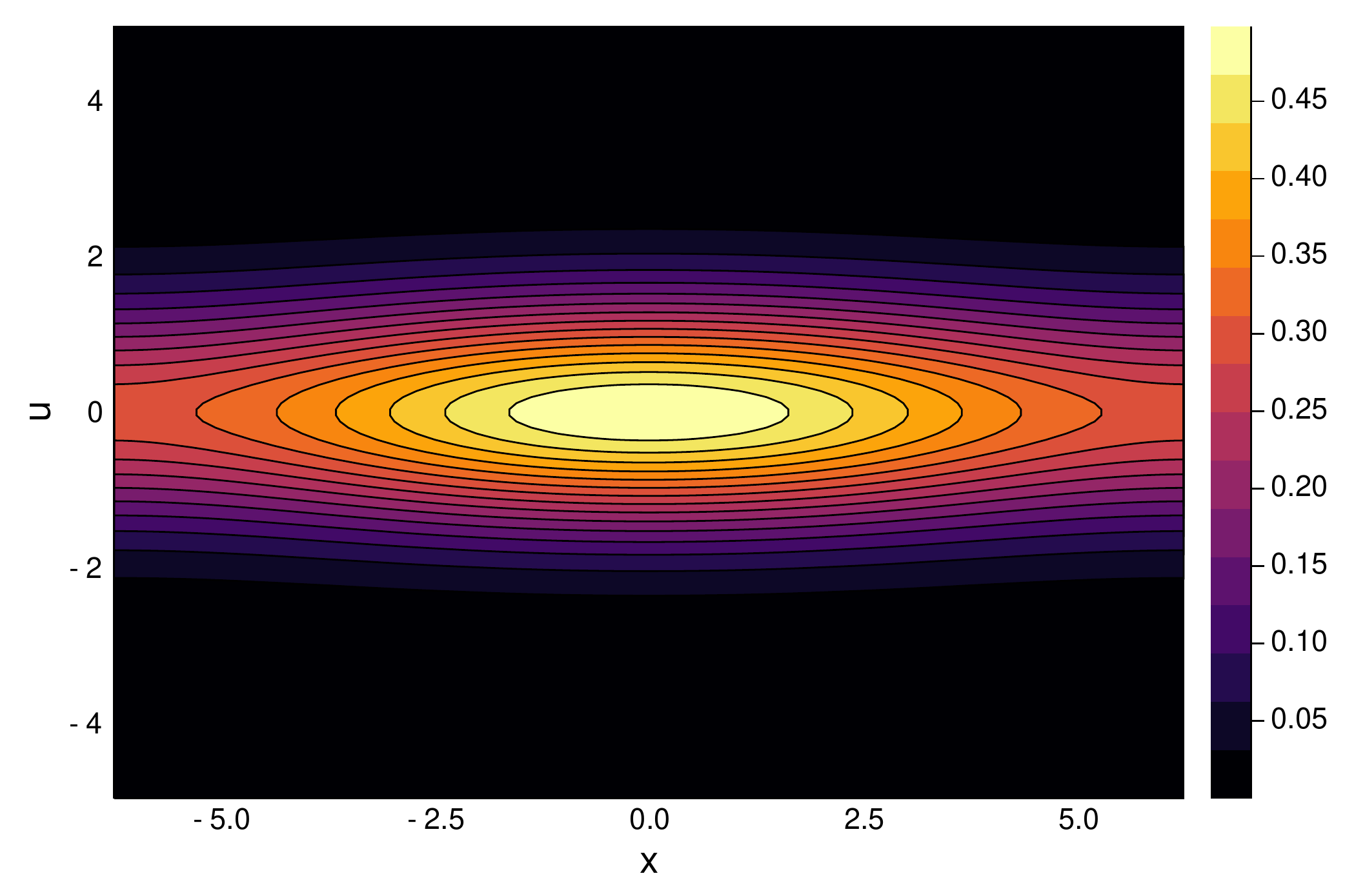}
	}
	\subfigure[$t=5$]{
		\includegraphics[width=0.3\textwidth]{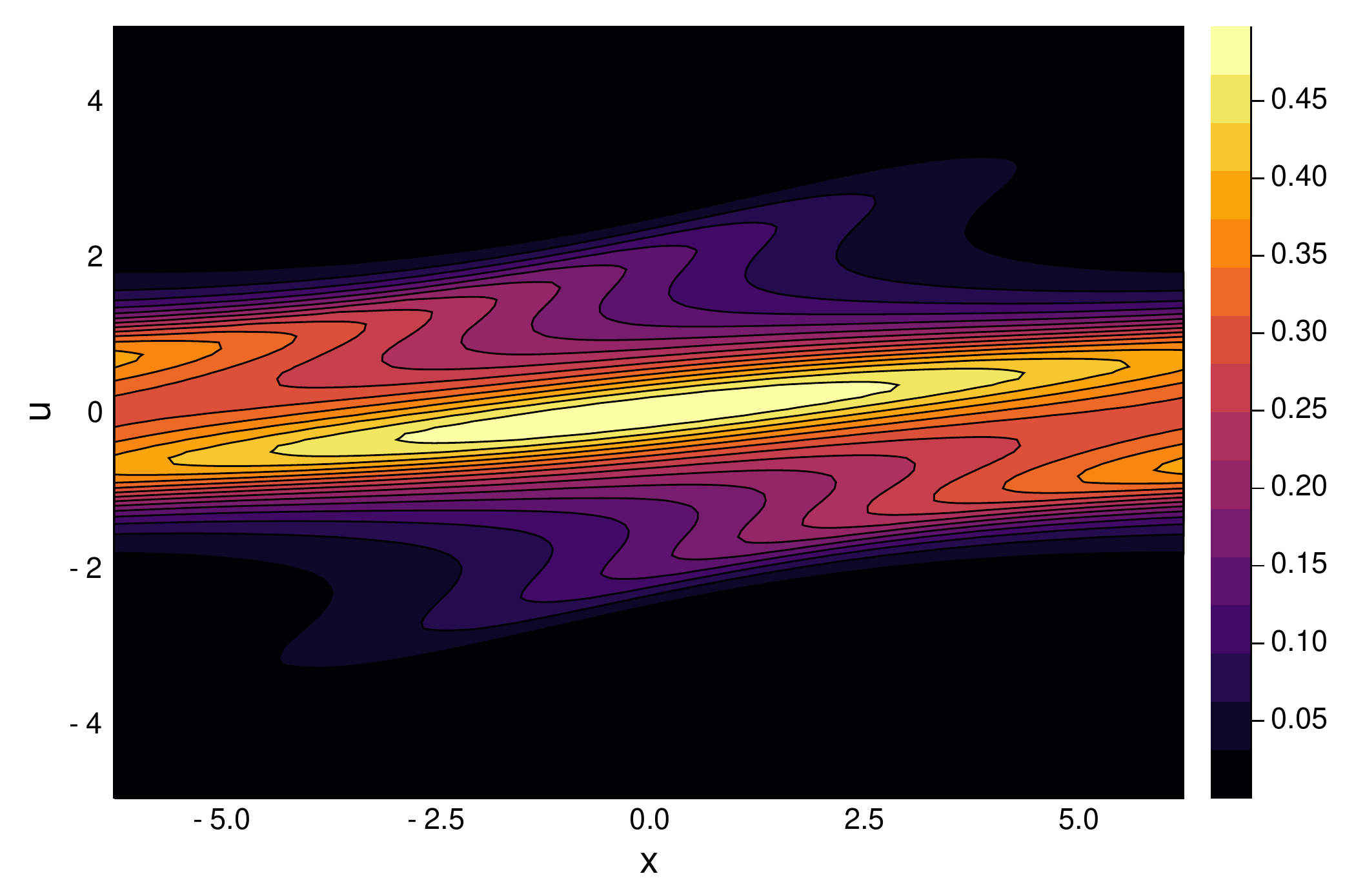}
	}
	\subfigure[$t=10$]{
		\includegraphics[width=0.3\textwidth]{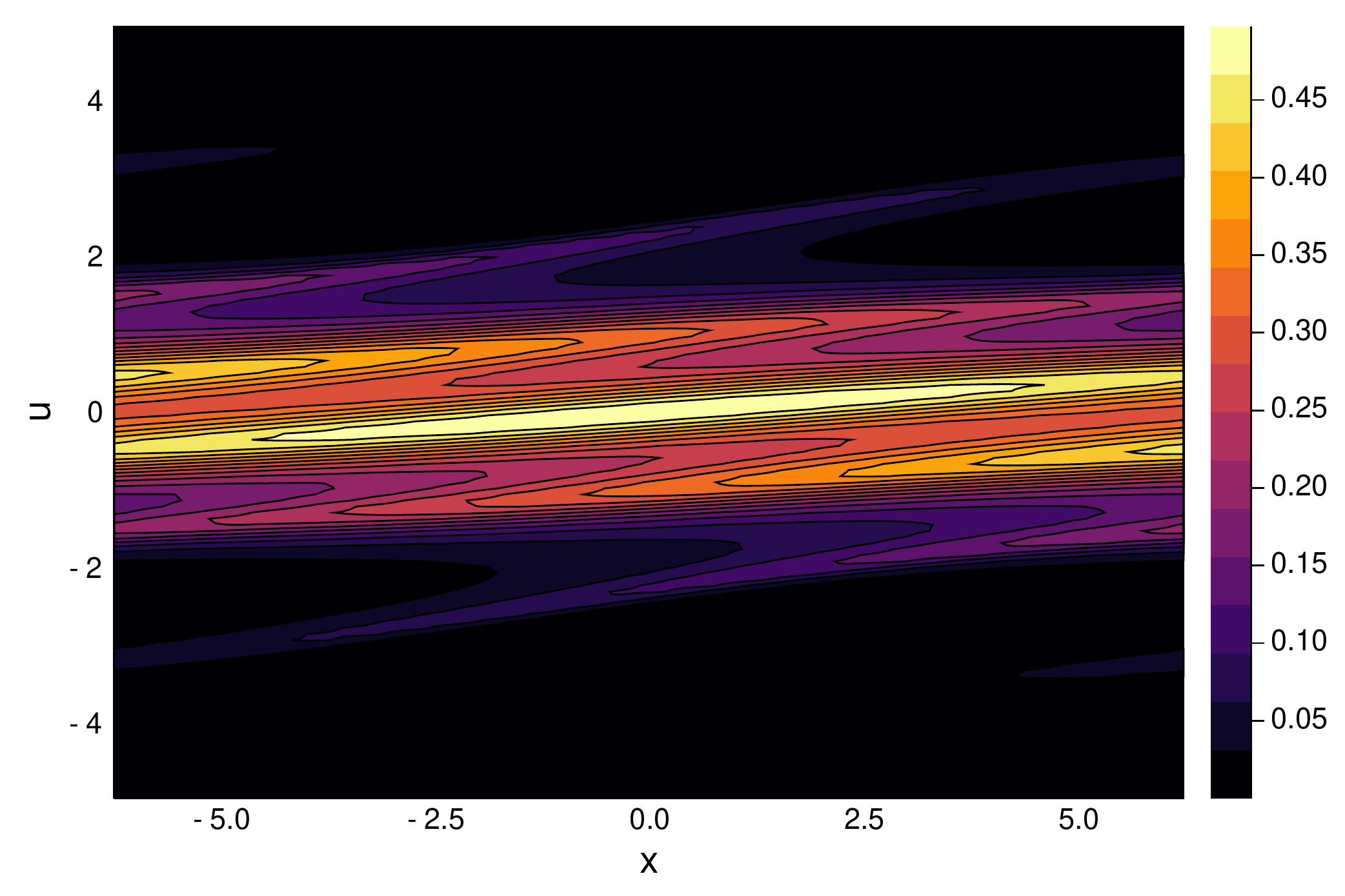}
	}
	\subfigure[$t=20$]{
		\includegraphics[width=0.3\textwidth]{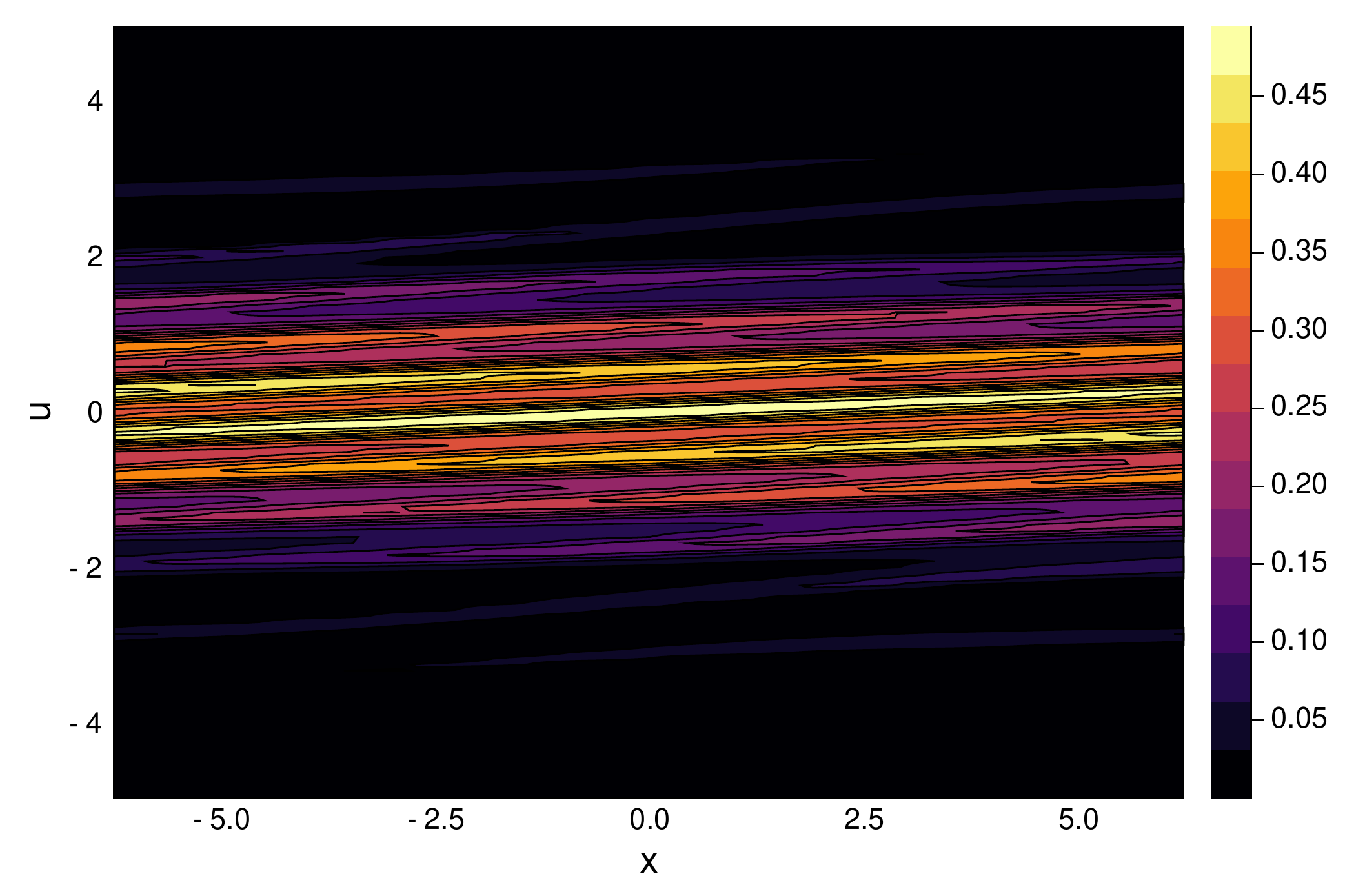}
	}
	\subfigure[$t=30$]{
		\includegraphics[width=0.3\textwidth]{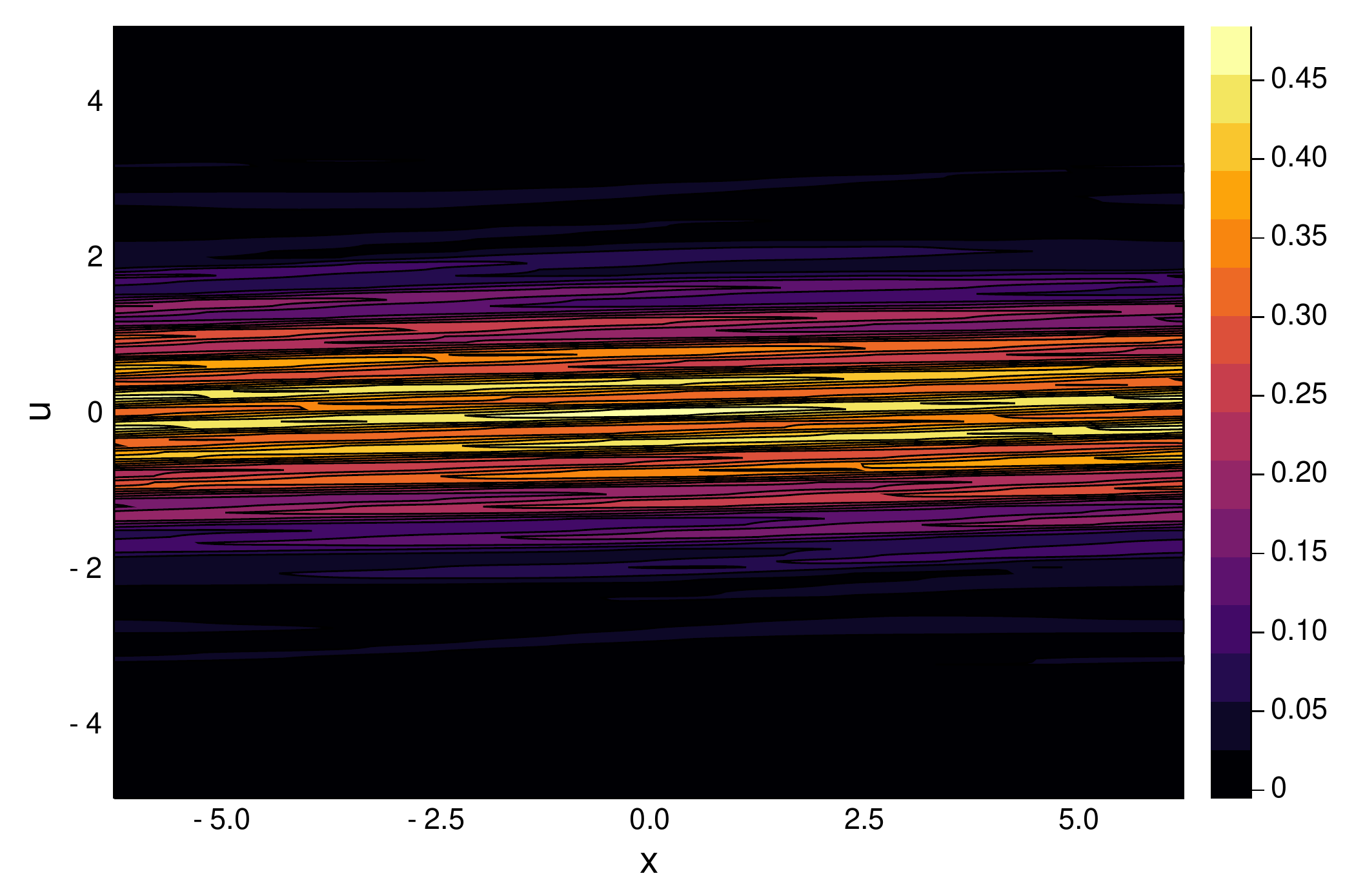}
	}
	\subfigure[$t=50$]{
		\includegraphics[width=0.3\textwidth]{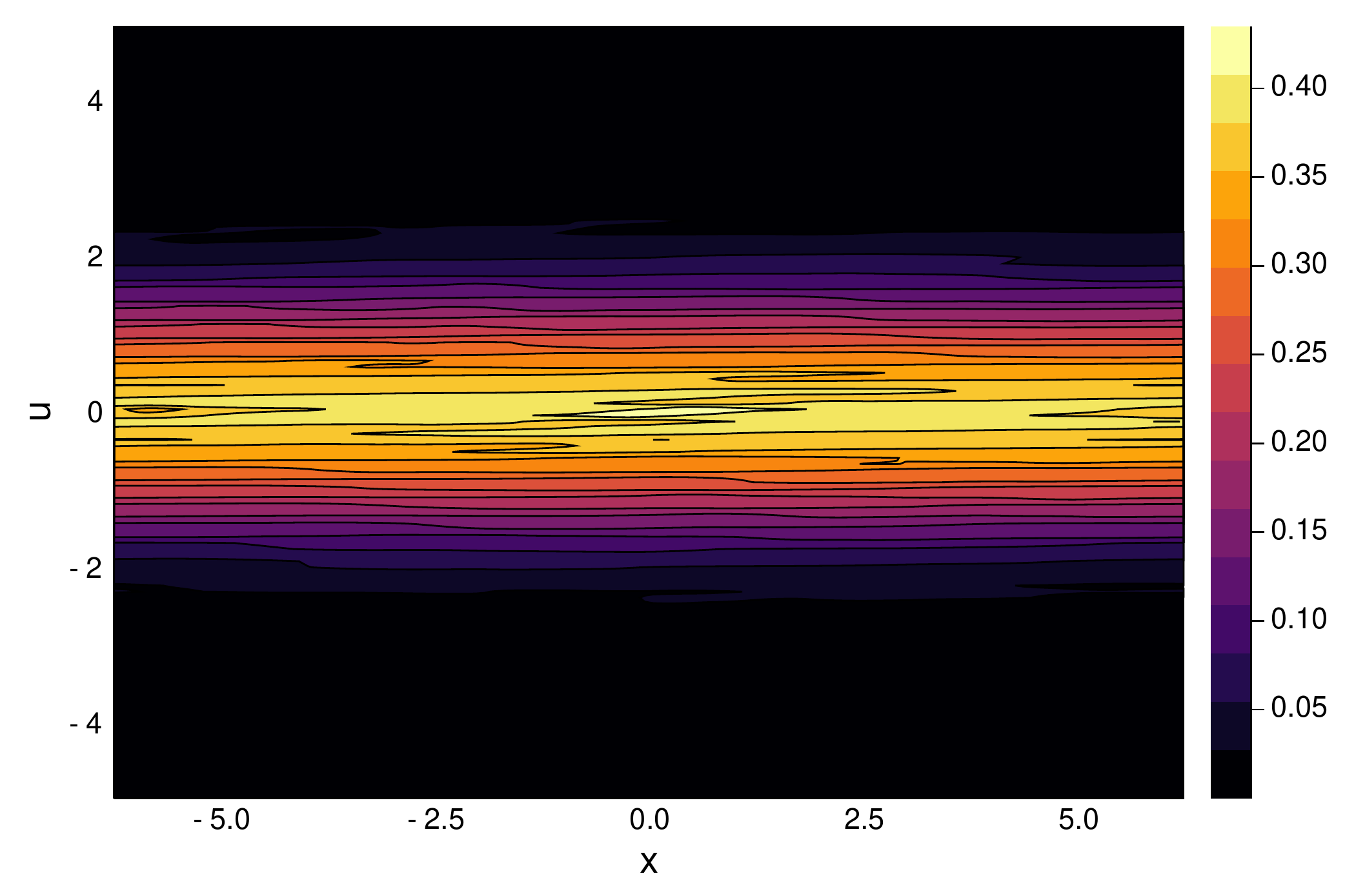}
	}
	\caption{Time evolved expectation value of particle distribution function over phase space $(x,u)$ in nonlinear landau damping.}
	\label{pic:damping nonlinear contour}
\end{figure}

\begin{figure}[htb!]
	\centering
	\subfigure[Expectation value]{
		\includegraphics[width=0.45\textwidth]{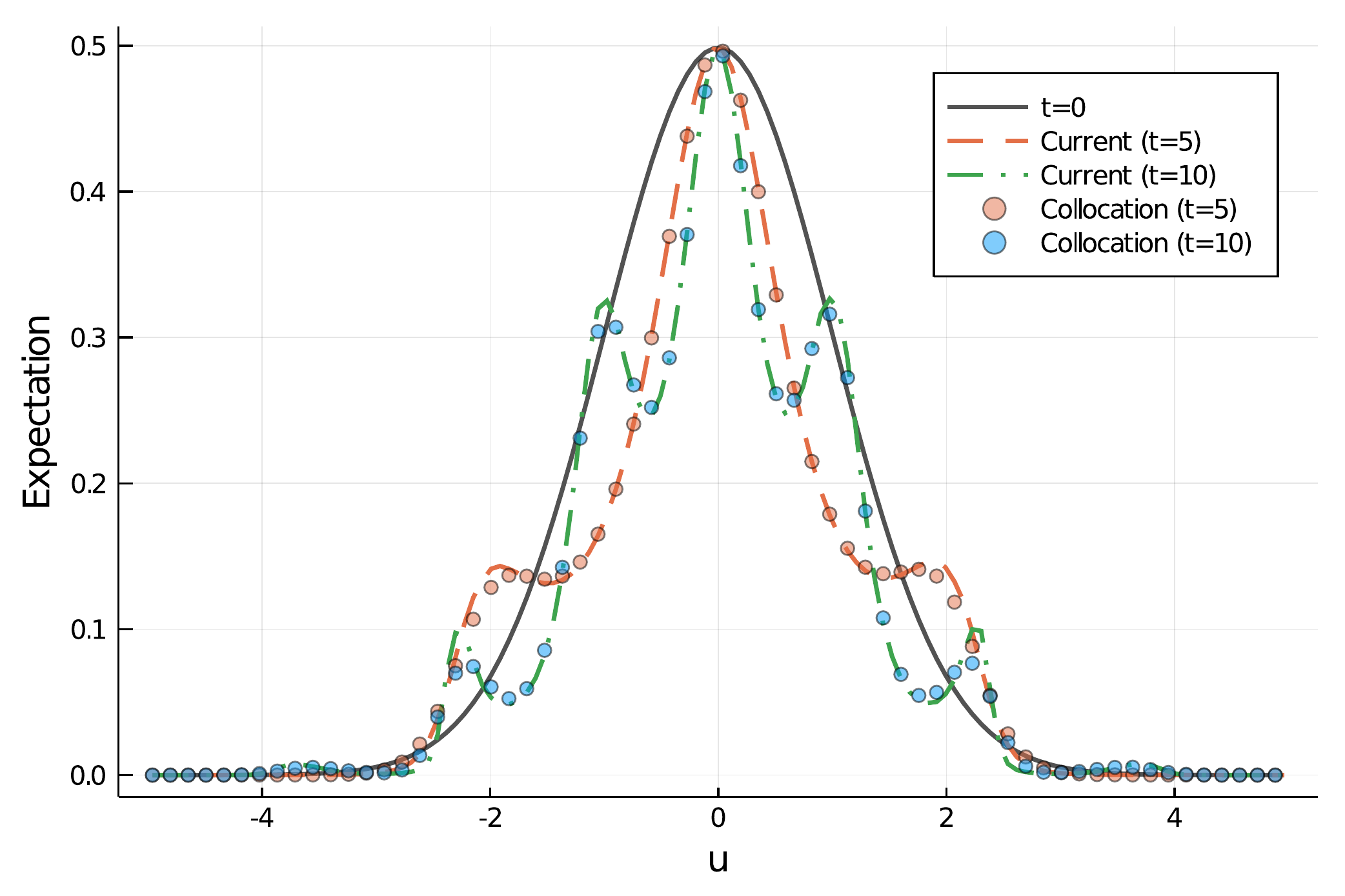}
	}
	\subfigure[Standard deviation]{
		\includegraphics[width=0.45\textwidth]{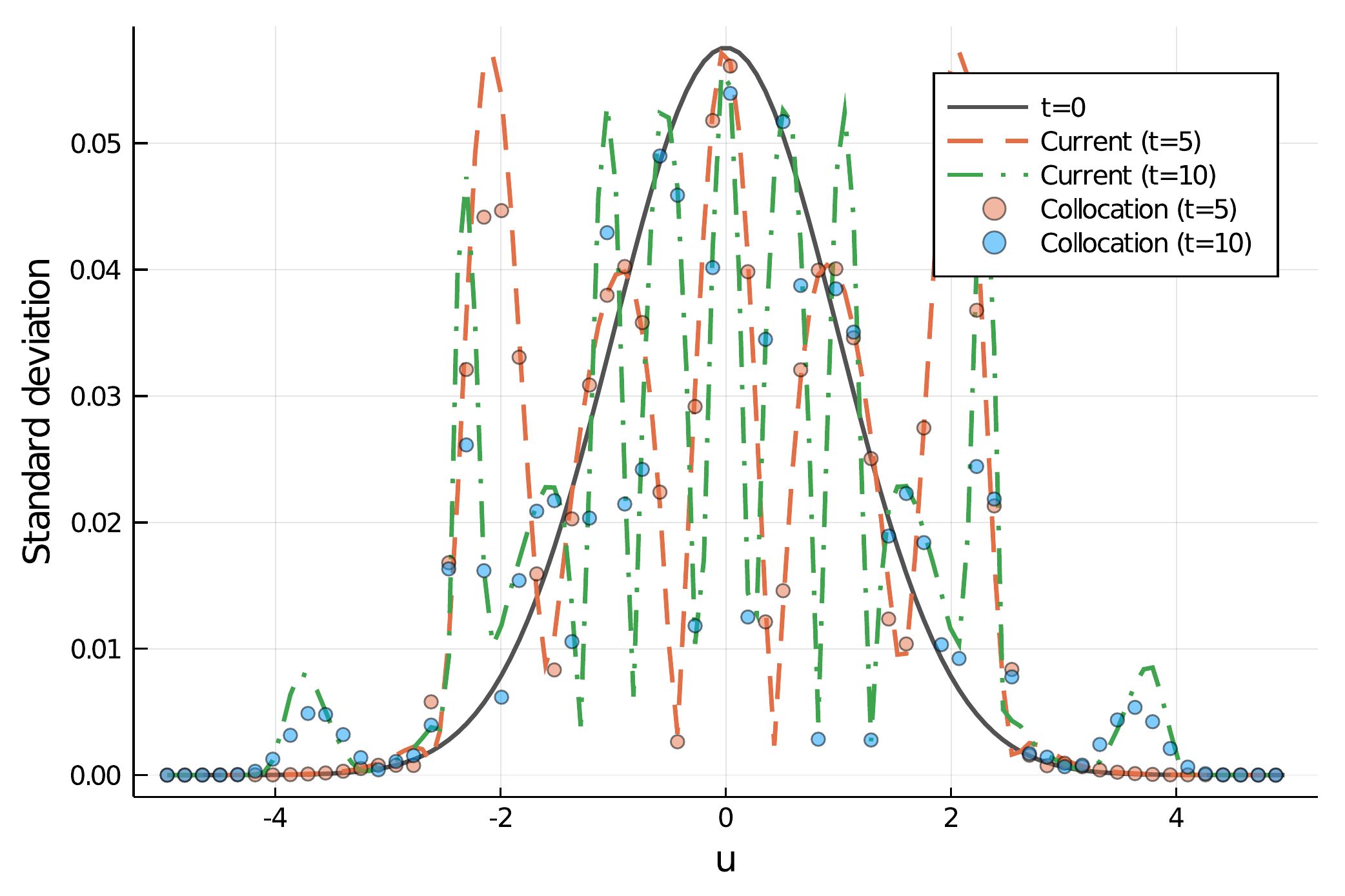}
	}
	\caption{Time evolved expectation value and standard deviation of particle distribution function at $x=0$ in nonlinear landau damping.}
	\label{pic:damping nonlinear pdf}
\end{figure}

Fig. \ref{pic:damping nonlinear contour} shows the time evolution of expected particle distribution over the phase space $(x,u)$,
and Fig. \ref{pic:damping nonlinear pdf} picks out the expectation and variance of particle distribution at physical domain center $x=0$.
Given the increasing intensity of radio field, here the particle distribution function is deformed in phase space.
Consistent with the linear case, the change of variance here is more significant than expectation value, indicating a stronger sensibility.

\subsection{Two-stream instability}

The two-stream instability is another typical phenomenon in for collisionless plasmas. 
To a certain extent, it can also be regarded as an inverse phenomenon of Landau damping.

\subsubsection{Linear case}

Consider the following initial system,
\begin{equation*}
\left[\begin{array}{c}
n_{ion} \\
n_{ele} \\
U \\
T \\
f_{ion} \\
f_{ele} \\
\end{array}\right]_{\text {t=0}}=\left[\begin{array}{c}
1 \\
\frac{2}{7} \left[1+\alpha \xi \left(\frac{\cos (2 k x)+\cos (3 k x)}{1.2}+\cos (k x)\right)\right] \\
0 \\
1 \\
n_{ion} \left(\frac{m_{ion}}{2\pi T}\right)^{1/2} \exp\left(-\frac{m_{ion}}{2 T} (u-U)^2\right) \\
n_{ele} \left(\frac{m_{ele}}{2\pi T}\right)^{1/2} \left(1+5 u^{2}\right) \exp\left(-\frac{m_{ele}}{2 T} (u-U)^2\right) \\
\end{array}\right],
\end{equation*}
where the computational setup is listed in Table \ref{tab:linear twostream}.
Same as the Landau damping case, we fix the slow motions of ions as background, and solve the evolution of electrons.
\begin{table}[htbp]
	\caption{Computational setup of linear two-stream instability.} 
	\centering
	\begin{tabular}{lllllll} 
		\toprule 
		$t$ & $x$ & $N_x$ & $m_{ion}/m_{ele}$ & $u_{ele}$ & $N_u$ & $\alpha$ \\ 
		\midrule 
		$(0,70]$ & $[0,2\pi/k]$ & 128 & $1836$ & $[-5,5]$ & 128 & 0.001 \\ 
		\midrule 
		$k$ & $\xi$ & Polynomial & $N_z$(gPC) & $N_z$(quad) & Kn & cfl \\ 
		\midrule 
		0.5 & $\mathcal U(0,1)$ & Legendre & $5$ & $9$ & $100$  & 0.2   \\ 
		\bottomrule 
	\end{tabular} 
	\label{tab:linear twostream}
\end{table}

\begin{figure}[htb!]
	\centering
	\subfigure[Expectation value]{
		\includegraphics[width=0.45\textwidth]{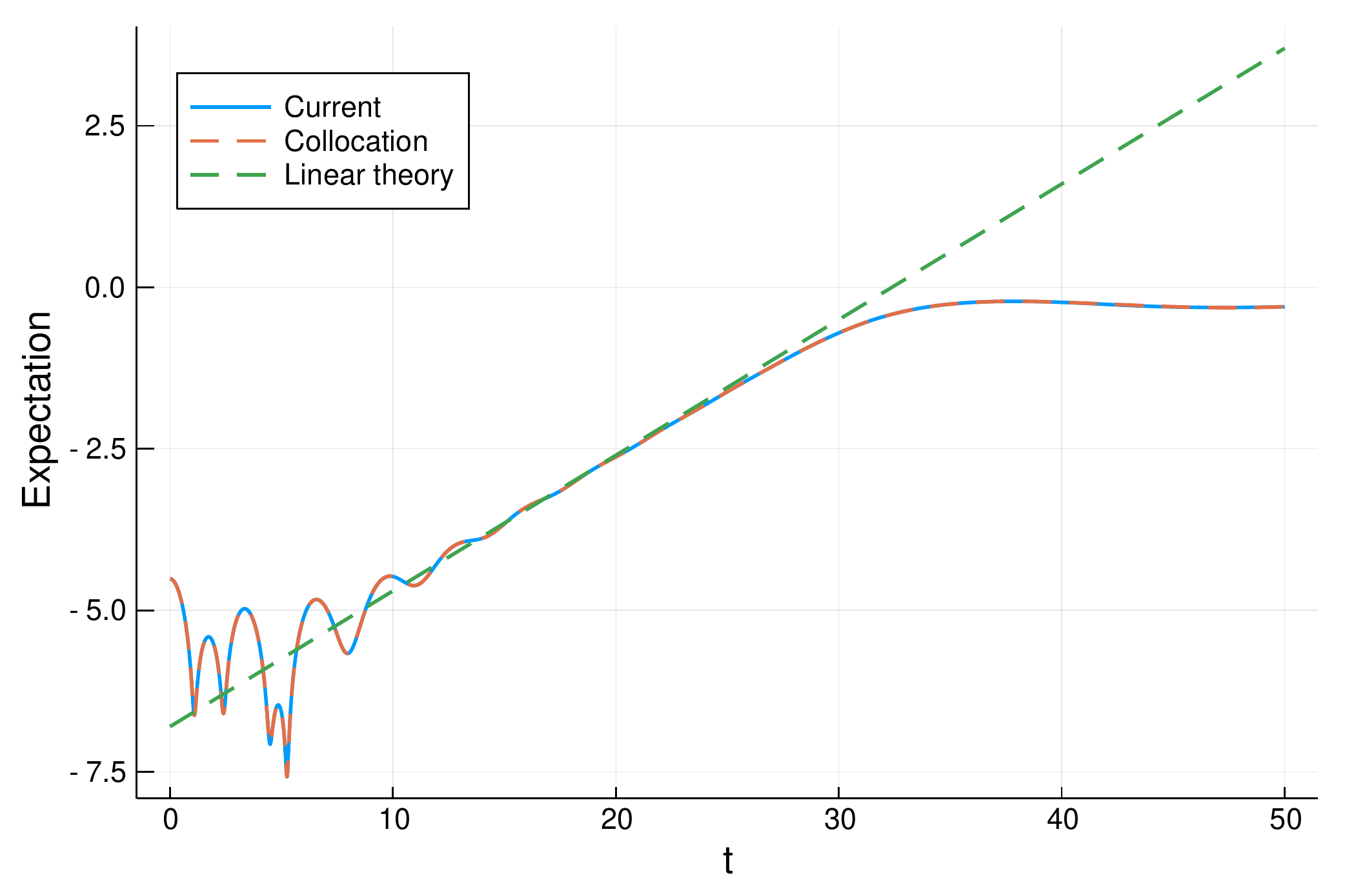}
	}
	\subfigure[Standard deviation]{
		\includegraphics[width=0.45\textwidth]{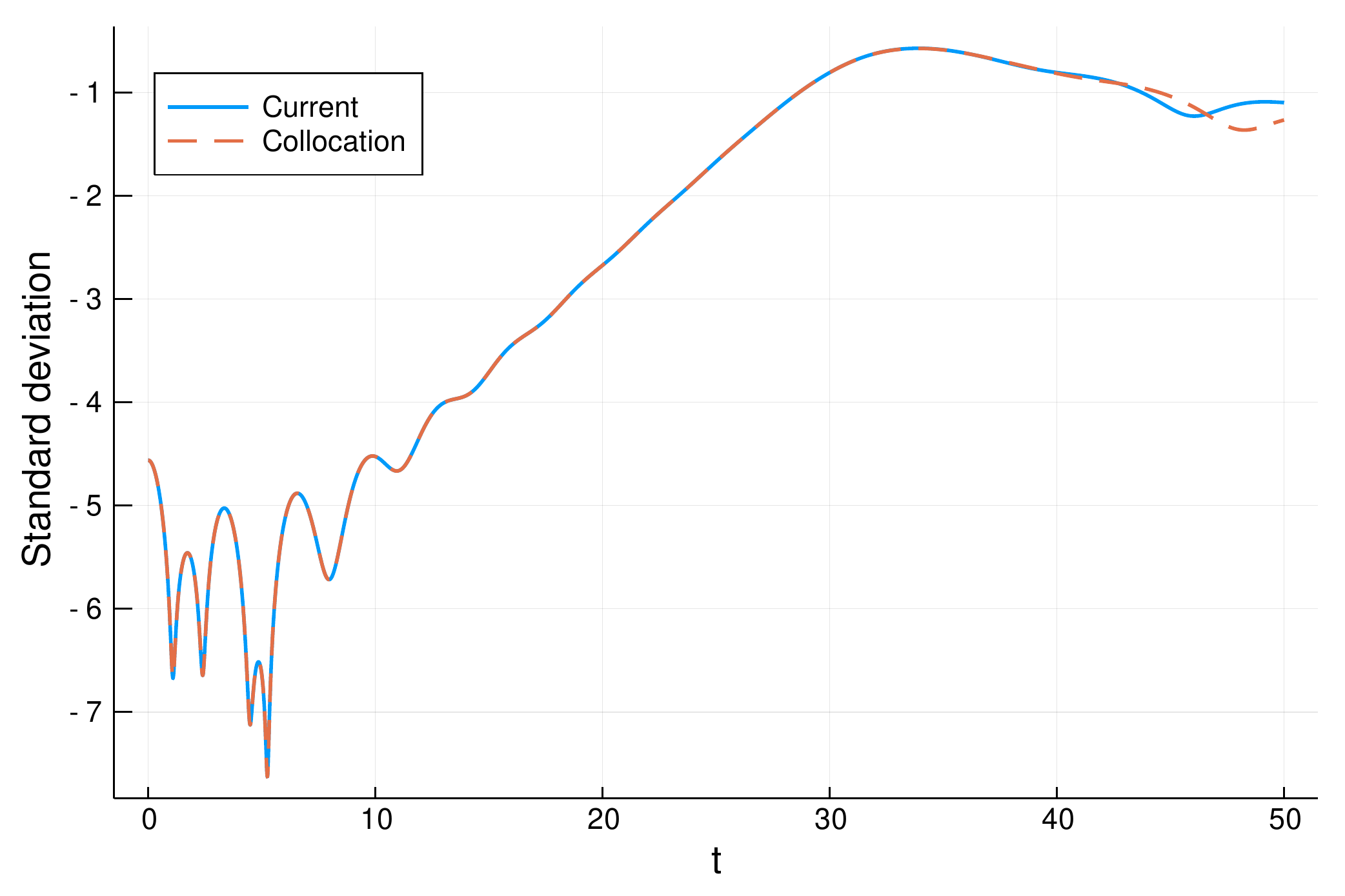}
	}
	\caption{Expectation value and standard deviation of electric field energy (logarithmic) in linear two-stream instability.}
	\label{pic:twostream linear}
\end{figure}

\begin{figure}[htb!]
	\centering
	\subfigure[Expectation value]{
		\includegraphics[width=0.45\textwidth]{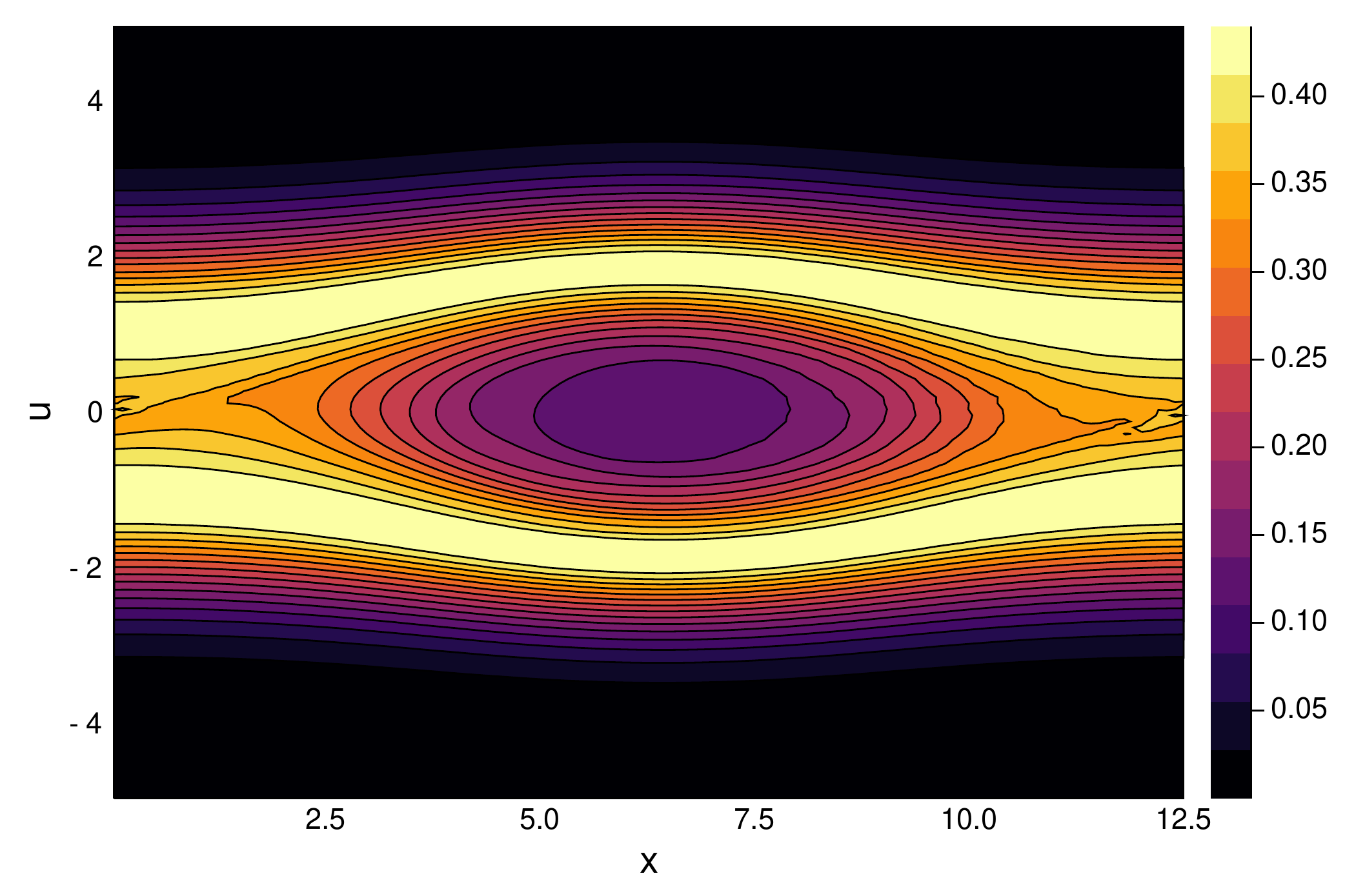}
	}
	\subfigure[Standard deviation]{
		\includegraphics[width=0.45\textwidth]{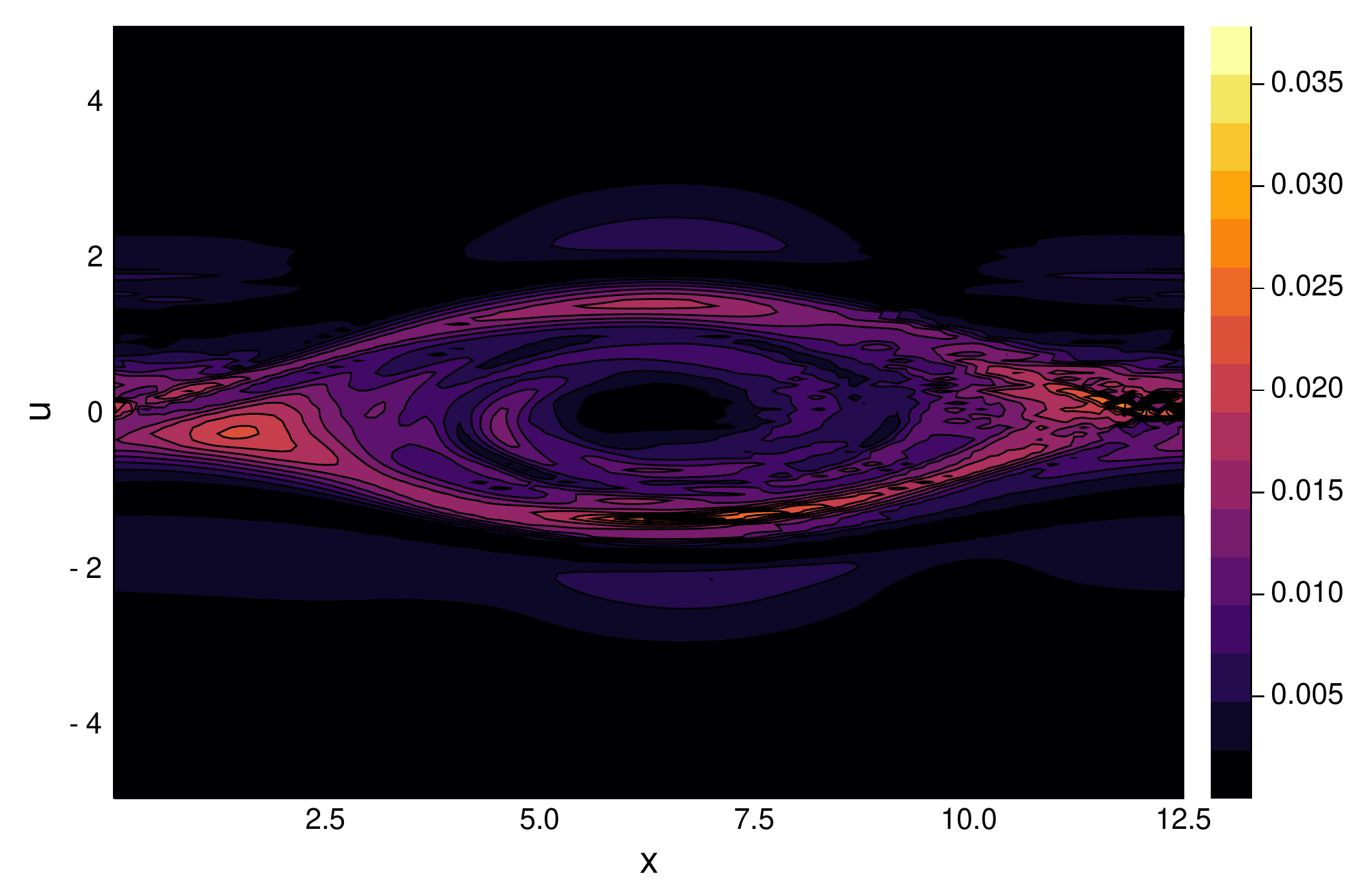}
	}
	\caption{Expectation value and standard deviation of particle distribution function over phase space $(x,u)$ in linear two-stream instability.}
	\label{pic:twostream linear contour}
\end{figure}

The evolution of electric field energy is shown in Fig. \ref{pic:twostream linear}, from which it can be seen that the numerical and theoretical solutions \cite{chen2012introduction} fit well,
and the uncertainties of electric energy propagate in the similar pattern as mean field.
Fig. \ref{pic:twostream linear contour} provides the particle distribution function over the phase space $(x,u)$ at $t=70$.
The swirling pattern is clearly identified in both expectation and variance values.
More fine structures can be seen in the standard deviation, which provides a clearer way to quantify the stochastic evolution of particles.

\subsubsection{Nonlinear case}

In the following let us consider the nonlinear case.
The initial system is given as,
\begin{equation*}
\left[\begin{array}{c}
n_{ion} \\
n_{ele} \\
U \\
T \\
f_{ion} \\
f_{ele} \\
\end{array}\right]_{\text {t=0}}=\left[\begin{array}{c}
1 \\
\frac{1}{2} \left( 1 + \alpha \xi * \cos(kx) \right) \\
0 \\
0.09 \\
n_{ion} \left(\frac{m_{ion}}{2\pi T}\right)^{1/2} \exp\left(-\frac{m_{ion}}{2 T} (u-U)^2\right) \\
\frac{1}{2} n_{ele} \left(\frac{m_{ele}}{2\pi T}\right)^{1/2} 
\left( \exp\left(-\frac{m_{ele}}{2 T} (u-0.99)^2\right) + \exp\left(-\frac{m_{ele}}{2 T} (u+0.99)^2\right) \right) \\
\end{array}\right].
\end{equation*}
The computational setups adopted here is shown in Table \ref{tab:nonlinear twostream}.
\begin{table}[htbp]
	\caption{Computational setup of nonlinear two-stream instability.} 
	\centering
	\begin{tabular}{lllllll} 
		\toprule 
		$t$ & $x$ & $N_x$ & $m_{ion}/m_{ele}$ & $u_{ele}$ & $N_u$ & $\alpha$ \\ 
		\midrule 
		$(0,70]$ & $[0,2\pi/k]$ & 256 & $1836$ & $[-5,5]$ & 256 & 0.001 \\ 
		\midrule 
		$k$ & $\xi$ & Polynomial & $N_z$(gPC) & $N_z$(quad) & Kn & cfl \\ 
		\midrule 
		0.5 & $\mathcal U(0,1)$ & Legendre & $5$ & $9$ & $100$  & 0.2   \\ 
		\bottomrule 
	\end{tabular} 
	\label{tab:nonlinear twostream}
\end{table}

\begin{figure}[htb!]
	\centering
	\subfigure[Expectation value]{
		\includegraphics[width=0.45\textwidth]{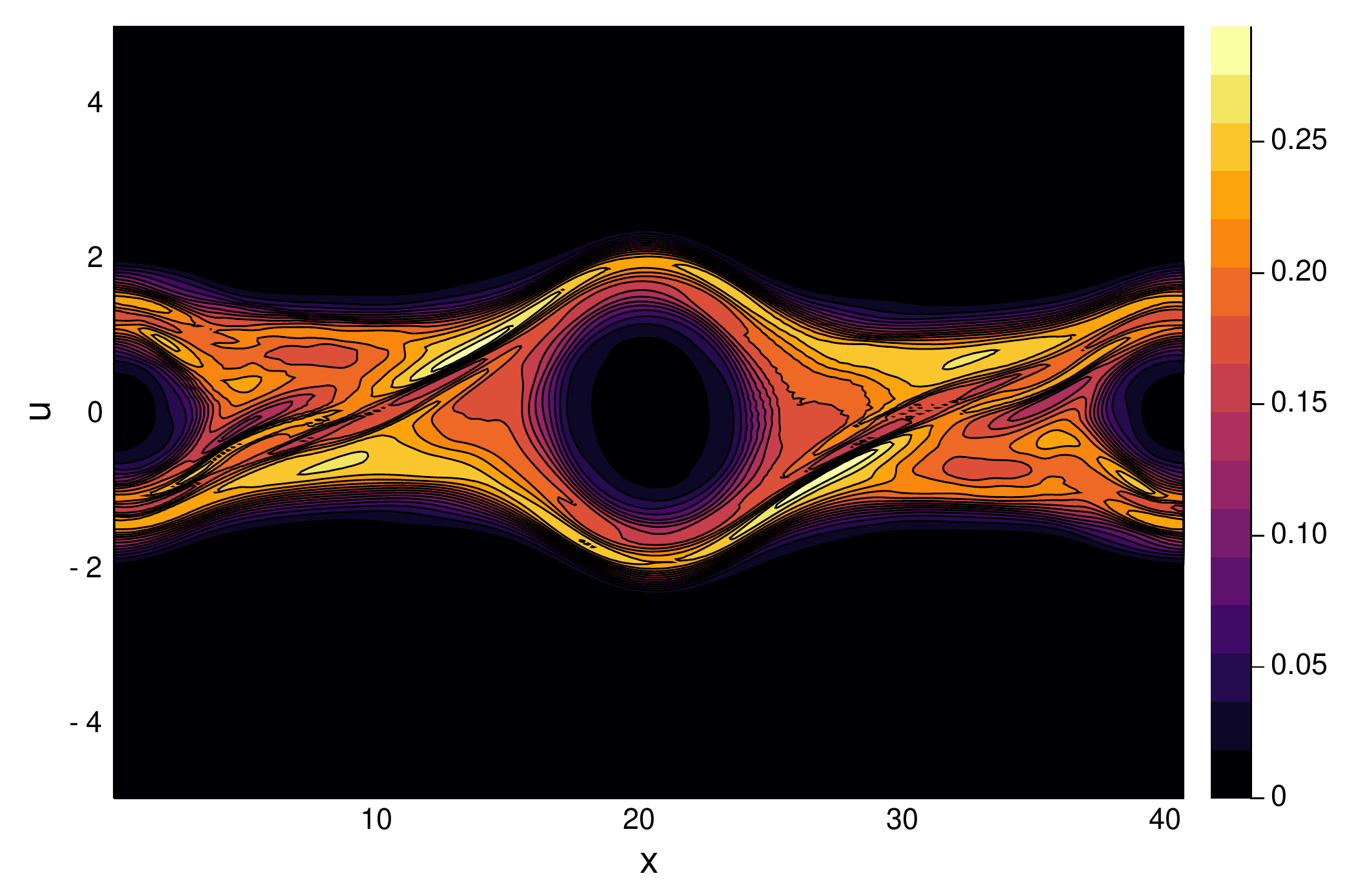}
	}
	\subfigure[Standard deviation]{
		\includegraphics[width=0.45\textwidth]{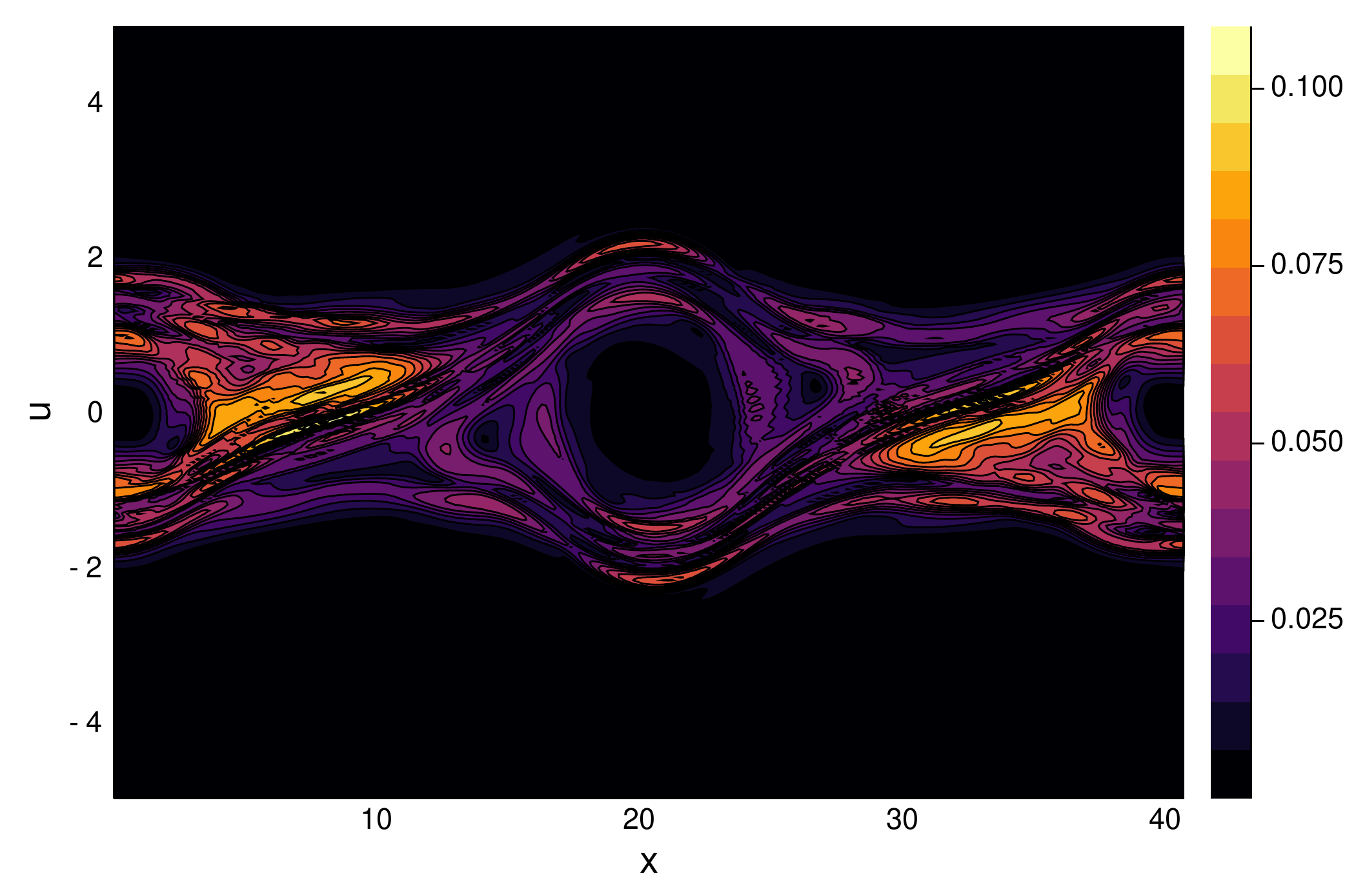}
	}
	\caption{Expectation value and standard deviation of particle distribution function over phase space $(x,u)$ in nonlinear two-stream instability.}
	\label{pic:twostream nonlinear contour}
\end{figure}

Fig. \ref{pic:twostream nonlinear contour} shows the time evolution of expected particle distribution over the phase space $(x,u)$.
Due to the increasing kinetic energy from two particle streams, the particle distribution function is stretched and warped in the phase space.
The pattern of standard deviation is similar as mean field, yet presenting more fine-scale structures, indicating a stronger sensitivity with respect to stochastic parameters.

\subsection{Brio-Wu shock tube} \label{sec:sod}

After the validation of current scheme in the Vlasov limit, let us turn to another limiting regime, i.e. the magnetohydrodynamic (MHD) transport problem under continuum assumption.
In this case, we employ the Brio-Wu shock tube \cite{brio1988upwind} as benchmark case for ideal MHD solutions.

In the simulation, the initial background of deterministic solutions is consistent with the Sod problem, with an additional magnetic discontinuity, i.e.,
\begin{equation*}
\left[\begin{array}{c}
n_{ion} \\
n_{ele} \\
U \\
p \\
E_x \\
E_y \\
E_z \\
B_x \\
B_y \\
B_z \\
\phi \\
\psi \\
\end{array}\right]_{{L}}=\left[\begin{array}{c}
1 \\
1  \\
0 \\
1 \\
0\\
0\\
0\\
0.75\\
1\\
0\\
0\\
0\\
\end{array}\right],
\end{equation*}
for the left half, and
\begin{equation*}
\left[\begin{array}{c}
n_{ion} \\
n_{ele} \\
U \\
p \\
E_x \\
E_y \\
E_z \\
B_x \\
B_y \\
B_z \\
\phi \\
\psi \\
\end{array}\right]_{{R}}=\left[\begin{array}{c}
0.125 \\
0.125  \\
0 \\
0.8 \\
0\\
0\\
0\\
0.75\\
-1\\
0\\
0\\
0\\
\end{array}\right],
\end{equation*}
for the right half.
The computational setup is listed in Table \ref{tab:briowu}.
To recover the ideal MHD solutions, here the interspecies coefficients $\nu_{ie}$ is set as zero.
\begin{table}[htbp]
	\caption{Computational setup of Brio-Wu shock tube.} 
	\centering
	\begin{tabular}{lllllll} 
		\toprule 
		$t$ & $x$ & $N_x$ & $m_{ion}/m_{ele}$ & $u_{ion}$ & $u_{ele}$ & cfl \\ 
		\midrule 
		$(0,0.1]$ & $[0,1]$ & 400 & $1836$ & $[-5,5]$ & $[-5,5]\times \sqrt{ \frac{m_{ion}}{m_{ele}} } $ & 0.3 \\ 
		\midrule 
		$\xi$ & Polynomial & $N_z$(gPC) & $N_z$(quad) & Kn & $r_g$ & $\lambda_D$ \\ 
		\midrule 
		$\mathcal U(0.95,1.05)$ & Legendre & $5$ & $9$ & $1.0 \times 10^{-6}$ & $[0.003,100]$  & 0.01   \\ 
		\bottomrule 
	\end{tabular} 
	\label{tab:briowu}
\end{table}

\subsubsection{Case 1: stochastic flow field}

\begin{figure}[htb!]
	\centering
	\subfigure[$\mathbb E(N)$]{
		\includegraphics[width=0.31\textwidth]{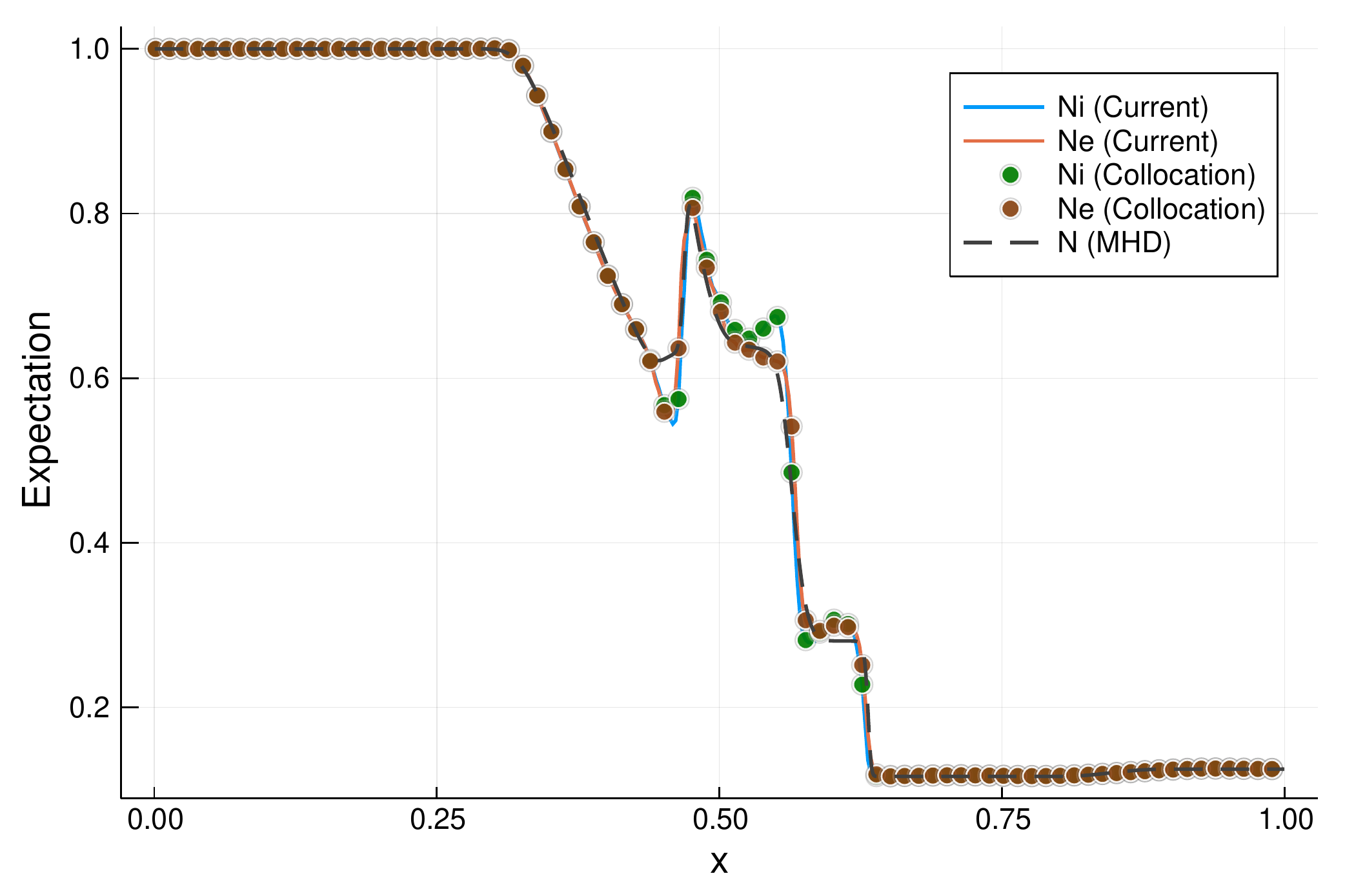}
	}
	\subfigure[$\mathbb E(U)$]{
		\includegraphics[width=0.31\textwidth]{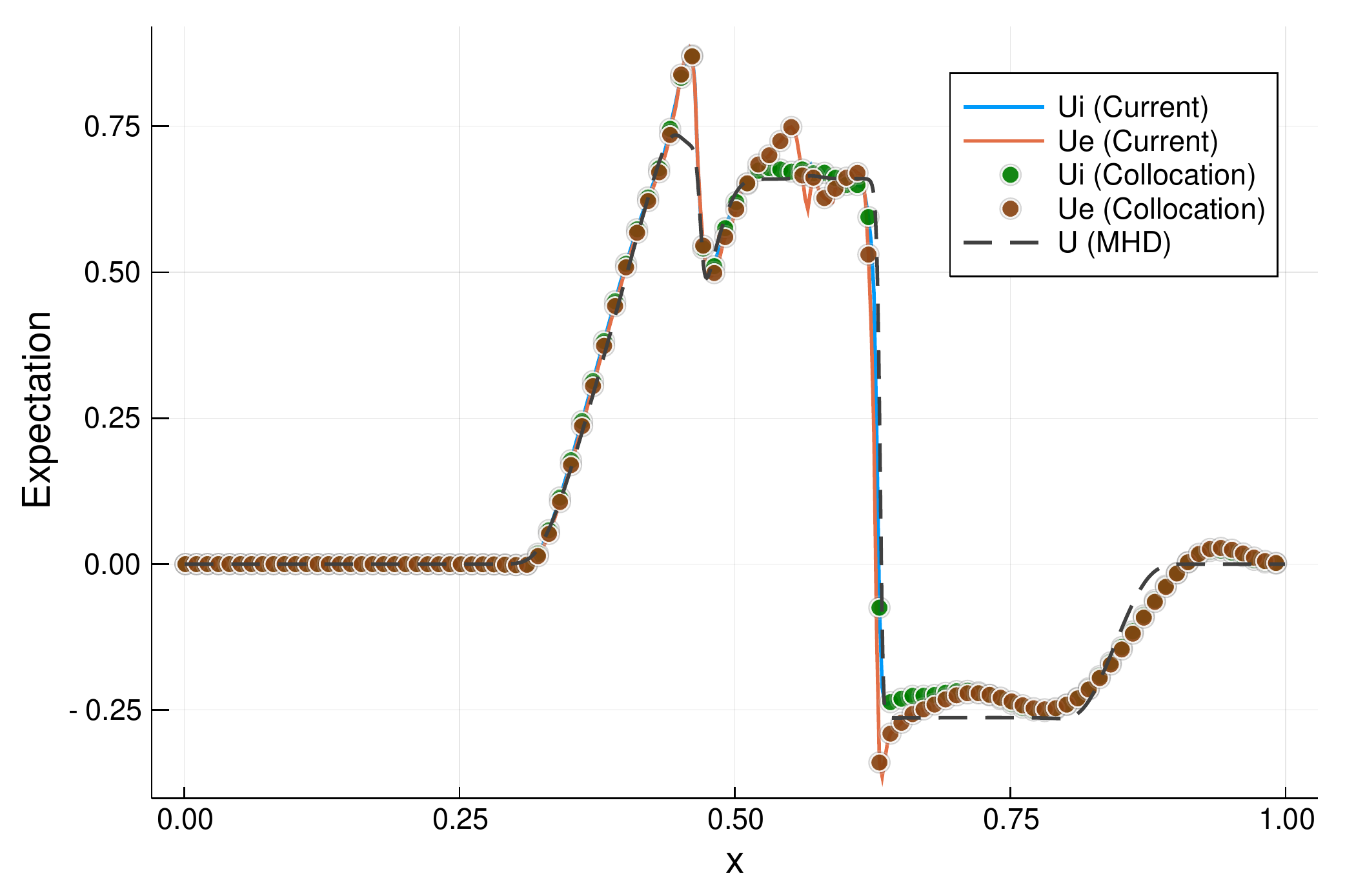}
	}
	\subfigure[$\mathbb E(B_y)$]{
		\includegraphics[width=0.31\textwidth]{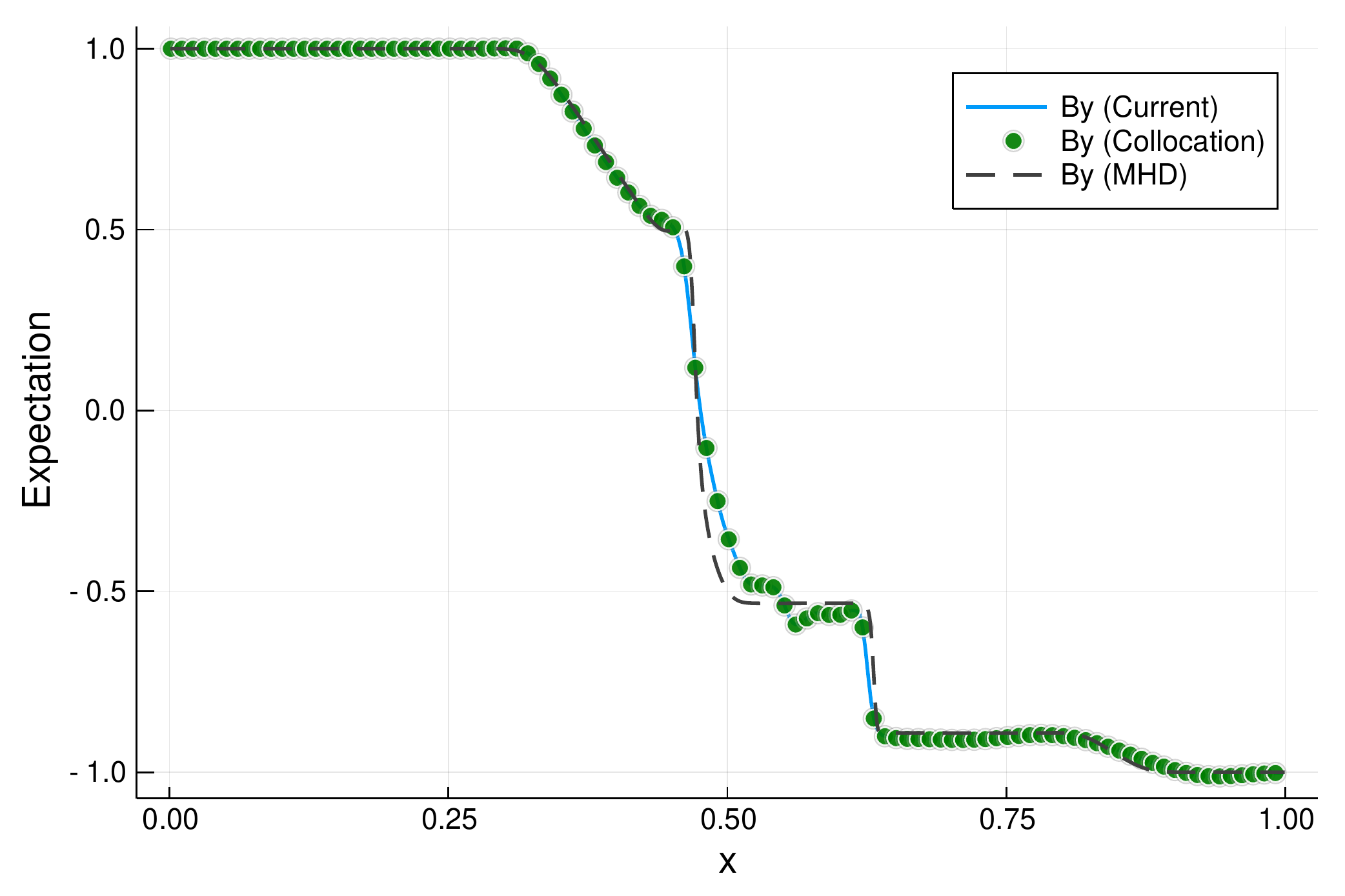}
	}
	\subfigure[$\mathbb E(N)$]{
		\includegraphics[width=0.31\textwidth]{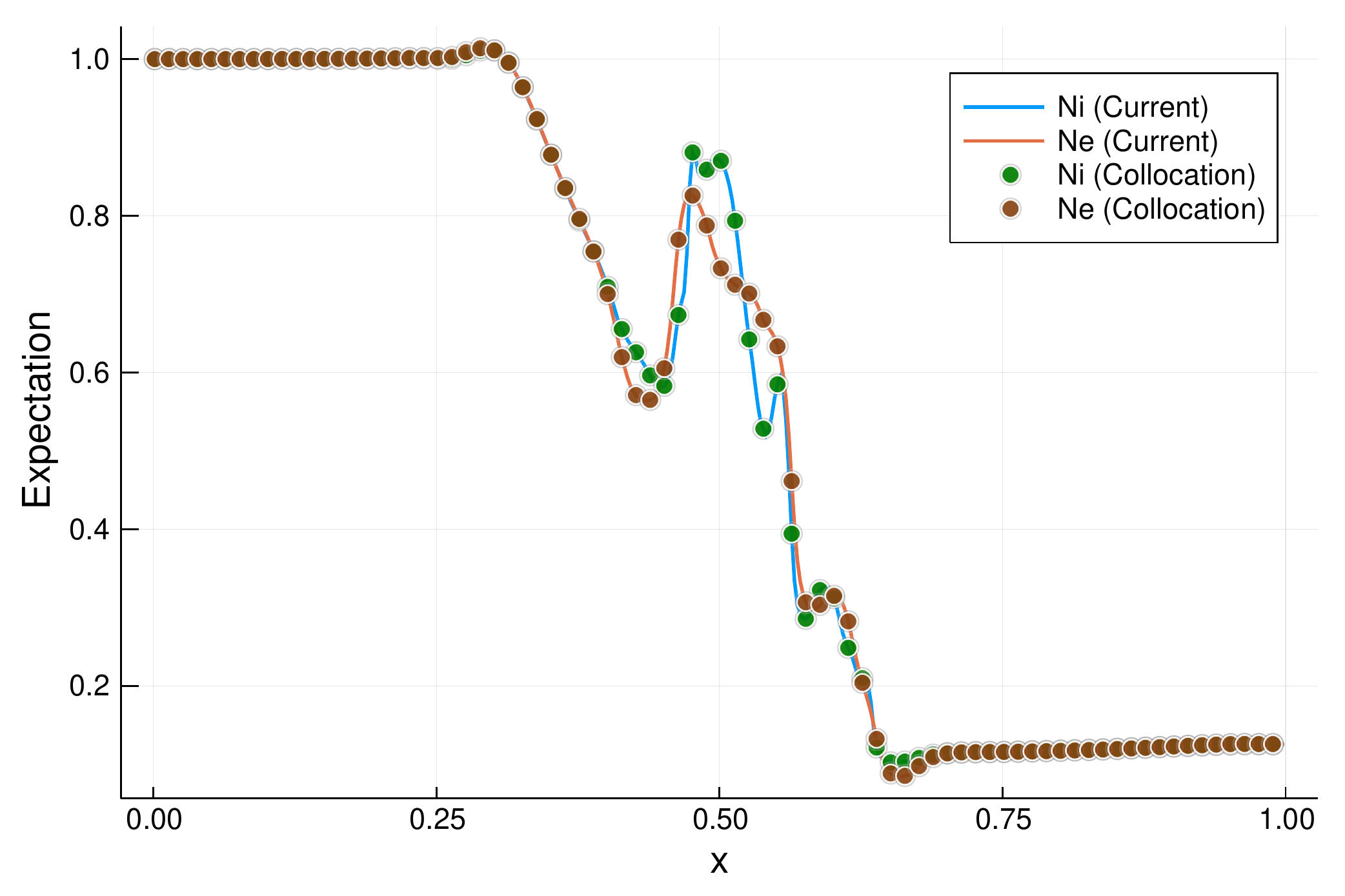}
	}
	\subfigure[$\mathbb E(U)$]{
		\includegraphics[width=0.31\textwidth]{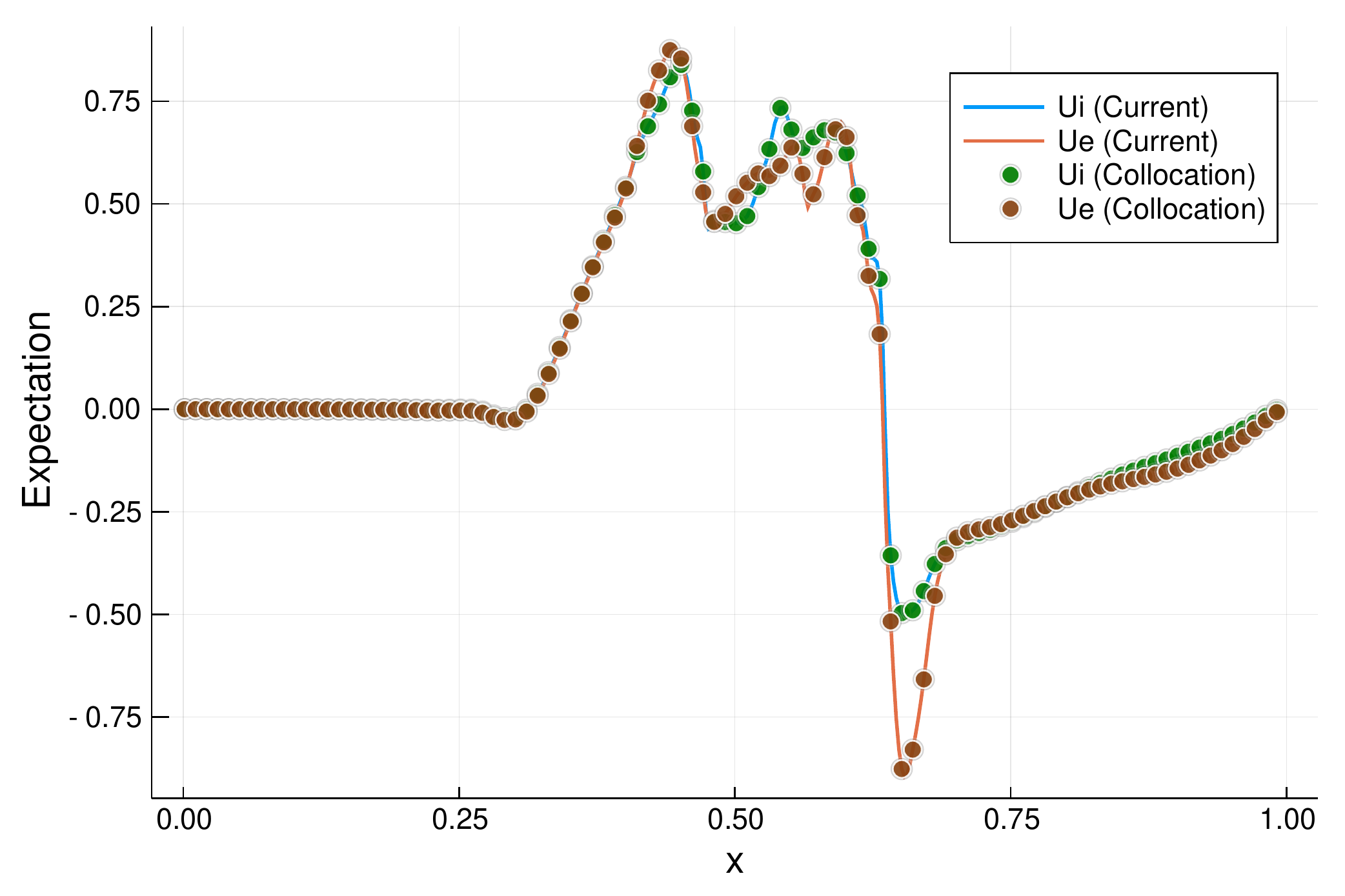}
	}
	\subfigure[$\mathbb E(B_y)$]{
		\includegraphics[width=0.31\textwidth]{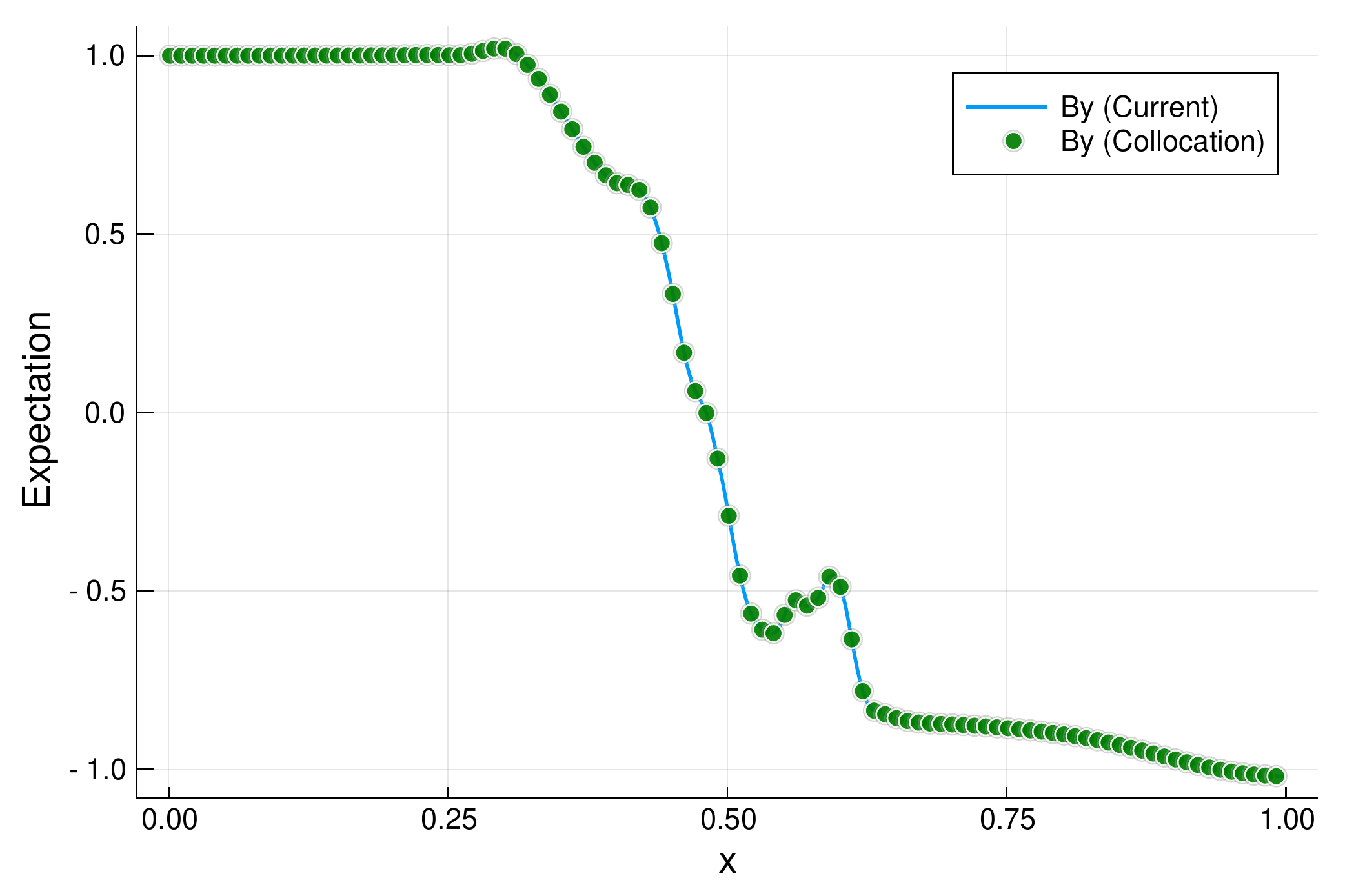}
	}
	\subfigure[$\mathbb E(N)$]{
		\includegraphics[width=0.31\textwidth]{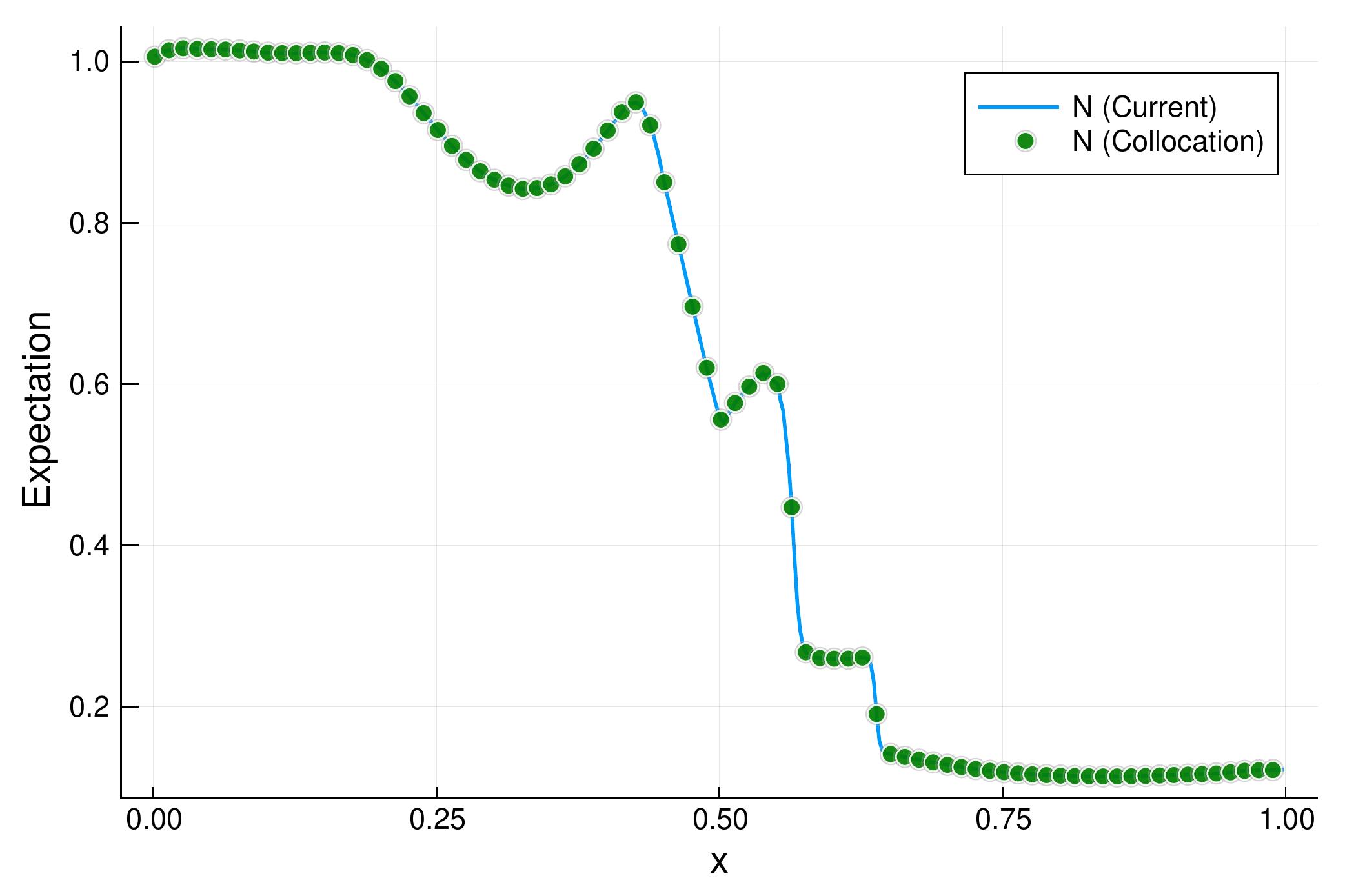}
	}
	\subfigure[$\mathbb E(U)$]{
		\includegraphics[width=0.31\textwidth]{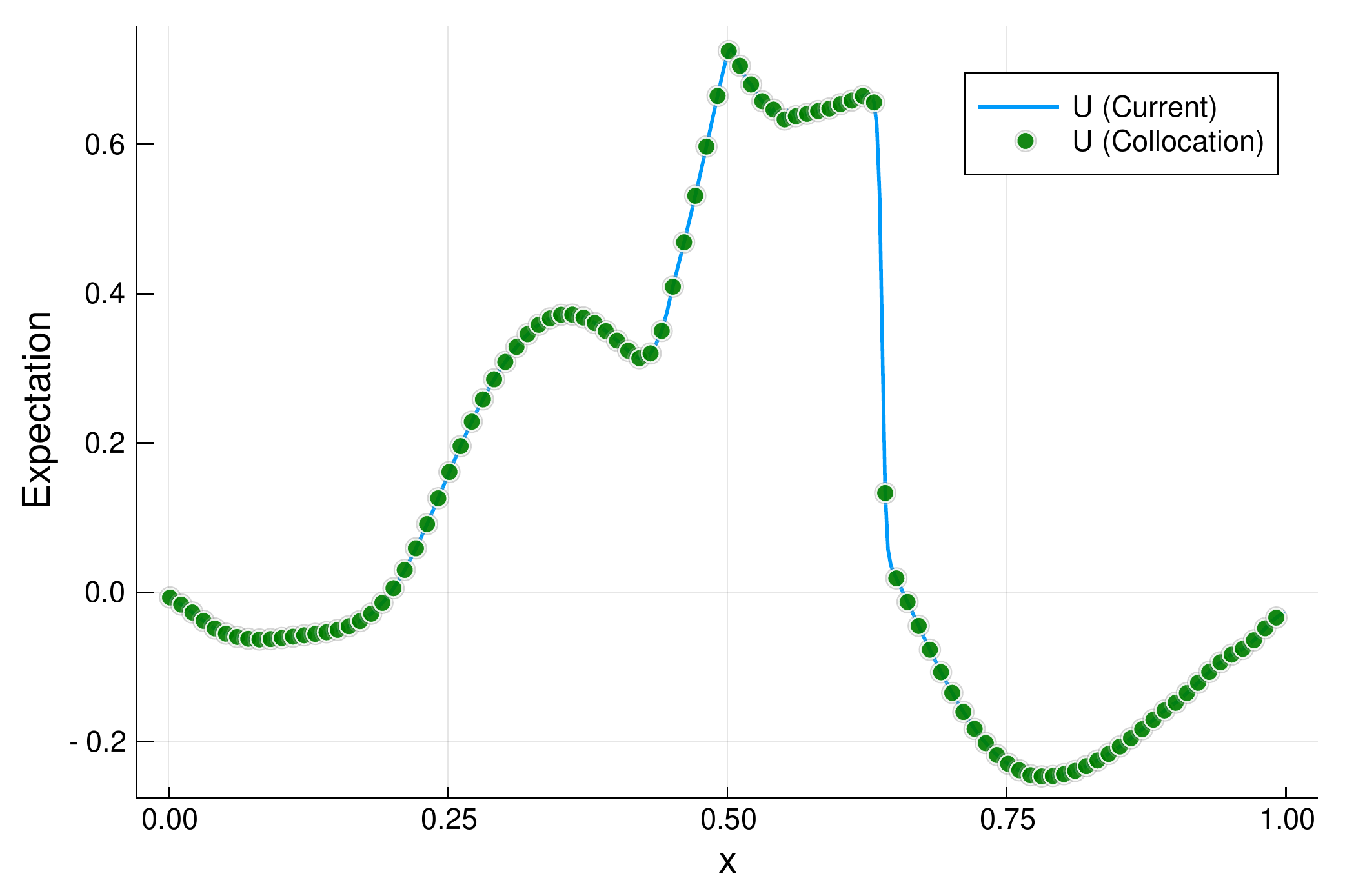}
	}
	\subfigure[$\mathbb E(B_y)$]{
		\includegraphics[width=0.31\textwidth]{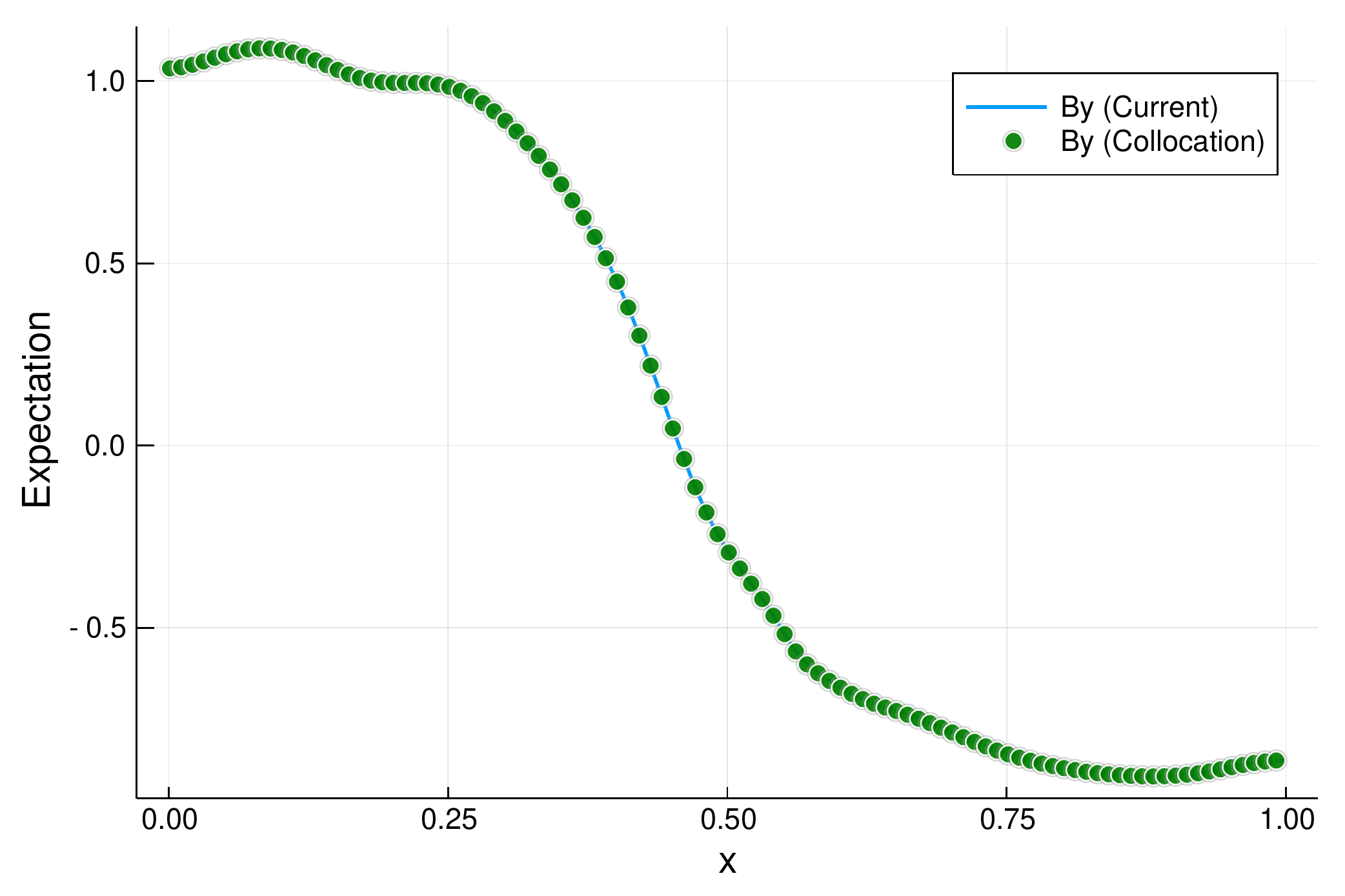}
	}
	\subfigure[$\mathbb E(N)$]{
		\includegraphics[width=0.31\textwidth]{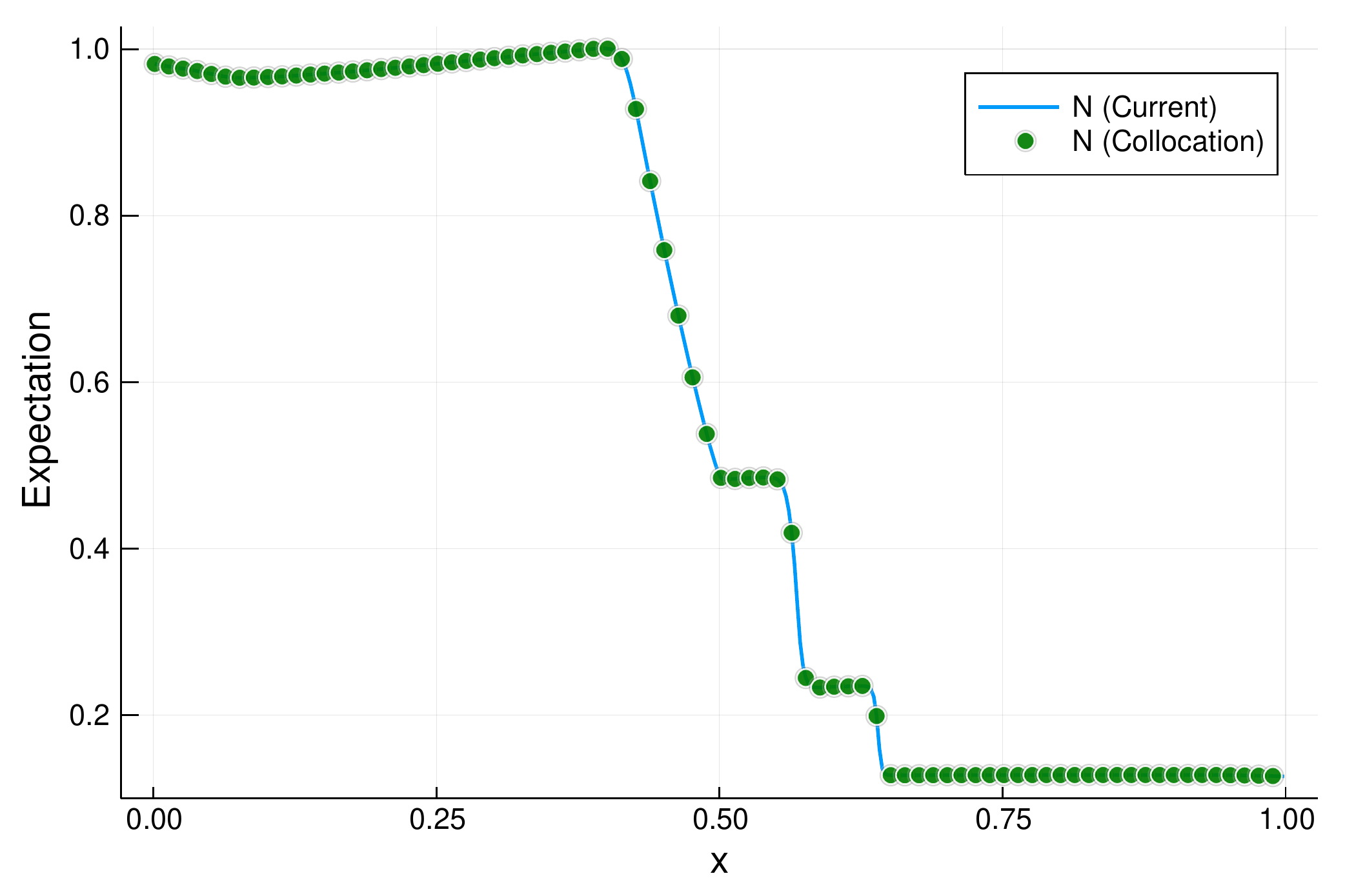}
	}
	\subfigure[$\mathbb E(U)$]{
		\includegraphics[width=0.31\textwidth]{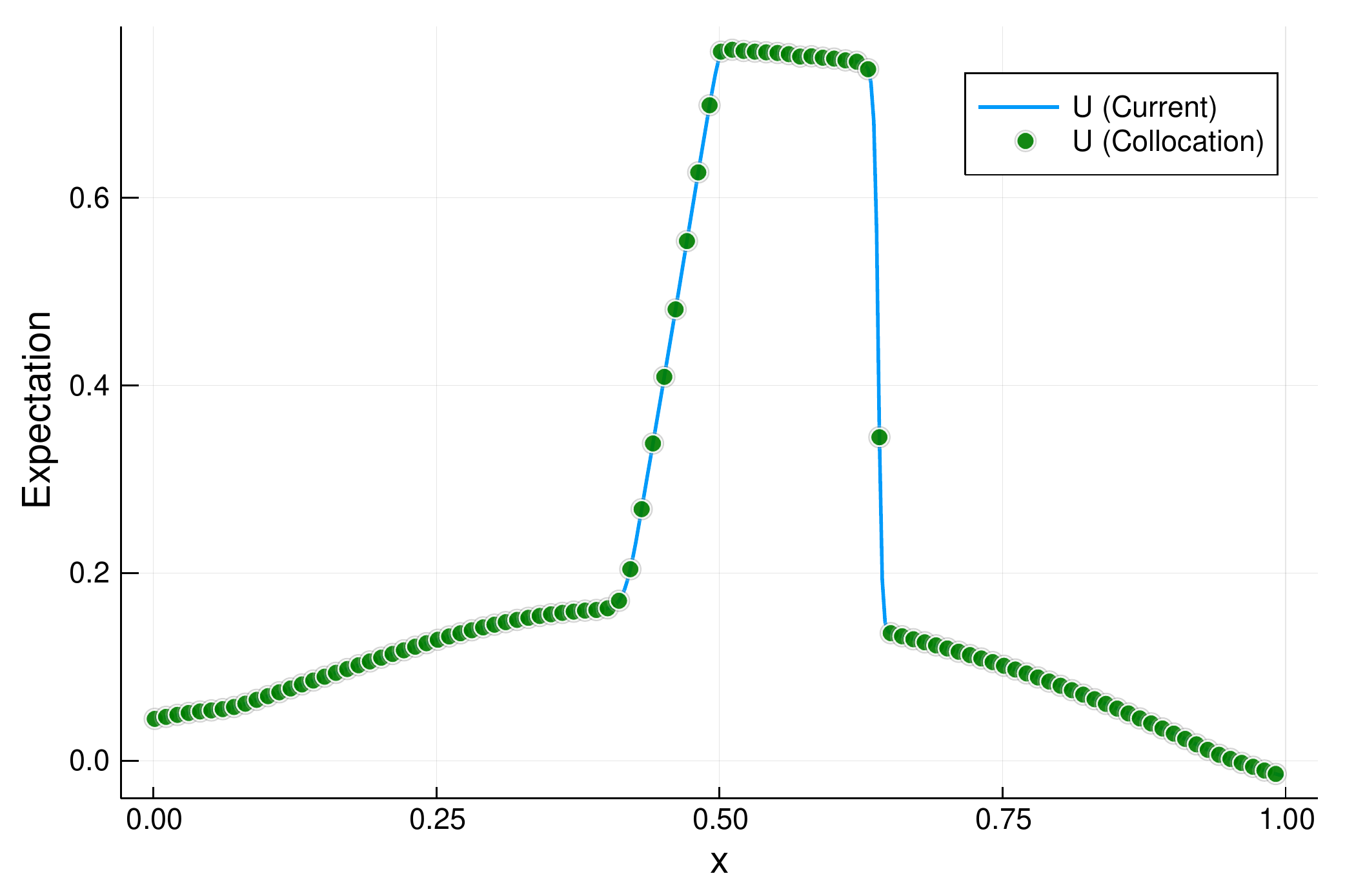}
	}
	\subfigure[$\mathbb E(B_y)$]{
		\includegraphics[width=0.31\textwidth]{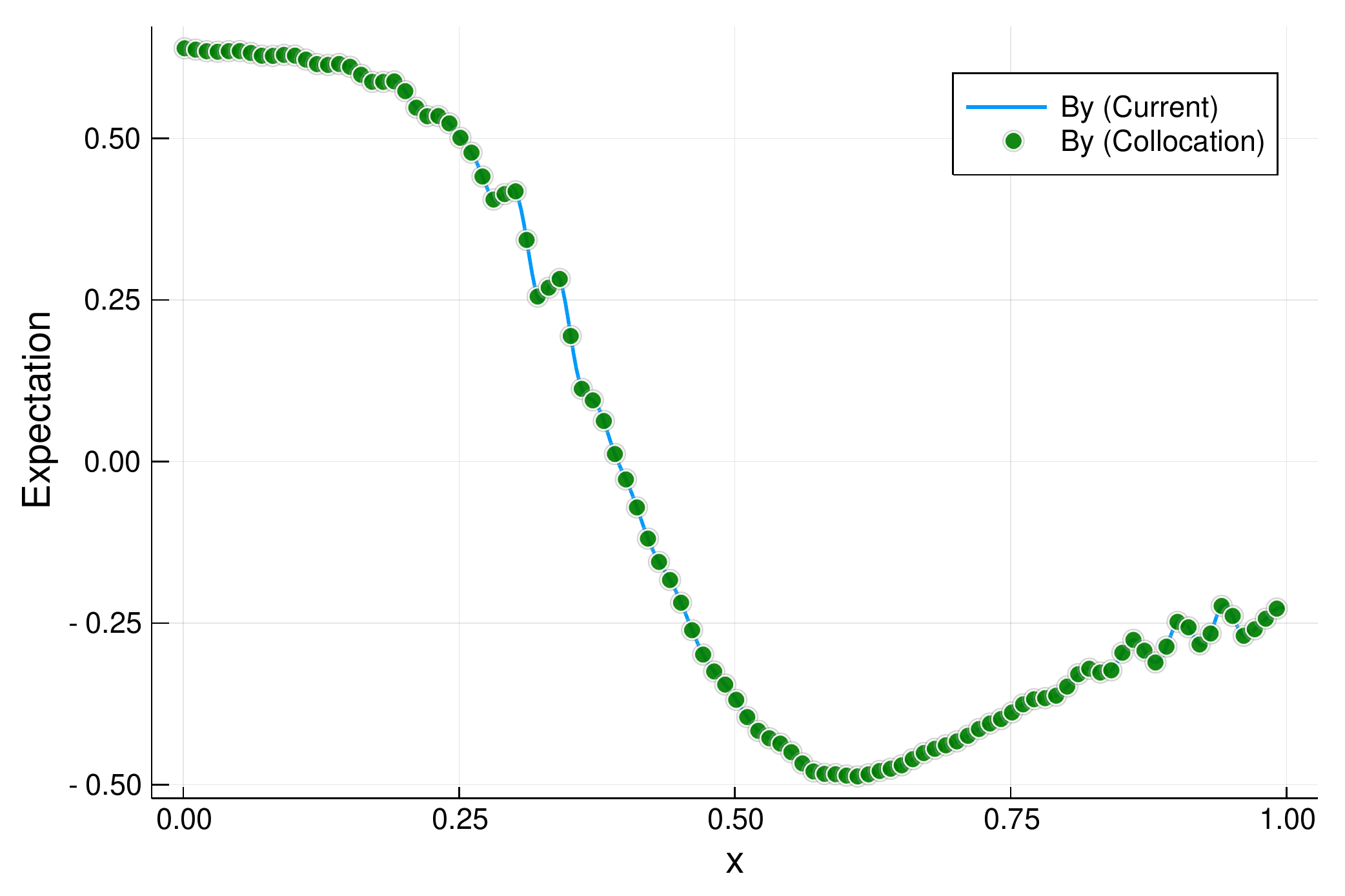}
	}
	\subfigure[$\mathbb E(N)$]{
		\includegraphics[width=0.31\textwidth]{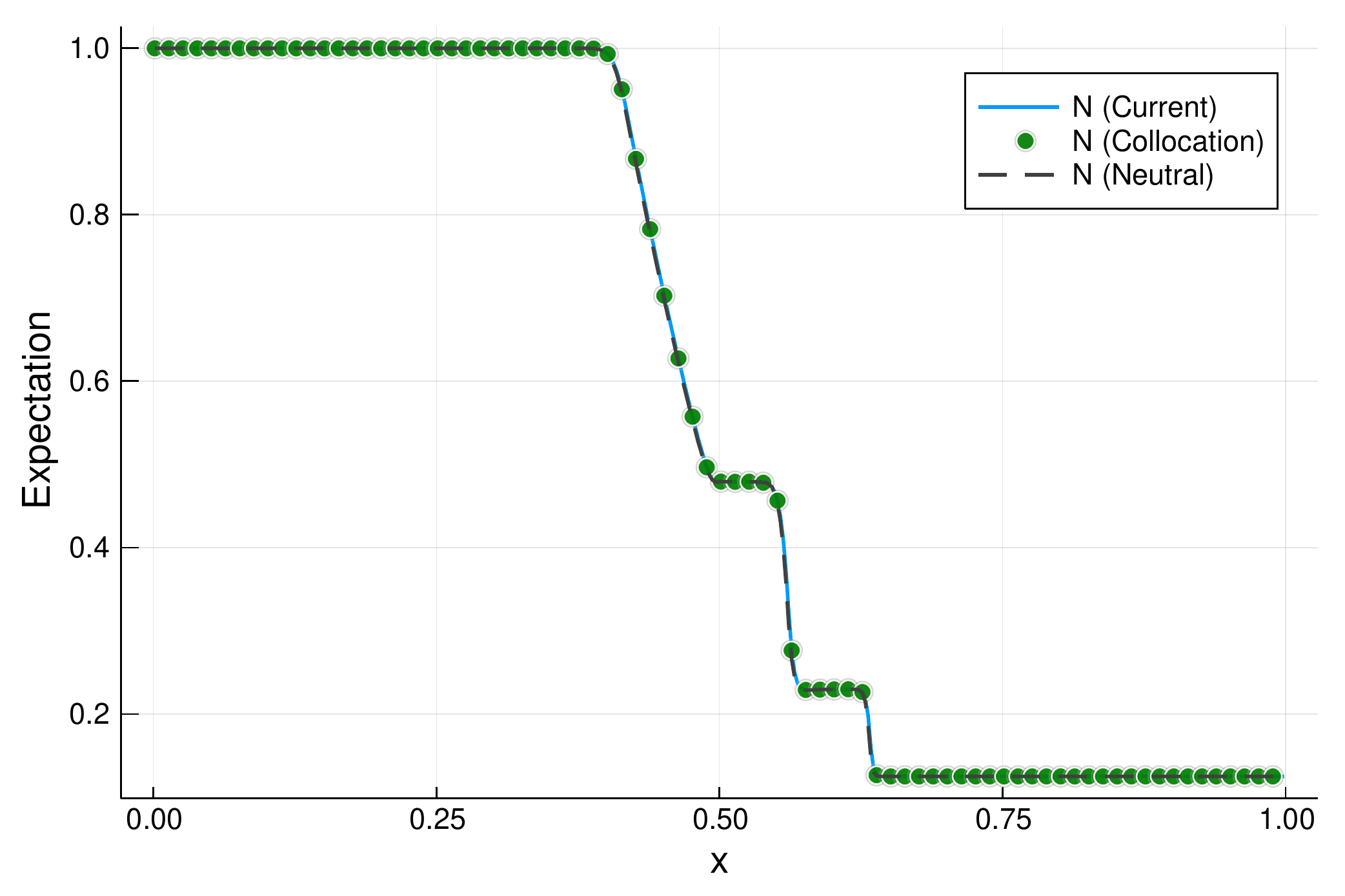}
	}
	\subfigure[$\mathbb E(U)$]{
		\includegraphics[width=0.31\textwidth]{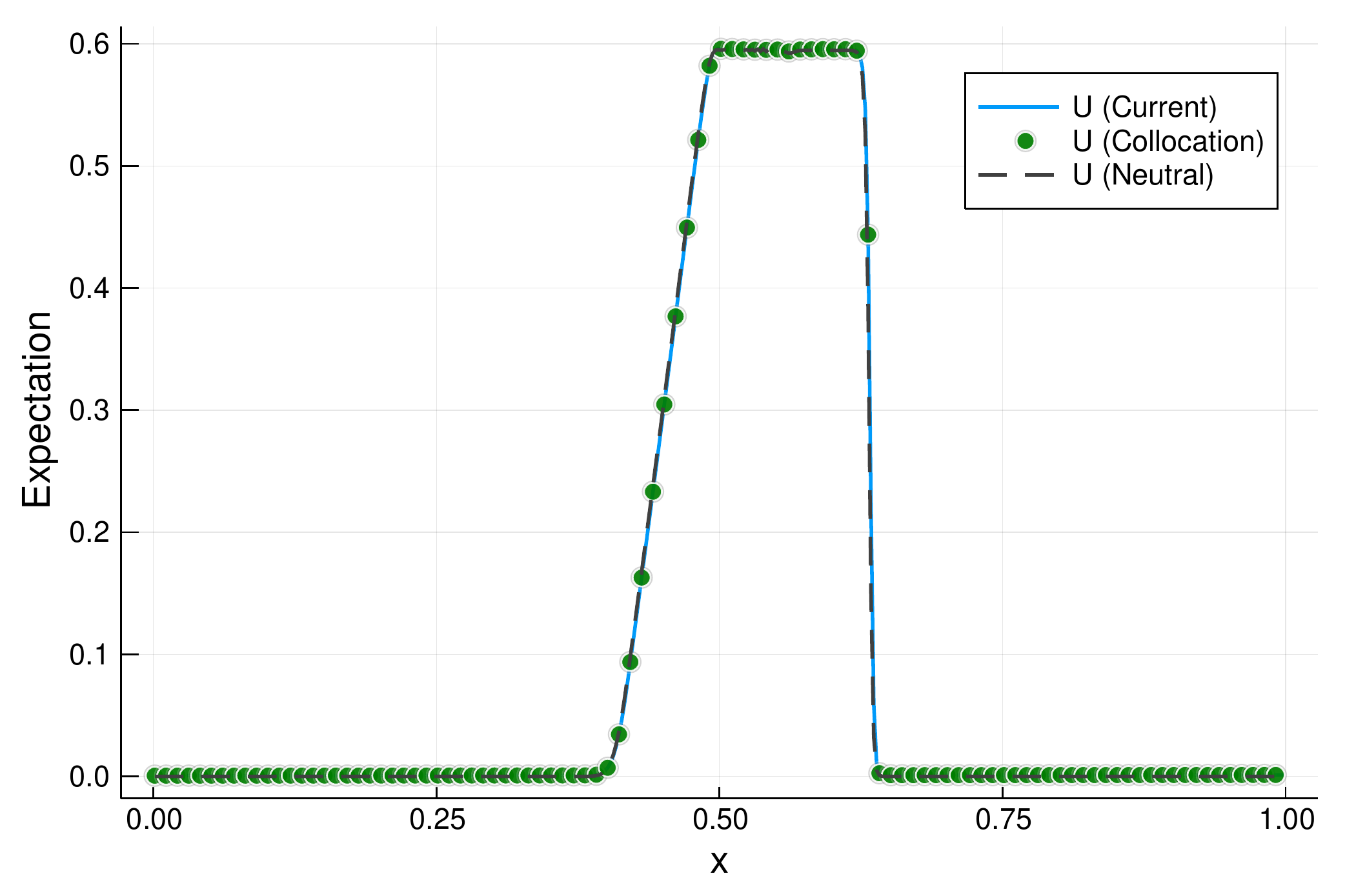}
	}
	\subfigure[$\mathbb E(B_y)$]{
		\includegraphics[width=0.31\textwidth]{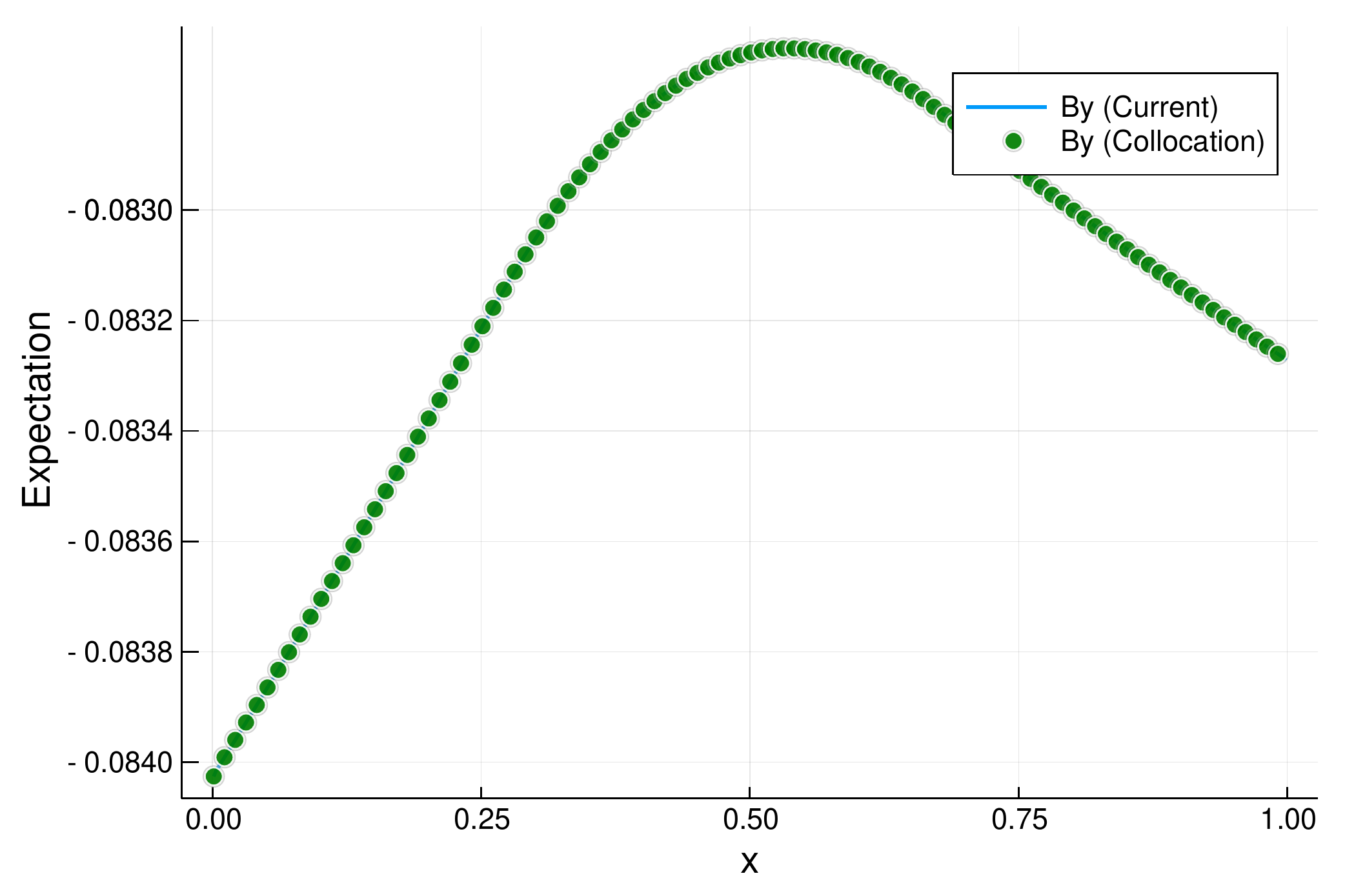}
	}
	\caption{Expectation values of $N$, $U$ and $B_y$ in Brio-Wu shock tube with density uncertainty at $t=0.1$ (row 1: $r_g=0.003$, row 2: $r_g=0.01$, row 3: $r_g=0.1$, row 4: $r_g=1$, row 5: $r_g=100$).}
	\label{pic:briowu case1 mean}
\end{figure}

First, we consider the uncertainties of plasma density in the left half tube, with $n_{ion,L}=n_{ele,L}=\xi$.
The numerical expected solutions of macroscopic variables under different reference gyroradius at $t = 0.1$ are shown in Fig. \ref{pic:briowu case1 mean}.
It is known that the evolution of flow and electromagnetic fields can be well described by MHD equations when $r_g$ is small.
At $r_g = 0.003$, besides the typical wave structures in the classic Sod problem, a compound wave emerges around the tube center. 
As illustrated by Brio and Wu \cite{brio1988upwind}, with the nonconvexity of ideal MHD equation, this wave structure is induced from initial magnetic discontinuity.
From Fig. \ref{pic:briowu case1 mean}(a)-(c), we see the current scheme from the BGK-Maxwell system provides equivalent numerical solutions as ideal MHD equations,
which identifies the asymptotic preserving (AP) property of the current scheme in the MHD limit.

As the gyroradius $r_g$ increases, the electric and Lorentz force decreases accordingly.
The effects of charge separation are enhanced, and the omitted terms, e.g. the Hall current term, become important.
The plasma gradually manifests itself as dielectric material.
As a result, the magnetic diffusion is enhanced, and the wave structures from discontinuous magnetic field $B_y$ are flattened.
When $r_g = 100$, at this point the gyroradius is much larger than the characteristic length of the shock tube, and thus plasma behaves as neutral gas, and the Brio-Wu shock tube degenerates into a standard Sod case. 
In Fig. \ref{pic:briowu case1 mean}(m) and (n), we see the current scheme provides equivalent solutions as Euler equations, indicating the AP property of the current scheme in the Euler limit.

\begin{figure}[htb!]
	\centering
	\subfigure[$\mathbb S(N)$]{
		\includegraphics[width=0.31\textwidth]{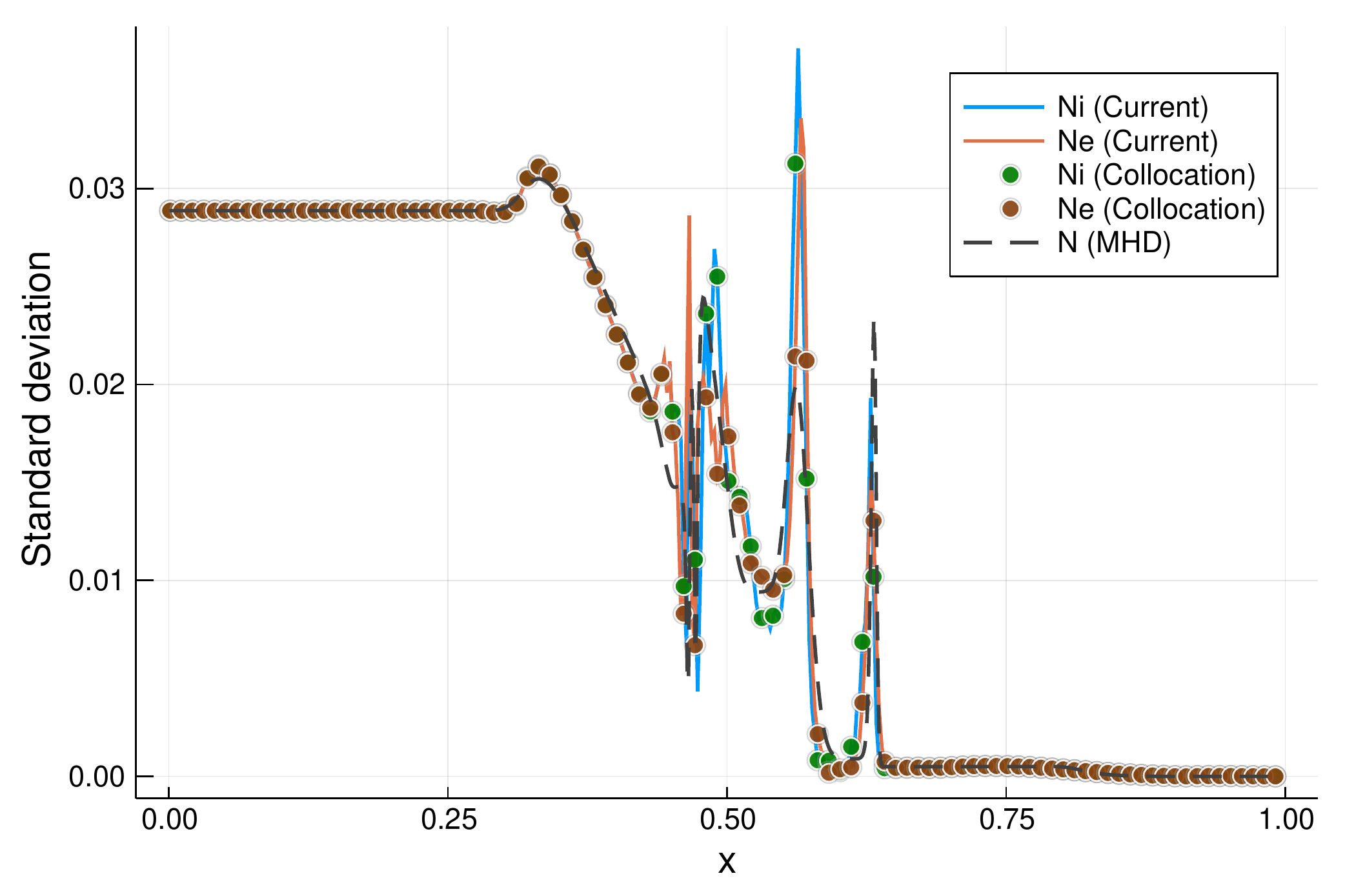}
	}
	\subfigure[$\mathbb S(U)$]{
		\includegraphics[width=0.31\textwidth]{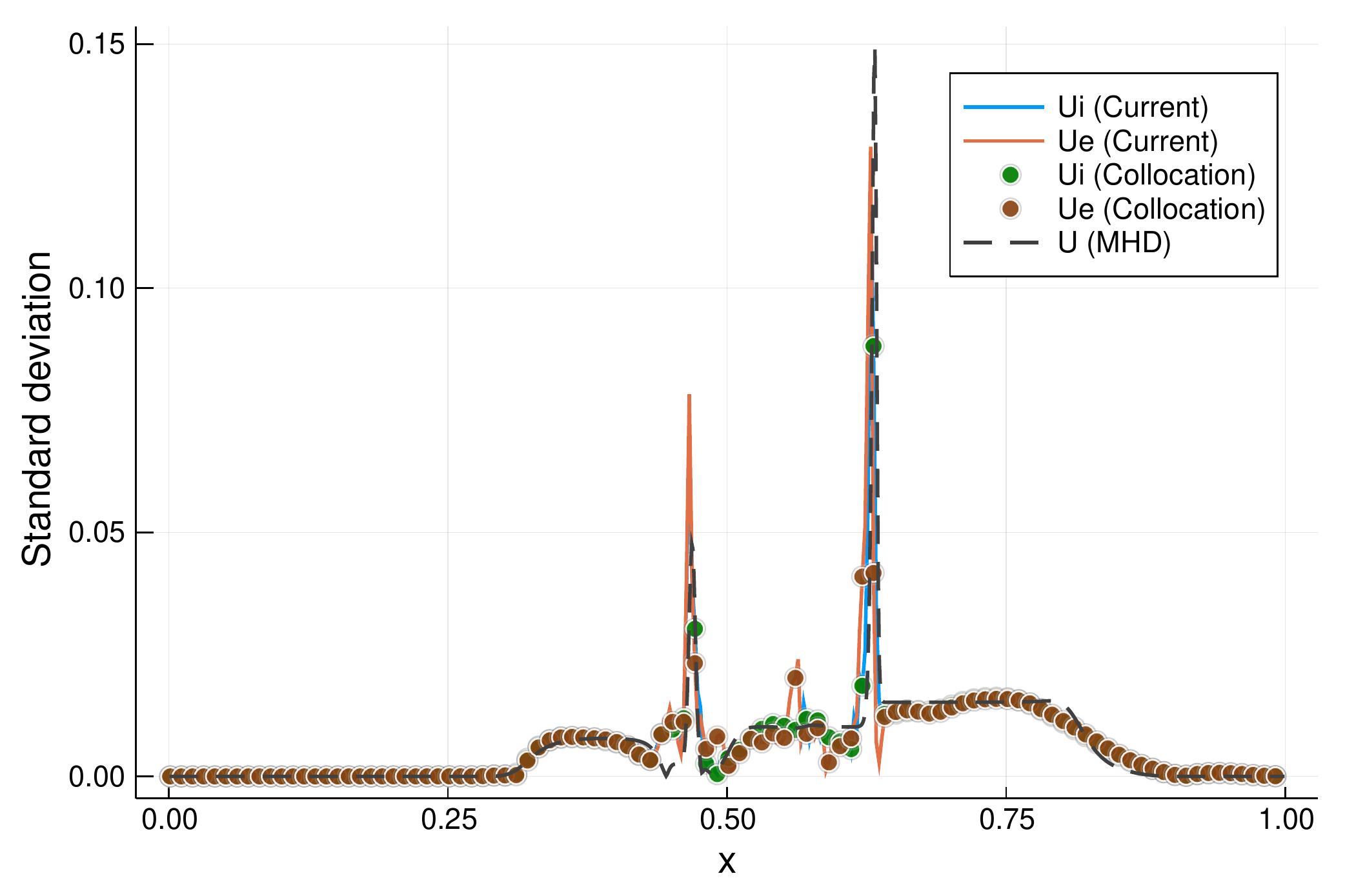}
	}
	\subfigure[$\mathbb S(B_y)$]{
		\includegraphics[width=0.31\textwidth]{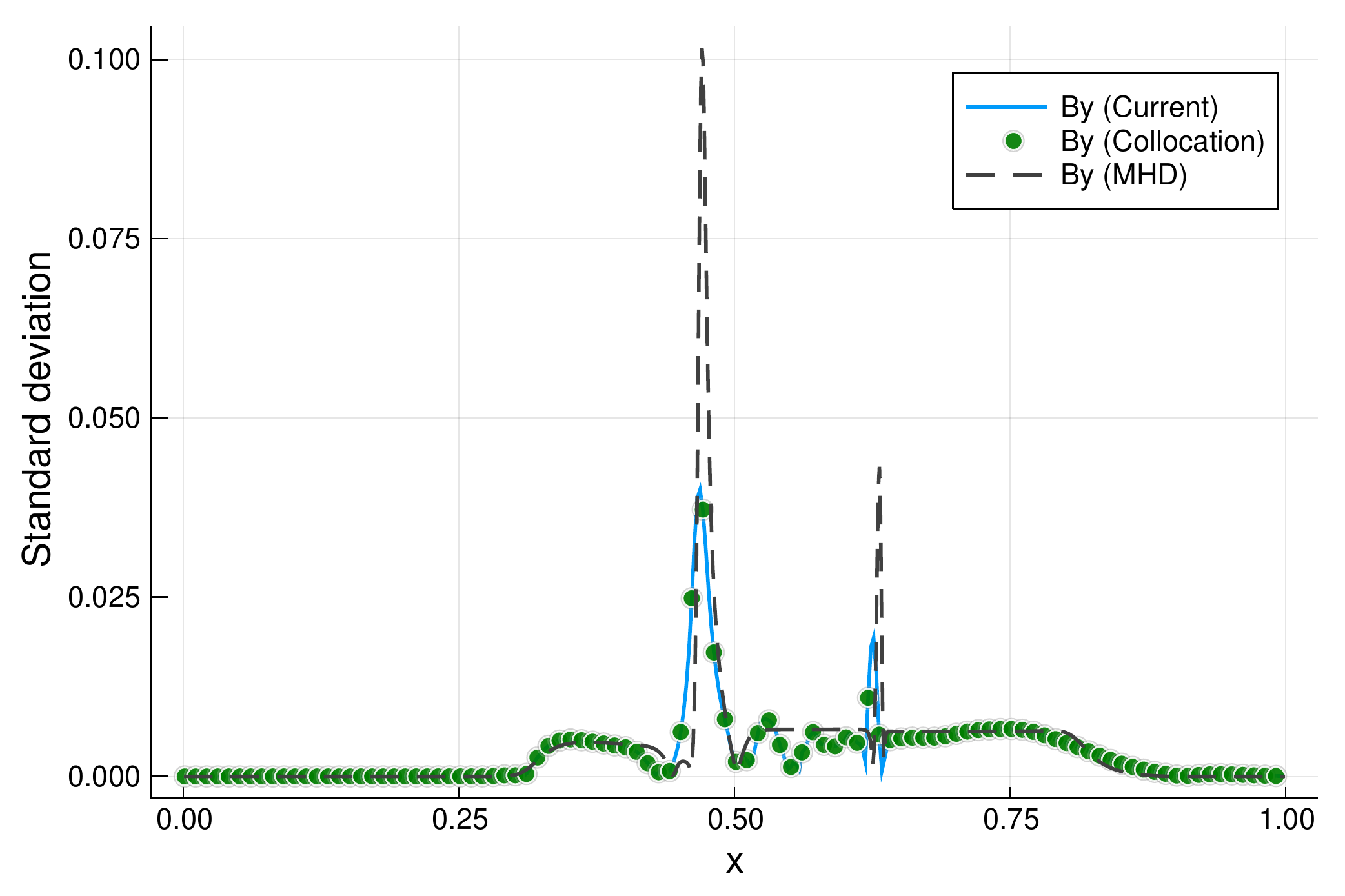}
	}
	\subfigure[$\mathbb S(N)$]{
		\includegraphics[width=0.31\textwidth]{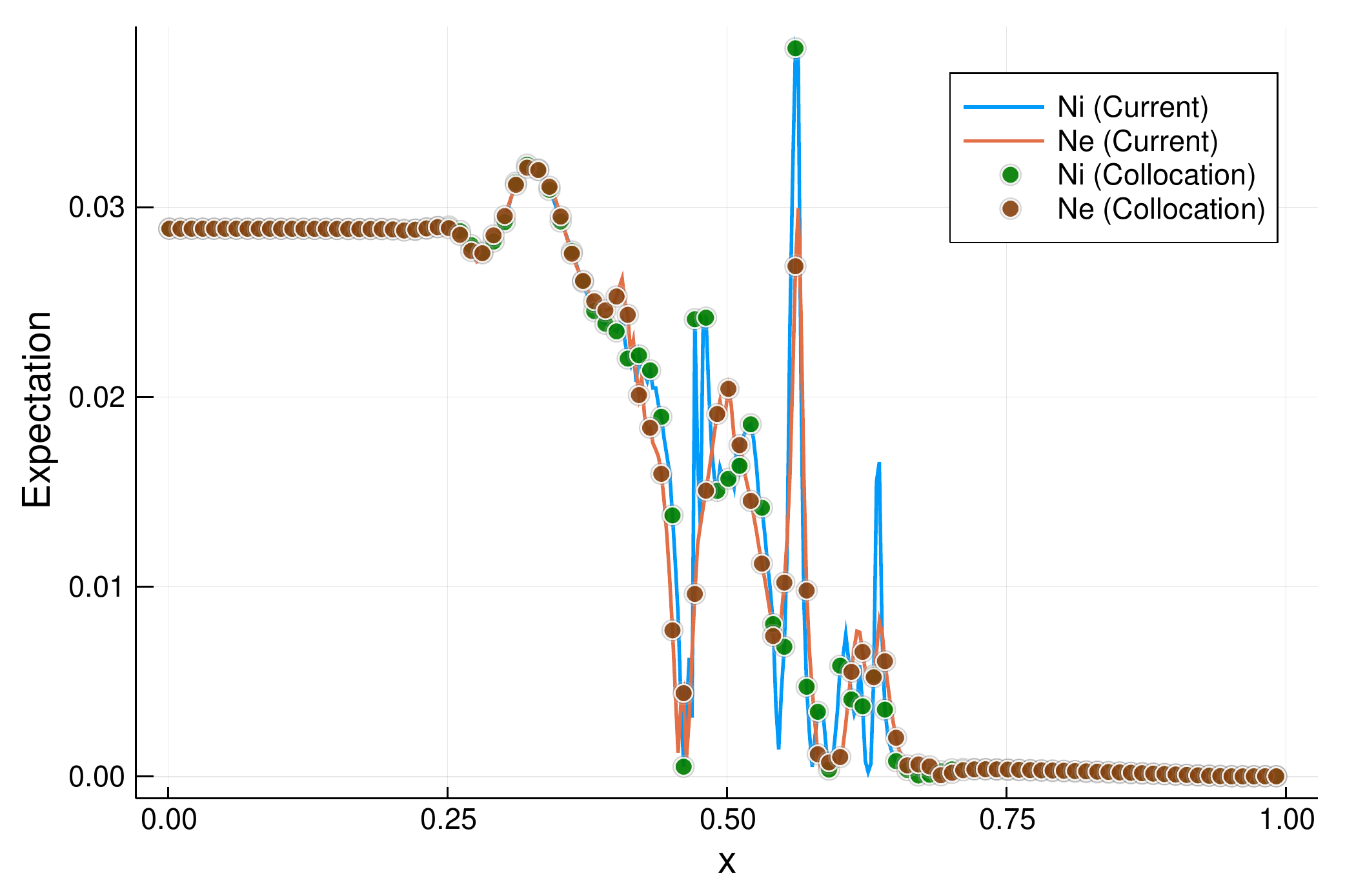}
	}
	\subfigure[$\mathbb S(U)$]{
		\includegraphics[width=0.31\textwidth]{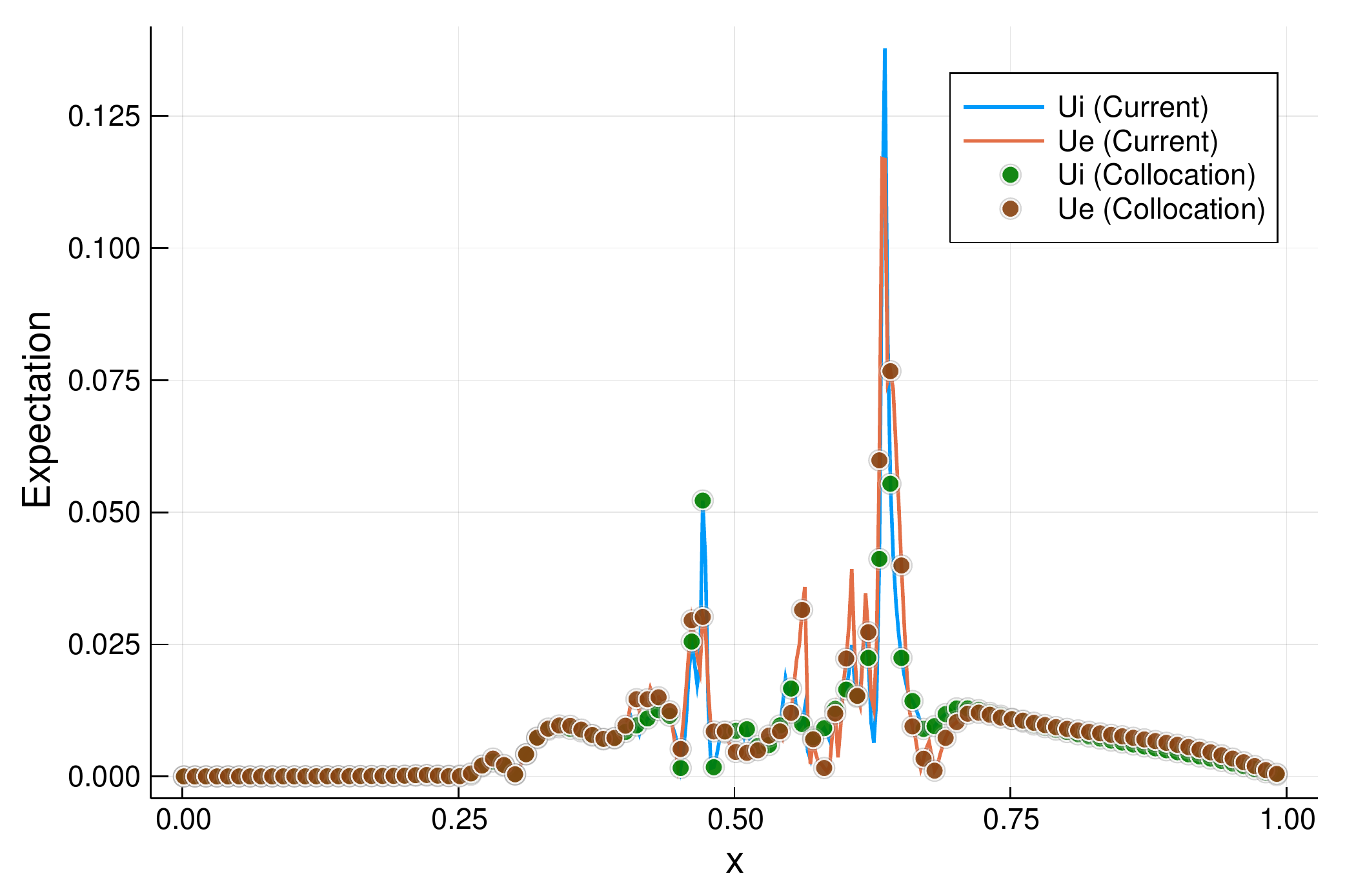}
	}
	\subfigure[$\mathbb S(B_y)$]{
		\includegraphics[width=0.31\textwidth]{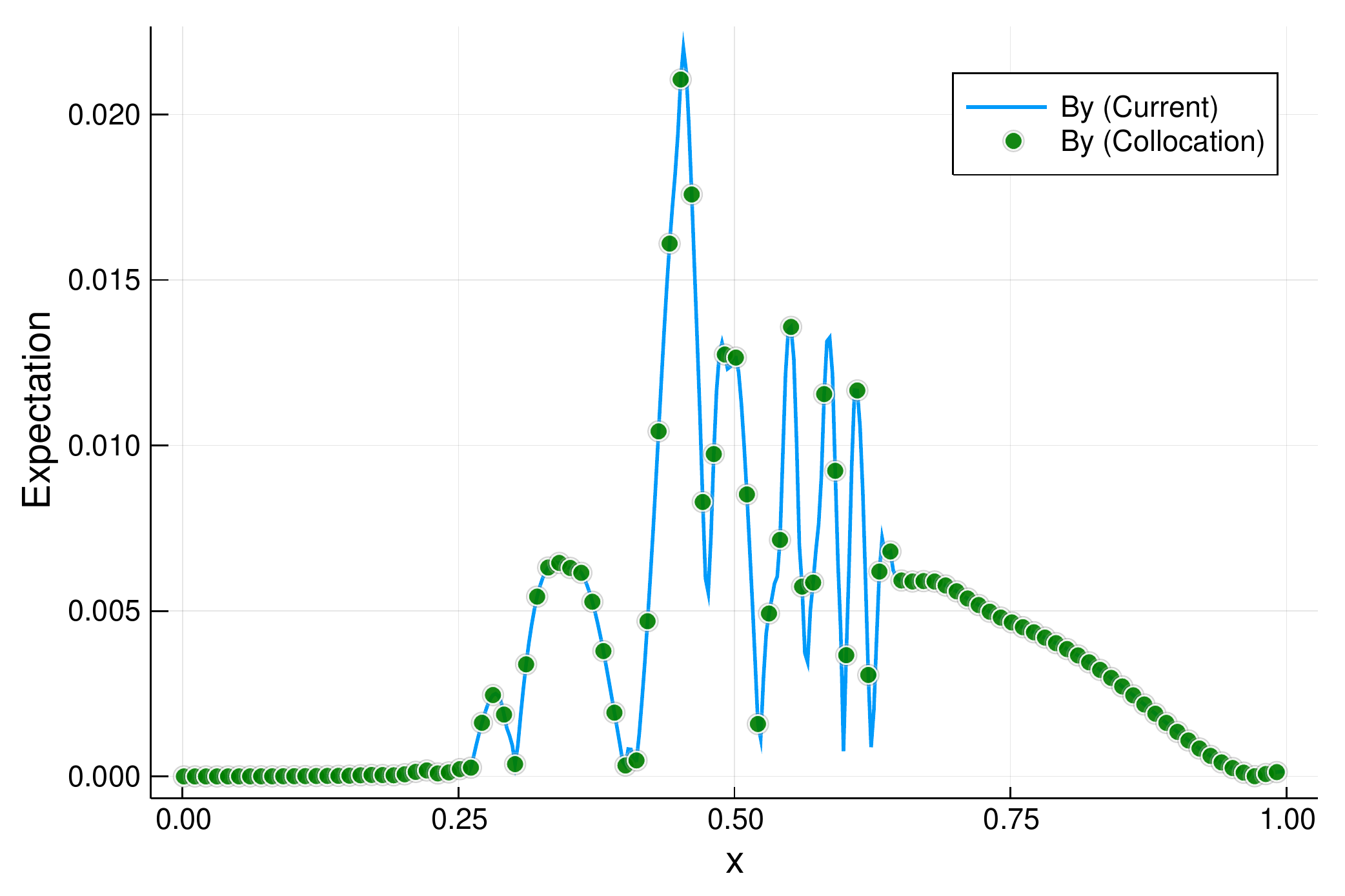}
	}
	\subfigure[$\mathbb S(N)$]{
		\includegraphics[width=0.31\textwidth]{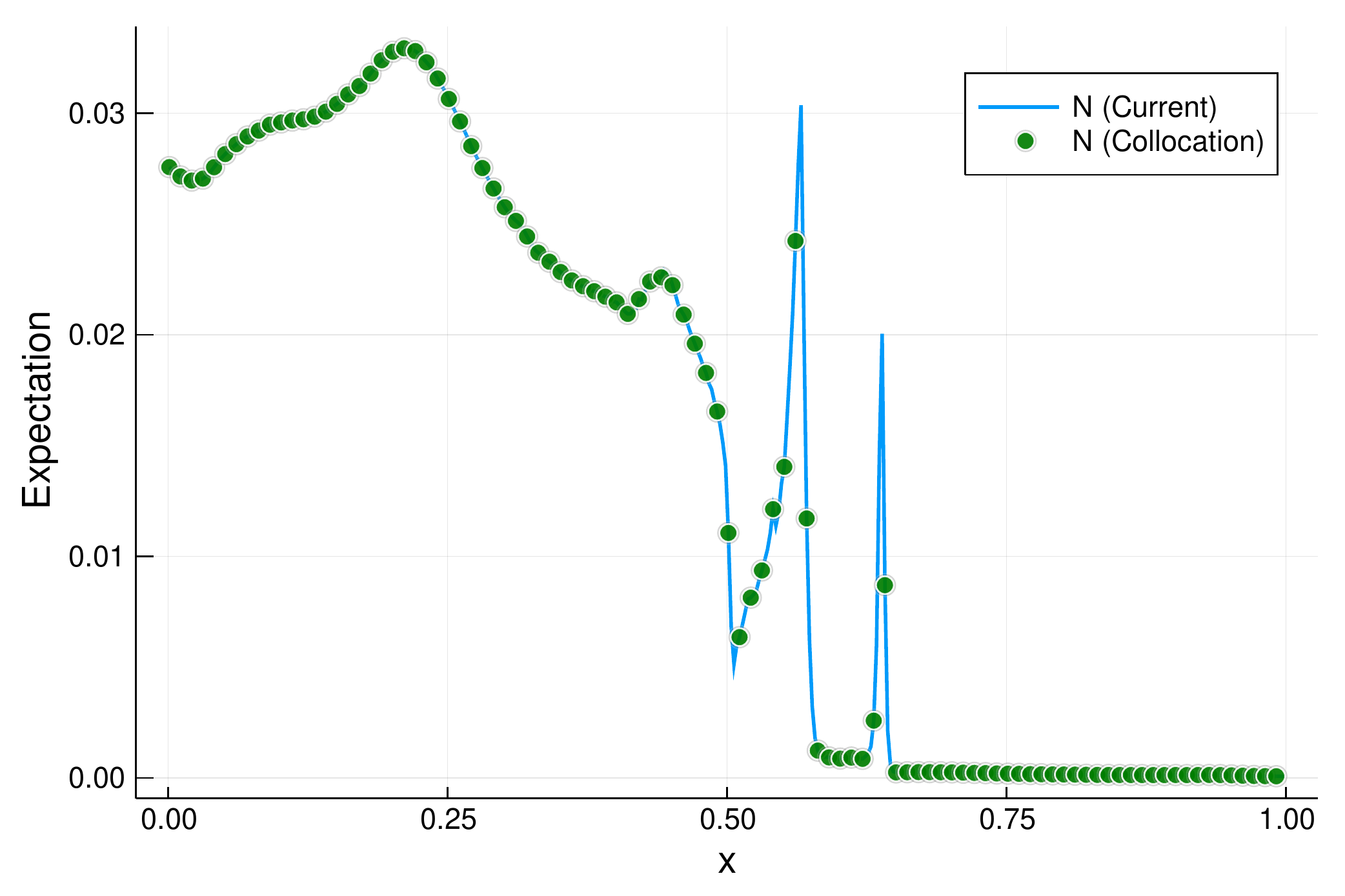}
	}
	\subfigure[$\mathbb S(U)$]{
		\includegraphics[width=0.31\textwidth]{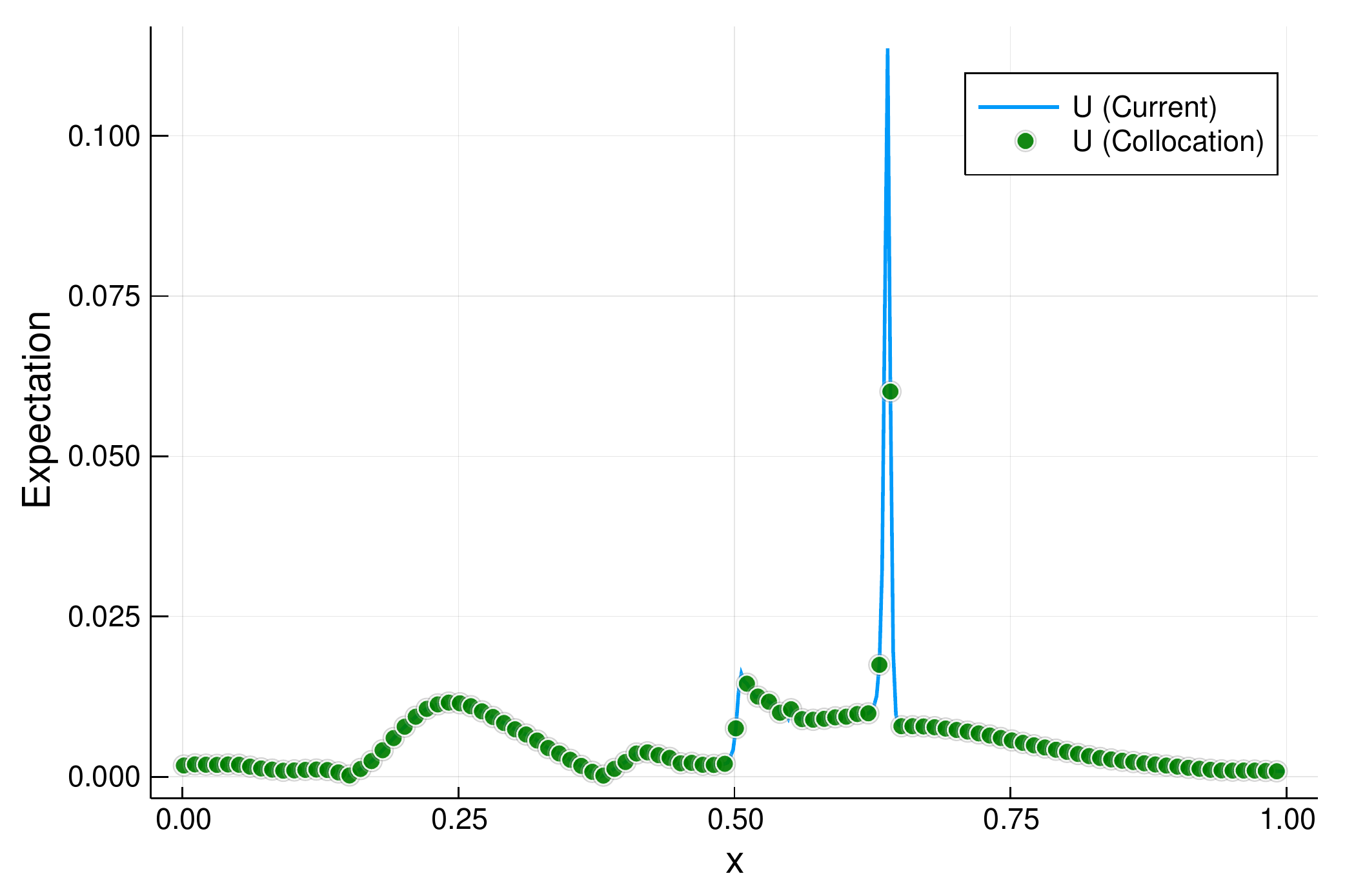}
	}
	\subfigure[$\mathbb S(B_y)$]{
		\includegraphics[width=0.31\textwidth]{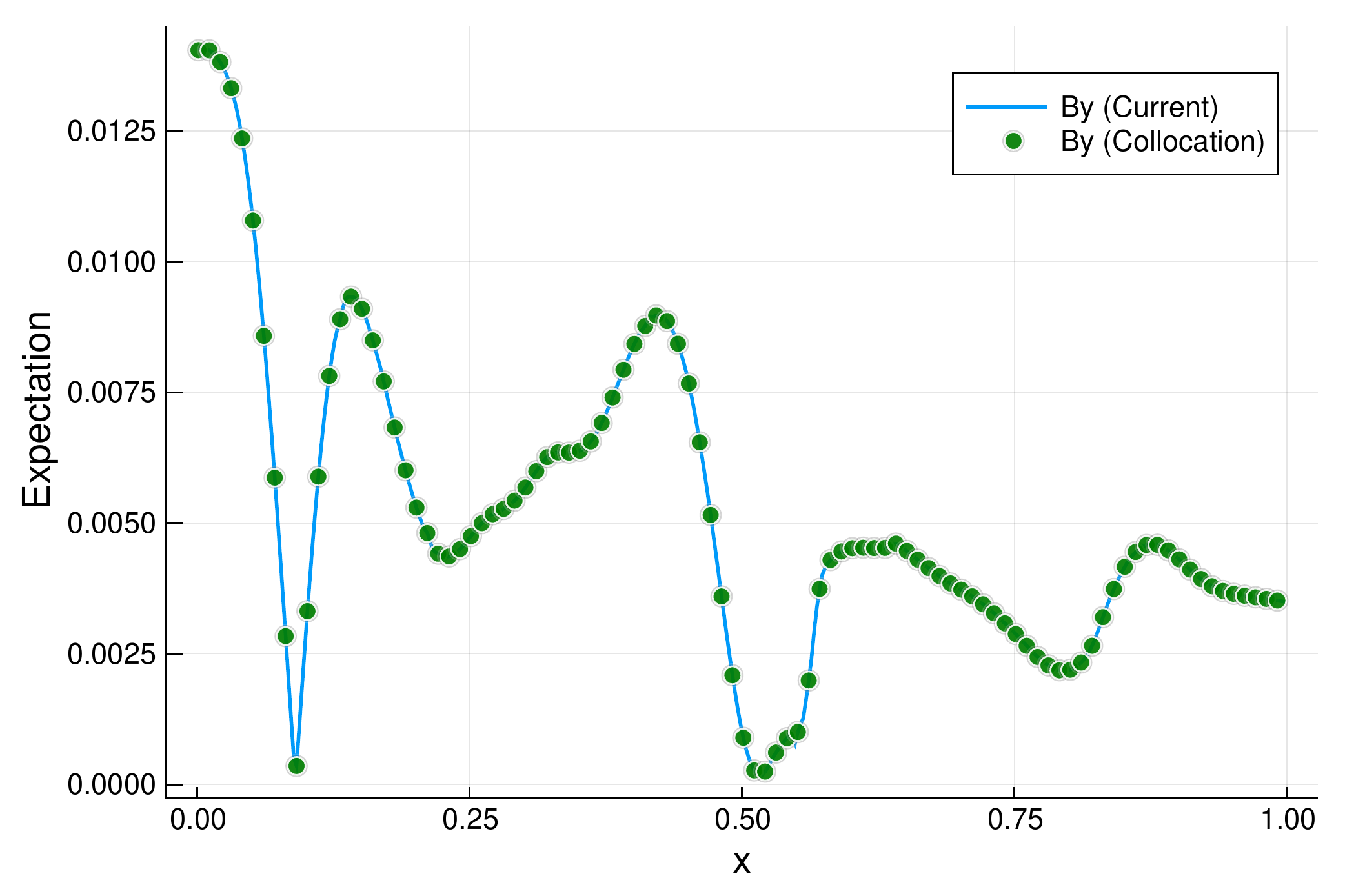}
	}
	\subfigure[$\mathbb S(N)$]{
		\includegraphics[width=0.31\textwidth]{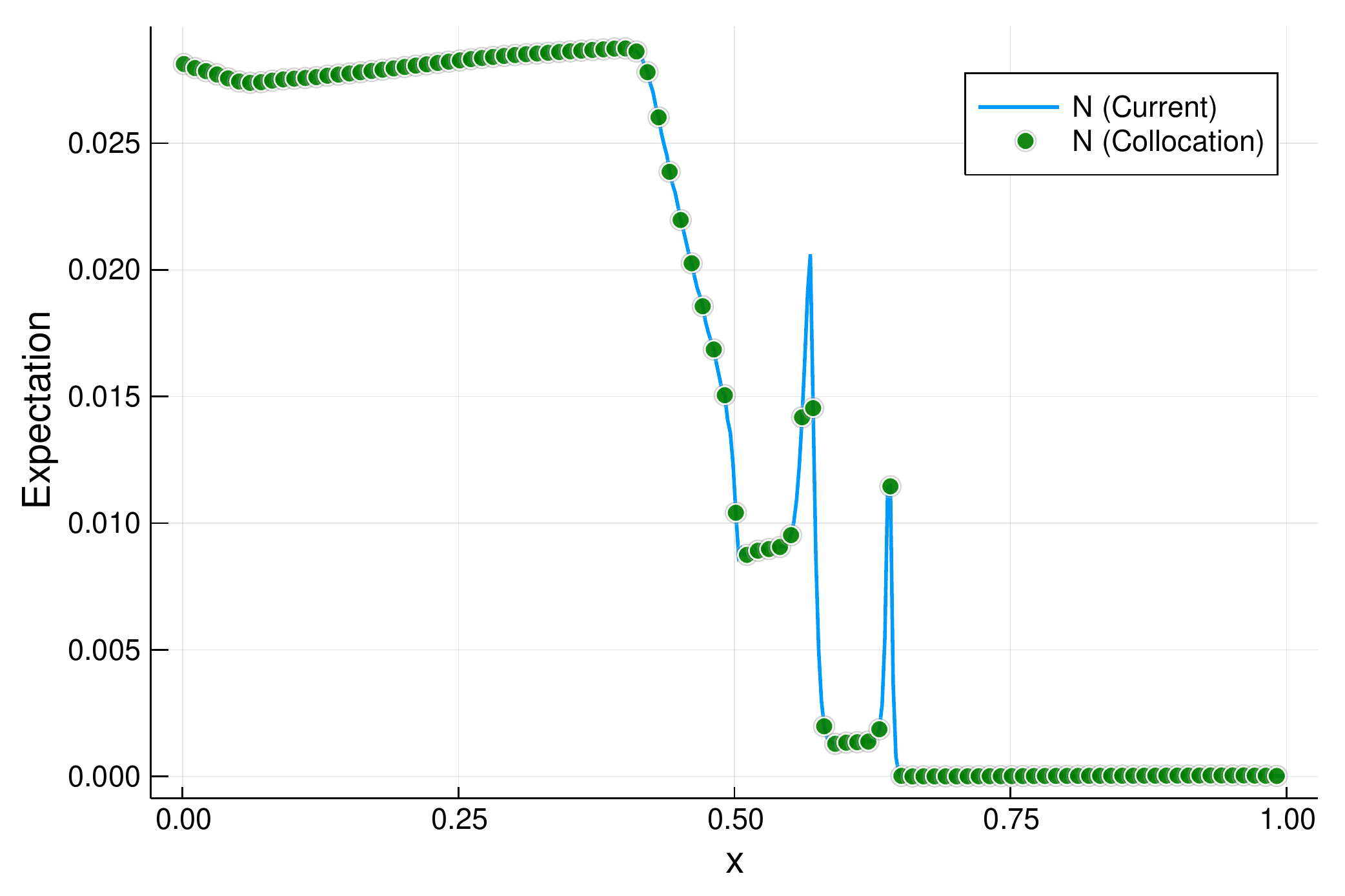}
	}
	\subfigure[$\mathbb S(U)$]{
		\includegraphics[width=0.31\textwidth]{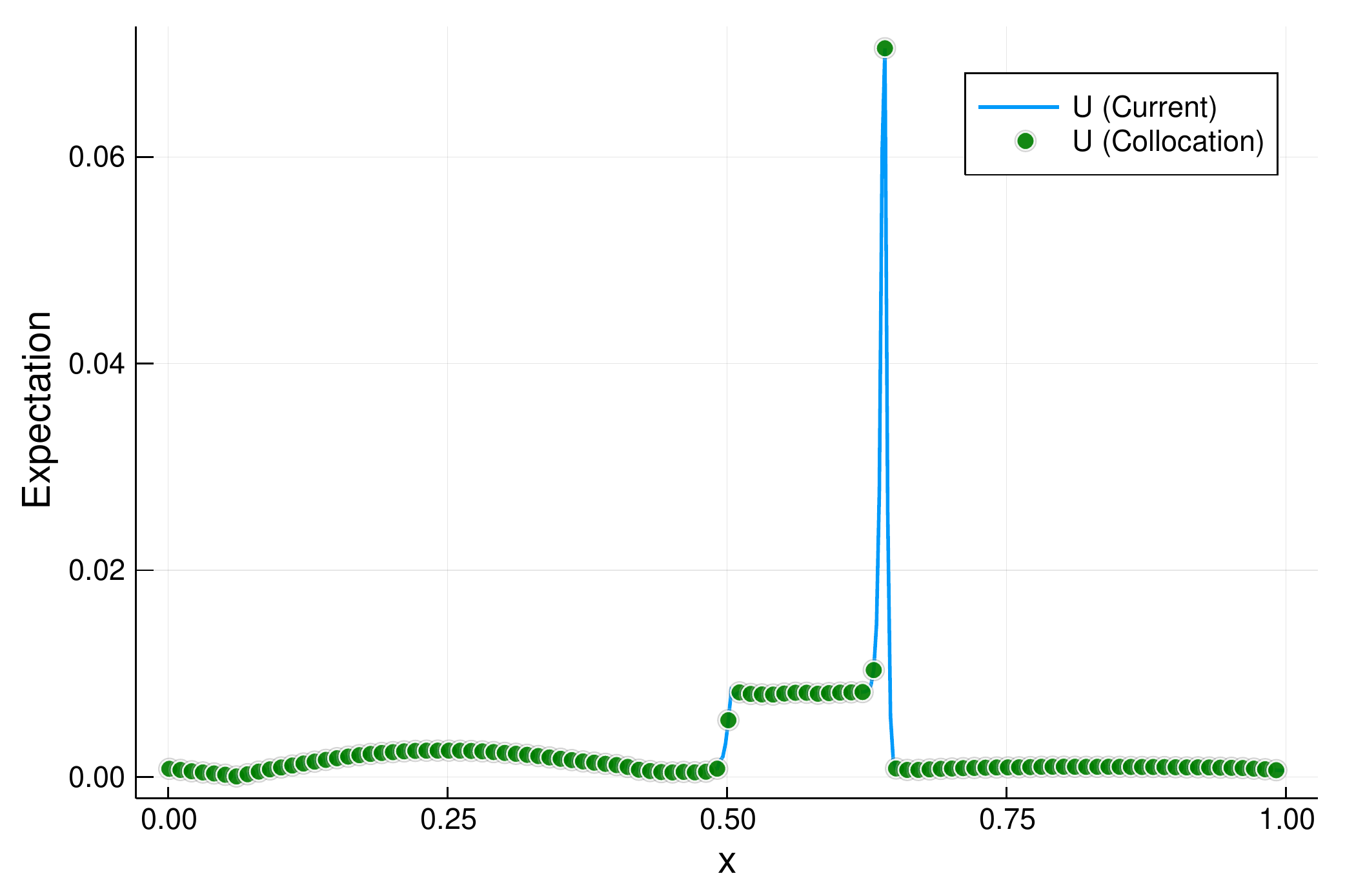}
	}
	\subfigure[$\mathbb S(B_y)$]{
		\includegraphics[width=0.31\textwidth]{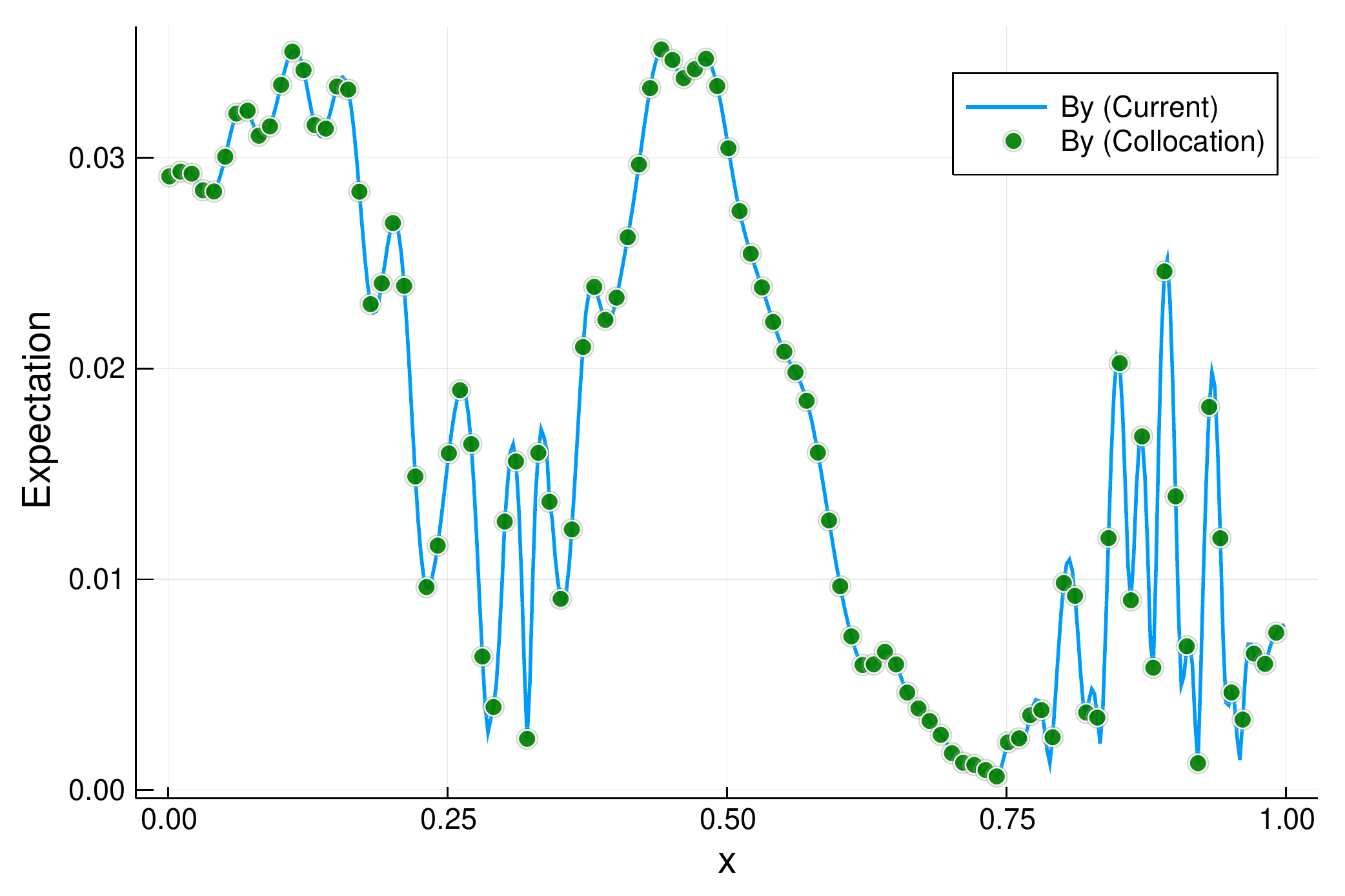}
	}
	\subfigure[$\mathbb S(N)$]{
		\includegraphics[width=0.31\textwidth]{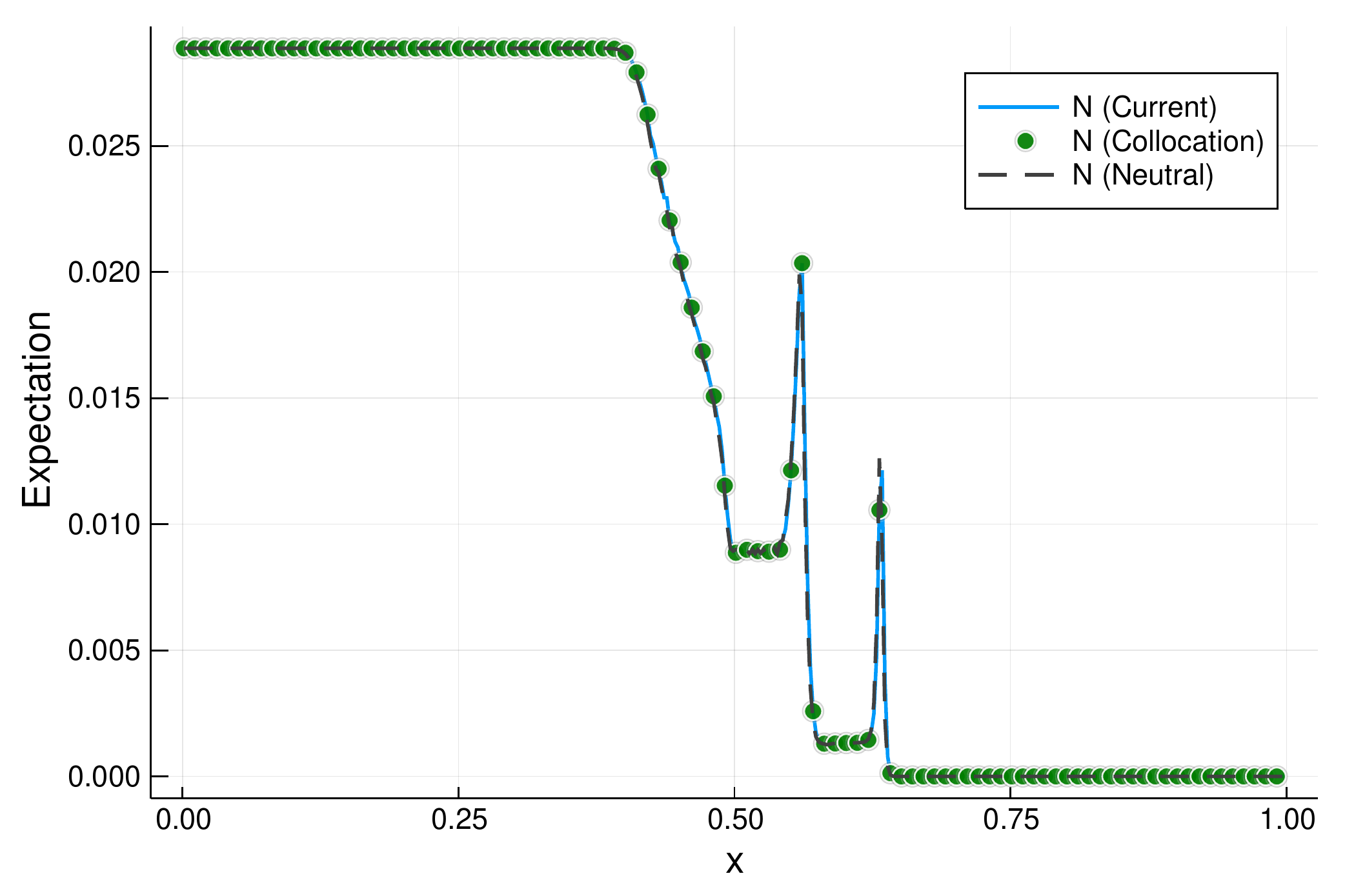}
	}
	\subfigure[$\mathbb S(U)$]{
		\includegraphics[width=0.31\textwidth]{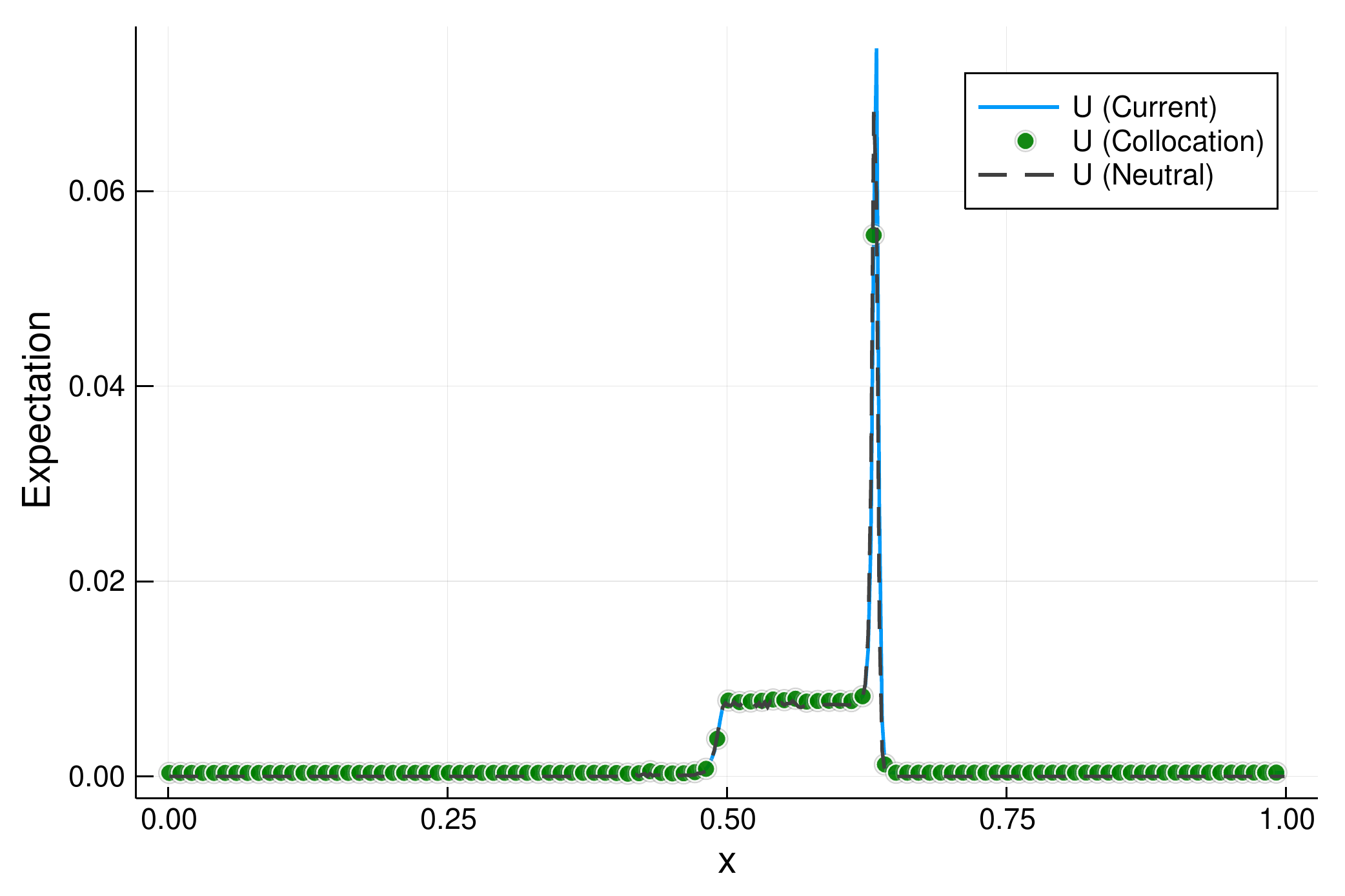}
	}
	\subfigure[$\mathbb S(B_y)$]{
		\includegraphics[width=0.31\textwidth]{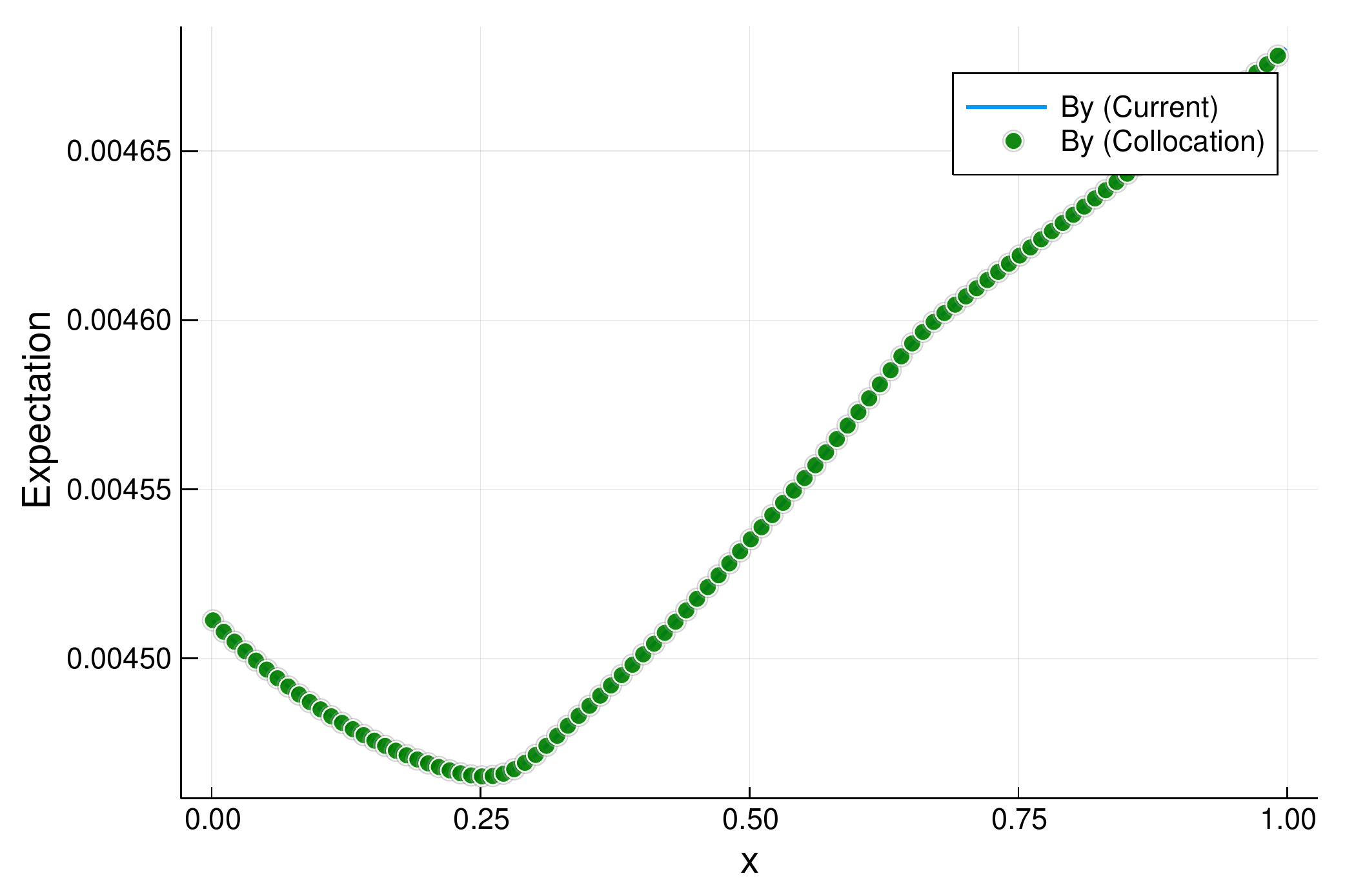}
	}
	\caption{Standard deviations of $N$, $U$ and $B_y$ in Brio-Wu shock tube with density uncertainty at $t=0.1$ (row 1: $r_g=0.003$, row 2: $r_g=0.01$, row 3: $r_g=0.1$, row 4: $r_g=1$, row 5: $r_g=100$).}
	\label{pic:briowu case1 std}
\end{figure}

Fig. \ref{pic:briowu case1 std} presents the standard deviations at the same output instant. 
Generally, the uncertainties travel along with the wave structure of expectation values and present similar propagating patterns. 
Typical wave structures serve as sources of local maximums of variance.
Compared with the expected value, its variance is more sensitive to physical discontinuities and holds finer-scale structures due to the spectrum formulation in the random space. 
As a result, the overshoots near contact discontinuity and shock, as well as the oscillations around central compound wave are observed.

\begin{figure}[htb!]
	\centering
	\subfigure[$\mathbb E(h_{ion})$]{
		\includegraphics[width=0.4\textwidth]{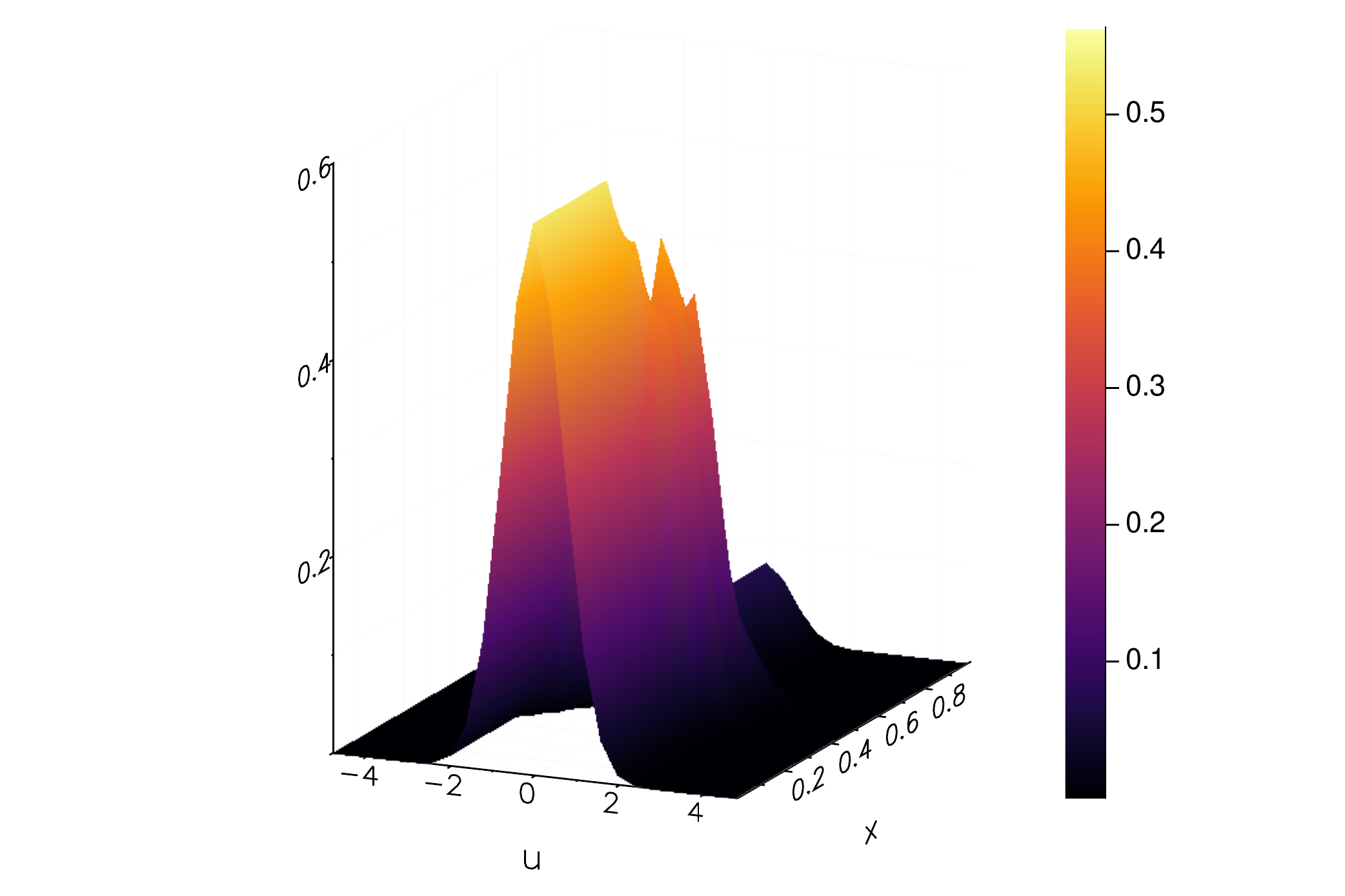}
	}
	\subfigure[$\mathbb S(h_{ion})$]{
		\includegraphics[width=0.4\textwidth]{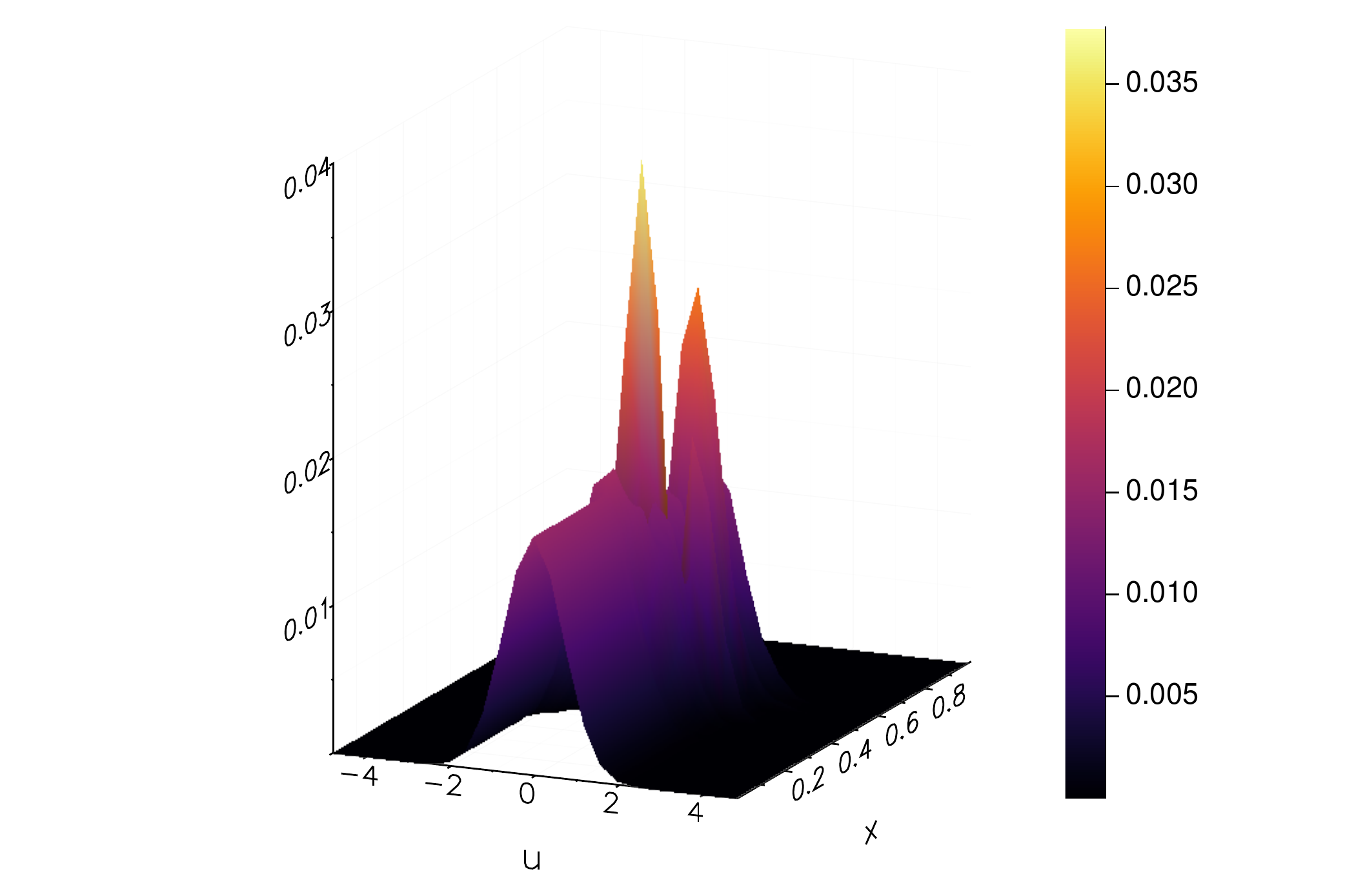}
	}
	\subfigure[$\mathbb E(h_{ion})$]{
		\includegraphics[width=0.4\textwidth]{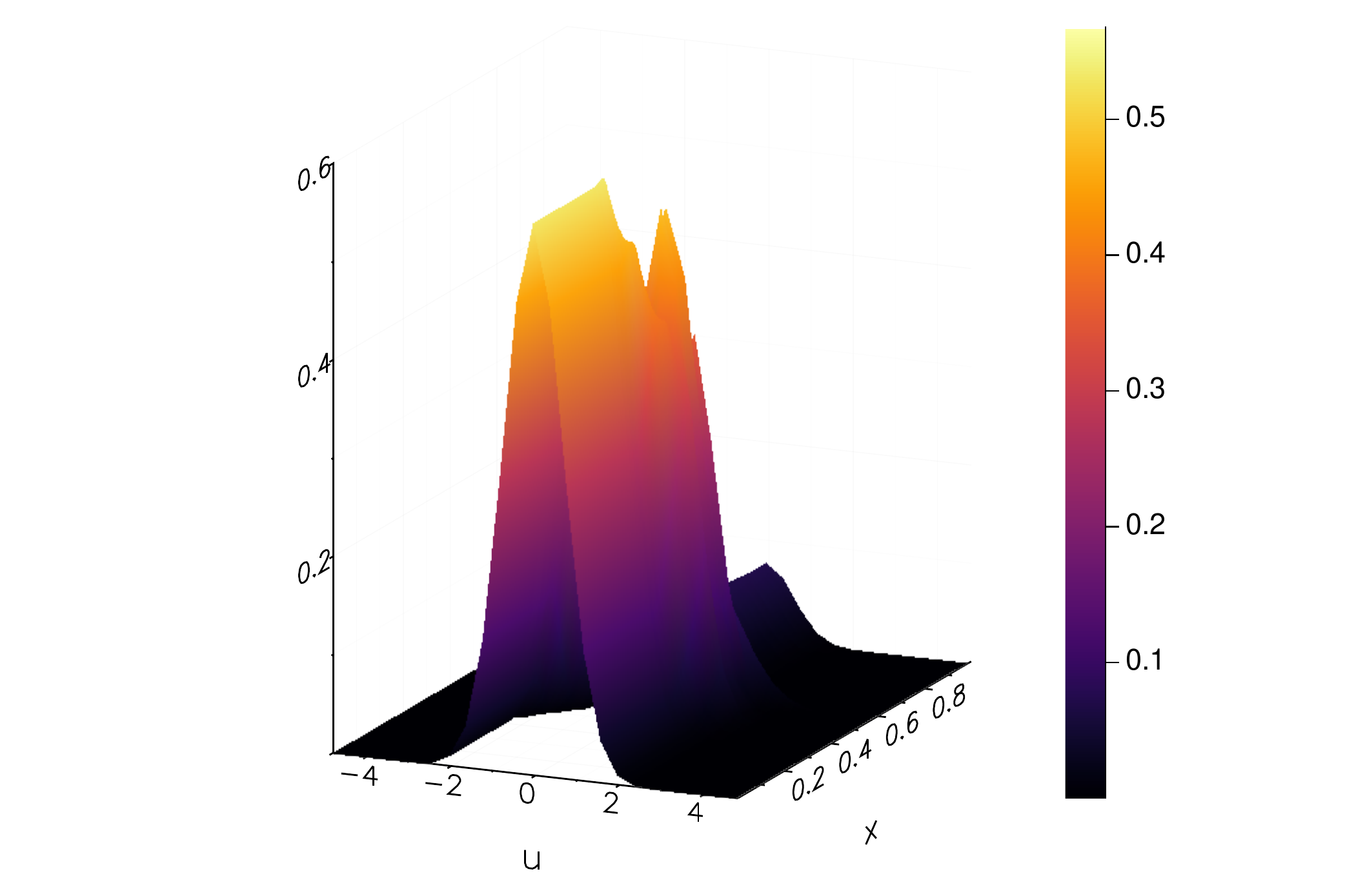}
	}
	\subfigure[$\mathbb S(h_{ion})$]{
		\includegraphics[width=0.4\textwidth]{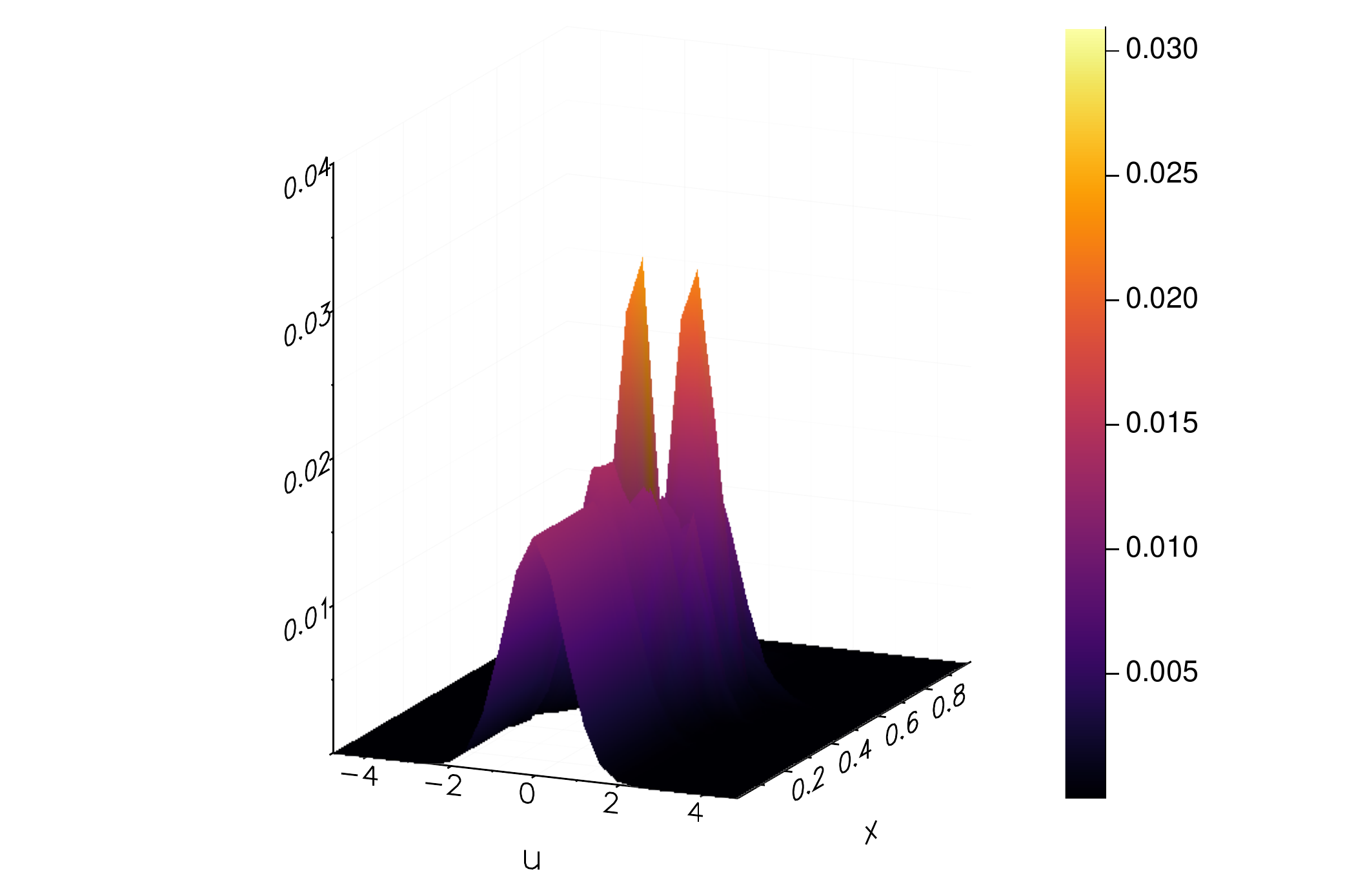}
	}
	\subfigure[$\mathbb E(h_{ion})$]{
		\includegraphics[width=0.4\textwidth]{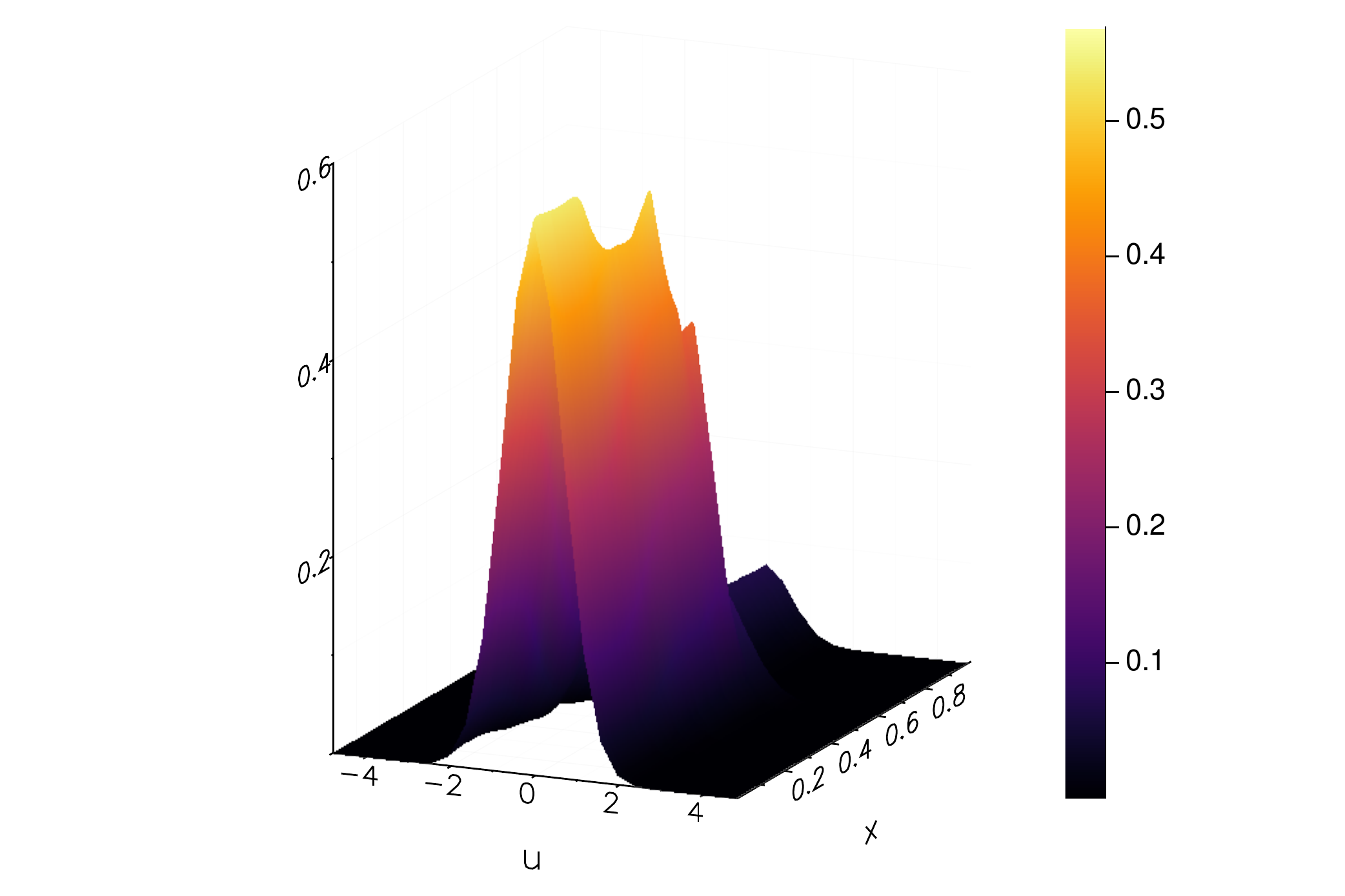}
	}
	\subfigure[$\mathbb S(h_{ion})$]{
		\includegraphics[width=0.4\textwidth]{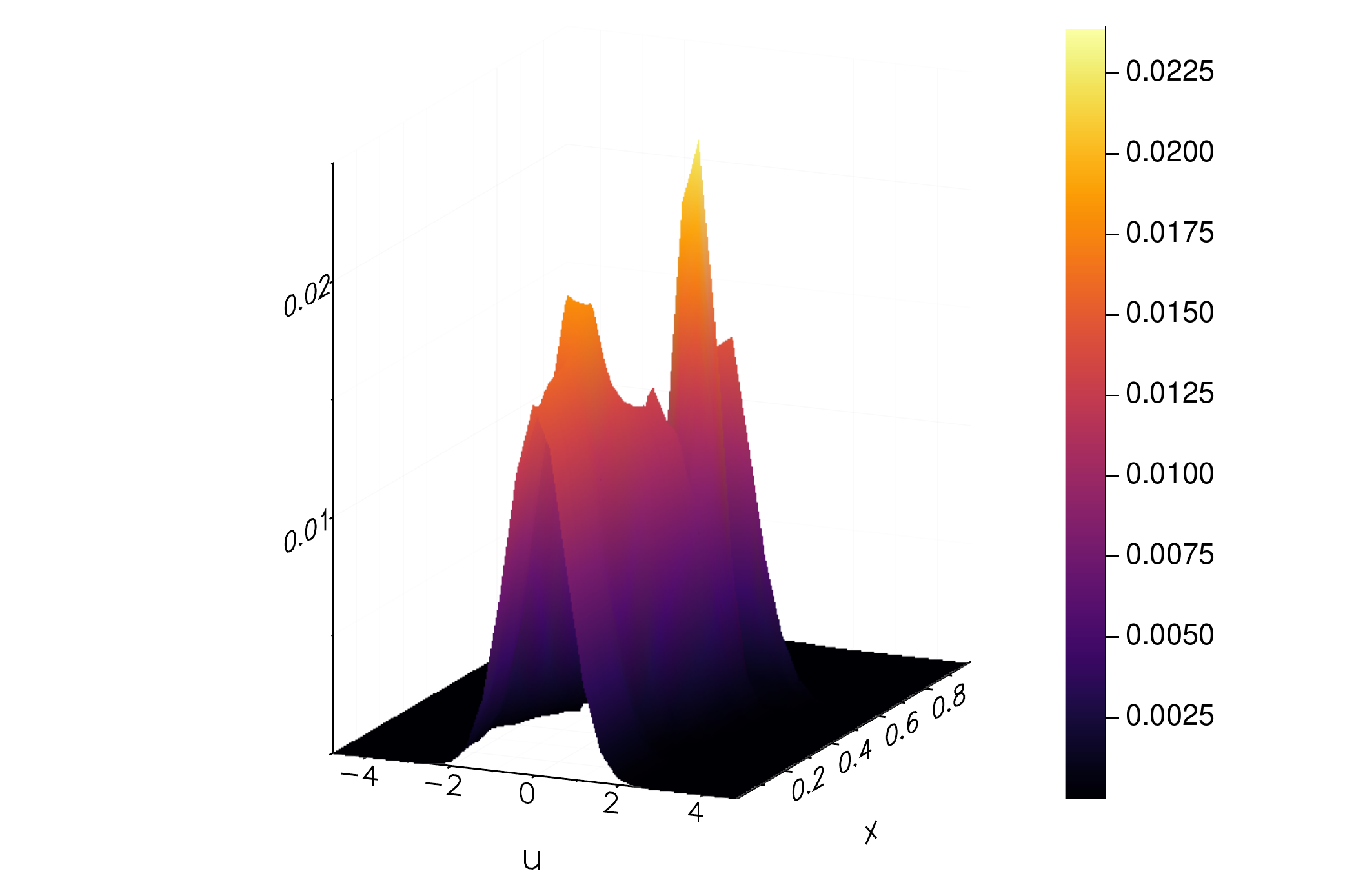}
	}
	\subfigure[$\mathbb E(h_{ion})$]{
		\includegraphics[width=0.4\textwidth]{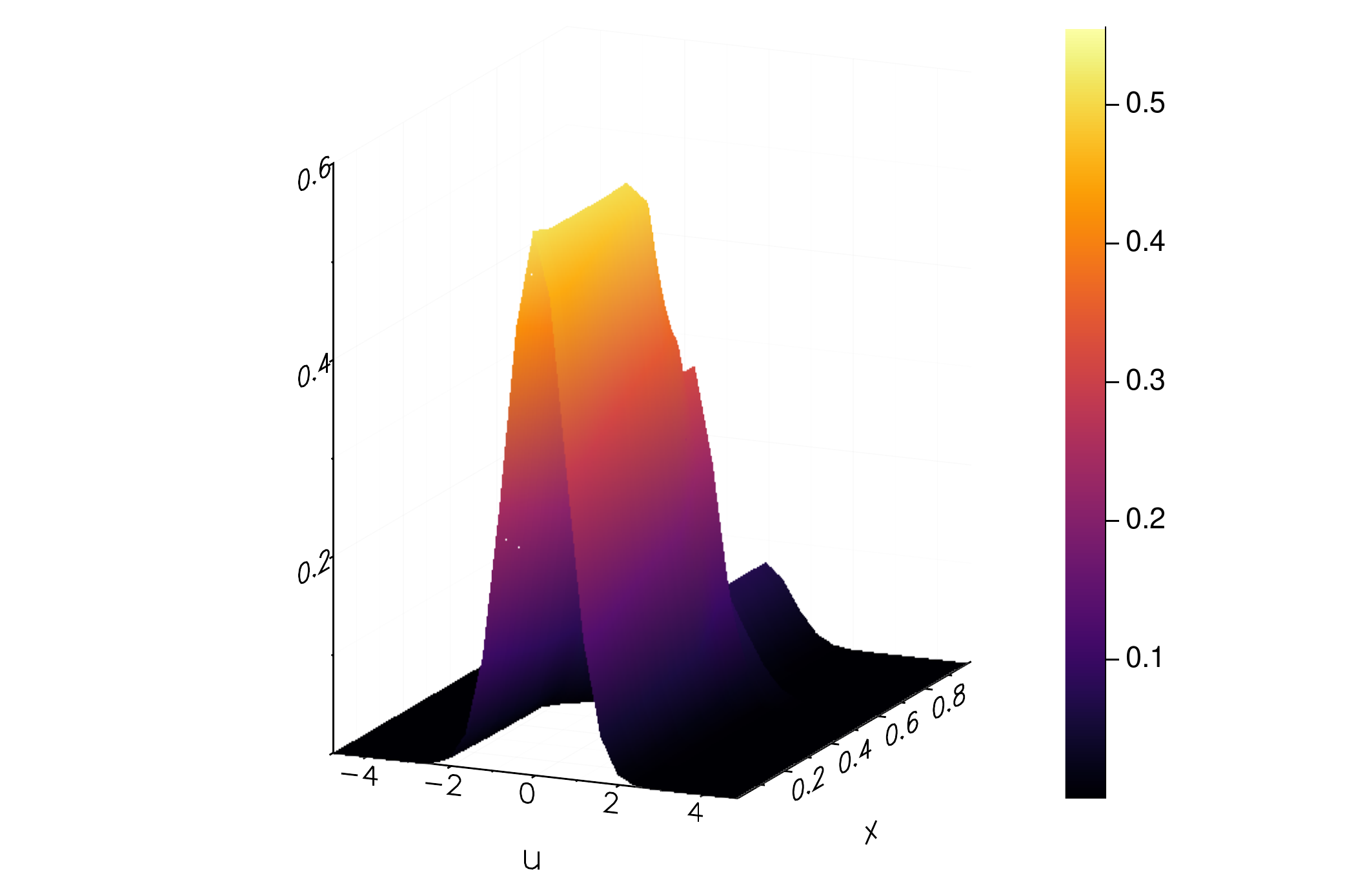}
	}
	\subfigure[$\mathbb S(h_{ion})$]{
		\includegraphics[width=0.4\textwidth]{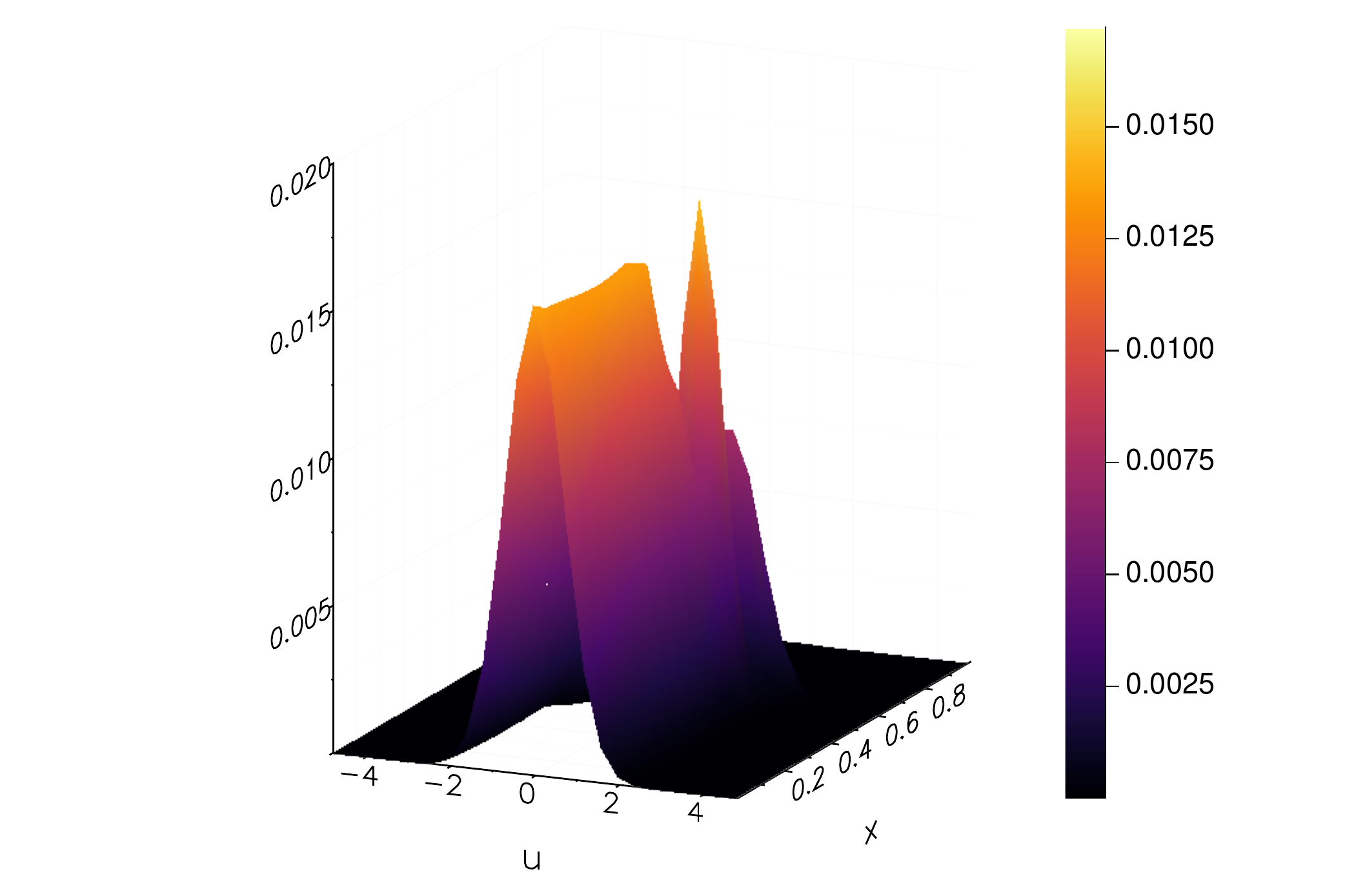}
	}
	\caption{Expectations and variances of reduced ion distribution $h_{ion}$, in Brio-Wu shock tube with density uncertainty at $t=0.1$ (row 1: $r_g=0.003$, row 2: $r_g=0.01$, row 3: $r_g=0.1$, row 4: $r_g=1$).}
	\label{pic:briowu case1 pdf}
\end{figure}

Fig. \ref{pic:briowu case1 pdf} presents the expectations and standard deviations of particle distribution function at different reference gyrorarius.
As is shown, the discontinuities in macroscopic expectations and overshoots in standard deviations come from the uncertainties contained in the particle distribution function near the center of velocity space. 
From MHD to Euler regimes, the randomness on particles get reduced and smoothed, resulting in gentle profiles of macroscopic quantities.

\subsubsection{Case 2: stochastic magnetic field}

\begin{figure}[htb!]
	\centering
	\subfigure[$\mathbb E(N)$]{
		\includegraphics[width=0.31\textwidth]{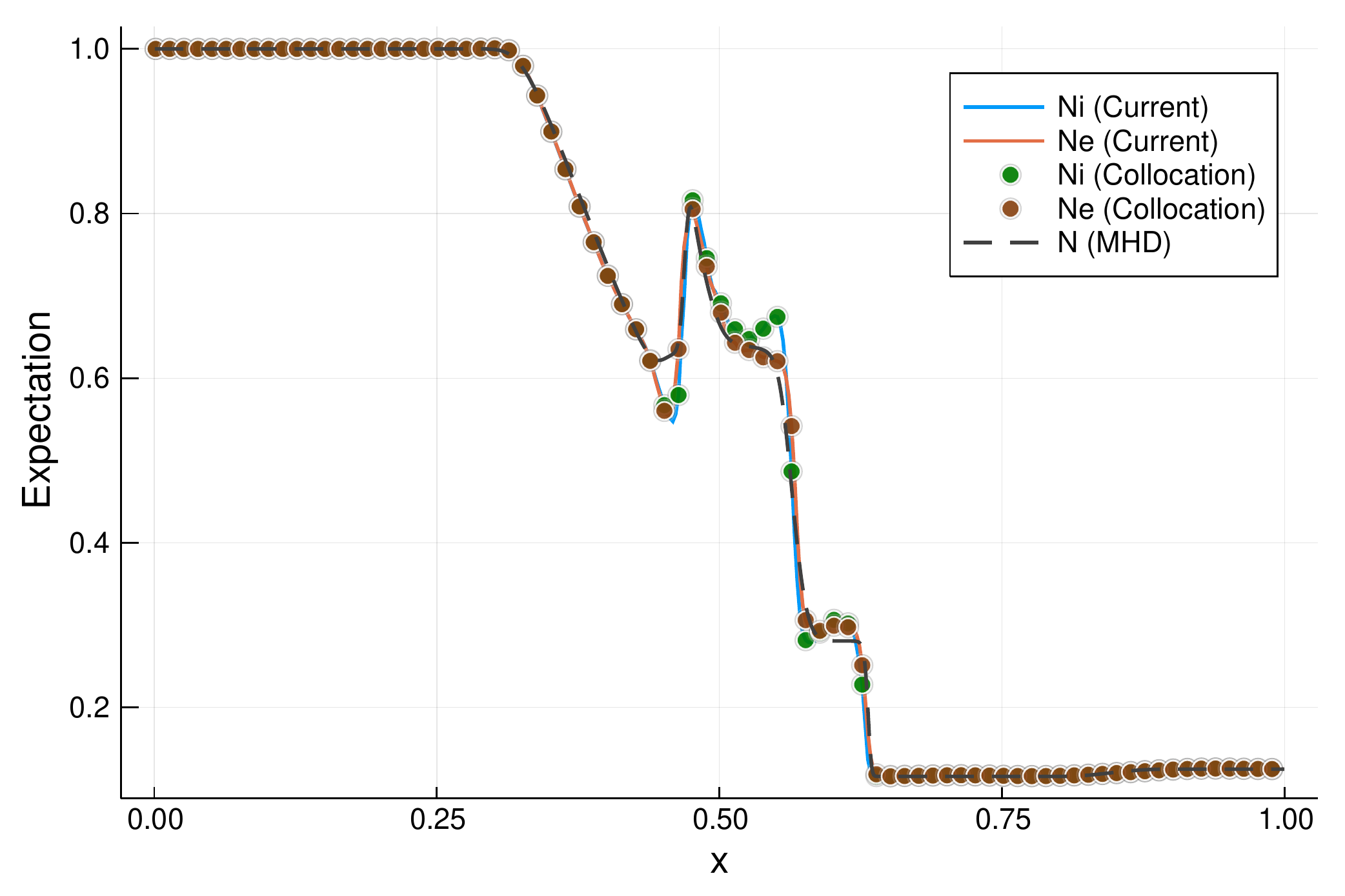}
	}
	\subfigure[$\mathbb E(U)$]{
		\includegraphics[width=0.31\textwidth]{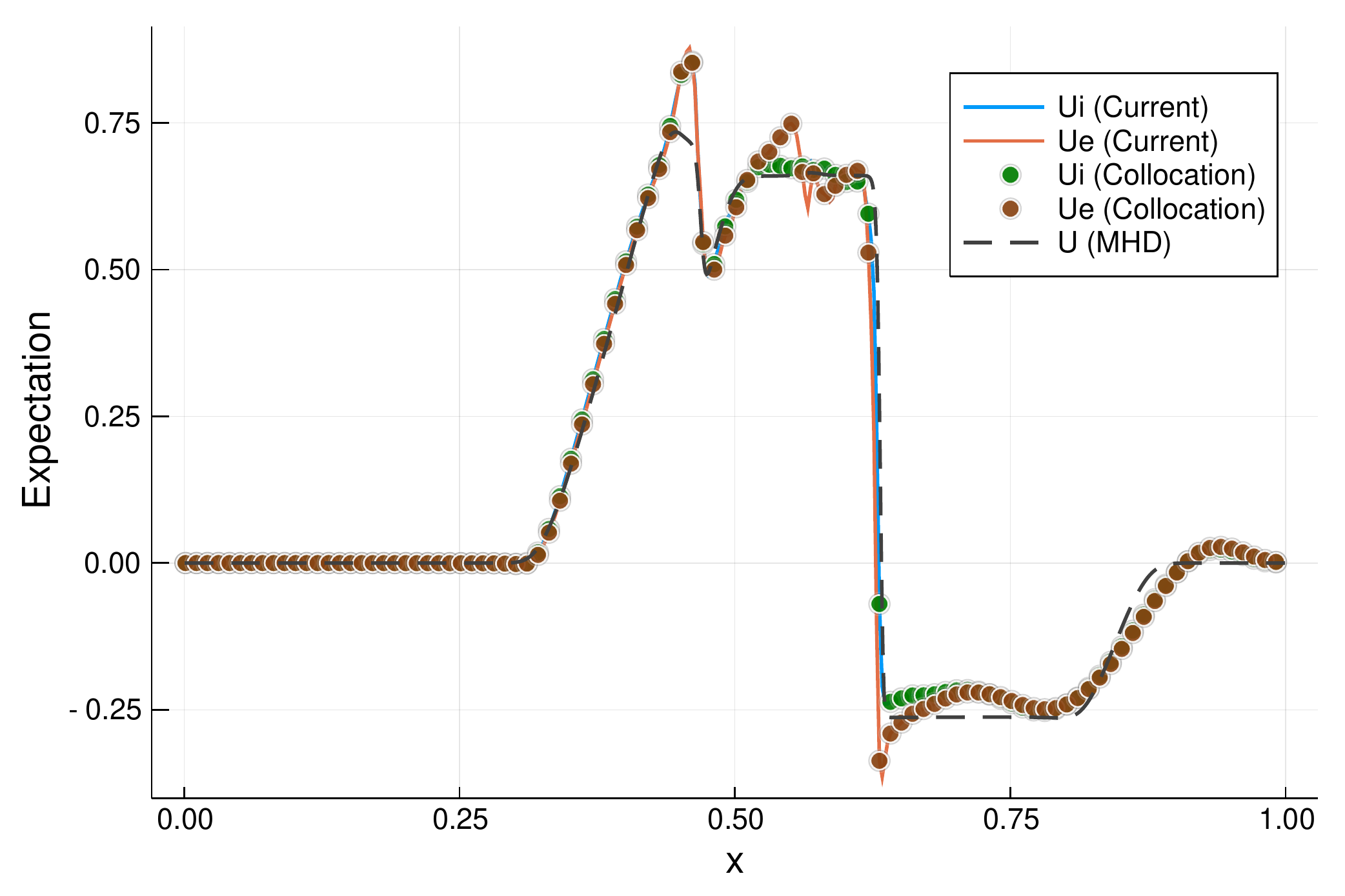}
	}
	\subfigure[$\mathbb E(B_y)$]{
		\includegraphics[width=0.31\textwidth]{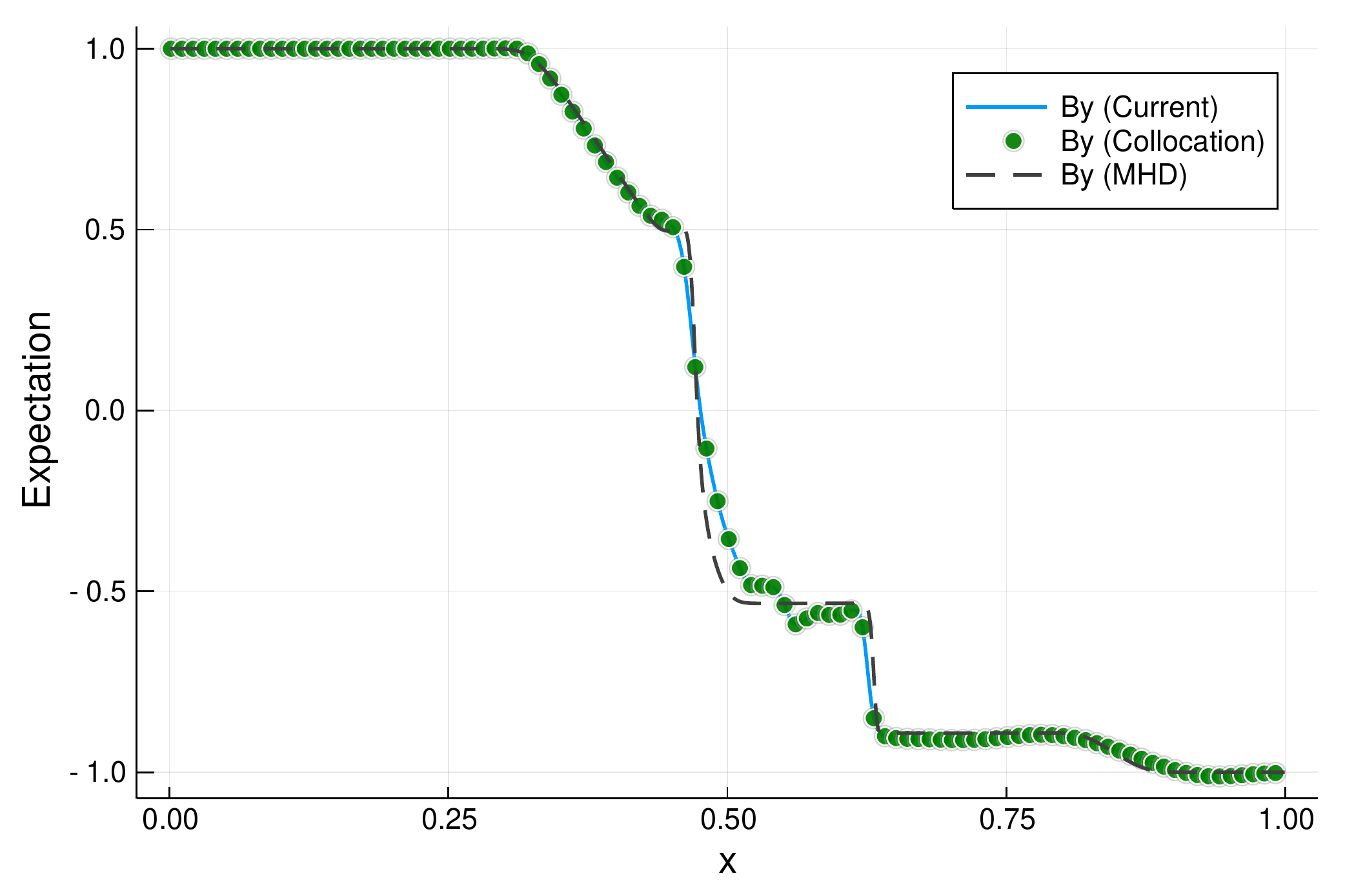}
	}
	\subfigure[$\mathbb E(N)$]{
		\includegraphics[width=0.31\textwidth]{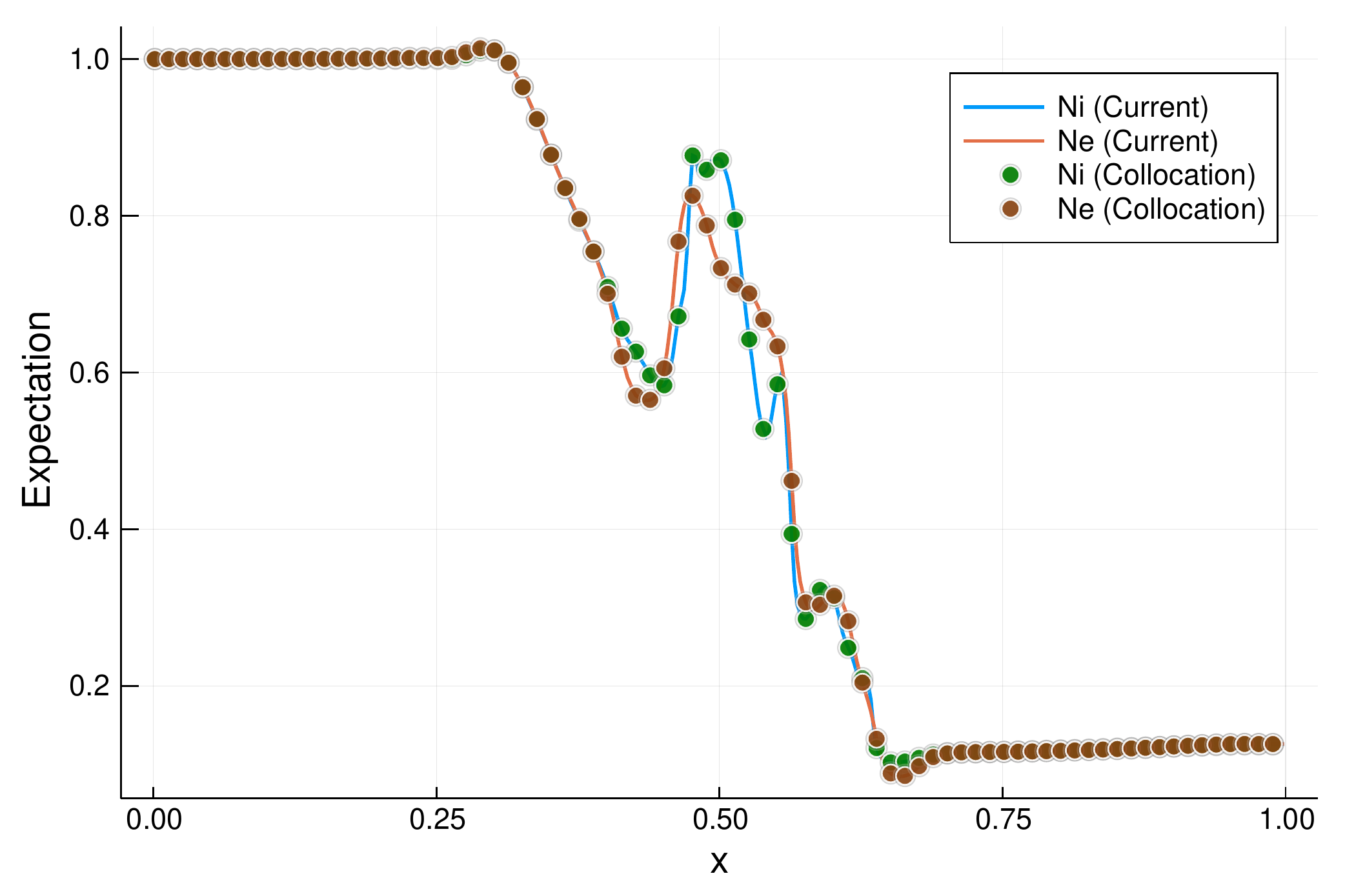}
	}
	\subfigure[$\mathbb E(U)$]{
		\includegraphics[width=0.31\textwidth]{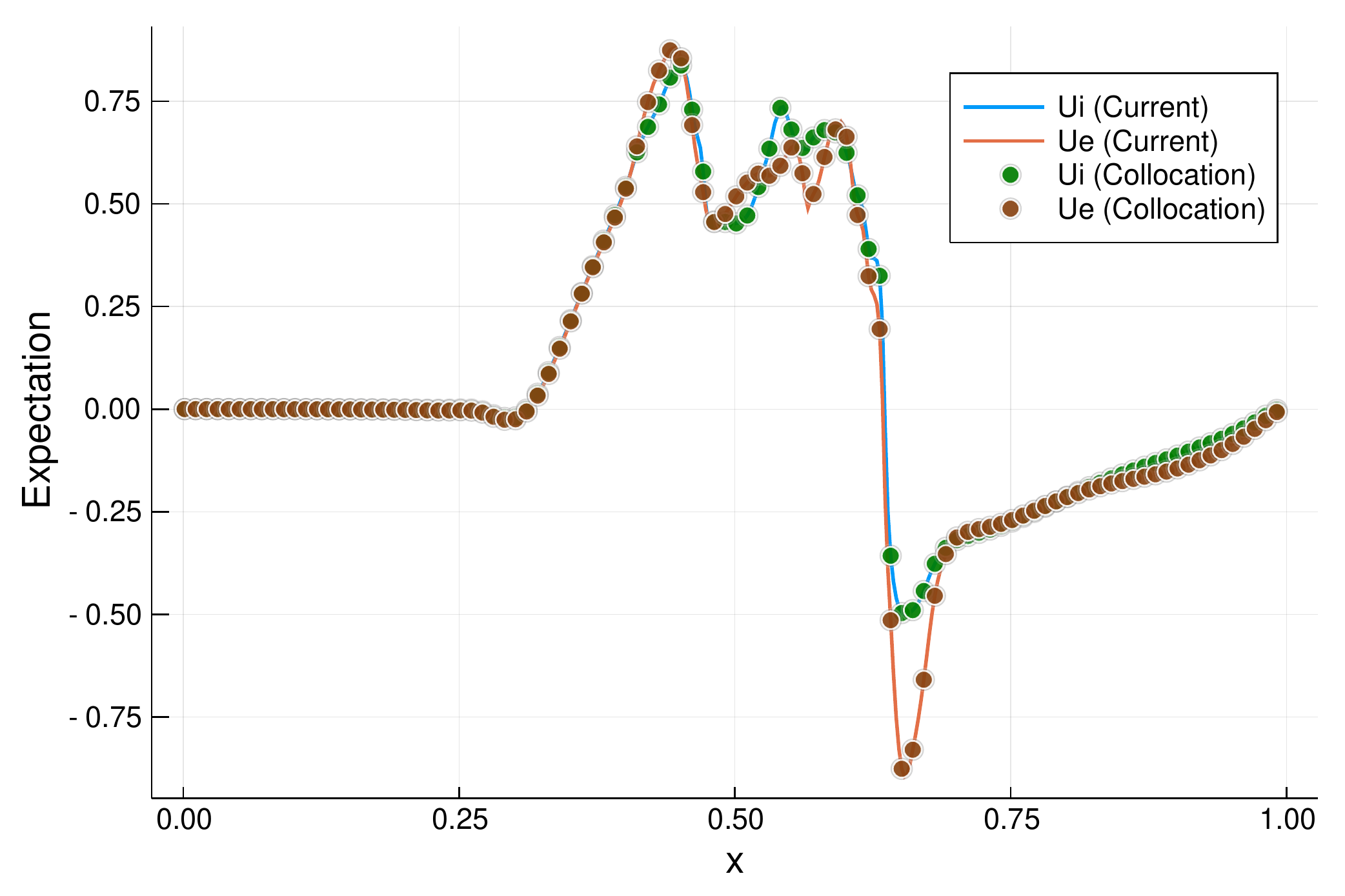}
	}
	\subfigure[$\mathbb E(B_y)$]{
		\includegraphics[width=0.31\textwidth]{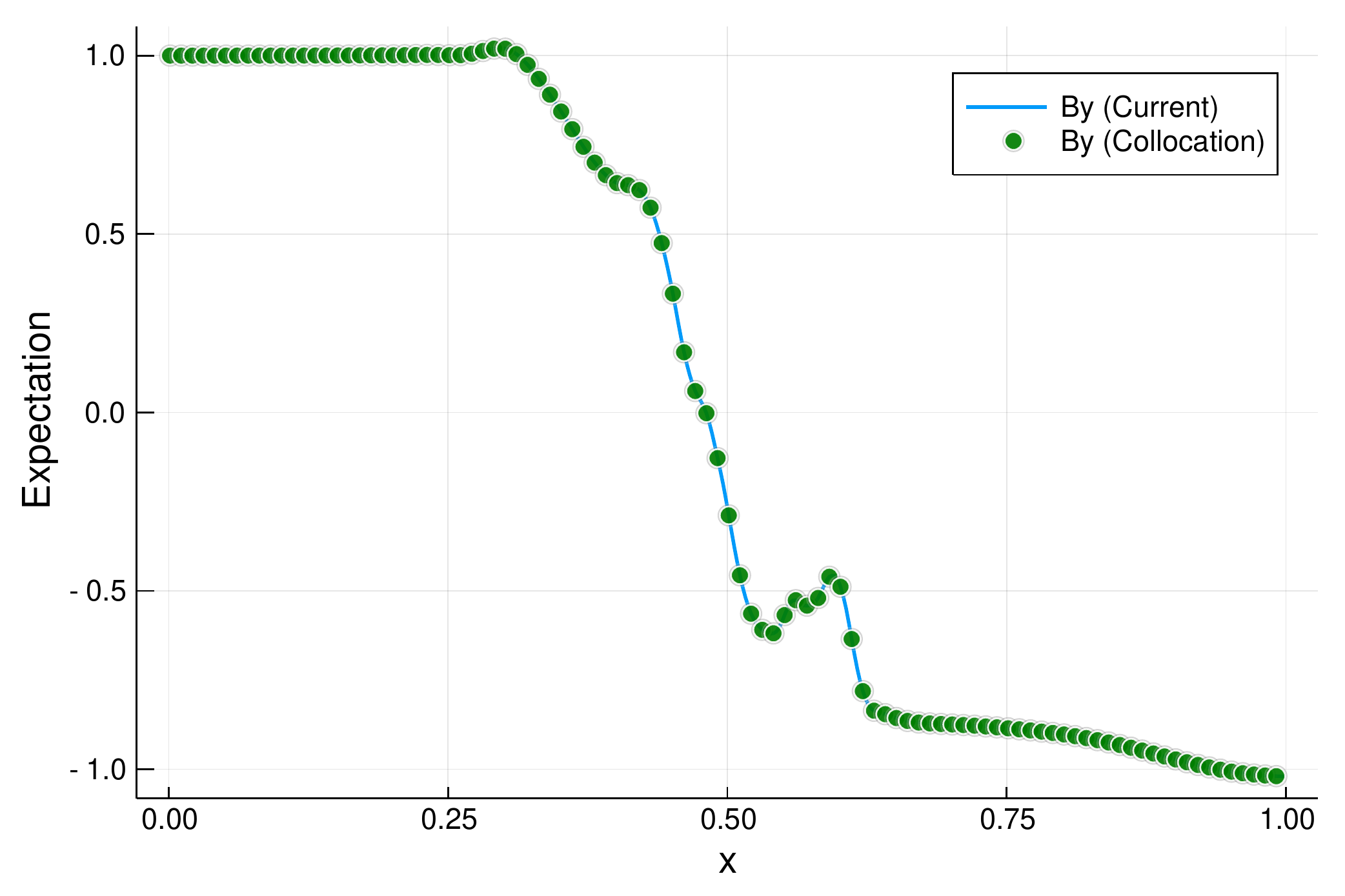}
	}
	\subfigure[$\mathbb E(N)$]{
		\includegraphics[width=0.31\textwidth]{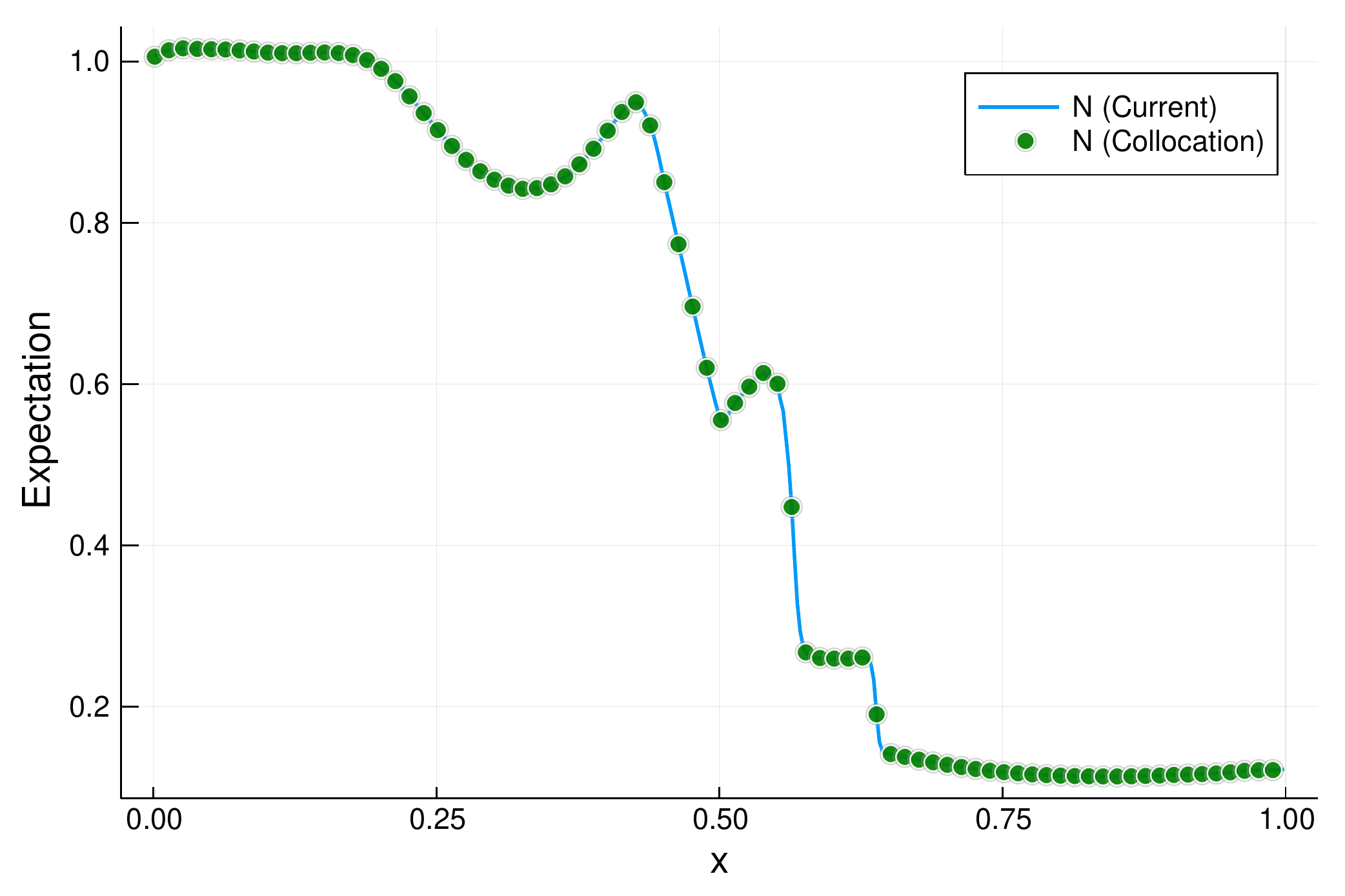}
	}
	\subfigure[$\mathbb E(U)$]{
		\includegraphics[width=0.31\textwidth]{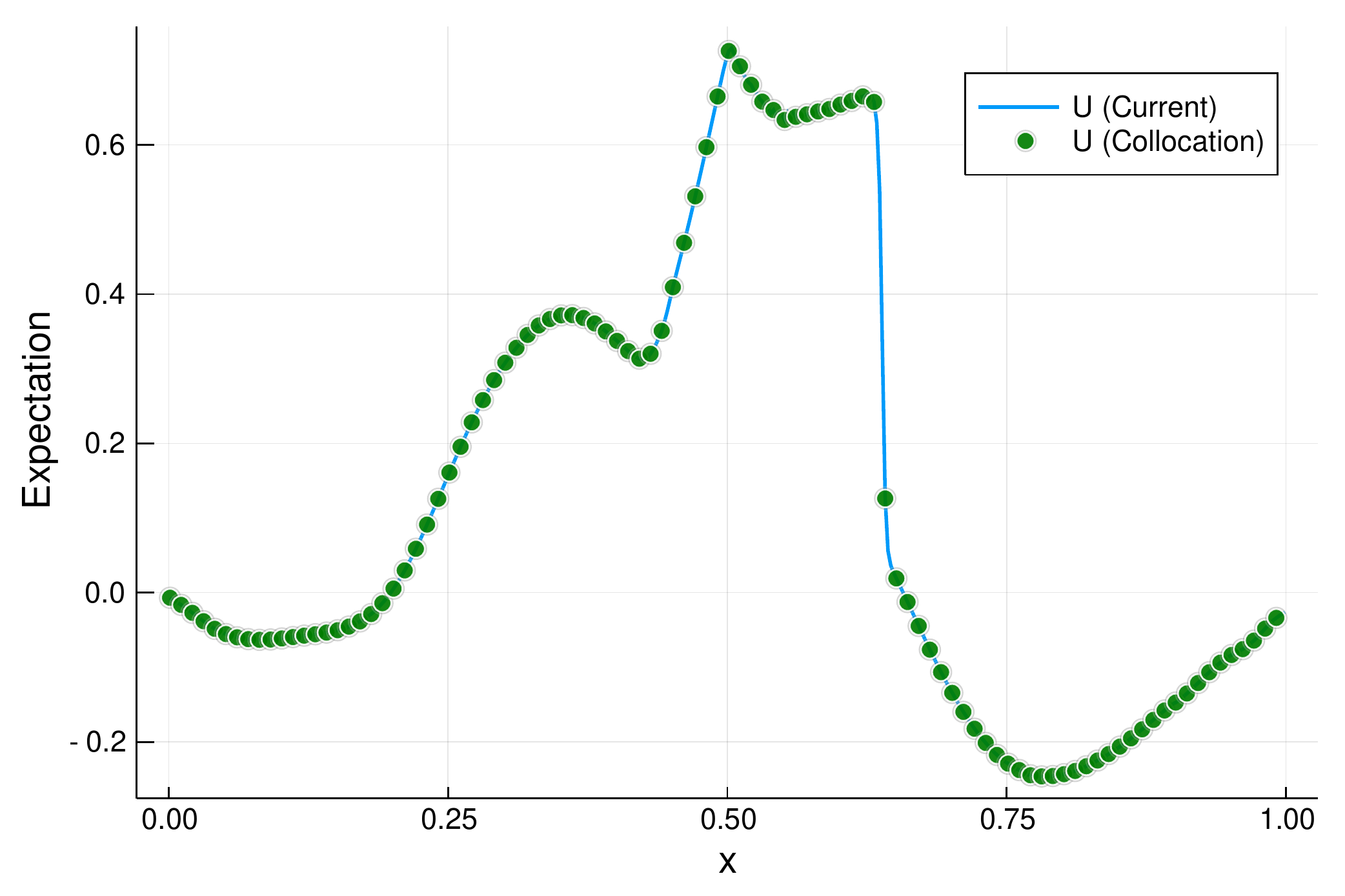}
	}
	\subfigure[$\mathbb E(B_y)$]{
		\includegraphics[width=0.31\textwidth]{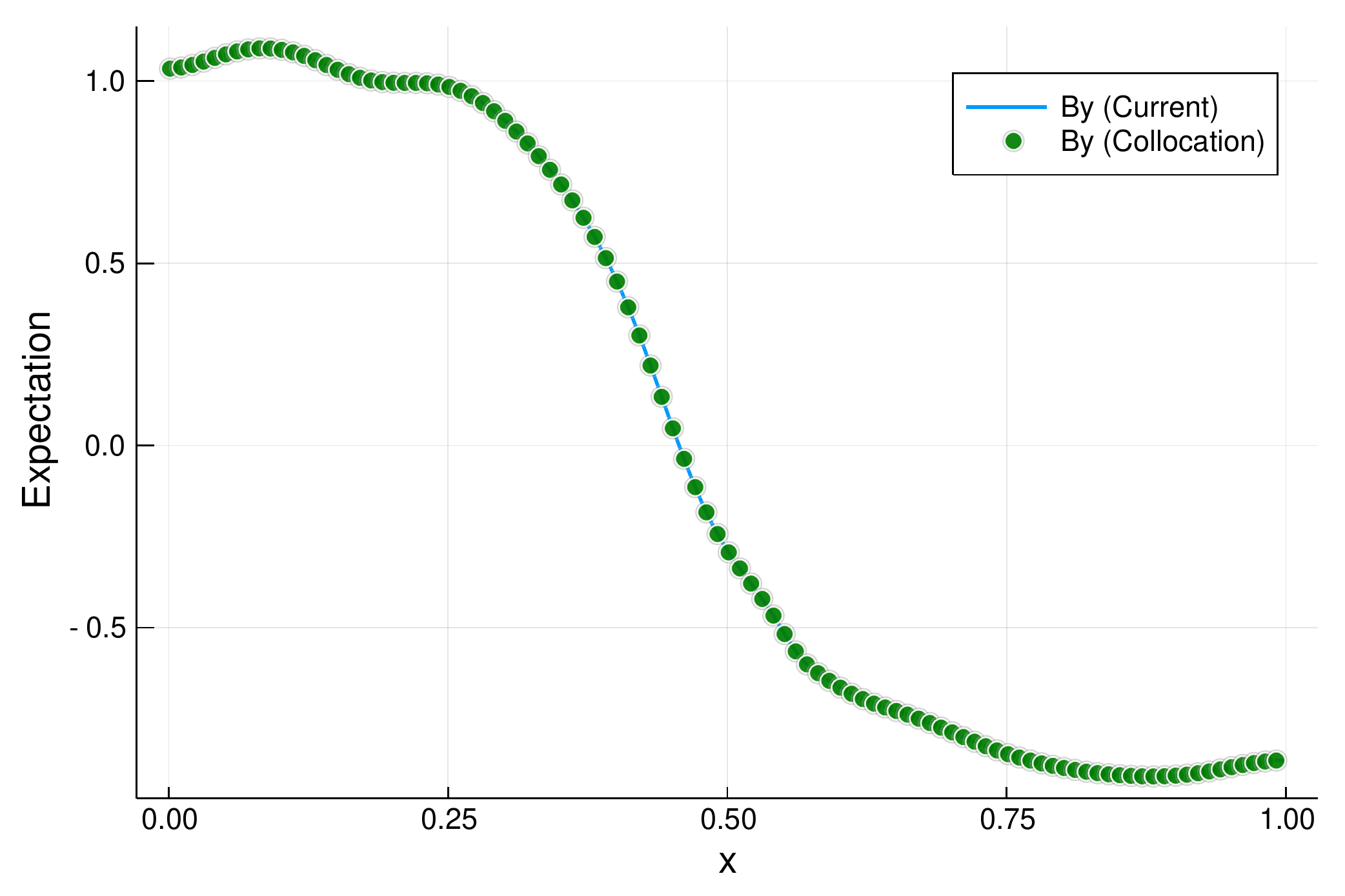}
	}
	\subfigure[$\mathbb E(N)$]{
		\includegraphics[width=0.31\textwidth]{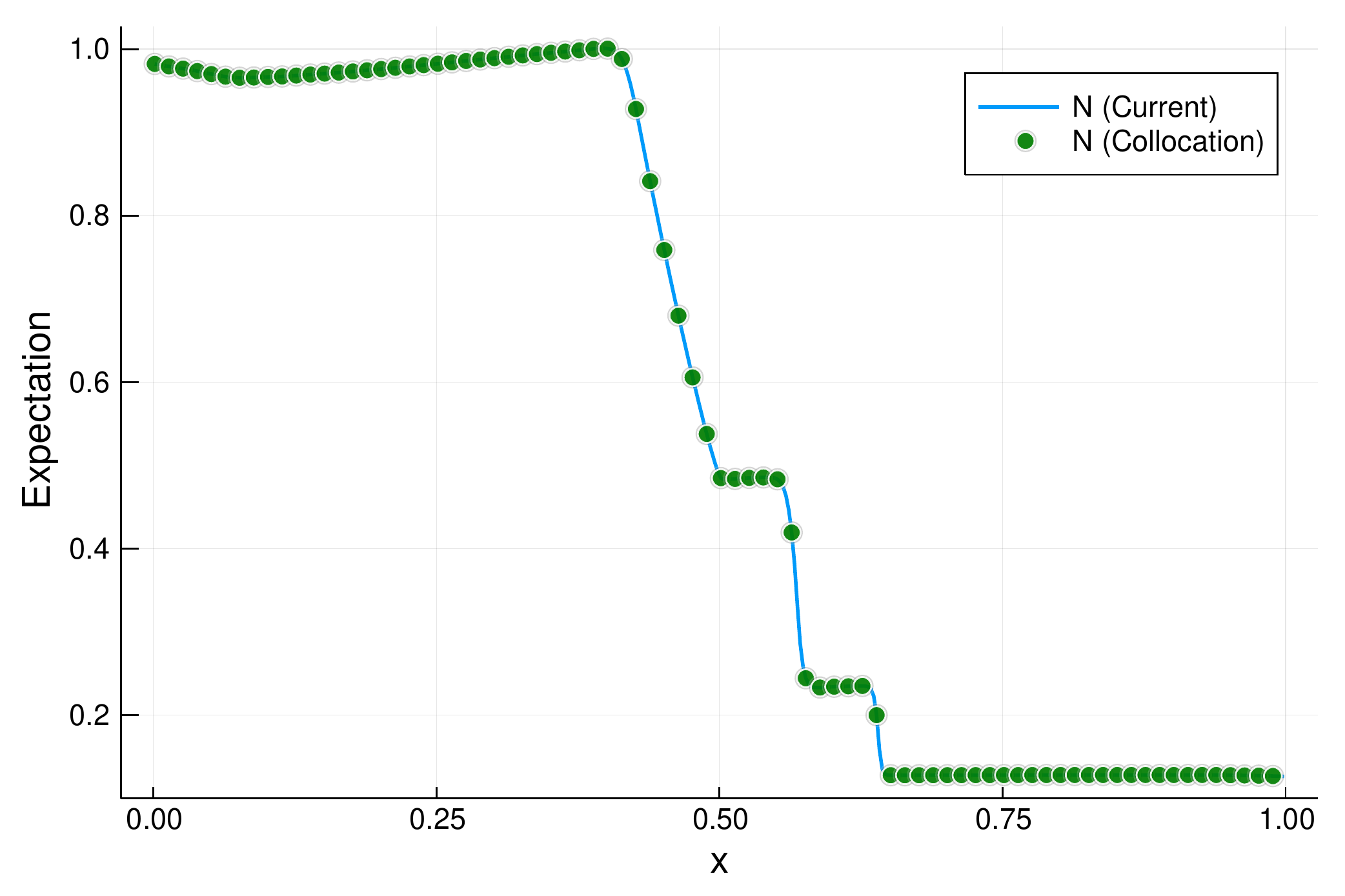}
	}
	\subfigure[$\mathbb E(U)$]{
		\includegraphics[width=0.31\textwidth]{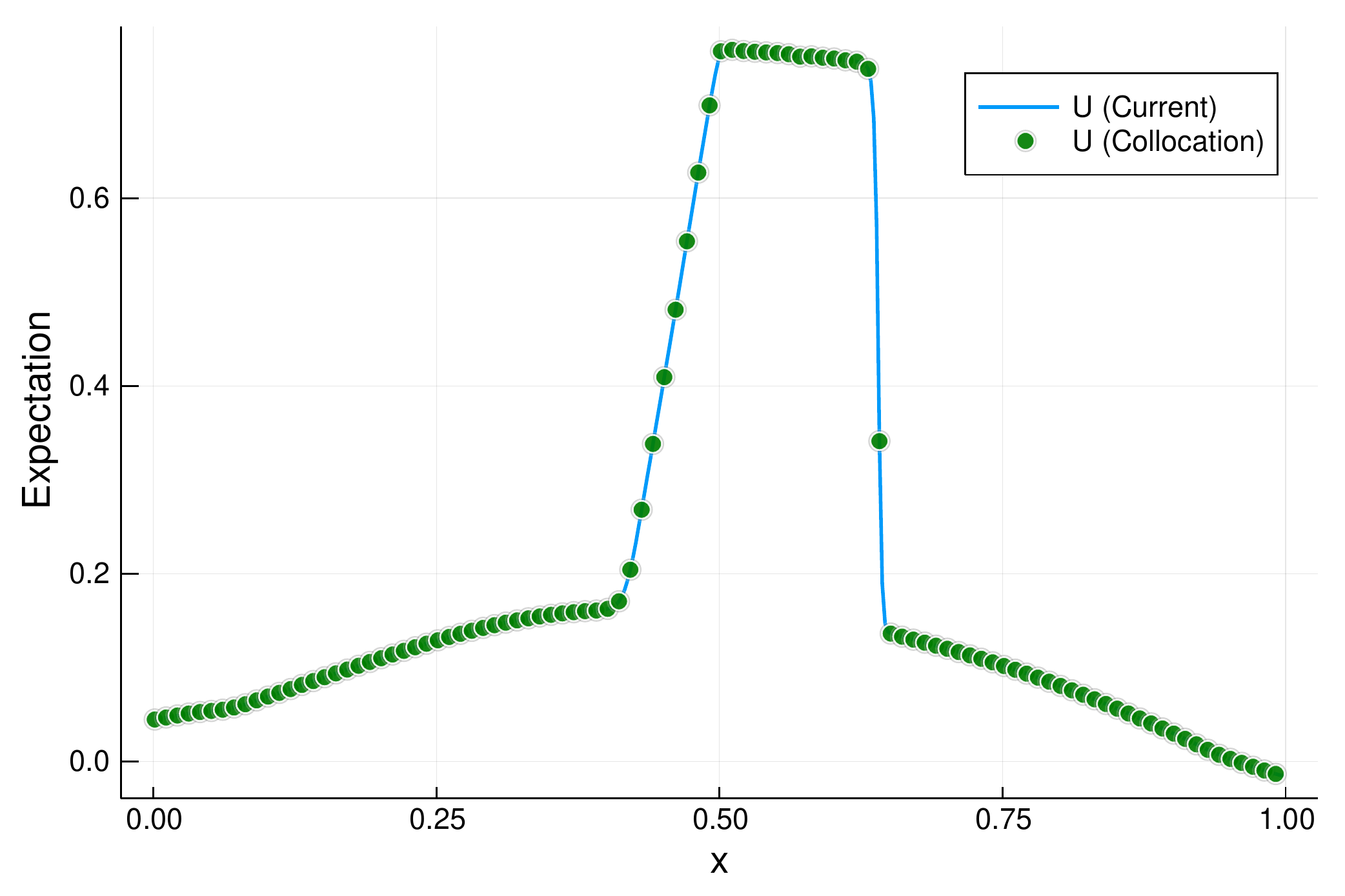}
	}
	\subfigure[$\mathbb E(B_y)$]{
		\includegraphics[width=0.31\textwidth]{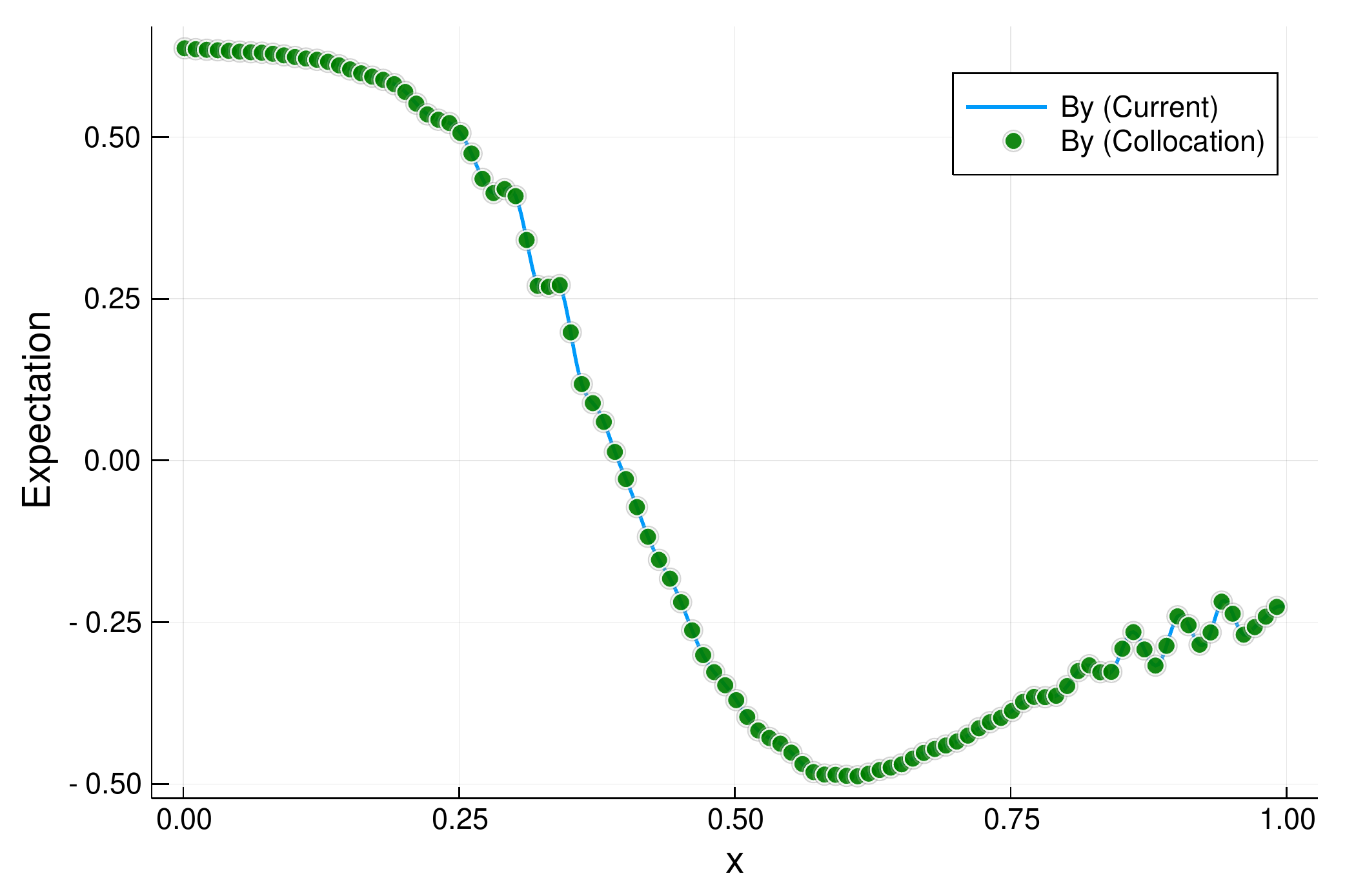}
	}
	\subfigure[$\mathbb E(N)$]{
		\includegraphics[width=0.31\textwidth]{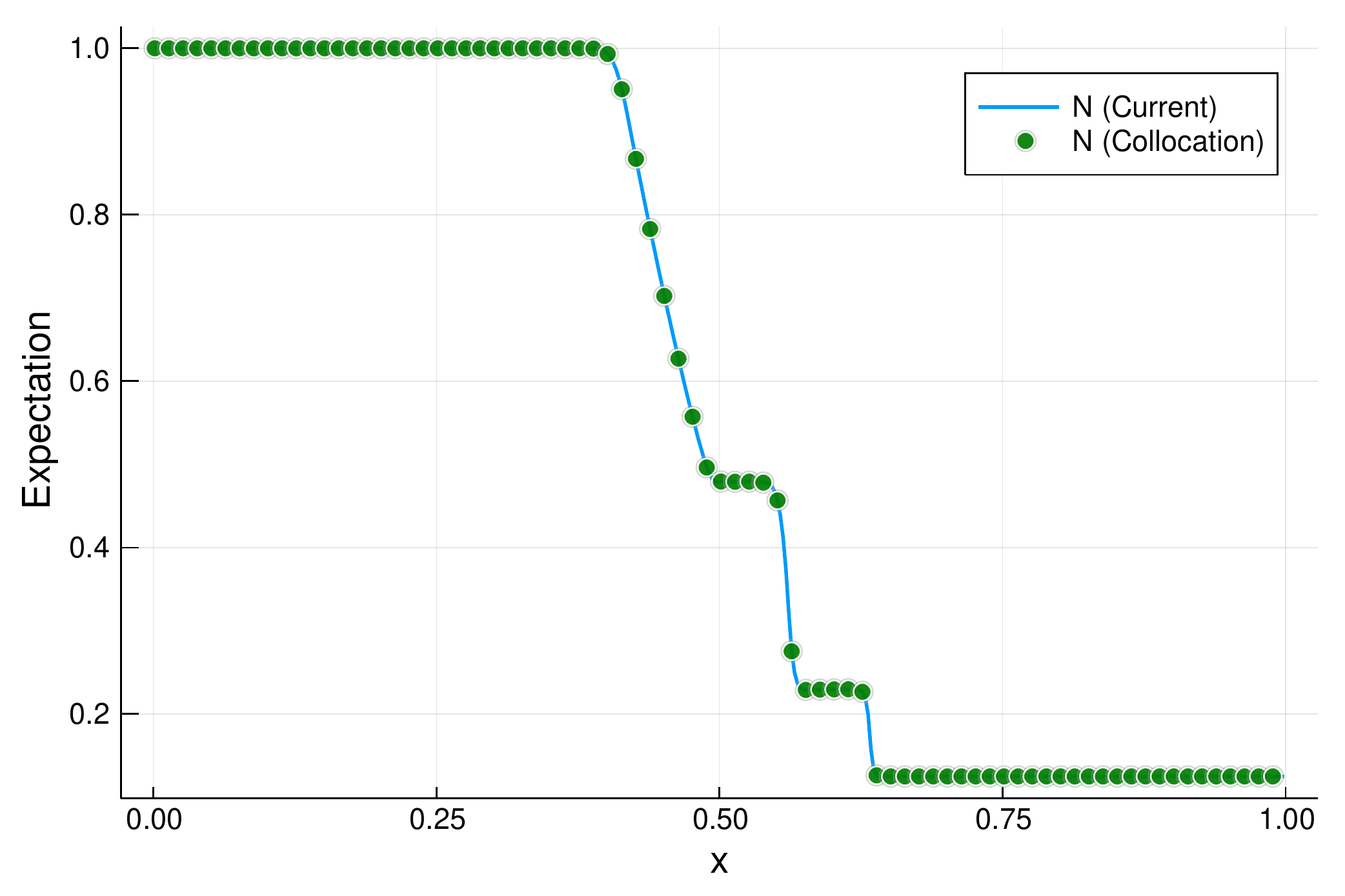}
	}
	\subfigure[$\mathbb E(U)$]{
		\includegraphics[width=0.31\textwidth]{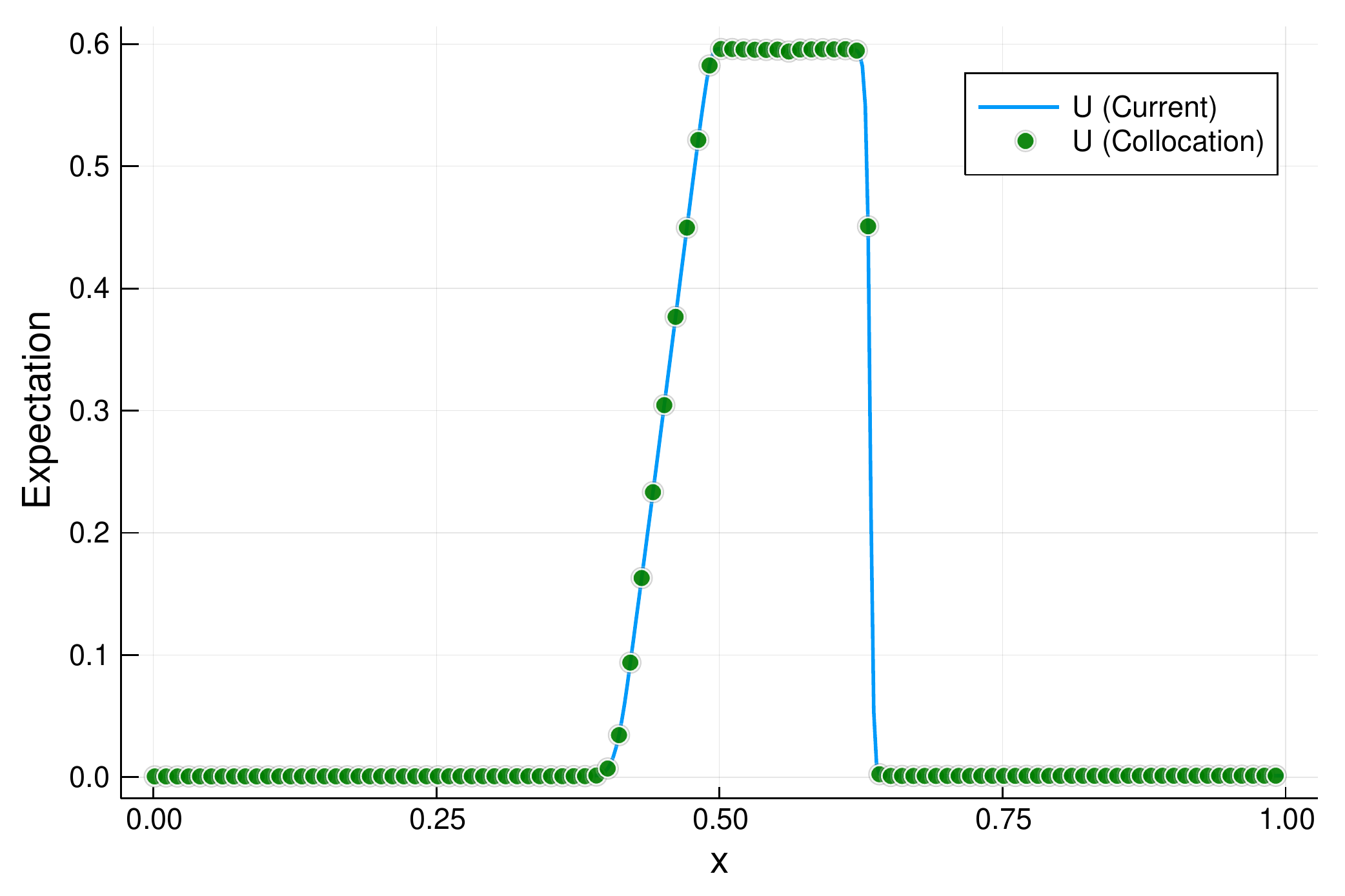}
	}
	\subfigure[$\mathbb E(B_y)$]{
		\includegraphics[width=0.31\textwidth]{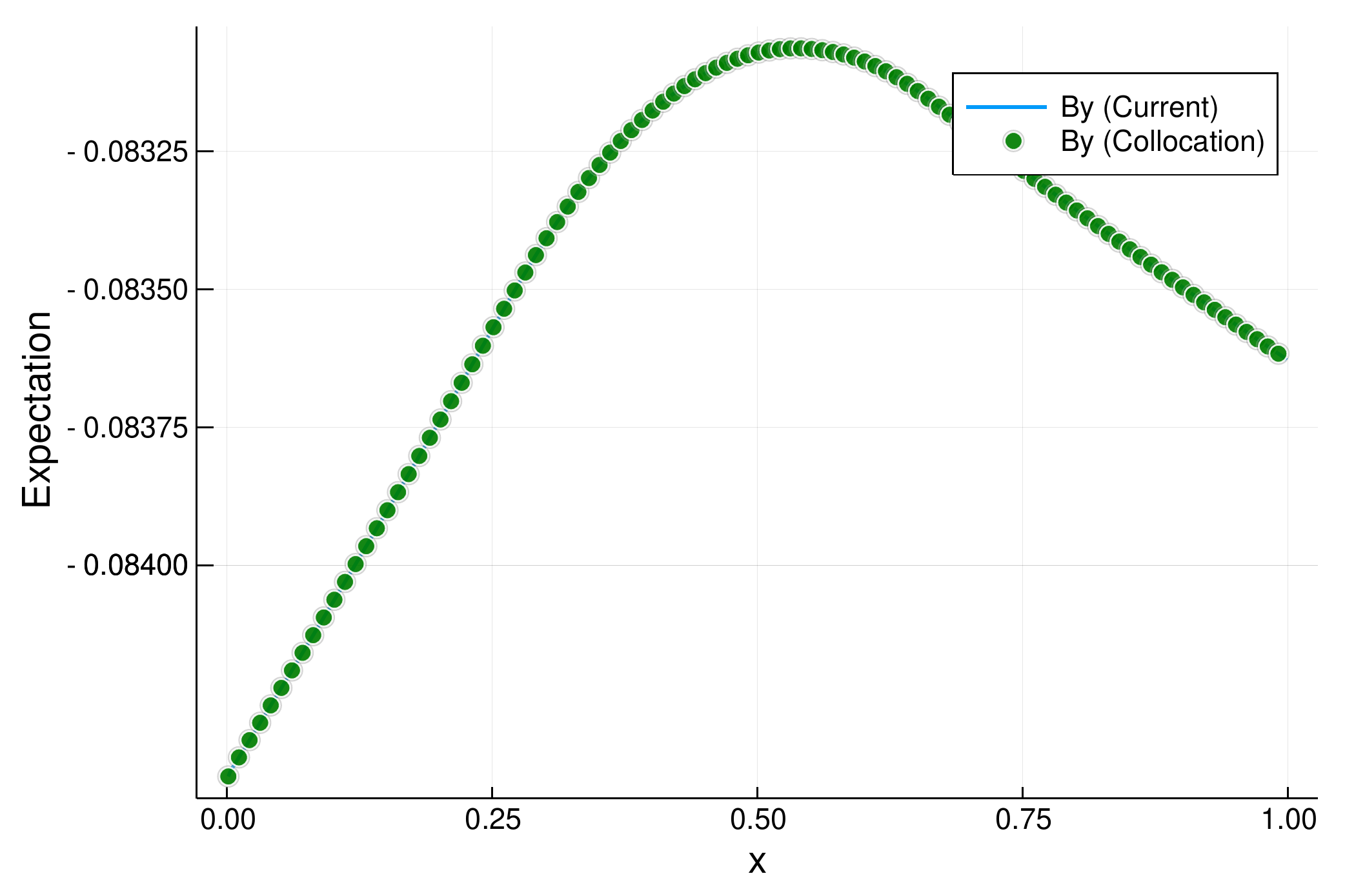}
	}
	\caption{Expectation values of $N$, $U$ and $B_y$ in Brio-Wu shock tube with magnetic uncertainty at $t=0.1$ (row 1: $r_g=0.003$, row 2: $r_g=0.01$, row 3: $r_g=0.1$, row 4: $r_g=1$, row 5: $r_g=100$).}
	\label{pic:briowu case2 mean}
\end{figure}

\begin{figure}[htb!]
	\centering
	\subfigure[$\mathbb S(N)$]{
		\includegraphics[width=0.31\textwidth]{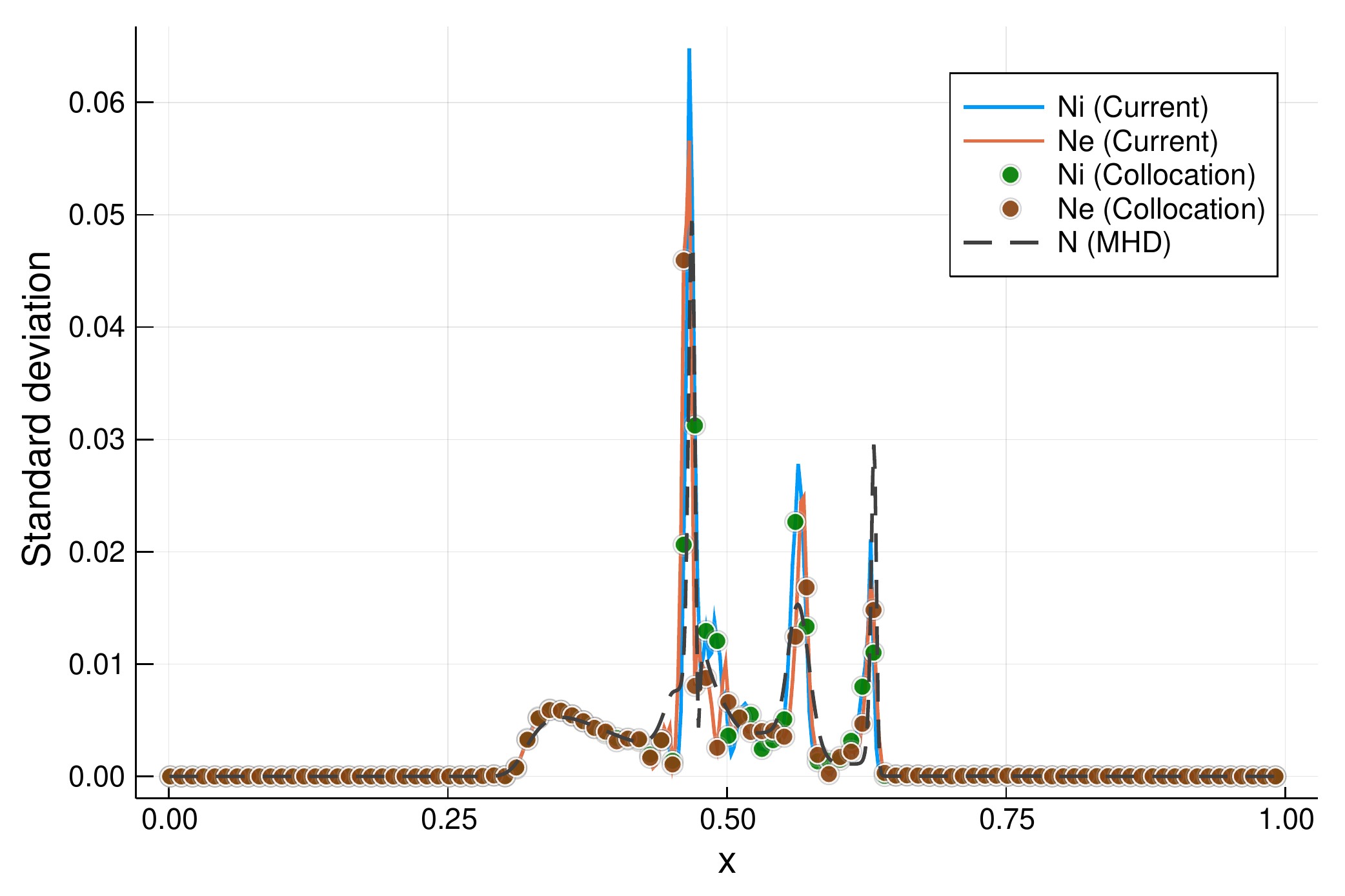}
	}
	\subfigure[$\mathbb S(U)$]{
		\includegraphics[width=0.31\textwidth]{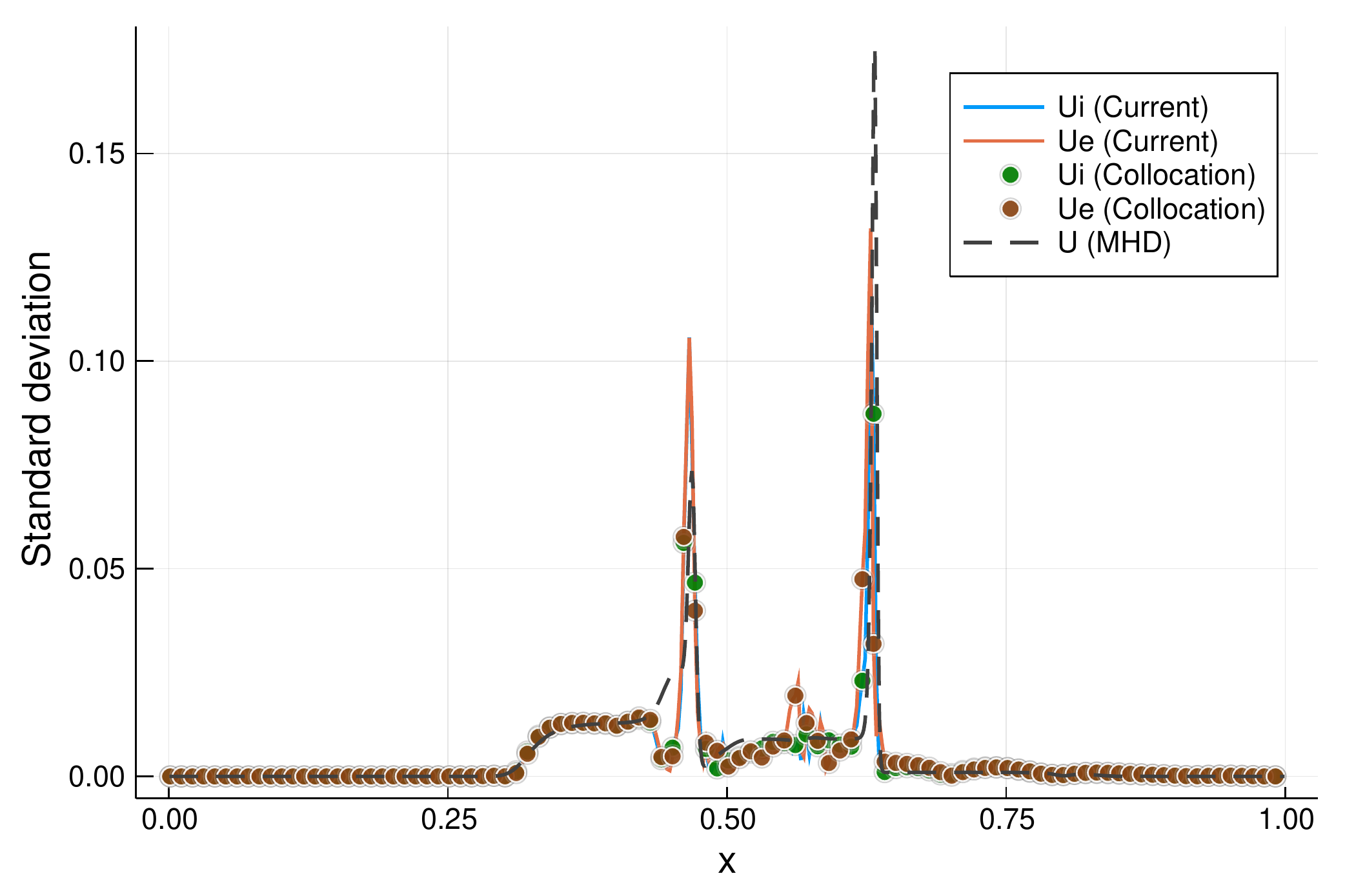}
	}
	\subfigure[$\mathbb S(B_y)$]{
		\includegraphics[width=0.31\textwidth]{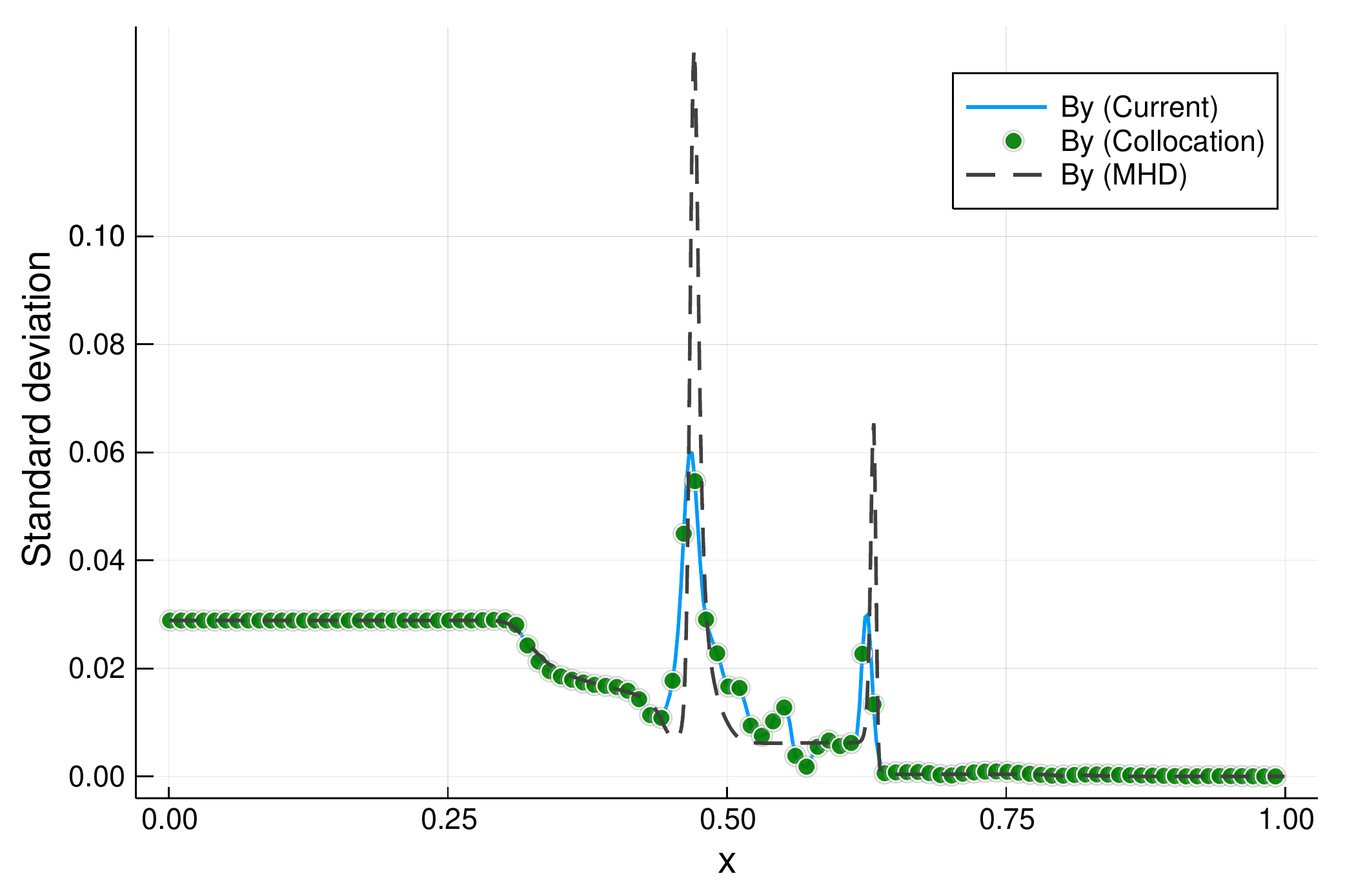}
	}
	\subfigure[$\mathbb S(N)$]{
		\includegraphics[width=0.31\textwidth]{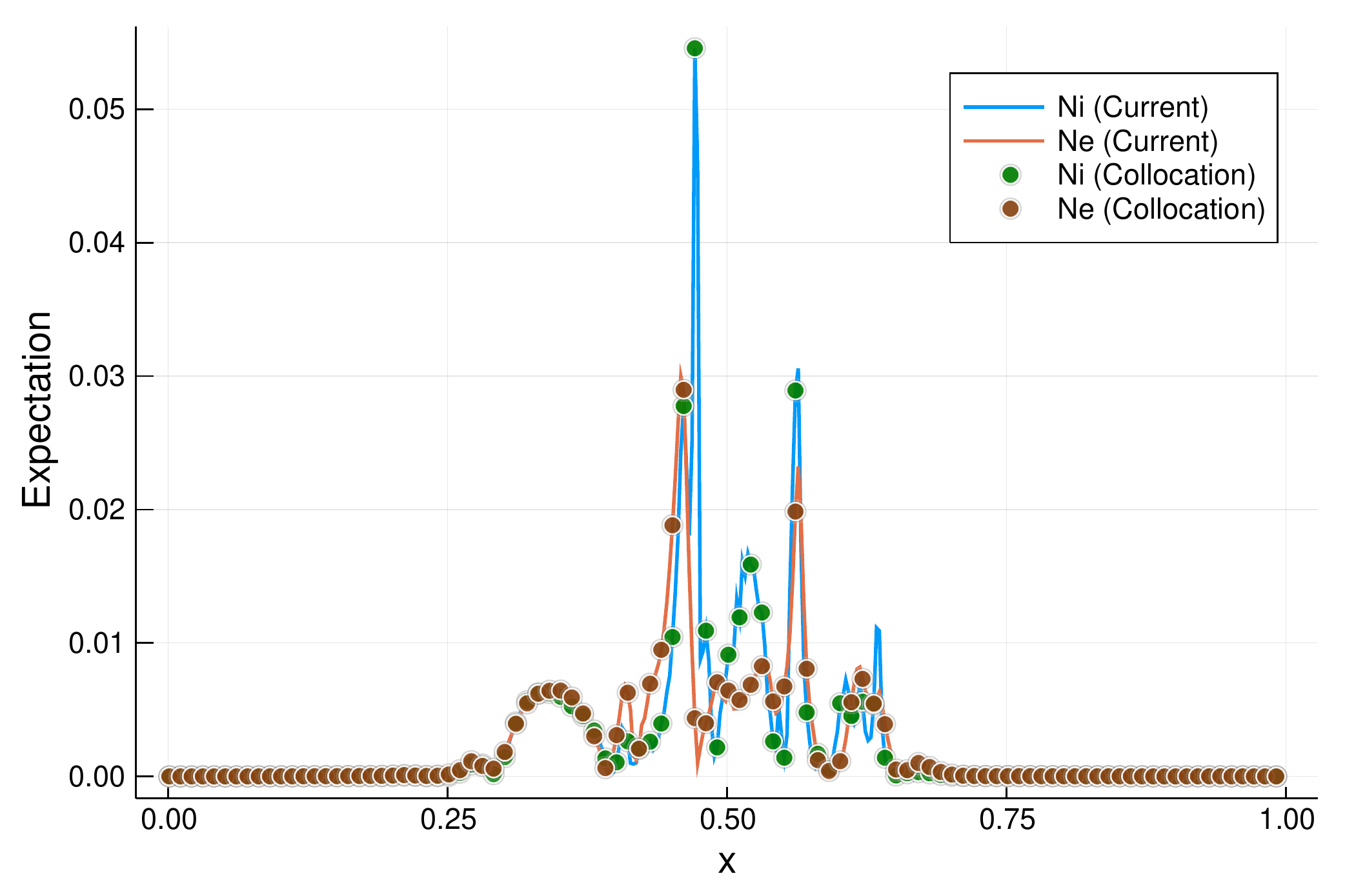}
	}
	\subfigure[$\mathbb S(U)$]{
		\includegraphics[width=0.31\textwidth]{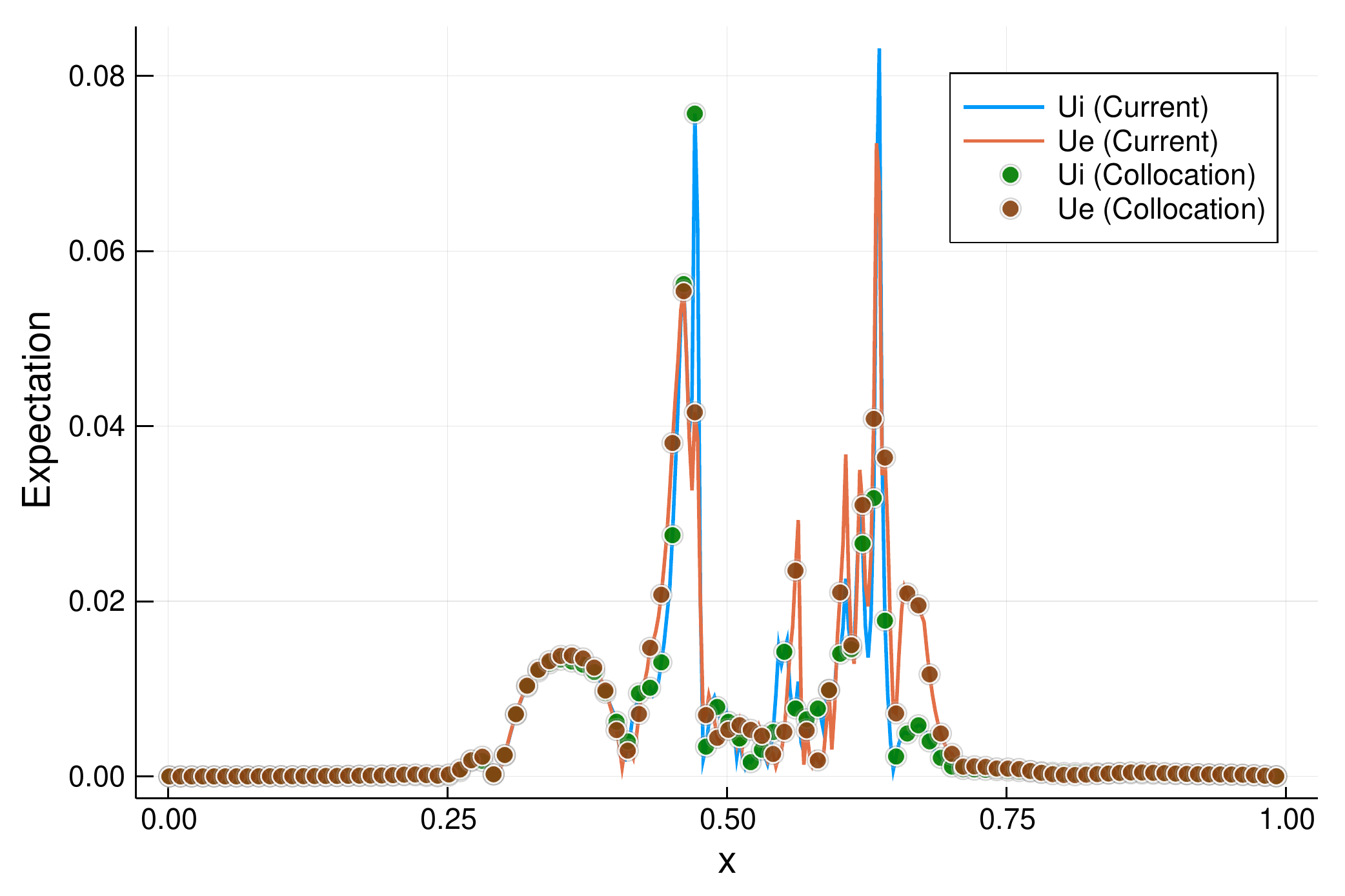}
	}
	\subfigure[$\mathbb S(B_y)$]{
		\includegraphics[width=0.31\textwidth]{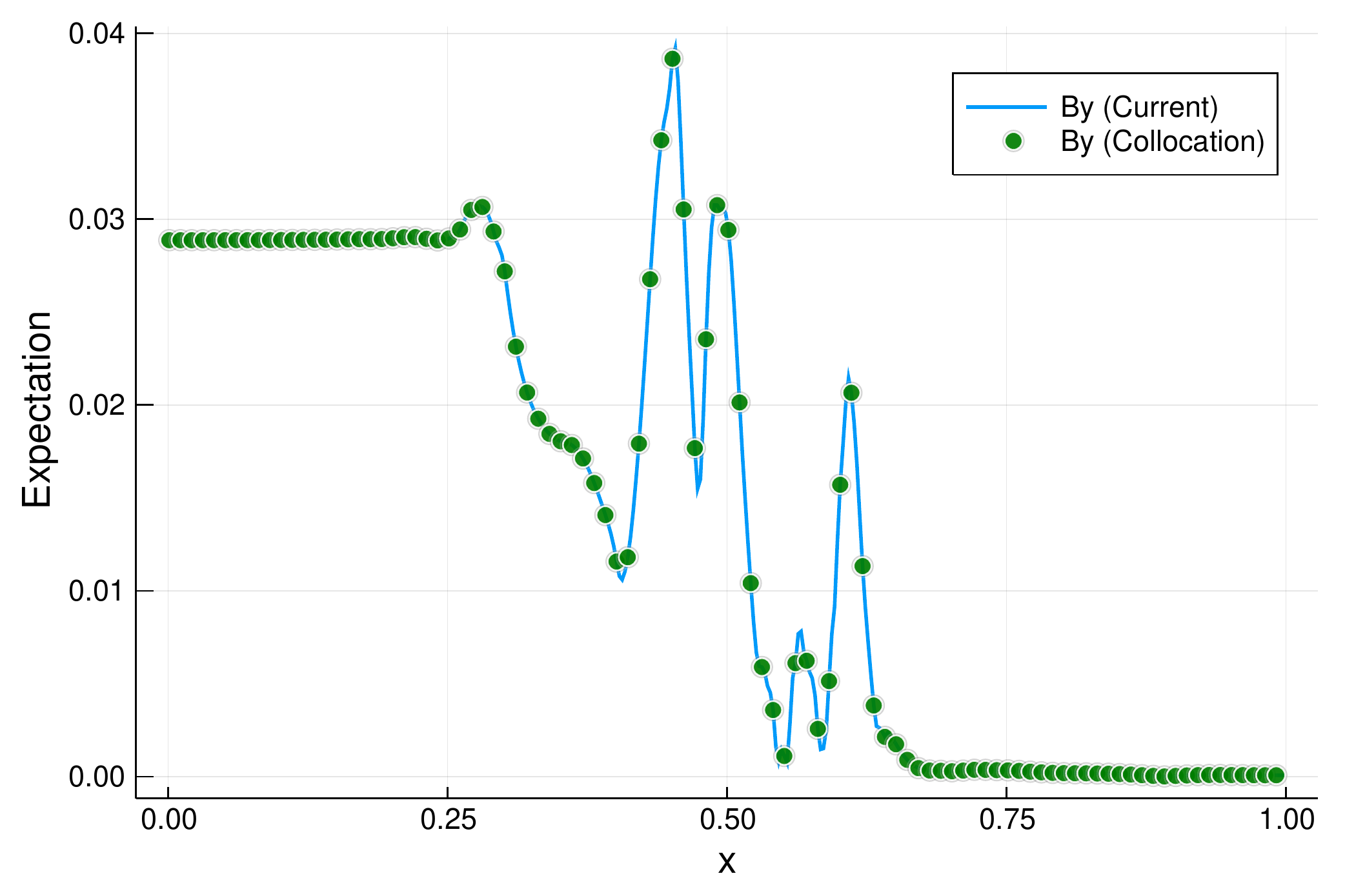}
	}
	\subfigure[$\mathbb S(N)$]{
		\includegraphics[width=0.31\textwidth]{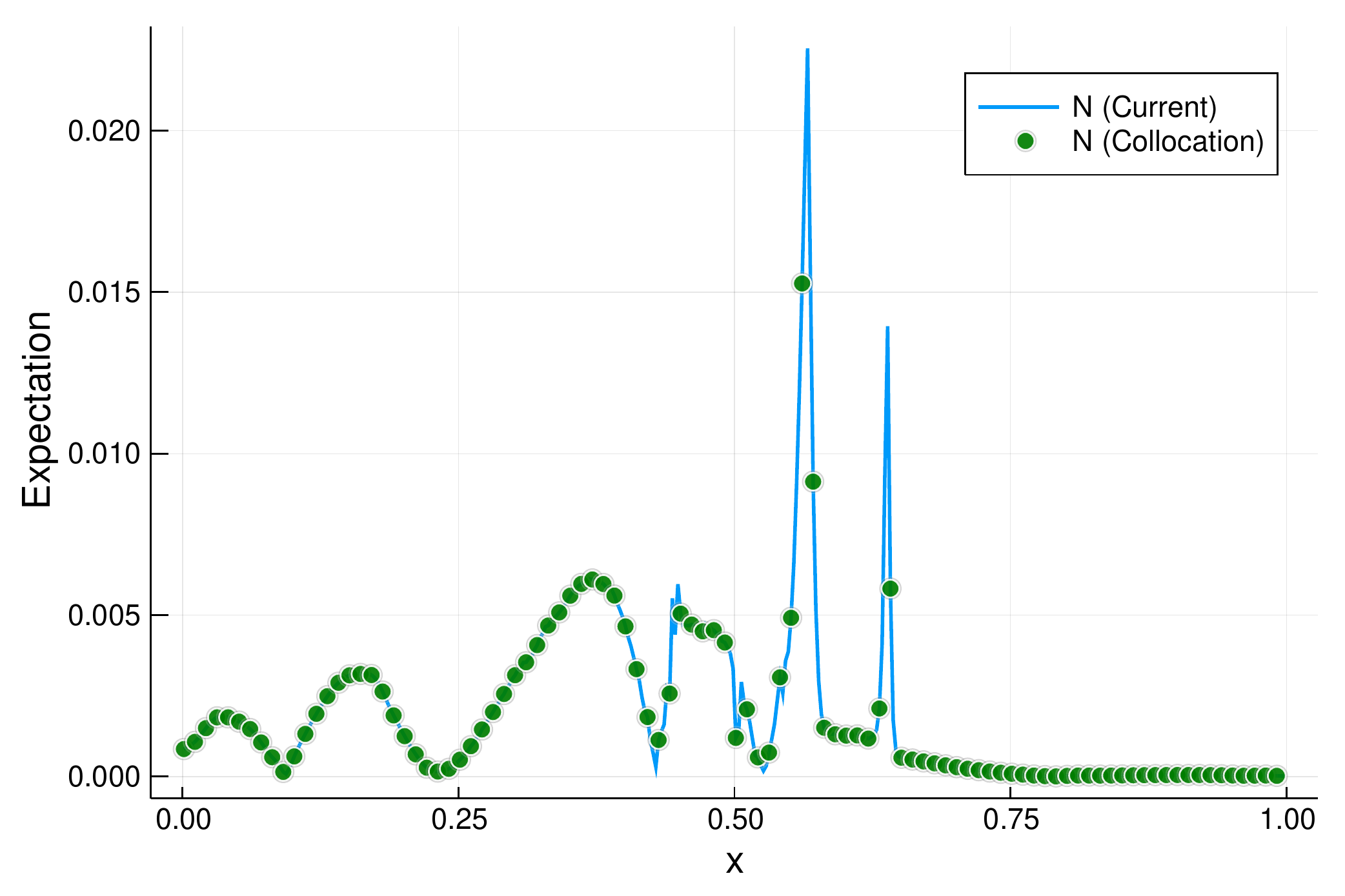}
	}
	\subfigure[$\mathbb S(U)$]{
		\includegraphics[width=0.31\textwidth]{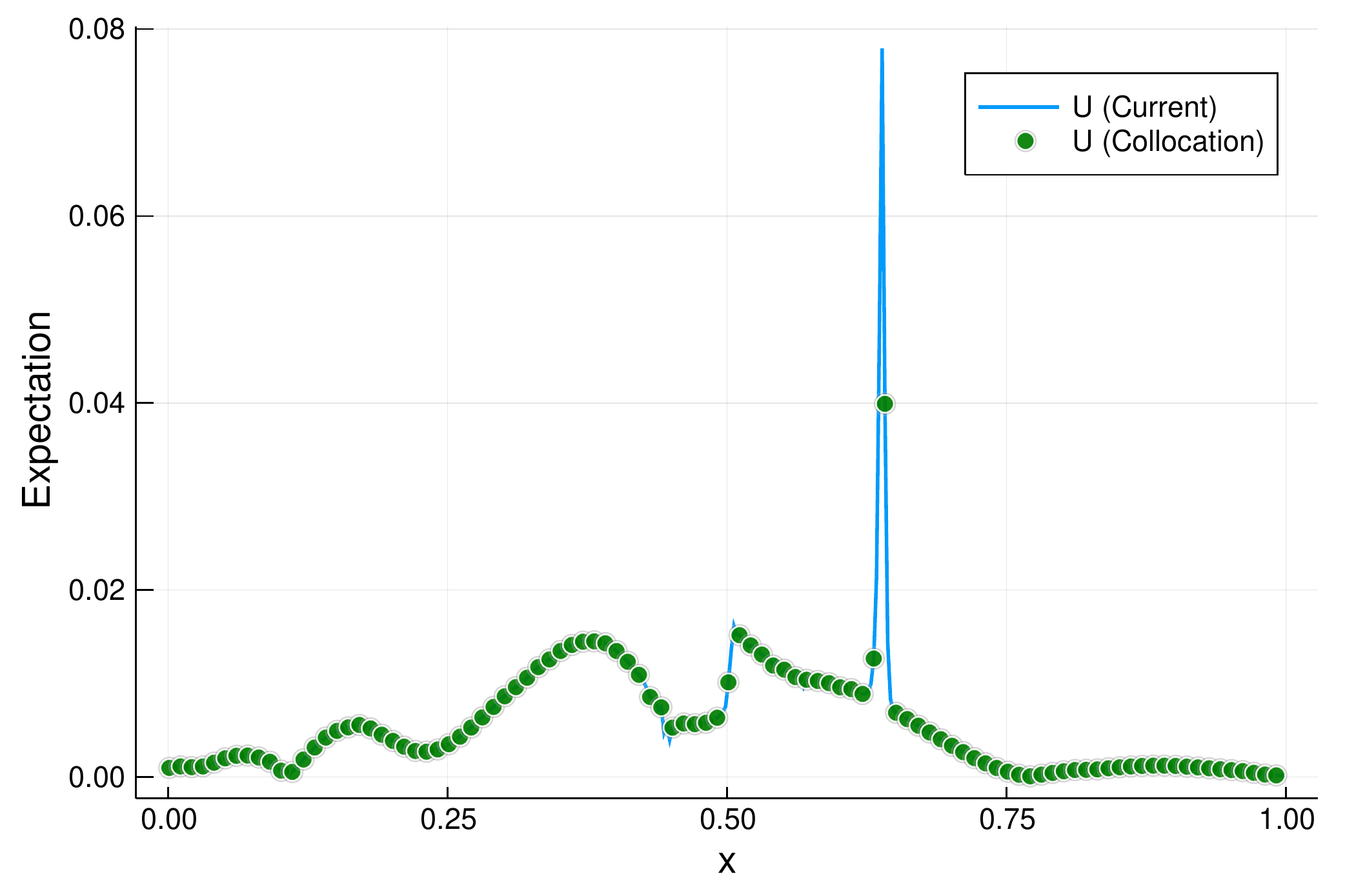}
	}
	\subfigure[$\mathbb S(B_y)$]{
		\includegraphics[width=0.31\textwidth]{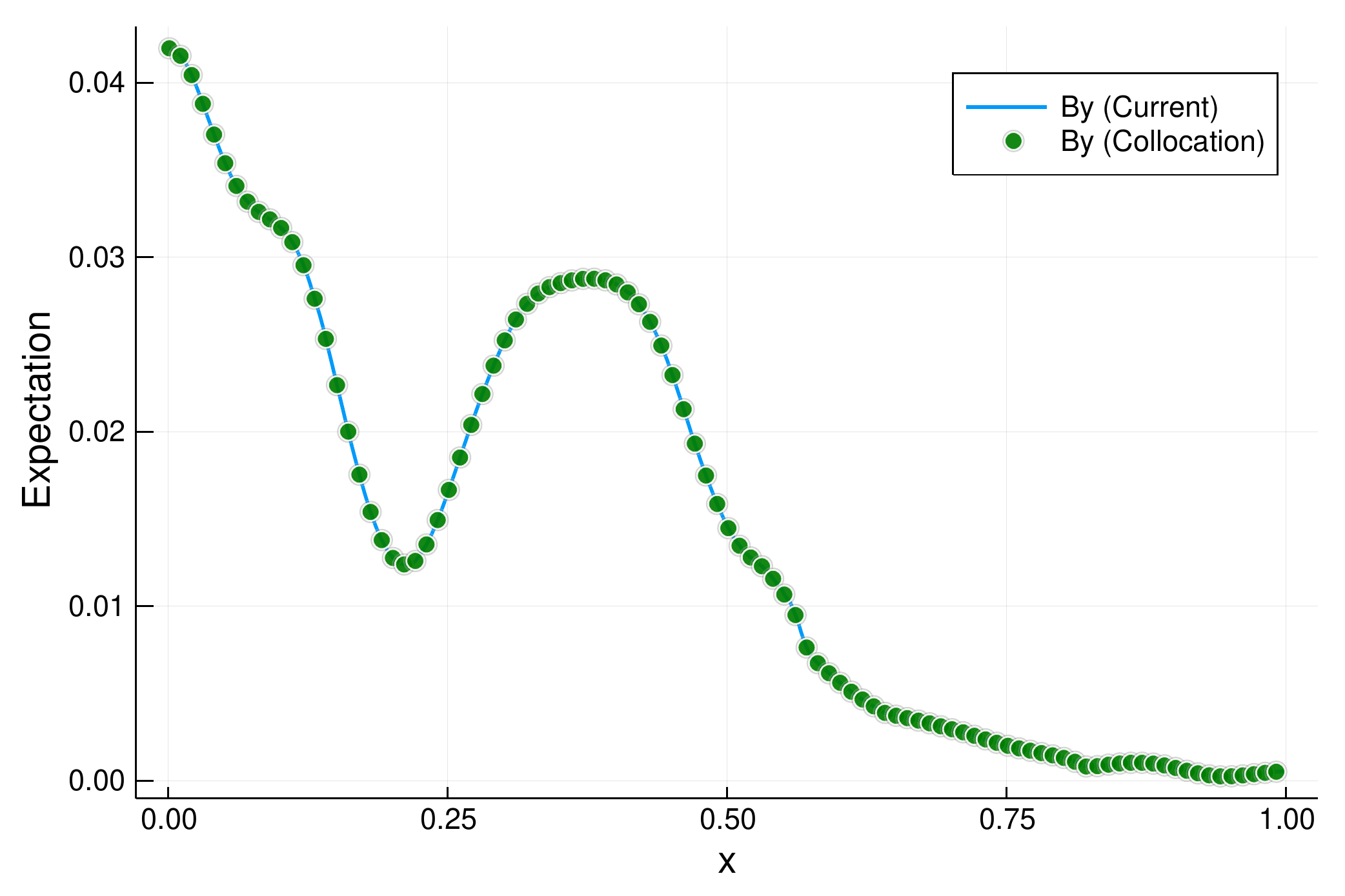}
	}
	\subfigure[$\mathbb S(N)$]{
		\includegraphics[width=0.31\textwidth]{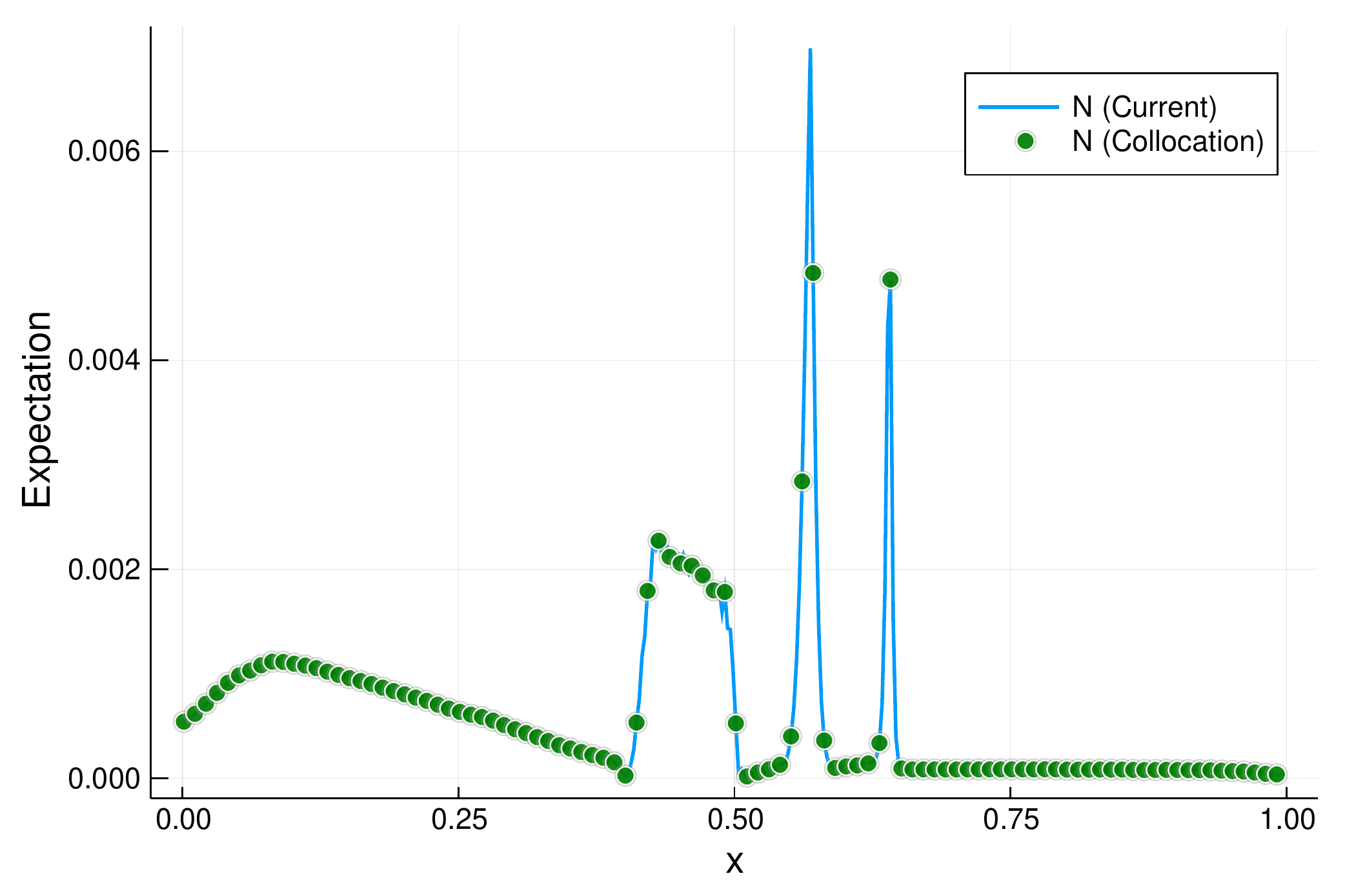}
	}
	\subfigure[$\mathbb S(U)$]{
		\includegraphics[width=0.31\textwidth]{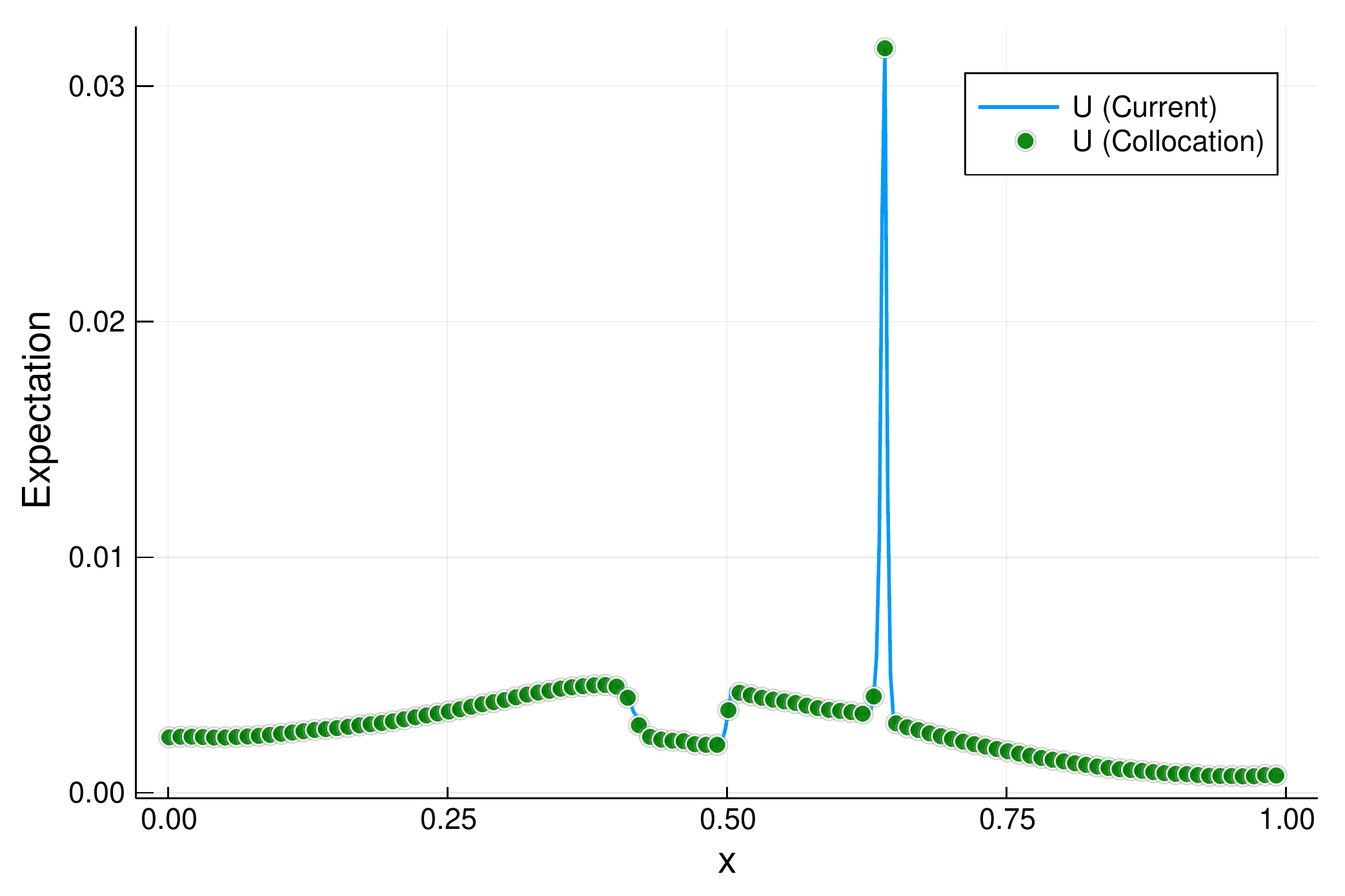}
	}
	\subfigure[$\mathbb S(B_y)$]{
		\includegraphics[width=0.31\textwidth]{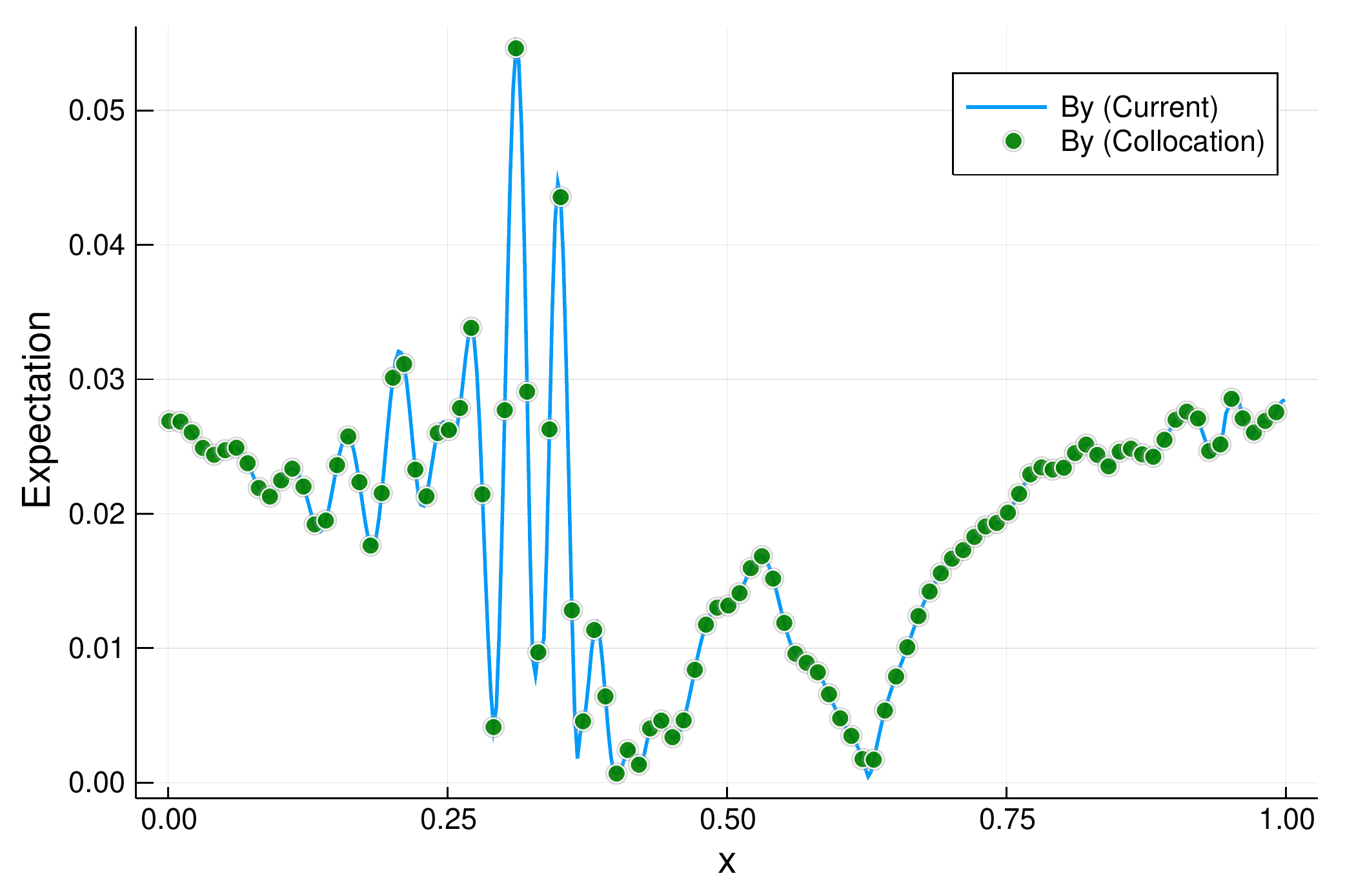}
	}
	\subfigure[$\mathbb S(N)$]{
		\includegraphics[width=0.31\textwidth]{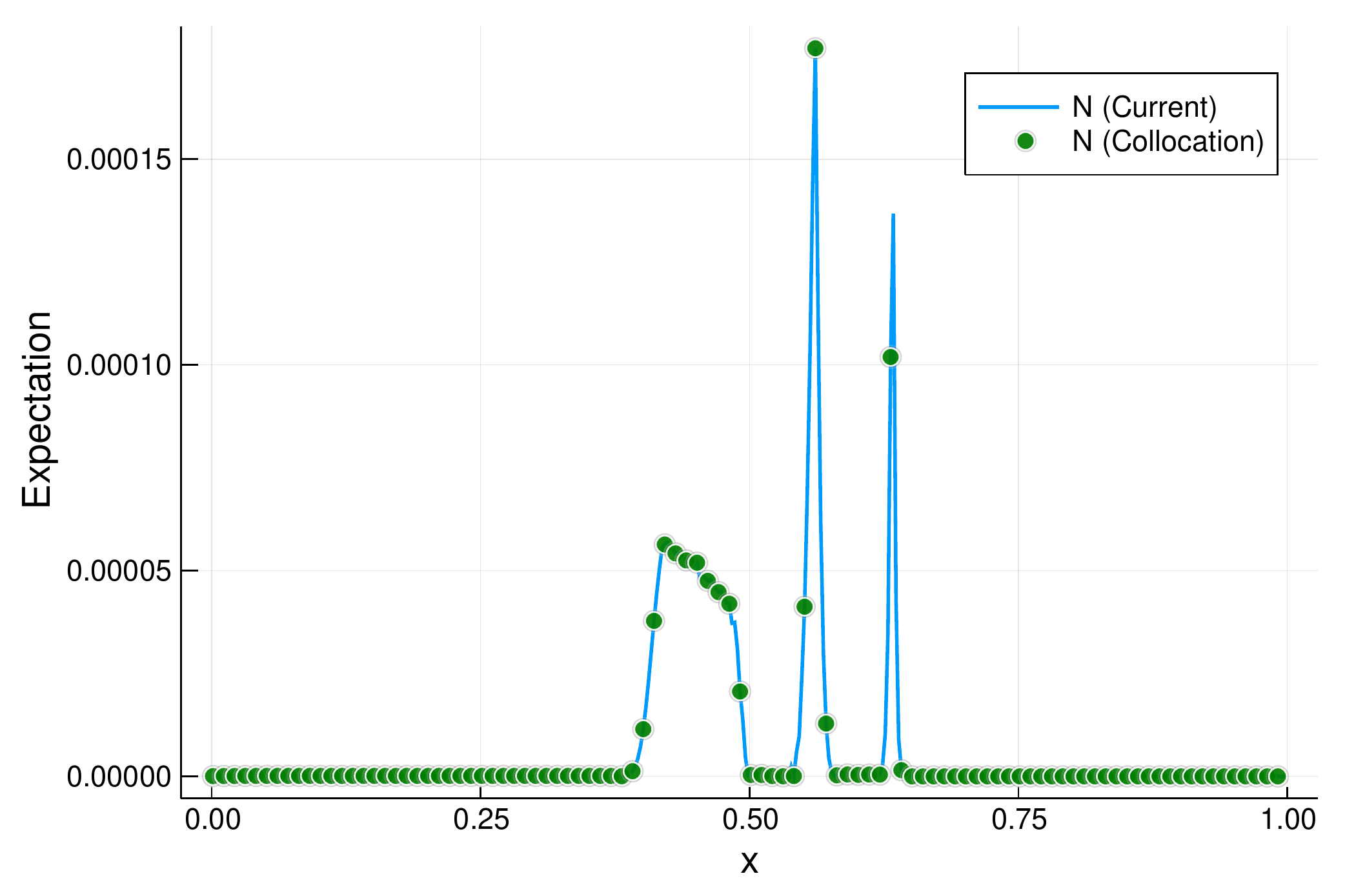}
	}
	\subfigure[$\mathbb S(U)$]{
		\includegraphics[width=0.31\textwidth]{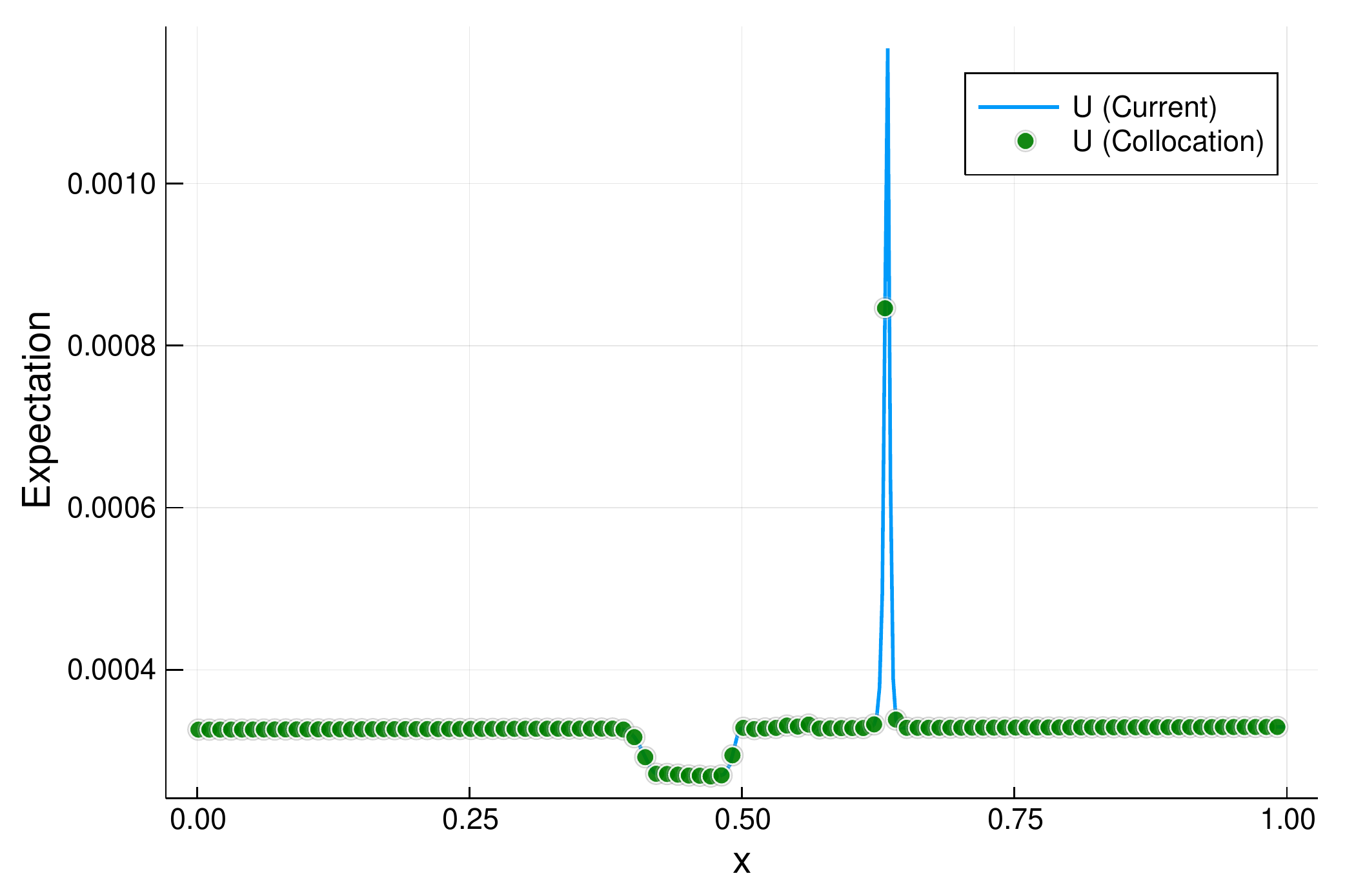}
	}
	\subfigure[$\mathbb S(B_y)$]{
		\includegraphics[width=0.31\textwidth]{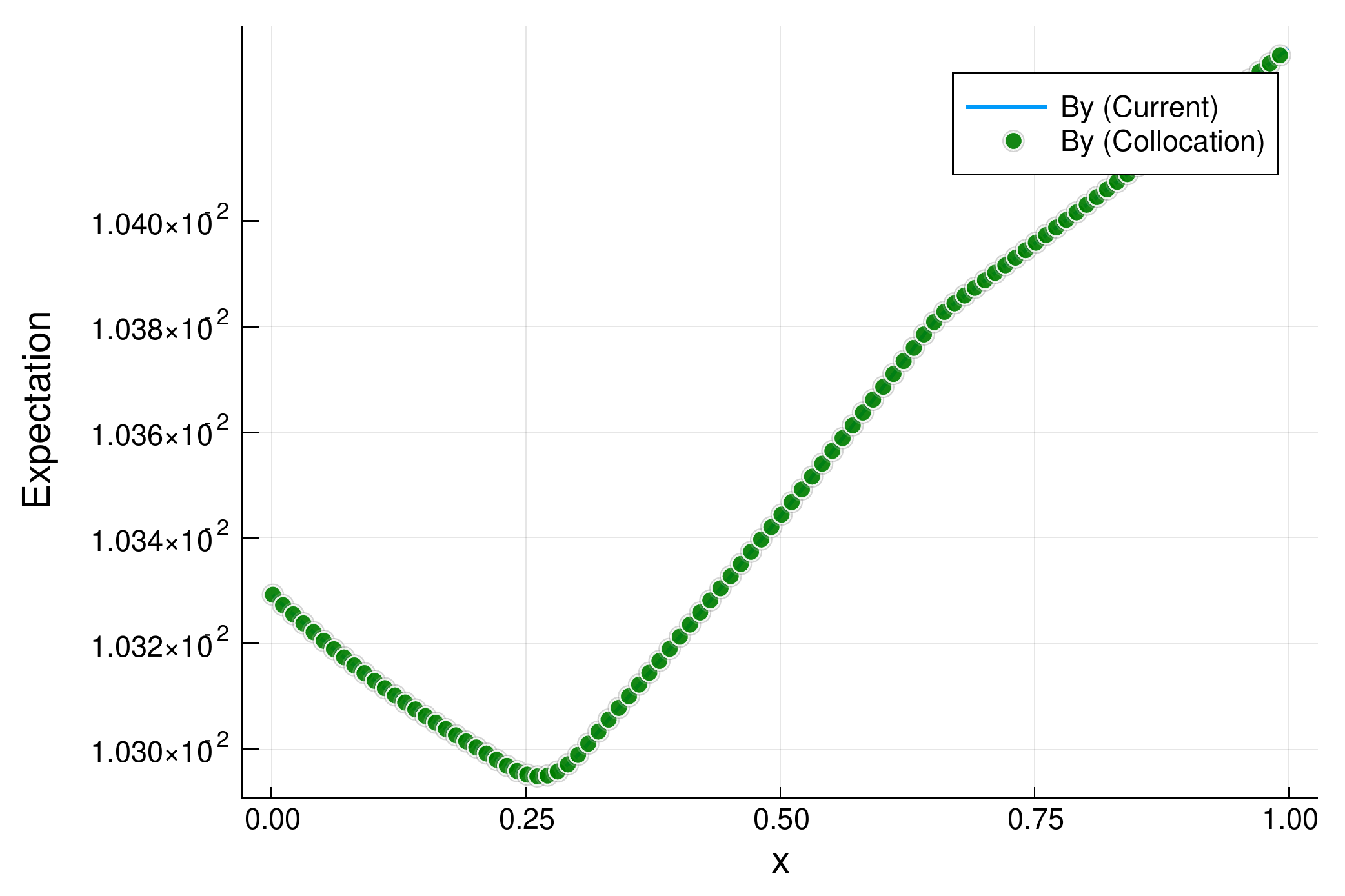}
	}
	\caption{Standard deviations of $N$, $U$ and $B_y$ in Brio-Wu shock tube with magnetic uncertainty at $t=0.1$ (row 1: $r_g=0.003$, row 2: $r_g=0.01$, row 3: $r_g=0.1$, row 4: $r_g=1$, row 5: $r_g=100$).}
	\label{pic:briowu case2 std}
\end{figure}

In the second case, we turn to the uncertainties from magnetic field in the left half tube, with $B_{y,L}=\xi$.
The numerical solutions of macroscopic variables under different reference gyroradius at $t = 0.1$ are shown in Fig. \ref{pic:briowu case2 mean} and \ref{pic:briowu case2 std}.
For the expected values in Fig. \ref{pic:briowu case2 mean}, we see that different initial stochastic conditions don't significantly change the wave structures of mean field.
Both MHD and Euler limits are precisely preserved under stochastic magnetic field.

Fig. \ref{pic:briowu case2 std} presents the corresponding standard deviations of macroscopic system.
As can be seen, the uncertainties manifest the consistent propagating patterns of expected values.
Different from the previous case, here the high-frequency variance of density in compound wave region is much reduced, which indicates this phenomenon comes from the uncertain densities.
As $r_g$ increases, the wave structures around tube center becomes even more complicated in the transition regime $r_g=0.01$, and simplify again when it comes to $r_g=0.1$, which demonstrates the complex nonlinearity from Hall current, two-fluid effects, etc.
The increasing gyroradius doesn't lead to a simple process of monotonic variation.

This test case clearly shows the consistency and distinction of propagation modes between expectation value and variance. It also illustrates the capacity of current scheme to simulate multi-scale and multi-physics plasma transports, and capture the propagation of uncertainties in different regimes.

\section{Conclusion}

Plasma dynamics is associated with an intrinsic multi-scale nature due to the large variations of particle density and temperature, as well as the characteristic scales of the local structures. 
Based on the multi-component BGK model and Maxwell's equations, a stochastic kinetic scheme with hybrid Galerkin-collocation strategy has been constructed in this paper, which allows for a unified numerical simulation for multi-scale plasma physics. 
Based on the cross-scale modeling, the solution algorithm is able to capture both equilibrium magnetohyrodynamics and non-equilibrium gyrations of charged particles simultaneously, and recover the scale-dependent plasma physics along with the emergence, propagation, and evolution of randomness. 
The asymptotic-preserving property of the scheme is validated through theoretical analysis and numerical tests.

\section*{Acknowledgement}

The current research is funded by the Alexander von Humboldt Foundation.

\clearpage
\newpage

\bibliographystyle{unsrt}
\bibliography{v1}

\begin{thebibliography}{10}

\bibitem{national2008plasma}
National~Research Council, Plasma~Science Committee, Plasma~2010 Committee,
  et~al.
\newblock {\em Plasma science: advancing knowledge in the national interest},
  volume~3.
\newblock National Academies Press, 2008.

\bibitem{chen2012introduction}
Francis~F Chen.
\newblock {\em Introduction to plasma physics}.
\newblock Springer Science \& Business Media, 2012.

\bibitem{vahedi1995monte}
Vahid Vahedi and Maheswaran Surendra.
\newblock A monte carlo collision model for the particle-in-cell method:
  applications to argon and oxygen discharges.
\newblock {\em Computer Physics Communications}, 87(1-2):179--198, 1995.

\bibitem{filbet2001conservative}
Francis Filbet, Eric Sonnendr{\"u}cker, and Pierre Bertrand.
\newblock Conservative numerical schemes for the vlasov equation.
\newblock {\em Journal of Computational Physics}, 172(1):166--187, 2001.

\bibitem{crouseilles2009forward}
Nicolas Crouseilles, Thomas Respaud, and Eric Sonnendr{\"u}cker.
\newblock A forward semi-lagrangian method for the numerical solution of the
  vlasov equation.
\newblock {\em Computer Physics Communications}, 180(10):1730--1745, 2009.

\bibitem{qiu2010conservative}
Jing-Mei Qiu and Andrew Christlieb.
\newblock A conservative high order semi-lagrangian weno method for the vlasov
  equation.
\newblock {\em Journal of Computational Physics}, 229(4):1130--1149, 2010.

\bibitem{brio1988upwind}
Moysey Brio and Cheng~Chin Wu.
\newblock An upwind differencing scheme for the equations of ideal
  magnetohydrodynamics.
\newblock {\em Journal of computational physics}, 75(2):400--422, 1988.

\bibitem{powell1999solution}
Kenneth~G Powell, Philip~L Roe, Timur~J Linde, Tamas~I Gombosi, and Darren~L
  De~Zeeuw.
\newblock A solution-adaptive upwind scheme for ideal magnetohydrodynamics.
\newblock {\em Journal of Computational Physics}, 154(2):284--309, 1999.

\bibitem{degond2010asymptotic}
Pierre Degond, Fabrice Deluzet, Laurent Navoret, An-Bang Sun, and
  Marie-H{\'e}l{\`e}ne Vignal.
\newblock Asymptotic-preserving particle-in-cell method for the vlasov--poisson
  system near quasineutrality.
\newblock {\em Journal of Computational Physics}, 229(16):5630--5652, 2010.

\bibitem{degond2017asymptotic}
Pierre Degond and Fabrice Deluzet.
\newblock Asymptotic-preserving methods and multiscale models for plasma
  physics.
\newblock {\em Journal of Computational Physics}, 336:429--457, 2017.

\bibitem{nistrefer2019}
National Institute of Standards and Technology.
\newblock {\em The NIST Reference on Constants, Units, and Uncertainty}, 5
  2019.

\bibitem{xiu2010numerical}
Dongbin Xiu.
\newblock {\em Numerical methods for stochastic computations: a spectral method
  approach}.
\newblock Princeton university press, 2010.

\bibitem{canuto1982approximation}
Claudio Canuto and Alfio Quarteroni.
\newblock Approximation results for orthogonal polynomials in sobolev spaces.
\newblock {\em Mathematics of Computation}, 38(157):67--86, 1982.

\bibitem{xiu2005high}
Dongbin Xiu and Jan~S Hesthaven.
\newblock High-order collocation methods for differential equations with random
  inputs.
\newblock {\em SIAM Journal on Scientific Computing}, 27(3):1118--1139, 2005.

\bibitem{hu2016stochastic}
Jingwei Hu, Shi Jin, and Ruiwen Shu.
\newblock A stochastic galerkin method for the fokker--planck--landau equation
  with random uncertainties.
\newblock In {\em XVI International Conference on Hyperbolic Problems: Theory,
  Numerics, Applications}, pages 1--19. Springer, 2016.

\bibitem{hu2017uncertainty}
Jingwei Hu and Shi Jin.
\newblock Uncertainty quantification for kinetic equations.
\newblock In {\em Uncertainty Quantification for Hyperbolic and Kinetic
  Equations}, pages 193--229. Springer, 2017.

\bibitem{jin2018hypocoercivity}
Shi Jin and Yuhua Zhu.
\newblock Hypocoercivity and uniform regularity for the
  vlasov--poisson--fokker--planck system with uncertainty and multiple scales.
\newblock {\em SIAM Journal on Mathematical Analysis}, 50(2):1790--1816, 2018.

\bibitem{ding2019random}
Zhiyan Ding and Shi Jin.
\newblock Random regularity of a nonlinear landau damping solution for the
  vlasov-poisson equations with random inputs.
\newblock {\em International Journal for Uncertainty Quantification}, 9(2),
  2019.

\bibitem{phillips2015stochastic}
Edward~G Phillips and Howard~C Elman.
\newblock A stochastic approach to uncertainty in the equations of mhd
  kinematics.
\newblock {\em Journal of Computational Physics}, 284:334--350, 2015.

\bibitem{yamazaki2017stochastic}
Kazuo Yamazaki.
\newblock Stochastic hall-magneto-hydrodynamics system in three and two and a
  half dimensions.
\newblock {\em Journal of Statistical Physics}, 166(2):368--397, 2017.

\bibitem{xiao2020stochastic}
Tianbai Xiao and Martin Frank.
\newblock A stochastic kinetic scheme for multi-scale flow transport with
  uncertainty quantification.
\newblock {\em arXiv preprint arXiv:2002.00277}, 2020.

\bibitem{andries2002consistent}
Pierre Andries, Kazuo Aoki, and Benoit Perthame.
\newblock A consistent bgk-type model for gas mixtures.
\newblock {\em Journal of Statistical Physics}, 106(5):993--1018, 2002.

\bibitem{morse1963energy}
TF~Morse.
\newblock Energy and momentum exchange between nonequipartition gases.
\newblock {\em The Physics of Fluids}, 6(10):1420--1427, 1963.

\bibitem{braginskii1965transport}
SI~Braginskii.
\newblock Transport processes in a plasma.
\newblock {\em Reviews of plasma physics}, 1, 1965.

\bibitem{munz2000divergence}
C-D Munz, Pascal Omnes, Rudolf Schneider, Eric Sonnendr{\"u}cker, and Ursula
  Voss.
\newblock Divergence correction techniques for maxwell solvers based on a
  hyperbolic model.
\newblock {\em Journal of Computational Physics}, 161(2):484--511, 2000.

\bibitem{koganrarefied}
Mikhail~N Kogan.
\newblock {\em Rarefied Gas Dynamics}.
\newblock Plenum Press, 1969.

\bibitem{xu2010unified}
Kun Xu and Juan-Chen Huang.
\newblock A unified gas-kinetic scheme for continuum and rarefied flows.
\newblock {\em Journal of Computational Physics}, 229(20):7747--7764, 2010.

\bibitem{xiao2017well}
Tianbai Xiao, Qingdong Cai, and Kun Xu.
\newblock A well-balanced unified gas-kinetic scheme for multiscale flow
  transport under gravitational field.
\newblock {\em Journal of Computational Physics}, 332:475 -- 491, 2017.

\bibitem{liu2017unified}
Chang Liu and Kun Xu.
\newblock A unified gas kinetic scheme for continuum and rarefied flows v:
  Multiscale and multi-component plasma transport.
\newblock {\em Communications in Computational Physics}, 22(5):1175--1223,
  2017.

\bibitem{xiao2019unified}
Tianbai Xiao, Kun Xu, and Qingdong Cai.
\newblock A unified gas-kinetic scheme for multiscale and multicomponent flow
  transport.
\newblock {\em Applied Mathematics and Mechanics}, pages 1--18, 2019.

\bibitem{xiao2020velocity}
Tianbai Xiao, Chang Liu, Kun Xu, and Qingdong Cai.
\newblock A velocity-space adaptive unified gas kinetic scheme for continuum
  and rarefied flows.
\newblock {\em Journal of Computational Physics}, page 109535, 2020.

\bibitem{hakim2006high}
Ammar Hakim, John Loverich, and Uri Shumlak.
\newblock A high resolution wave propagation scheme for ideal two-fluid plasma
  equations.
\newblock {\em Journal of Computational Physics}, 219(1):418--442, 2006.

\bibitem{shampine1982implementation}
Lawrence~F Shampine.
\newblock Implementation of rosenbrock methods.
\newblock {\em ACM Transactions on Mathematical Software (TOMS)}, 8(2):93--113,
  1982.

\bibitem{landau1946vibration}
L.~Landau.
\newblock On the vibration of the electronic plasma.
\newblock {\em Journal of Physics}, 10(1):25 -- 34, 1946.

\end{thebibliography}
\newpage

\begin{table}
	\centering
	\caption{Nomenclature of stochastic kinetic scheme.}
	\begin{tabular*}{16cm}{lll}
		\hline
		$k_B$ & Boltzmann constant \\
		$\varepsilon_0$ & vacuum permittivity \\
		$\mu_0$ & vacuum permeability \\
		$e$ & elementary charge \\
		$c$ & speed of light   \\
		$q_\alpha$ & charge of species $\alpha$, with $q_\alpha=\pm e$ ($\alpha = ion,\ ele$) \\
		$m_\alpha$ & particle mass of species $\alpha$ \\
		$n_\alpha$ & number density of species $\alpha$ \\
		$\rho_\alpha$ & density of species $\alpha$ \\
		$\mathbf U_\alpha$ & macroscopic velocity of species $\alpha$ \\
		$T_\alpha$ & temperature of species $\alpha$ \\
		$\mathbf P_\alpha$ & stress tensor of species $\alpha$ \\
		$p_\alpha$ & pressure of species $\alpha$ \\
		$\mathbf q_\alpha$ & heat flux of species $\alpha$ \\
		$\mathbf W_\alpha$ & macroscopic conservative variables of species $\alpha$ (density, momentum, energy) \\
		$f_\alpha$ & particle distribution function of species $\alpha$  \\
		$\mathbf u$ & particle velocity \\
		$\mathbf a_\alpha$ & electromagnetic force acting on species $\alpha$ \\
		$Q_\alpha$ & kinetic collision operator of species $\alpha$ \\
		$\mathcal M_\alpha$ & equilibrium distribution function of species $\alpha$ \\
		$\mathbf E$ & electric field \\
		$\mathbf B$ & magnetic field \\
		$\mathbf \phi$ & correction potential for $\mathbf E$ in perfectly hyperbolic Maxwell's equations \\
		$\mathbf \psi$ & correction potential for $\mathbf B$ in perfectly hyperbolic Maxwell's equations \\
		$\sigma$ & charge density \\
		$\mathbf J$ & current density \\
		$\mathbf M$ & abbreviation of electromagnetic variables ($\mathbf E, \mathbf B, \phi, \psi$) \\
		$\varpi$ & vector of collision invariants  \\
		$\nu_\alpha$ & collision frequency of species $\alpha$ \\
		$\nu_{ie}$ & interaction frequency between ion and electron \\
		$\tau_\alpha$ & Collision time of species $\alpha$ with $\tau_\alpha=1/\nu_{\alpha}$ \\
		$\mathbf F^W_\alpha$ & flux for macroscopic conservative variables of species $\alpha$ \\
		$\mathbf F^f_\alpha$ & flux for particle distribution function of species $\alpha$ \\
		$\mathbf F^M$ & flux for electromagnetic variables \\
		${L}_N$ & generalized polynomial chaos expansion of any stochastic variable $L$ with degree $N$ \\
		$\hat{\boldsymbol l}$ & vector of generalized polynomial chaos coefficients of any stochastic variable $L$ \\
		$\Phi$ & orthogonal polynomials in random space $\mathbf z$ \\
		$\varrho$ & probability density function of random variable $\mathbf z$ \\
		$\rm Kn$ & Knudsen number  \\
		$r_L$ & gyroradius  \\
		$\lambda_D$ & Debye length  \\
		$H[x]$ & Heaviside step function \\
		\hline
	\end{tabular*}
	\label{table:nomenclature}
\end{table}

\end{document}